\pdfoutput=1
\documentclass[11pt,letterpaper]{amsart}
\usepackage{graphicx}
\usepackage{hyperref}
\theoremstyle{plain}
\newtheorem{theorem}{Theorem}[subsection]
\newtheorem{proposition}{Proposition}[subsection]
\newtheorem{lemma}{Lemma}[subsection]
\newtheorem{corollary}{Corollary}[subsection]
\theoremstyle{remark}
\newtheorem{remark}{Remark}[subsection]
\theoremstyle{definition}
\newtheorem{definition}{Definition}[subsection]
\begin{document}
\def\omathbf#1{\overline{\mathbf{#1}}}
\newcommand\qreflink[2]{#2}
\renewcommand\descriptionlabel[1]{#1}
\numberwithin{equation}{subsection}
\title[Squeezing gravitational energy to arbitrarily small scales]{Squeezing a fixed amount of gravitational energy to arbitrarily small scales, in $U(1)$ symmetry}
\author[S. Alexakis]{Spyros Alexakis}\address{Mathematics Department, University of Toronto}\email{alexakis@math.toronto.edu}
\author[N. T. Carruth]{Nathan Thomas Carruth}\address{Mathematics Department, University of Toronto}\curraddr{Yau Mathematical Sciences Center, Tsinghua University}\email{lutianci@mail.tsinghua.edu.cn}
\begin{abstract}
We prove uniform finite-time existence of solutions to the vacuum Einstein equations in polarized $U(1)$ symmetry which have uniformly positive incoming $H^1$ energy supported on an arbitrarily small set in the $2 + 1$ spacetime obtained by quotienting by the $U(1)$ symmetry. We also construct a subclass of solutions for which the energy remains concentrated (along a $U(1)$ family of geodesics) throughout its evolution. These results rely on three innovations: a direct treatment of the $2 + 1$ Einstein equations in a null geodesic gauge, a novel parabolic scaling of the Einstein equations in this gauge, and a new Klainerman-Sobolev inequality on rectangular strips.
\end{abstract}
\maketitle
\font\mbbt=msbm10
\font\mbbs=msbm7
\font\mbbss=msbm5
\newfam\mbbfam\textfont\mbbfam=\mbbt\scriptfont\mbbfam=\mbbs\scriptscriptfont\mbbfam=\mbbss
\def\mbb{\fam\mbbfam}
\font\rust=wncyr10
\font\russ=wncyr7
\font\russs=wncyr5
\newfam\rusfam\textfont\rusfam=\rust\scriptfont\rusfam=\russ\scriptscriptfont\rusfam=\russs
\def\rus{\fam\rusfam}
\font\mbft=cmbx10
\font\mbfs=cmbx7
\font\mbfss=cmbx5
\newfam\mbffam\textfont\mbffam=\mbft\scriptfont\mbffam=\mbfs\scriptscriptfont\mbffam=\mbfss
\def\mbff{\fam\mbffam }
\font\msat=msam10
\font\msas=msam7
\font\msass=msam5
\textfont15=\msat\scriptfont15=\msas\scriptscriptfont15=\msass
\def\msaf{\fam15 }
\font\frakt=eufm10
\font\fraks=eufm7
\font\frakss=eufm5
\newfam\frakfam\textfont\frakfam=\frakt\scriptfont\frakfam=\fraks\scriptscriptfont\frakfam=\frakss
\def\frakf{\fam\frakfam }
\font\twfi=cmmi17	scaled	\magstep0
\font\twfsy=cmsy17	scaled	\magstep0
\font\twfex=cmex17	scaled	\magstep0
\def\twff{\textfont0=\twfrm\textfont1=\twfi\textfont2=\twfsy\twfrm}
\font\txtgrk=grmn1000	scaled	\magstep0
\font\mbf=cmmib10	scaled \magstep0
\newcommand{\gambznorm}{N}
\newcommand{\quupsz}{\quupsilon_0}
\newcommand{\quusigu}{\widehat{\quusig}}
\newcommand{\qusbnmin}{7}
\newcommand{\gambC}{C_G}
\newcommand{\gambCdx}{\gambC'}
\newcommand{\gambCp}{\overline{\gambC}}
\newcommand{\gambCdxp}{\overline{\gambCdx}}
\newcommand{\OmegbC}{C_O}
\newcommand{\OmegbCp}{\overline{\OmegbC}}
\newcommand{\enerC}{C^{\rm E}}
\newcommand{\ellbinvC}{C^{\rm I}}
\newcommand{\diffeqC}{C^{\rm d}}
\newcommand{\quOmegbs}{\quOmegbh'}
\newcommand{\quAppfr}{{F}}
\newcommand{\quAsigmatt}{\quwlntilde{A}_\qusigt}
\newcommand{\quAsigmat}{{A}_{\qusigmat}}
\newcommand{\quAsigmatp}{{A}_{\qusigmat'}}
\newcommand{\quAsigmu}{{A}_{\quupsilon}}
\newcommand{\quBsigfb}{{\overline{\quBsigf}}}
\newcommand{\quBsigffbt}{\quwlntilde{\quBsigffb}}
\newcommand{\quBsigffb}{{\overline{\quBsigff}}}
\newcommand{\quBsigff}{{{\frakf s}}}
\newcommand{\quBsigf}{{{\frakf S}}}
\newcommand{\quBups}{{B_\quupsilon}}
\newcommand{\quBxi}{{{\mathcal B}_{\quxi}}}
\newcommand{\quBzet}{{{\mathcal B}_{\quzeta}}}
\newcommand{\quXvec}{{\mathbf X}}
\newcommand{\quVvec}{{\mathbf V}}
\newcommand{\quCGeod}{{C}^{{\rm G}}}
\newcommand{\quCKS}{{C}_{{KS}}}
\newcommand{\quCOmegb}{{{\mathcal M}}_{{\overline{\Omega}}}}
\newcommand{\quCb}{\overline{C}}
\newcommand{\quChhbarC}{{C}^{\Gamma}}
\newcommand{\quChhbaro}{{C^{\rm h}_0}}
\newcommand{\quCnmi}{{C}^{{\rm M}}}
\newcommand{\quCopss}{C_\quopsset}
\newcommand{\quCosize}{C_{\rm fi}}
\newcommand{\quCpr}{{C}^{{\rm p}}}
\newcommand{\quCrip}{\quCri}
\newcommand{\quCri}{{C}_{{\rm i}}}
\newcommand{\quCsb}{{C}^{{\rm S}}}
\newcommand{\quCte}{576C_0^2 (\quuCnmi)^2}
\newcommand{\quCty}{{\overline C}_0}
\newcommand{\quC}{\mathbb{C}}
\newcommand{\quDHb}{{\overline{H}}_{{\!\qudifset}}}
\newcommand{\quDH}{{H}_{{\qudifset}}}
\newcommand{\quDb}{\overline{\Delta}}
\newcommand{\quEEEAsembeq}{\eqref{Asembeq}}
\newcommand{\quEEEAsobolevemb}{Lemma \ref{Asobolevemb}}
\newcommand{\quEEEAziboginit}{\eqref{Aziboginit}}
\newcommand{\quEEEAzinitbound}{Proposition \ref{Azinitbound}}
\newcommand{\quEEEAzstuffdelbound}{\eqref{Azstuffdelbound}}
\newcommand{\quEEEBsigfbeeqyi}{\eqref{Bsigfbeeqyi}}
\newcommand{\quEEEBsigfbexist}{Proposition \ref{Bsigfbexist}}
\newcommand{\quEEEBsigffbdefeq}{\eqref{Bsigffbdefeq}}
\newcommand{\quEEEBsigffbupb}{\eqref{Bsigffbupb}}
\newcommand{\quEEEBsigffdefeq}{\eqref{Bsigffdefeq}}
\newcommand{\quEEEBsigffdef}{\eqref{Bsigffdef}}
\newcommand{\quEEEBsigfpcond}{\eqref{Bsigfpcond}}
\newcommand{\quEEEBxizetdef}{\eqref{Bxizetdef}}
\newcommand{\quEEECKSdefeq}{\eqref{CKSdefeq}}
\newcommand{\quEEECinfzQSdef}{\eqref{CinfzQSdef}}
\newcommand{\quEEECnmidefeq}{\eqref{Cnmidefeq}}
\newcommand{\quEEECprmdef}{\eqref{Cprmdef}}
\newcommand{\quEEECsbdef}{\eqref{Csbdef}}
\newcommand{\quEEECtydefeq}{\eqref{Ctydefeq}}
\newcommand{\quEEECtydef}{\eqref{Ctydef}}
\newcommand{\quEEECtysdef}{\eqref{Ctysdef}}
\newcommand{\quEEECyGeodboundep}{\eqref{CyGeodboundep}}
\newcommand{\quEEECyGeodbounde}{\eqref{CyGeodbounde}}
\newcommand{\quEEECyGeodbound}{\eqref{CyGeodbound}}
\newcommand{\quEEECzGeodbounde}{\eqref{CzGeodbounde}}
\newcommand{\quEEECzGeodbound}{\eqref{CzGeodbound}}
\newcommand{\quEEEDESdomset}{\eqref{DESdomset}}
\newcommand{\quEEEDGeodsbound}{\eqref{DGeodsbound}}
\newcommand{\quEEEDHbnormdef}{\eqref{DHbnormdef}}
\newcommand{\quEEEDHbprodA}{\eqref{DHbprodA}}
\newcommand{\quEEEDHbspacdef}{\eqref{DHbspacdef}}
\newcommand{\quEEEDHenerinitgambb}{\eqref{DHenerinitgambb}}
\newcommand{\quEEEDHenerinitomegbb}{\eqref{DHenerinitomegbb}}
\newcommand{\quEEEDHenerinit}{Lemma \ref{DHenerinit}}
\newcommand{\quEEEDboxcommeqii}{\eqref{Dboxcommeqii}}
\newcommand{\quEEEDboxcommeqi}{\eqref{Dboxcommeqi}}
\newcommand{\quEEEDboxcomm}{Proposition \ref{Dboxcomm}}
\newcommand{\quEEEDboxstatbound}{\eqref{Dboxstatbound}}
\newcommand{\quEEEDfboxcommreser}{\eqref{Dfboxcommreser}}
\newcommand{\quEEEDfboxcommressan}{\eqref{Dfboxcommressan}}
\newcommand{\quEEEDfboxcommresyi}{\eqref{Dfboxcommresyi}}
\newcommand{\quEEEDfboxcomm}{Proposition \ref{Dfboxcomm}}
\newcommand{\quEEEEEbwtEEb}{\eqref{EEbwtEEb}}
\newcommand{\quEEEEbmsim}{\eqref{Ebmsim}}
\newcommand{\quEEEEbnusineqyi}{\eqref{Ebnusineqyi}}
\newcommand{\quEEEEbnusq}{\eqref{Ebnusq}}
\newcommand{\quEEEEboot}{\eqref{Eboot}}
\newcommand{\quEEEEeqgauge}{Subsection \ref{Eeqgauge}}
\newcommand{\quEEEEsbneTpbound}{\eqref{EsbneTpbound}}
\newcommand{\quEEEEsbnedef}{\eqref{Esbnedef}}
\newcommand{\quEEEEsbpeqyi}{\eqref{Esbpeqyi}}
\newcommand{\quEEEFyineqz}{\eqref{Fyineqz}}
\newcommand{\quEEEGSBdef}{\eqref{GSBdef}}
\newcommand{\quEEEGammaeqns}{\eqref{Gammaeqns}}
\newcommand{\quEEEGeodAfin}{\eqref{GeodAfin}}
\newcommand{\quEEEGeodSigz}{\eqref{GeodSigz}}
\newcommand{\quEEEGeodbootstrap}{\eqref{Geodbootstrap}}
\newcommand{\quEEEGeodbyi}{\eqref{Geodbyi}}
\newcommand{\quEEEGeoddefcond}{\eqref{Geoddefcond}}
\newcommand{\quEEEGeoddefeq}{\eqref{Geoddefeq}}
\newcommand{\quEEEGeodinitling}{\eqref{Geodinitling}}
\newcommand{\quEEEGeodsdefeq}{\eqref{Geodsdefeq}}
\newcommand{\quEEEGeodtsbxbvbdef}{\eqref{Geodtsbxbvbdef}}
\newcommand{\quEEEHSinequal}{Lemma \ref{HSinequal}}
\newcommand{\quEEEHbarspace}{Definition \ref{Hbarspace}}
\newcommand{\quEEEHyptsbdef}{\eqref{Hyptsbdef}}
\newcommand{\quEEEHyptsbfluxineq}{\eqref{Hyptsbfluxineq}}
\newcommand{\quEEEIbSigsbdef}{\eqref{IbSigsbdef}}
\newcommand{\quEEEIbSigsbhbarfbb}{\eqref{IbSigsbhbarfbb}}
\newcommand{\quEEEIbSigsbhbarfb}{\eqref{IbSigsbhbarfb}}
\newcommand{\quEEEIbsim}{\eqref{Ibsim}}
\newcommand{\quEEEIdefeq}{\eqref{Idefeq}}
\newcommand{\quEEEIjWmnorm}{\eqref{IjWmnorm}}
\newcommand{\quEEEIjeq}{\eqref{Ijeq}}
\newcommand{\quEEEIoperator}{Definition \ref{Ioperator}}
\newcommand{\quEEEIopprop}{Proposition \ref{Iopprop}}
\newcommand{\quEEEKdef}{\eqref{Kdef}}
\newcommand{\quEEEKeq}{Proposition \ref{Keq}}
\newcommand{\quEEEKriceqer}{\eqref{Kriceqer}}
\newcommand{\quEEEKriceq}{\eqref{Kriceq}}
\newcommand{\quEEELlinezederyi}{\eqref{Llinezederyi}}
\newcommand{\quEEELlinezeder}{\eqref{Llinezeder}}
\newcommand{\quEEELlinezedyi}{\eqref{Llinezedyi}}
\newcommand{\quEEENbXorthnull}{\eqref{NbXorthnull}}
\newcommand{\quEEENveczdef}{\eqref{Nveczdef}}
\newcommand{\quEEEOmegbtdefin}{\eqref{Omegbtdefin}}
\newcommand{\quEEEOmegdefin}{\eqref{Omegdefin}}
\newcommand{\quEEEPSti}{Subsection \ref{PSti}}
\newcommand{\quEEEQSdefeq}{\eqref{QSdefeq}}
\newcommand{\quEEEQTaphdef}{\eqref{QTaphdef}}
\newcommand{\quEEEQTdhbound}{Proposition \ref{QTdhbound}}
\newcommand{\quEEERiccomegbineq}{\eqref{Riccomegbineq}}
\newcommand{\quEEERicconstgaugecond}{\eqref{Ricconstgaugecond}}
\newcommand{\quEEERicconst}{Corollary \ref{Ricconst}}
\newcommand{\quEEERicexp}{\eqref{Ricexp}}
\newcommand{\quEEERicwav}{\eqref{Ricwav}}
\newcommand{\quEEEScalc}{\eqref{Scalc}}
\newcommand{\quEEEScalmdefeq}{\eqref{Scalmdefeq}}
\newcommand{\quEEESdefeqn}{\eqref{Sdefeqn}}
\newcommand{\quEEESgambseqone}{\eqref{Sgambseqone}}
\newcommand{\quEEESigsbfSigsbmoteq}{\eqref{SigsbfSigsbmoteq}}
\newcommand{\quEEESigsbfboundineq}{\eqref{Sigsbfboundineq}}
\newcommand{\quEEESigsbfdefeqc}{\eqref{Sigsbfdefeqc}}
\newcommand{\quEEESigsbfdefeq}{\eqref{Sigsbfdefeq}}
\newcommand{\quEEESigsbfmotcond}{\eqref{Sigsbfmotcond}}
\newcommand{\quEEESigsbfmotdefeq}{\eqref{Sigsbfmotdefeq}}
\newcommand{\quEEESigsurf}{\eqref{Sigsurf}}
\newcommand{\quEEESigzzdef}{\eqref{Sigzzdef}}
\newcommand{\quEEESijSeq}{\eqref{SijSeq}}
\newcommand{\quEEESitwoeq}{\eqref{Sitwoeq}}
\newcommand{\quEEESjelldef}{\eqref{Sjelldef}}
\newcommand{\quEEESsyseq}{\eqref{Ssyseq}}
\newcommand{\quEEESztp}{\eqref{Sztp}}
\newcommand{\quEEETSSbdef}{\eqref{TSSbdef}}
\newcommand{\quEEETstardef}{\eqref{Tstardef}}
\newcommand{\quEEEUconteqyi}{\eqref{Uconteqyi}}
\newcommand{\quEEEUcont}{Proposition \ref{Ucont}}
\newcommand{\quEEEUe}{\eqref{Ue}}
\newcommand{\quEEEUtvbfluxineq}{\eqref{Utvbfluxineq}}
\newcommand{\quEEEUy}{\eqref{Uy}}
\newcommand{\quEEEUztrderiv}{\eqref{Uztrderiv}}
\newcommand{\quEEEWOOhXz}{\eqref{WOOhXz}}
\newcommand{\quEEEWinftyalg}{Lemma \ref{Winftyalg}}
\newcommand{\quEEEWinftydef}{\eqref{Winftydef}}
\newcommand{\quEEEabcK}{\eqref{abcK}}
\newcommand{\quEEEablowboundeq}{\eqref{ablowboundeq}}
\newcommand{\quEEEacknowledgements}{Subsection \ref{acknowledgements}}
\newcommand{\quEEEadmnonlinLinfty}{Lemma \ref{admnonlinLinfty}}
\newcommand{\quEEEadmnonlinpropcor}{Corollary \ref{admnonlinpropcor}}
\newcommand{\quEEEadmnonlinpropmultmult}{\eqref{admnonlinpropmultmult}}
\newcommand{\quEEEadmnonlinprop}{Proposition \ref{admnonlinprop}}
\newcommand{\quEEEalgres}{Subsection \ref{algres}}
\newcommand{\quEEEalpensricone}{\eqref{alpensricone}}
\newcommand{\quEEEalpensricthree}{\eqref{alpensricthree}}
\newcommand{\quEEEalpensrictwo}{\eqref{alpensrictwo}}
\newcommand{\quEEEamsmalldet}{Lemma \ref{amsmalldet}}
\newcommand{\quEEEappfrgambUzbd}{\eqref{appfrgambUzbd}}
\newcommand{\quEEEappfrlowbound}{\eqref{appfrlowbound}}
\newcommand{\quEEEapproxsoldef}{Definition \ref{approxsoldef}}
\newcommand{\quEEEapproxsol}{Subsection \ref{approxsol}}
\newcommand{\quEEEarccscbound}{\eqref{arccscbound}}
\newcommand{\quEEEasdparkeqii}{\eqref{asdparkeqii}}
\newcommand{\quEEEbIbootinfii}{\eqref{bIbootinfii}}
\newcommand{\quEEEbIbootinfmod}{\eqref{bIbootinfmod}}
\newcommand{\quEEEbIbootinf}{\eqref{bIbootinf}}
\newcommand{\quEEEbasecommexp}{\eqref{basecommexp}}
\newcommand{\quEEEbaseests}{\eqref{baseests}}
\newcommand{\quEEEbasegambbound}{\eqref{basegambbound}}
\newcommand{\quEEEbaseinitboundii}{\eqref{baseinitboundii}}
\newcommand{\quEEEbaseinitbound}{\eqref{baseinitbound}}
\newcommand{\quEEEbaseres}{Subsection \ref{baseres}}
\newcommand{\quEEEbbener}{Subsection \ref{bbener}}
\newcommand{\quEEEbbsbdone}{\eqref{bbsbdone}}
\newcommand{\quEEEbbsbvbbound}{\eqref{bbsbvbbound}}
\newcommand{\quEEEbbsderbounds}{\eqref{bbsderbounds}}
\newcommand{\quEEEbbsdonei}{\eqref{bbsdonei}}
\newcommand{\quEEEbbseqone}{\eqref{bbseqone}}
\newcommand{\quEEEbigambb}{\eqref{bigambb}}
\newcommand{\quEEEbiomegbb}{\eqref{biomegbb}}
\newcommand{\quEEEbootstrapimpLinfgamb}{Lemma \ref{bootstrapimpLinfgamb}}
\newcommand{\quEEEbootstrapimpcor}{Corollary \ref{bootstrapimpcor}}
\newcommand{\quEEEbootstrapimp}{Lemma \ref{bootstrapimp}}
\newcommand{\quEEEboundsinitdatgamb}{Subsection \ref{approxsol}}
\newcommand{\quEEEbulkFfcdefi}{\eqref{bulkFfcdefi}}
\newcommand{\quEEEbulkFfcdef}{\eqref{bulkFfcdef}}
\newcommand{\quEEEbulkGtdefe}{\eqref{bulkGtdefe}}
\newcommand{\quEEEbulkGtdef}{\eqref{bulkGtdef}}
\newcommand{\quEEEbulkdef}{\eqref{bulkdef}}
\newcommand{\quEEEbulkexdefeqe}{\eqref{bulkexdefeqe}}
\newcommand{\quEEEbulkexdefeq}{\eqref{bulkexdefeq}}
\newcommand{\quEEEbulksdef}{\eqref{bulksdef}}
\newcommand{\quEEEcEboot}{\eqref{cEboot}}
\newcommand{\quEEEcalSdef}{\eqref{calSdef}}
\newcommand{\quEEEcbsbdonei}{\eqref{cbsbdonei}}
\newcommand{\quEEEcbsbdone}{\eqref{cbsbdone}}
\newcommand{\quEEEcbsderbounds}{\eqref{cbsderbounds}}
\newcommand{\quEEEcbseqone}{\eqref{cbseqone}}
\newcommand{\quEEEchihatkr}{\eqref{chihatkr}}
\newcommand{\quEEEchiichihatbound}{\eqref{chiichihatbound}}
\newcommand{\quEEEchintchihatinf}{\eqref{chintchihatinf}}
\newcommand{\quEEEchsymbound}{\eqref{chsymbound}}
\newcommand{\quEEEchsymdbound}{\eqref{chsymdbound}}
\newcommand{\quEEEclinitsuppcondgamb}{\eqref{clinitsuppcondgamb}}
\newcommand{\quEEEclinitsuppcondomegb}{\eqref{clinitsuppcondomegb}}
\newcommand{\quEEEcnstEqchoice}{\eqref{cnstEqchoice}}
\newcommand{\quEEEcommnormreseq}{\eqref{commnormreseq}}
\newcommand{\quEEEcommnorm}{Proposition \ref{commnorm}}
\newcommand{\quEEEcomonefsf}{\eqref{comonefsf}}
\newcommand{\quEEEcompsemi}{Subsection \ref{compsemi}}
\newcommand{\quEEEconsistencycond}{\eqref{consistencycond}}
\newcommand{\quEEEconsthinvexp}{\eqref{consthinvexp}}
\newcommand{\quEEEconstintro}{Subsection \ref{constintro}}
\newcommand{\quEEEconstpres}{Proposition \ref{constpres}}
\newcommand{\quEEEconstraints}{Section \ref{constraints}}
\newcommand{\quEEEconstronel}{\eqref{constronel}}
\newcommand{\quEEEconstrone}{\eqref{constrone}}
\newcommand{\quEEEconstroowave}{\eqref{constroowave}}
\newcommand{\quEEEconstroricone}{\eqref{constroricone}}
\newcommand{\quEEEconstroricthree}{\eqref{constroricthree}}
\newcommand{\quEEEconstrorictwo}{\ref{constrorictwo}}
\newcommand{\quEEEconstrsoln}{Lemma \ref{constrsoln}}
\newcommand{\quEEEconstrthree}{\eqref{constrthree}}
\newcommand{\quEEEconstrtwo}{\eqref{constrtwo}}
\newcommand{\quEEEcontboundord}{Definition \ref{contboundord}}
\newcommand{\quEEEcrsinitdat}{\eqref{crsinitdat}}
\newcommand{\quEEEcsbBsigffbdefeq}{\eqref{csbBsigffbdefeq}}
\newcommand{\quEEEcsbBsigffbdef}{\eqref{csbBsigffbdef}}
\newcommand{\quEEEcsbEboot}{\eqref{csbEboot}}
\newcommand{\quEEEcsbisoEboot}{\eqref{csbisoEboot}}
\newcommand{\quEEEcsbsbulkdef}{\eqref{csbsbulkdef}}
\newcommand{\quEEEcscbound}{\eqref{cscbound}}
\newcommand{\quEEEcudifsetfix}{\eqref{cudifsetfix}}
\newcommand{\quEEEdAccommF}{\eqref{dAccommF}}
\newcommand{\quEEEdAcexpwaveq}{\eqref{dAcexpwaveq}}
\newcommand{\quEEEdAcexp}{Proposition \ref{dAcexp}}
\newcommand{\quEEEdAcfunclist}{\eqref{dAcfunclist}}
\newcommand{\quEEEdAcpdAchXz}{\eqref{dAcpdAchXz}}
\newcommand{\quEEEdAcwboxeq}{\eqref{dAcwboxeq}}
\newcommand{\quEEEdGeodbootstrap}{\eqref{dGeodbootstrap}}
\newcommand{\quEEEdGeodinitbound}{\eqref{dGeodinitbound}}
\newcommand{\quEEEddAcFbbyi}{\eqref{ddAcFbbyi}}
\newcommand{\quEEEddAcFbb}{\eqref{ddAcFbb}}
\newcommand{\quEEEdefadmnonlin}{Definition \ref{defadmnonlin}}
\newcommand{\quEEEdfunzdef}{\eqref{dfunzdef}}
\newcommand{\quEEEdgambappdef}{\eqref{dgambappdef}}
\newcommand{\quEEEdgambpbbound}{\eqref{dgambpbbound}}
\newcommand{\quEEEdhhd}{\eqref{dhhd}}
\newcommand{\quEEEdhtildedef}{\eqref{dhtildedef}}
\newcommand{\quEEEdifAset}{\eqref{difAset}}
\newcommand{\quEEEdiffwave}{\eqref{diffwave}}
\newcommand{\quEEEdifopsetdef}{\eqref{difopsetdef}}
\newcommand{\quEEEdifopsettprop}{\eqref{difopsettprop}}
\newcommand{\quEEEdifraset}{Definition \ref{difraset}}
\newcommand{\quEEEdivthSigy}{Proposition \ref{divthSigy}}
\newcommand{\quEEEdivth}{Proposition \ref{divth}}
\newcommand{\quEEEdlbbbetclinecond}{\eqref{dlbbbetclinecond}}
\newcommand{\quEEEdlbbbinitzed}{\eqref{dlbbbinitzed}}
\newcommand{\quEEEdlbderbounds}{\eqref{dlbderbounds}}
\newcommand{\quEEEdlbeqone}{\eqref{dlbeqone}}
\newcommand{\quEEEdlbgambsuppi}{\eqref{dlbgambsuppi}}
\newcommand{\quEEEdlbgambsupp}{\eqref{dlbgambsupp}}
\newcommand{\quEEEdlblocbound}{\eqref{dlblocbound}}
\newcommand{\quEEEdlbsbdonei}{\eqref{dlbsbdonei}}
\newcommand{\quEEEdlbsbdone}{\eqref{dlbsbdone}}
\newcommand{\quEEEdlbsbvbbound}{\eqref{dlbsbvbbound}}
\newcommand{\quEEEdlbsderbounds}{\eqref{dlbsderbounds}}
\newcommand{\quEEEdlbseqone}{\eqref{dlbseqone}}
\newcommand{\quEEEdlbvby}{\eqref{dlbvby}}
\newcommand{\quEEEdsopelldef}{Definition \ref{dsopelldef}}
\newcommand{\quEEEeTeTpdef}{\eqref{eTeTpdef}}
\newcommand{\quEEEeTeTpnuscaldef}{\eqref{eTeTpnuscaldef}}
\newcommand{\quEEEeTpeTpark}{\eqref{eTpeTpark}}
\newcommand{\quEEEeenerdefii}{\eqref{eenerdefii}}
\newcommand{\quEEEeenerdefi}{\eqref{eenerdefi}}
\newcommand{\quEEEellbinssmall}{Proposition \ref{ellbinssmall}}
\newcommand{\quEEEellbinvlbs}{\eqref{ellbinvlbs}}
\newcommand{\quEEEellbinvlb}{\eqref{ellbinvlb}}
\newcommand{\quEEEellbinvsmall}{Corollary \ref{ellbinvsmall}}
\newcommand{\quEEEenerbound}{Subsection \ref{enerbound}}
\newcommand{\quEEEenerdef}{Subsection \ref{enerdef}}
\newcommand{\quEEEenerineqerrtermexp}{\eqref{enerineqerrtermexp}}
\newcommand{\quEEEenerineq}{Section \ref{enerineq}}
\newcommand{\quEEEenerperboot}{Subsection \ref{enerperboot}}
\newcommand{\quEEEensricone}{\eqref{ensricone}}
\newcommand{\quEEEensricthree}{\eqref{ensricthree}}
\newcommand{\quEEEensrictwo}{\eqref{ensrictwo}}
\newcommand{\quEEEepsDIedefm}{\eqref{epsDIedefm}}
\newcommand{\quEEEepsbound}{\eqref{epsbound}}
\newcommand{\quEEEepsdef}{\eqref{epsdef}}
\newcommand{\quEEEepsenerboundeqi}{\eqref{epsenerboundeqi}}
\newcommand{\quEEEepsenerboundresineqyi}{\eqref{epsenerboundresineqyi}}
\newcommand{\quEEEepsenerbound}{Proposition \ref{epsenerbound}}
\newcommand{\quEEEepsnetaappfr}{\eqref{epsnetaappfr}}
\newcommand{\quEEEepssim}{\eqref{epssim}}
\newcommand{\quEEEeswave}{\eqref{eswave}}
\newcommand{\quEEEexistintro}{Subsection \ref{existintro}}
\newcommand{\quEEEfHTU}{\eqref{fHTU}}
\newcommand{\quEEEfLinfUln}{\eqref{fLinfUln}}
\newcommand{\quEEEfeparrbyi}{\eqref{feparrbyi}}
\newcommand{\quEEEfeparrb}{\eqref{feparrb}}
\newcommand{\quEEEfgaugechoice}{Figure \ref{fgaugechoice}}
\newcommand{\quEEEfhXzsizb}{\eqref{fhXzsizb}}
\newcommand{\quEEEfindivth}{\eqref{findivth}}
\newcommand{\quEEEfindiv}{\eqref{findiv}}
\newcommand{\quEEEfineqerver}{\eqref{fineqerver}}
\newcommand{\quEEEfineqyiver}{\eqref{fineqyiver}}
\newcommand{\quEEEfirmbdeq}{\eqref{firmbdeq}}
\newcommand{\quEEEfirmbdleq}{\eqref{firmbdleq}}
\newcommand{\quEEEfluxboundsone}{\eqref{fluxboundsone}}
\newcommand{\quEEEfluxboundstwo}{\eqref{fluxboundstwo}}
\newcommand{\quEEEfocapp}{Subsection \ref{focapp}}
\newcommand{\quEEEfoccoordmetbderdi}{\eqref{foccoordmetbderdi}}
\newcommand{\quEEEfoccoordmetbderd}{\eqref{foccoordmetbderd}}
\newcommand{\quEEEfoccoordmetbdi}{\eqref{foccoordmetbdi}}
\newcommand{\quEEEfoccoordmetbd}{\eqref{foccoordmetbd}}
\newcommand{\quEEEfoccoordparkcondi}{\eqref{foccoordparkcondi}}
\newcommand{\quEEEfoccoordparkcond}{\eqref{foccoordparkcond}}
\newcommand{\quEEEfoccoordsmth}{Proposition \ref{foccoordsmth}}
\newcommand{\quEEEfocsol}{Section \ref{focsol}}
\newcommand{\quEEEfrJKLeq}{\eqref{frJKLeq}}
\newcommand{\quEEEfrwboxJKLeq}{\eqref{frwboxJKLeq}}
\newcommand{\quEEEfsbvbbound}{\eqref{fsbvbbound}}
\newcommand{\quEEEfscalcdefiii}{\eqref{fscalcdefiii}}
\newcommand{\quEEEfscalcdefii}{\eqref{fscalcdefii}}
\newcommand{\quEEEfscalcdef}{\eqref{fscalcdef}}
\newcommand{\quEEEfscaldef}{\eqref{fscaldef}}
\newcommand{\quEEEfscalgamdef}{\eqref{fscalgamdef}}
\newcommand{\quEEEfsintro}{Subsection \ref{fsintro}}
\newcommand{\quEEEfullgmetdef}{\eqref{fullgmetdef}}
\newcommand{\quEEEgambGappr}{Proposition \ref{gambGappr}}
\newcommand{\quEEEgambIdiff}{\eqref{gambIdiff}}
\newcommand{\quEEEgambNNapprox}{Corollary \ref{gambNNapprox}}
\newcommand{\quEEEgambNbaseprop}{Lemma \ref{gambNbaseprop}}
\newcommand{\quEEEgambNdefeq}{\eqref{gambNdefeq}}
\newcommand{\quEEEgambNsbz}{\eqref{gambNsbz}}
\newcommand{\quEEEgambPloc}{\eqref{gambPloc}}
\newcommand{\quEEEgambUzSigzsbz}{Lemma \ref{gambUzSigzsbz}}
\newcommand{\quEEEgambadmnonlin}{Proposition \ref{gambadmnonlin}}
\newcommand{\quEEEgambappfrSzeq}{\eqref{gambappfrSzeq}}
\newcommand{\quEEEgambdecbound}{\eqref{gambdecbound}}
\newcommand{\quEEEgambkinitbound}{Proposition \ref{gambkinitbound}}
\newcommand{\quEEEgambrmbdleq}{\eqref{gambrmbdleq}}
\newcommand{\quEEEgambsbdone}{\eqref{gambsbdone}}
\newcommand{\quEEEgambsderbounds}{\eqref{gambsderbounds}}
\newcommand{\quEEEgambseqone}{\eqref{gambseqone}}
\newcommand{\quEEEgambsinitzed}{\eqref{gambsinitzed}}
\newcommand{\quEEEgambsuppyi}{\eqref{gambsuppyi}}
\newcommand{\quEEEgambsupp}{Figure \ref{gambsupp}}
\newcommand{\quEEEgambvboundmainbound}{\eqref{gambvboundmainbound}}
\newcommand{\quEEEgambvbound}{Proposition \ref{gambvbound}}
\newcommand{\quEEEgambveqone}{\eqref{gambveqone}}
\newcommand{\quEEEgambzdef}{\eqref{gambzdef}}
\newcommand{\quEEEgammaenboundshighenerineq}{\eqref{gammaenboundshighenerineq}}
\newcommand{\quEEEgammaenbounds}{Lemma \ref{gammaenbounds}}
\newcommand{\quEEEgammainitdat}{\eqref{gammainitdat}}
\newcommand{\quEEEgaugechoicech}{Section \ref{gaugechoicech}}
\newcommand{\quEEEgaugechoice}{Subsection \ref{gaugechoice}}
\newcommand{\quEEEgaugemetcondii}{\eqref{gaugemetcondii}}
\newcommand{\quEEEgaugemetcond}{\eqref{gaugemetcond}}
\newcommand{\quEEEgaugemetform}{\eqref{gaugemetform}}
\newcommand{\quEEEgaugemetii}{Proposition \ref{gaugemetii}}
\newcommand{\quEEEgaugemeti}{Proposition \ref{gaugemeti}}
\newcommand{\quEEEgcintro}{Subsection \ref{gcintro}}
\newcommand{\quEEEgdiffeq}{\eqref{gdiffeq}}
\newcommand{\quEEEgeomoptNapp}{\eqref{geomoptNapp}}
\newcommand{\quEEEgeooptvarpib}{\eqref{geooptvarpib}}
\newcommand{\quEEEgeooptvarpi}{\eqref{geooptvarpi}}
\newcommand{\quEEEgestinq}{Lemma \ref{gestinq}}
\newcommand{\quEEEgibmaininfbd}{\eqref{gibmaininfbd}}
\newcommand{\quEEEgronwallcont}{Proposition \ref{gronwallcont}}
\newcommand{\quEEEgronwallde}{\eqref{gronwallde}}
\newcommand{\quEEEgronwalle}{Proposition \ref{gronwalle}}
\newcommand{\quEEEhXzPhicond}{\eqref{hXzPhicond}}
\newcommand{\quEEEhXznormdef}{\eqref{hXznormdef}}
\newcommand{\quEEEhatfuncdef}{\eqref{hatfuncdef}}
\newcommand{\quEEEhbardefii}{\eqref{hbardefii}}
\newcommand{\quEEEhbardef}{\eqref{hbardef}}
\newcommand{\quEEEhbarspatdef}{\eqref{hbarspatdef}}
\newcommand{\quEEEhinvexp}{\eqref{hinvexp}}
\newcommand{\quEEEiSgambseqone}{\eqref{iSgambseqone}}
\newcommand{\quEEEibounddef}{\eqref{ibounddef}}
\newcommand{\quEEEiconstrone}{\eqref{iconstrone}}
\newcommand{\quEEEiconstrthree}{\eqref{iconstrthree}}
\newcommand{\quEEEiconstrtwo}{\eqref{iconstrtwo}}
\newcommand{\quEEEidscalconste}{\eqref{idscalconste}}
\newcommand{\quEEEidscalconsts}{\eqref{idscalconsts}}
\newcommand{\quEEEidscalconsty}{\eqref{idscalconsty}}
\newcommand{\quEEEidsolnsupp}{\eqref{idsolnsupp}}
\newcommand{\quEEEifluxboundsone}{\eqref{ifluxboundsone}}
\newcommand{\quEEEifluxboundstwo}{\eqref{ifluxboundstwo}}
\newcommand{\quEEEigBBsigffcond}{\eqref{igBBsigffcond}}
\newcommand{\quEEEigBcond}{\eqref{igBcond}}
\newcommand{\quEEEigScond}{\eqref{igScond}}
\newcommand{\quEEEigVcond}{\eqref{igVcond}}
\newcommand{\quEEEigambdsuppeq}{\eqref{igambdsuppeq}}
\newcommand{\quEEEigambsuppeq}{\eqref{igambsuppeq}}
\newcommand{\quEEEihbarexp}{\eqref{ihbarexp}}
\newcommand{\quEEEiibsuppgambcond}{\eqref{iibsuppgambcond}}
\newcommand{\quEEEiichihatbound}{\eqref{iichihatbound}}
\newcommand{\quEEEiihbarexp}{\eqref{iihbarexp}}
\newcommand{\quEEEimethexp}{\eqref{imethexp}}
\newcommand{\quEEEindhypstat}{\eqref{indhypstat}}
\newcommand{\quEEEingasymp}{\eqref{ingasymp}}
\newcommand{\quEEEingexp}{\eqref{ingexp}}
\newcommand{\quEEEinitdatbounds}{Proposition \ref{initdatbounds}}
\newcommand{\quEEEinitdatintro}{Subsection \ref{initdatintro}}
\newcommand{\quEEEinitdat}{Section \ref{initdat}}
\newcommand{\quEEEinitenernormeq}{Proposition \ref{initenernormeq}}
\newcommand{\quEEEinitnormsener}{Definition \ref{initnormsener}}
\newcommand{\quEEEinjfuncHScond}{\eqref{injfuncHScond}}
\newcommand{\quEEEinjfunc}{Lemma \ref{injfunc}}
\newcommand{\quEEEinseqbound}{Theorem \ref{inseqbound}}
\newcommand{\quEEEinsricone}{\eqref{insricone}}
\newcommand{\quEEEinsricthree}{\eqref{insricthree}}
\newcommand{\quEEEinsrictwo}{\eqref{insrictwo}}
\newcommand{\quEEEintGeodineq}{\eqref{intGeodineq}}
\newcommand{\quEEEintRicwav}{\eqref{intRicwav}}
\newcommand{\quEEEintchihatinf}{\eqref{intchihatinf}}
\newcommand{\quEEEinterintro}{Subsection \ref{interintro}}
\newcommand{\quEEEinterlude}{Section \ref{interlude}}
\newcommand{\quEEEintfoccoordsmth}{Proposition \ref{intfoccoordsmth}}
\newcommand{\quEEEintgagniren}{\eqref{intgagniren}}
\newcommand{\quEEEintgammaform}{\eqref{intgammaform}}
\newcommand{\quEEEintgaugecond}{\eqref{intgaugecond}}
\newcommand{\quEEEintklainsobol}{\eqref{intklainsobol}}
\newcommand{\quEEEintroduction}{Section \ref{introduction}}
\newcommand{\quEEEinverse}{Lemma \ref{inverse}}
\newcommand{\quEEEioowave}{\eqref{ioowave}}
\newcommand{\quEEEioricone}{\eqref{ioricone}}
\newcommand{\quEEEioricthree}{\eqref{ioricthree}}
\newcommand{\quEEEiorictwo}{\eqref{iorictwo}}
\newcommand{\quEEEiotandef}{\eqref{iotandef}}
\newcommand{\quEEEiover}{Subsection \ref{iover}}
\newcommand{\quEEEirbthmgammastip}{\eqref{irbthmgammastip}}
\newcommand{\quEEEirbthmscalfrac}{\eqref{irbthmscalfrac}}
\newcommand{\quEEEirbthm}{Theorem \ref{irbthm}}
\newcommand{\quEEEiscalcdef}{\eqref{iscalcdef}}
\newcommand{\quEEEiscalconste}{\eqref{iscalconste}}
\newcommand{\quEEEiscalconsts}{\eqref{iscalconsts}}
\newcommand{\quEEEiscalconsty}{\eqref{iscalconsty}}
\newcommand{\quEEEiscalvdef}{\eqref{iscalvdef}}
\newcommand{\quEEEiseqboundconsyi}{\eqref{iseqboundconsyi}}
\newcommand{\quEEEiseqbound}{Theorem \ref{iseqbound}}
\newcommand{\quEEEishopbaseester}{\eqref{ishopbaseester}}
\newcommand{\quEEEishopbaseestyi}{\eqref{ishopbaseestyi}}
\newcommand{\quEEEishopbaseest}{\eqref{ishopbaseest}}
\newcommand{\quEEEishuffleops}{Lemma \ref{ishuffleops}}
\newcommand{\quEEEisummary}{Subsection \ref{isummary}}
\newcommand{\quEEEiswave}{\eqref{iswave}}
\newcommand{\quEEEklainsobolevfige}{Figure \ref{klainsobolevfige}}
\newcommand{\quEEEklainsobolevfigy}{Figure \ref{klainsobolevfigy}}
\newcommand{\quEEEklainsobolv}{Proposition \ref{klainsobolv}}
\newcommand{\quEEEksUsdef}{\eqref{ksUsdef}}
\newcommand{\quEEEkseqero}{\eqref{kseqero}}
\newcommand{\quEEEkseqer}{\eqref{kseqer}}
\newcommand{\quEEEkseqsan}{\eqref{kseqsan}}
\newcommand{\quEEEkspbUdef}{\eqref{kspbUdef}}
\newcommand{\quEEElambframedef}{\eqref{lambframedef}}
\newcommand{\quEEElinfbounds}{Subsection \ref{linfbounds}}
\newcommand{\quEEElishopIconder}{\eqref{lishopIconder}}
\newcommand{\quEEElishopIcond}{\eqref{lishopIcond}}
\newcommand{\quEEElishopresest}{\eqref{lishopresest}}
\newcommand{\quEEElishoptpicond}{\eqref{lishoptpicond}}
\newcommand{\quEEElocener}{\eqref{locener}}
\newcommand{\quEEElwuegle}{\eqref{lwuegle}}
\newcommand{\quEEElwuegl}{\eqref{lwuegl}}
\newcommand{\quEEEmWinftydef}{\eqref{mWinftydef}}
\newcommand{\quEEEmaineTsqrtest}{\eqref{maineTsqrtest}}
\newcommand{\quEEEmainginitupperbound}{\eqref{mainginitupperbound}}
\newcommand{\quEEEmainprkestiii}{\eqref{mainprkestiii}}
\newcommand{\quEEEmainprkestii}{\eqref{mainprkestii}}
\newcommand{\quEEEmainprkest}{\eqref{mainprkest}}
\newcommand{\quEEEmainresultprf}{Subsection \ref{mainresultprf}}
\newcommand{\quEEEmetazbdef}{\eqref{metazbdef}}
\newcommand{\quEEEmethChristoffel}{Lemma \ref{methChristoffel}}
\newcommand{\quEEEmethasymp}{\eqref{methasymp}}
\newcommand{\quEEEmetinitform}{\eqref{metinitform}}
\newcommand{\quEEEmffdef}{\eqref{mffdef}}
\newcommand{\quEEEmffetasb}{\eqref{mffetasb}}
\newcommand{\quEEEmffetatau}{\eqref{mffetatau}}
\newcommand{\quEEEmffthetazdiff}{\eqref{mffthetazdiff}}
\newcommand{\quEEEmffthmffetasmallepsn}{\eqref{mffthmffetasmallepsn}}
\newcommand{\quEEEmfgagniren}{Proposition \ref{mfgagniren}}
\newcommand{\quEEEminkdef}{\eqref{minkdef}}
\newcommand{\quEEEmotconst}{Subsection \ref{motconst}}
\newcommand{\quEEEnDHbnormdef}{\eqref{nDHbnormdef}}
\newcommand{\quEEEnHyptsbdef}{\eqref{nHyptsbdef}}
\newcommand{\quEEEnLcompdefeq}{\eqref{nLcompdefeq}}
\newcommand{\quEEEnLcompdef}{\eqref{nLcompdef}}
\newcommand{\quEEEnablaT}{\eqref{nablaT}}
\newcommand{\quEEEnatchardef}{Definition \ref{natchardef}}
\newcommand{\quEEEndiffwave}{\eqref{ndiffwave}}
\newcommand{\quEEEnfoccoordmetbderd}{\eqref{nfoccoordmetbderd}}
\newcommand{\quEEEnfoccoordmetbd}{\eqref{nfoccoordmetbd}}
\newcommand{\quEEEningexp}{\eqref{ningexp}}
\newcommand{\quEEEnintgammaform}{\eqref{nintgammaform}}
\newcommand{\quEEEnirbthmgambstip}{\eqref{nirbthmgambstip}}
\newcommand{\quEEEnocoordRic}{\eqref{nocoordRic}}
\newcommand{\quEEEnormlesssim}{\eqref{normlesssim}}
\newcommand{\quEEEnormsim}{\eqref{normsim}}
\newcommand{\quEEEnotations}{Subsection \ref{notations}}
\newcommand{\quEEEnsCOdef}{\eqref{nsCOdef}}
\newcommand{\quEEEnsricone}{\eqref{nsricone}}
\newcommand{\quEEEnsricthree}{\eqref{nsricthree}}
\newcommand{\quEEEnsrictwo}{\eqref{nsrictwo}}
\newcommand{\quEEEnstuffdelbound}{\eqref{nstuffdelbound}}
\newcommand{\quEEEnullHbspace}{Definition \ref{nullHbspace}}
\newcommand{\quEEEnullfolfigyi}{Figure \ref{nullfolfigyi}}
\newcommand{\quEEEnullfolfig}{Figure \ref{nullfolfig}}
\newcommand{\quEEEnullsurf}{\eqref{nullsurf}}
\newcommand{\quEEEnwfzyi}{\eqref{nwfzyi}}
\newcommand{\quEEEoHspace}{Definition \ref{oHspace}}
\newcommand{\quEEEomegbdecbound}{\eqref{omegbdecbound}}
\newcommand{\quEEEomegbenboundseqi}{\eqref{omegbenboundseqi}}
\newcommand{\quEEEomegbenboundseqling}{\eqref{omegbenboundseqling}}
\newcommand{\quEEEomegbenbounds}{Lemma \ref{omegbenbounds}}
\newcommand{\quEEEomegbenerbound}{Lemma \ref{omegbenerbound}}
\newcommand{\quEEEomegbgambfocboundi}{Proposition \ref{omegbgambfocboundi}}
\newcommand{\quEEEomegbsupp}{\eqref{omegbsupp}}
\newcommand{\quEEEonedintineq}{\eqref{onedintineq}}
\newcommand{\quEEEonedlineineq}{\eqref{onedlineineq}}
\newcommand{\quEEEonedsobolev}{Lemma \ref{onedsobolev}}
\newcommand{\quEEEoowave}{\eqref{oowave}}
\newcommand{\quEEEopssetbrackid}{\eqref{opssetbrackid}}
\newcommand{\quEEEoriconep}{\eqref{oriconep}}
\newcommand{\quEEEoricone}{\eqref{oricone}}
\newcommand{\quEEEoricthree}{\eqref{oricthree}}
\newcommand{\quEEEorictwo}{\eqref{orictwo}}
\newcommand{\quEEEosginitquant}{\eqref{osginitquant}}
\newcommand{\quEEEoutconstr}{Subsection \ref{outconstr}}
\newcommand{\quEEEparkCKSnu}{\eqref{parkCKSnu}}
\newcommand{\quEEEparkparrbd}{\eqref{parkparrbd}}
\newcommand{\quEEEpartconstr}{\eqref{partconstr}}
\newcommand{\quEEEpbarAsigmtdef}{\eqref{pbarAsigmtdef}}
\newcommand{\quEEEpgambsderbounds}{\eqref{pgambsderbounds}}
\newcommand{\quEEEphysnullsurff}{Figure \ref{physnullsurff}}
\newcommand{\quEEEplseii}{\eqref{plseii}}
\newcommand{\quEEEplsei}{\eqref{plsei}}
\newcommand{\quEEEpolSineq}{\eqref{polSineq}}
\newcommand{\quEEEponeres}{\eqref{poneres}}
\newcommand{\quEEEprecprop}{Proposition \ref{precprop}}
\newcommand{\quEEEprfer}{\eqref{prfer}}
\newcommand{\quEEEprimdiv}{\eqref{primdiv}}
\newcommand{\quEEEprodrulebound}{\eqref{prodrulebound}}
\newcommand{\quEEEpsbJgambN}{\eqref{psbJgambN}}
\newcommand{\quEEEpsbbeqone}{\eqref{psbbeqone}}
\newcommand{\quEEEpsbderbounds}{\eqref{psbderbounds}}
\newcommand{\quEEEpsbgambhXz}{\eqref{psbgambhXz}}
\newcommand{\quEEEpsbpxbbbeq}{\eqref{psbpxbbbeq}}
\newcommand{\quEEEpsbtOmegsub}{\eqref{psbtOmegsub}}
\newcommand{\quEEEpsbusys}{\eqref{psbusys}}
\newcommand{\quEEEpsdderbounds}{\eqref{psdderbounds}}
\newcommand{\quEEEpsdlbeqone}{\eqref{psdlbeqone}}
\newcommand{\quEEEpsoricthree}{\eqref{psoricthree}}
\newcommand{\quEEEpsorictwo}{\eqref{psorictwo}}
\newcommand{\quEEEpstWOOpssgamb}{\eqref{pstWOOpssgamb}}
\newcommand{\quEEEpthetf}{\eqref{pthetf}}
\newcommand{\quEEEpvbexp}{\eqref{pvbexp}}
\newcommand{\quEEEpvbgambgeq}{\eqref{pvbgambgeq}}
\newcommand{\quEEEpvbgambtsb}{\eqref{pvbgambtsb}}
\newcommand{\quEEEpvbgamb}{\eqref{pvbgamb}}
\newcommand{\quEEEpvvbexp}{\eqref{pvvbexp}}
\newcommand{\quEEEpxbexper}{\eqref{pxbexper}}
\newcommand{\quEEEpxbexpyi}{\eqref{pxbexpyi}}
\newcommand{\quEEEpxbtgamb}{\eqref{pxbtgamb}}
\newcommand{\quEEEpxvbexper}{\eqref{pxvbexper}}
\newcommand{\quEEEpxvbexpyi}{\eqref{pxvbexpyi}}
\newcommand{\quEEEpxxbexper}{\eqref{pxxbexper}}
\newcommand{\quEEEpxxbexpyi}{\eqref{pxxbexpyi}}
\newcommand{\quEEEquadineq}{Lemma \ref{quadineq}}
\newcommand{\quEEEquotresbound}{\eqref{quotresbound}}
\newcommand{\quEEErbmainprkestii}{\eqref{rbmainprkestii}}
\newcommand{\quEEErbsbeTpcond}{\eqref{rbsbeTpcond}}
\newcommand{\quEEErbthmosizedef}{\eqref{rbthmosizedef}}
\newcommand{\quEEErbthm}{Theorem \ref{rbthm}}
\newcommand{\quEEEredwaveeq}{\eqref{redwaveeq}}
\newcommand{\quEEErefgronwallcont}{Proposition \ref{refgronwallcont}}
\newcommand{\quEEErefgronwallde}{\eqref{refgronwallde}}
\newcommand{\quEEErefgronwalleqyi}{\eqref{refgronwalleqyi}}
\newcommand{\quEEErelworks}{Subsection \ref{relworks}}
\newcommand{\quEEEremainbound}{Subsection \ref{remainbound}}
\newcommand{\quEEEricric}{Proposition \ref{ricric}}
\newcommand{\quEEErmbdeq}{\eqref{rmbdeq}}
\newcommand{\quEEErmbdleq}{\eqref{rmbdleq}}
\newcommand{\quEEEsCOdef}{\eqref{sCOdef}}
\newcommand{\quEEEsbHprodpbA}{\eqref{sbHprodpbA}}
\newcommand{\quEEEsbHyinormalg}{\eqref{sbHyinormalg}}
\newcommand{\quEEEsbWginitnorm}{\eqref{sbWginitnorm}}
\newcommand{\quEEEsbWsbxbvbdef}{\eqref{sbWsbxbvbdef}}
\newcommand{\quEEEsbWspace}{Definition \ref{sbWspace}}
\newcommand{\quEEEsbeTpcond}{\eqref{sbeTpcond}}
\newcommand{\quEEEsbeffcpc}{\eqref{sbeffcpc}}
\newcommand{\quEEEsbenerflux}{Corollary \ref{sbenerflux}}
\newcommand{\quEEEsbulkdef}{\eqref{sbulkdef}}
\newcommand{\quEEEscalbulk}{Figure \ref{scalbulk}}
\newcommand{\quEEEscalconste}{\eqref{scalconste}}
\newcommand{\quEEEscalconstsp}{\eqref{scalconstsp}}
\newcommand{\quEEEscalconsts}{\eqref{scalconsts}}
\newcommand{\quEEEscalconsty}{\eqref{scalconsty}}
\newcommand{\quEEEscalcut}{Subsection \ref{scalcut}}
\newcommand{\quEEEscaledeqintro}{Subsection \ref{scaledeqintro}}
\newcommand{\quEEEscaling}{Section \ref{scaling}}
\newcommand{\quEEEscalquantdef}{Definition \ref{scalquantdef}}
\newcommand{\quEEEscaltdef}{\eqref{scaltdef}}
\newcommand{\quEEEscaltransderiv}{\eqref{scaltransderiv}}
\newcommand{\quEEEscaltscaldef}{\eqref{scaltscaldef}}
\newcommand{\quEEEsclinitboundsallgamb}{Proposition \ref{sclinitboundsallgamb}}
\newcommand{\quEEEsclinitboundsallomeg}{Proposition \ref{sclinitboundsallomeg}}
\newcommand{\quEEEsclinitboundsi}{Proposition \ref{sclinitboundsi}}
\newcommand{\quEEEscoordsys}{Figure \ref{scoordsys}}
\newcommand{\quEEEseqboundcor}{Corollary \ref{seqboundcor}}
\newcommand{\quEEEseqbound}{Theorem \ref{seqbound}}
\newcommand{\quEEEsghupint}{\eqref{sghupint}}
\newcommand{\quEEEshopresester}{\eqref{shopresester}}
\newcommand{\quEEEshuffleops}{Corollary \ref{shuffleops}}
\newcommand{\quEEEshuffopseqyi}{\eqref{shuffopseqyi}}
\newcommand{\quEEEsigmatupsx}{\eqref{sigmatupsx}}
\newcommand{\quEEEsigubaseineq}{\eqref{sigubaseineq}}
\newcommand{\quEEEsiguintineq}{\eqref{siguintineq}}
\newcommand{\quEEEsiguintlem}{Lemma \ref{siguintlem}}
\newcommand{\quEEEsigupint}{\eqref{sigupint}}
\newcommand{\quEEEsiguprop}{Proposition \ref{siguprop}}
\newcommand{\quEEEsimpdifaset}{\eqref{simpdifaset}}
\newcommand{\quEEEsimpdifrset}{\eqref{simpdifrset}}
\newcommand{\quEEEsinitdate}{Subsection \ref{sinitdate}}
\newcommand{\quEEEsinitdat}{Subsection \ref{sinitdat}}
\newcommand{\quEEEsintroduction}{Subsection \ref{sintroduction}}
\newcommand{\quEEEsmoothnull}{Subsection \ref{smoothnull}}
\newcommand{\quEEEsobolevembeqi}{\eqref{sobolevembeqi}}
\newcommand{\quEEEsobolevemb}{Proposition \ref{sobolevemb}}
\newcommand{\quEEEsoowave}{\eqref{soowave}}
\newcommand{\quEEEspacetimephyss}{Figure \ref{spacetimephyss}}
\newcommand{\quEEEspacsurf}{\eqref{spacsurf}}
\newcommand{\quEEEspecsoboldefer}{\eqref{specsoboldefer}}
\newcommand{\quEEEspecsoboldefyi}{\eqref{specsoboldefyi}}
\newcommand{\quEEEssbIbootinfgamb}{\eqref{ssbIbootinfgamb}}
\newcommand{\quEEEssbIbootinfh}{\eqref{ssbIbootinfh}}
\newcommand{\quEEEstatres}{Subsection \ref{statres}}
\newcommand{\quEEEstuffdelbound}{\eqref{stuffdelbound}}
\newcommand{\quEEEswave}{\eqref{swave}}
\newcommand{\quEEEsymred}{Subsection \ref{symred}}
\newcommand{\quEEEtauSUSAdef}{\eqref{tauSUSAdef}}
\newcommand{\quEEEtauubound}{\eqref{tauubound}}
\newcommand{\quEEEtauzetadef}{\eqref{tauzetadef}}
\newcommand{\quEEEtdhaphnbound}{\eqref{tdhaphnbound}}
\newcommand{\quEEEtempGeodtboot}{\eqref{tempGeodtboot}}
\newcommand{\quEEEthinfshortdef}{\eqref{thinfshortdef}}
\newcommand{\quEEEtparmsbounds}{\eqref{tparmsbounds}}
\newcommand{\quEEEtregdef}{\eqref{tregdef}}
\newcommand{\quEEEttaudefeq}{\eqref{ttaudefeq}}
\newcommand{\quEEEudifsetpropeqer}{\eqref{udifsetpropeqer}}
\newcommand{\quEEEudifsetpropeqyi}{\eqref{udifsetpropeqyi}}
\newcommand{\quEEEudifsetprop}{Lemma \ref{udifsetprop}}
\newcommand{\quEEEudifsetyiordcom}{\eqref{udifsetyiordcom}}
\newcommand{\quEEEunGSBdef}{\eqref{unGSBdef}}
\newcommand{\quEEEupIdAcFii}{\eqref{upIdAcFii}}
\newcommand{\quEEEupIdAcFi}{\eqref{upIdAcFi}}
\newcommand{\quEEEuprtlIdef}{\eqref{uprtlIdef}}
\newcommand{\quEEEuprtldef}{Definition \ref{uprtldef}}
\newcommand{\quEEEupsbdef}{\eqref{upsbdef}}
\newcommand{\quEEEuzbbsbdone}{\eqref{uzbbsbdone}}
\newcommand{\quEEEuzcbsbdone}{\eqref{uzcbsbdone}}
\newcommand{\quEEEuzdlbsbdone}{\eqref{uzdlbsbdone}}
\newcommand{\quEEEuzgambsanb}{\eqref{uzgambsanb}}
\newcommand{\quEEEuzgambyib}{\eqref{uzgambyib}}
\newcommand{\quEEEuzindhypstat}{\eqref{uzindhypstat}}
\newcommand{\quEEEuzomegyib}{\eqref{uzomegyib}}
\newcommand{\quEEEvbjysigu}{\eqref{vbjysigu}}
\newcommand{\quEEEveqpquant}{\eqref{veqpquant}}
\newcommand{\quEEEvinitdat}{Subsection \ref{vinitdat}}
\newcommand{\quEEEwboxetaappfrb}{\eqref{wboxetaappfrb}}
\newcommand{\quEEEwboxetagambb}{\eqref{wboxetagambb}}
\newcommand{\quEEEwtbulkexpexp}{\eqref{wtbulkexpexp}}
\newcommand{\quEEEwtiotaniotan}{\eqref{wtiotaniotan}}
\newcommand{\quEEEwtpbHpbHgamb}{\eqref{wtpbHpbHgamb}}
\newcommand{\quEEEwtpbHpbH}{\eqref{wtpbHpbH}}
\newcommand{\quEEEyiCKSnpT}{\eqref{yiCKSnpT}}
\newcommand{\quEEEyidSemb}{\eqref{yidSemb}}
\newcommand{\quEEEyoucoordRic}{\eqref{youcoordRic}}
\newcommand{\quEb}{{{\overline{E}}}}
\newcommand{\quElinsetb}{X}
\newcommand{\quElinsetp}{Y}
\newcommand{\quElinset}{X}
\newcommand{\quElinsubsetp}{V}
\newcommand{\quElinsubset}{U}
\newcommand{\quElsp}{\qux}
\newcommand{\quE}{{E}}
\newcommand{\quGamb}{{\overline{\Gamma}}}
\newcommand{\quGeodend}{{\frakf f}}
\newcommand{\quGeodp}{{\widetilde{\quGeod}}}
\newcommand{\quGeodsg}{{\underline{\quGeod}}}
\newcommand{\quGeods}{{{\quGeodsg}}_{{\qutsbz}}}
\newcommand{\quGeodv}{{\underline{\underline{\quGeod}}}}
\newcommand{\quGeod}{{Z}}
\newcommand{\quHS}[1]{\vert #1\vert _{HS}}
\newcommand{\quHypt}[1]{\quwlntilde{\Sigma}_{#1}}
\newcommand{\quHyp}[1]{\qreflink{quantSigsb}{\Sigma_{#1}}}
\newcommand{\quIb}{{{\overline{I}}}}
\newcommand{\quI}{{{\mathcal I}}}
\newcommand{\quKSk}{\sigma}
\newcommand{\quKm}{\left(\matrix{0&0&0\cr 0&\alpha&\beta\cr 0&\beta&\delta\cr}\right)}
\newcommand{\quLambdabb}{{\overline\quLambdab}}
\newcommand{\quLambdab}{{\overline\Lambda}}
\newcommand{\quLb}{{N}}
\newcommand{\quLline}{{\mathcal L}}
\newcommand{\quLnb}{{L}}
\newcommand{\quL}{{\mathcal L}}
\newcommand{\quM}{{\bf M}}
\newcommand{\quNb}{{{\underline{L}}}}
\newcommand{\quN}{{\mathbb{N}}}
\newcommand{\quOmegbh}{\quOmegb}
\newcommand{\quOmegbt}{\quwlntilde{\quOmegb}}
\newcommand{\quOmegbz}{{\overline{\Omega}_0}}
\newcommand{\quOmegb}{{\overline{\Omega}}}
\newcommand{\quO}{O}
\newcommand{\quP}{{\mathcal P}}
\newcommand{\quQS}{{{\mathcal Q}}}
\newcommand{\quQTz}{\quQT_\queta}
\newcommand{\quQT}{{Q}}
\newcommand{\quQ}{{\mathcal Q}}
\newcommand{\quRe}{\hbox{\rm Re}}
\newcommand{\quRicwavsyseq}{\eqref{nsricone} -- \eqref{swave}}
\newcommand{\quRic}{{\hbox{\rm Ric}\,}}
\newcommand{\quRnn}{\quRpos_0}
\newcommand{\quRopset}{U}
\newcommand{\quRpos}{\quR^+}
\newcommand{\quR}{\mathbb{R}}
\newcommand{\quSb}{{\overline{S}}}
\newcommand{\quScalm}{\beta}
\newcommand{\quSigsbft}{\quwlntilde{\quSigsbf}}
\newcommand{\quSigsbf}{{S}}
\newcommand{\quSigsb}{{\Sigma_{\qutsb}}}
\newcommand{\quSigs}{{\Sigma'_\qusigmat}}
\newcommand{\quSigsp}{{\Sigma'_{\qusigmat'}}}
\newcommand{\quSigzt}{\quwlntilde{\Sigma}_{0}}
\newcommand{\quSigzz}{{\Sigma_0^0}}
\newcommand{\quSigz}{{{\Sigma}_{0}}}
\newcommand{\quSu}{U}
\newcommand{\quS}{{{\mathcal S}}}
\newcommand{\quTe}{T}
\newcommand{\quTu}{{\mathcal T}}
\newcommand{\quTy}{T'}
\newcommand{\quUln}{L_U}
\newcommand{\quUtvb}{{U_\qutvb}}
\newcommand{\quUub}{{\underline{U}}}
\newcommand{\quUzt}{\quwlntilde{U}_{0}}
\newcommand{\quUzz}{{U_0^0}}
\newcommand{\quUz}{{{U}_{0}}}
\newcommand{\quW}{{\mathcal W}}
\newcommand{\quXXbl}{\overline{\quXbl}}
\newcommand{\quXbl}{\overline{\Lambda}}
\newcommand{\quZ}{\mathbb{Z}}
\newcommand{\quaaph}{\quaph}
\newcommand{\quabma}{\sqrt{1 + \frac{\frac{1}{\theta_{2, 1}} \int_0^s \frac{1}{a_0}}{\sqrt{\bet^2 + \left(\frac{1}{\theta_{2, 1}} \int_0^s \frac{1}{a_0}\right)^2}}}}
\newcommand{\quabpa}{\sqrt{1 - \frac{\frac{1}{\theta_{2, 1}} \int_0^s \frac{1}{a_0}}{\sqrt{\bet^2 + \left(\frac{1}{\theta_{2, 1}} \int_0^s \frac{1}{a_0}\right)^2}}}}
\newcommand{\quab}{{\overline{a}}}
\newcommand{\quadk}{\qudk'{}}
\newcommand{\quadmterm}{small}
\newcommand{\qualgop}{{{\mathcal A}}}
\newcommand{\quaphdecnormexp}{{\quaphnorm{\quhbar}}}
\newcommand{\quaphdec}{\hbox{\txtgrk m}}
\newcommand{\quaphhbnum}{40\quCKS}
\newcommand{\quaphnorm}[1]{{\vert #1\vert _{\qreflink{quantaphdecnormexp}{\quaphdec}}}}
\newcommand{\quaph}{{{\frakf h}}}
\newcommand{\quappfr}{{f_\quparr}}
\newcommand{\quarccot}{\text{\rm arccot\,}}
\newcommand{\quarccsc}{{\rm arccsc}\,}
\newcommand{\quarcsin}{{\rm arcsin}\,}
\newcommand{\qubC}{\text{\rm \bf C}}
\newcommand{\qubHS}[1]{\left\vert #1\right\vert _{HS}}
\newcommand{\qubK}{{\hbox{\bf K}}}
\newcommand{\qubPhi}{\hbox{\mbf\cbPhi}}
\newcommand{\qubR}{{\hbox{\bf R}}}
\newcommand{\qubS}{\mathbb{S}}
\newcommand{\qubX}{\hbox{\bf X}}
\newcommand{\quba}{\overline{a}}
\newcommand{\qubbt}{\quwlntilde{\qubb}}
\newcommand{\qubbulk}{\widetilde{\qubulki}}
\newcommand{\qubb}{{\overline{b}}}
\newcommand{\qubc}{\overline{c}}
\newcommand{\qubdelta}{\hbox{\mbf\char'016}}
\newcommand{\qubdpol}{C(\queb_k[\qubgam], e_k[h], \queb_k[\qugamma], e_k[\qubh])}
\newcommand{\qubgam}{\overline{\qugamma}}
\newcommand{\qubh}{\overline{h}}
\newcommand{\qubln}{\left\| }
\newcommand{\qubootconst}{18432}
\newcommand{\qubphi}{\hbox{\mbf\cbphi}}
\newcommand{\qubrn}{\right\| }
\newcommand{\qubsighat}[1]{\setbox\hattmpbox=\hbox{$\displaystyle{#1}$}\ifdim\wd\hattmpbox>3em{\left(#1\right)^{\wedge}}\else\qusighat{#1}\fi\setbox\hattmpbox=\hbox{}}
\newcommand{\qubulkF}{{\Gamma_{\rm F}}}
\newcommand{\qubulkGt}{\widetilde{\qubulk}_\quGeod}
\newcommand{\qubulkG}{{\qubulk_\quGeod}}
\newcommand{\qubulkex}{\widetilde{\qubulk}}
\newcommand{\qubulki}{X}
\newcommand{\qubulks}{{\Gamma_\sigma}}
\newcommand{\qubulkz}{{\Gamma_0}}
\newcommand{\qubulk}{{\Gamma}}
\newcommand{\qubu}{\overline{u}}
\newcommand{\qubx}{{\mbff x}}
\newcommand{\qub}{{{\bf b}}}
\newcommand{\qucI}{I^\circ}
\newcommand{\qucbt}{\quwlntilde{\qucb}}
\newcommand{\qucb}{{\overline{c}}}
\newcommand{\quccvf}{{y}}
\newcommand{\qucds}{{s}}
\newcommand{\qucduuz}{\qucduu_0}
\newcommand{\qucduu}{{\underline{\qucdu}}}
\newcommand{\qucdu}{u}
\newcommand{\qucdv}{{v}}
\newcommand{\qucdxz}{\qucdx}
\newcommand{\qucdx}{{x}}
\newcommand{\qucdy}{{\phi}}
\newcommand{\quchihat}{{\widehat{\mathchar"011F}}}
\newcommand{\quconadj}{concentrated}
\newcommand{\quconcrat}{f}
\newcommand{\quconnoun}{concentration}
\newcommand{\qucosize}{c}
\newcommand{\qucvfLI}{{YLI}}
\newcommand{\qucvfL}{{BL}}
\newcommand{\qucvf}{{Y}}
\newcommand{\qudAc}{{{\frakf C}}}
\newcommand{\qudQ}{{\delta Q}}
\newcommand{\qudab}{{\overline{\delta a}}}
\newcommand{\qudaib}{{\overline{\delta^{-1} a}}}
\newcommand{\qudappfr}{\Delta_\quparr}
\newcommand{\qudb}{\dot{b}}
\newcommand{\qudc}{\dot{c}}
\newcommand{\quddo}{o}
\newcommand{\qudelb}{\qudlb}
\newcommand{\qudell}{\dot{\ell}}
\newcommand{\qudelth}{\mathchar"010E}
\newcommand{\qudepsymb}{\message{Warning: using deprecated symbol}{}}
\newcommand{\qudet}{{\rm det\,}}
\newcommand{\qudgambp}{\overline{\Delta}_\quparr}
\newcommand{\qudhi}{\delta^{-1} h}
\newcommand{\qudh}{\delta h}
\newcommand{\qudiffose}{{\qudiffos}}
\newcommand{\qudiffos}{\Delta}
\newcommand{\qudifopsett}{{[\quudifset, \qudifopset]}}
\newcommand{\qudifopset}{{{\mathcal D}}}
\newcommand{\qudifseta}{{{\mathcal X}_2}}
\newcommand{\qudifsetr}{{{\mathcal X}_1}}
\newcommand{\qudifset}{{\mathcal X}}
\newcommand{\qudk}{\delta}
\newcommand{\qudlbt}{\quwlntilde{\qudlb}}
\newcommand{\qudlb}{{\overline{\delta\ell}}}
\newcommand{\qudl}{\delta \ell}
\newcommand{\qudom}{\hbox{\rm dom}\,}
\newcommand{\qudsop}{{{\rus b}}}
\newcommand{\qudthet}{\epsilon}
\newcommand{\qudu}{\dot{\quu}}
\newcommand{\qudx}{\dot{\qux}}
\newcommand{\queTptupp}{1/\sqrt{2}}
\newcommand{\queTpt}{\quwlntilde{\queTp}}
\newcommand{\queTpupd}{\frac{1}{2\sqrt{2}}}
\newcommand{\queTpupp}{1/(2\sqrt{2})}
\newcommand{\queTp}{{T'}}
\newcommand{\queTt}{\quwlntilde{\queT}}
\newcommand{\queT}{{T}}
\newcommand{\queb}{\overline{e}}
\newcommand{\quellb}{{\overline{\ell}}}
\newcommand{\quemph}[1]{{\it #1}}
\newcommand{\quepsb}{{\mathchar"011C}_{*}}
\newcommand{\quepsn}{{\epsilon}}
\newcommand{\quetaus}{\qutau_0}
\newcommand{\queta}{{\mathchar"0111}}
\newcommand{\quez}{E_{k_0}}
\newcommand{\qufX}{{\frakf W}}
\newcommand{\qufbHS}[1]{\bigl\vert #1\bigr\vert }
\newcommand{\qufbar}{{\overline{f}}}
\newcommand{\qufb}{{\overline{f}}}
\newcommand{\qufs}{f_*}
\newcommand{\quft}{{\tilde{f}}}
\newcommand{\qugambGN}{{f_\quparr^{0, \quupN}}}
\newcommand{\qugambGNf}{{f_\quparr^\quupN}}
\newcommand{\qugambt}{\quwlntilde{\qugamb}}
\newcommand{\qugamb}{{{\overline{\qugamma}}}}
\newcommand{\qugamma}{{\mathchar"010D}}
\newcommand{\qugbabcseq}{\hbox{(\the\gbsyseqno--\advance\gbsyseqno by 2 \the\gbsyseqno)\advance\gbsyseqno by -2}}
\newcommand{\qugbeqnoct}{\gbeqno{\usecount{\gbeqcount}--\usecount{\gbeqcount}}}
\newcommand{\qugbeqnoc}{\gbeqno{\usecount{\gbeqcount}}}
\newcommand{\qugbgamseq}{\hbox{\advance\gbsyseqno by 3 (\the\gbsyseqno)\advance\gbsyseqno by -3}}
\newcommand{\qugblasteq}{\hbox{\advance\gbeqcount by -1(\the\gbeqcount)\advance\gbeqcount by 1}\relax}
\newcommand{\qugbo}{\varepsilon}
\newcommand{\qugbsyseqs}{\hbox{(\the\gbsyseqno--\advance\gbsyseqno by 3 \the\gbsyseqno)\advance\gbsyseqno by -3}}
\newcommand{\qugcvf}{X}
\newcommand{\quginit}{{\varpi}}
\newcommand{\qugss}{{\partial^2_\qucds \qugamma}}
\newcommand{\qugsv}{{\partial_\qucds \partial_\qucdv \qugamma}}
\newcommand{\qugsx}{{\partial_\qucds \partial_\qucdx \qugamma}}
\newcommand{\qugxx}{{\partial^2_\qucdx \qugamma}}
\newcommand{\quhL}{{\mathcal F}}
\newcommand{\quhSigmaz}{{\Sigma_0^{\frac{1}{2}}}}
\newcommand{\quhXb}{{\underline{\quhX}}}
\newcommand{\quhXz}{{{\hbox{\txtgrk D}}}}
\newcommand{\quhX}{{\widehat{X}}}
\newcommand{\quhbar}{{\overline{h}}}
\newcommand{\quhb}{\overline{h}}
\newcommand{\quhgo}{\hat{\qugamma}_1}
\newcommand{\quhgt}{\hat{\qugamma}_2}
\newcommand{\quhh}{\hat{h}}
\newcommand{\quhib}{\overline{h^{-1}}}
\newcommand{\quhinm}{\left[ \matrix{\frac{b^2}{a} - c & \frac{b}{a} & -1\cr\frac{b}{a}&\frac{1}{a}&0\cr -1&0&0\cr}\right]}
\newcommand{\quhl}{f}
\newcommand{\quhm}{\left[\matrix{0&0&-1\cr 0&a&b\cr -1&b&c\cr}\right]}
\newcommand{\quhtinm}{-\frac{1}{a}\left[\matrix{a\left(1 + \frac{1}{2}c\right) - \frac{1}{2} b^2 & -\frac{1}{\sqrt{2}} b & -\frac{1}{2} b^2 + \frac{1}{2} ac\cr -\frac{1}{\sqrt{2}} b & -1 & -\frac{1}{\sqrt{2}} b\cr -\frac{1}{2} b^2 + \frac{1}{2} ac & -\frac{1}{\sqrt{2}} b & a\left(-1 + \frac{1}{2}c\right) - \frac{1}{2}b^2\cr}\right]}
\newcommand{\quhtm}{\left[\matrix{ -1 + \frac{1}{2} c & \frac{1}{\sqrt{2}} b & -\frac{1}{2} c\cr \frac{1}{\sqrt{2}} b & a & -\frac{1}{\sqrt{2}} b\cr -\frac{1}{2} c & - \frac{1}{\sqrt{2}} b & 1 + \frac{1}{2} c\cr}\right]}
\newcommand{\quhug}{\hat{\underline{\qugamma}}}
\newcommand{\quhvb}{1}
\newcommand{\quiLline}{{\mathcal L}}
\newcommand{\quib}{\quIb}
\newcommand{\quidop}{{{\mathcal I}}}
\newcommand{\quido}{\hbox{\bf 1}}
\newcommand{\quigBz}{\quigB_0}
\newcommand{\quigB}{{{\mathcal B}}}
\newcommand{\quigSz}{\quigS_0}
\newcommand{\quigS}{{{\mathcal V}}}
\newcommand{\quigVz}{\quigV_0}
\newcommand{\quigV}{{{\mathcal S}}}
\newcommand{\quimage}{{\hbox{\rm Im}\,}}
\newcommand{\quim}{\hbox{im}\,}
\newcommand{\quinM}{{{\mathcal M}}}
\newcommand{\quinN}{{{\mathcal N}}}
\newcommand{\quinP}{{\mathcal P}}
\newcommand{\quinR}{{{\mathcal R}}}
\newcommand{\quindset}{{\mathcal I}}
\newcommand{\quingeod}{{\ell}}
\newcommand{\quing}{\qumetg}
\newcommand{\quinh}{{{{}_0\!h}}}
\newcommand{\quinin}{{\quSigz}}
\newcommand{\quinout}{{\quUz}}
\newcommand{\quinsA}{{\bf A}}
\newcommand{\quintrvlp}{J}
\newcommand{\quintrvlz}{\quintrvl_0}
\newcommand{\quintrvl}{I}
\newcommand{\quiny}{\qubx}
\newcommand{\quiotan}{{i}}
\newcommand{\quiota}{{\mathchar"0113}}
\newcommand{\quirad}{\qucdy}
\newcommand{\qui}{{\frakf i}}
\newcommand{\qulambdabb}{{\overline\qulambdab}}
\newcommand{\qulambdab}{{\overline\lambda}}
\newcommand{\qulb}{\quellb}
\newcommand{\qulee}{{q}}
\newcommand{\qulesssim}{\mathrel\mathchar"3F2E}
\newcommand{\qugreatsim}{\mathrel\mathchar"3F2F}
\newcommand{\qules}{{r}}
\newcommand{\quley}{{p}}
\newcommand{\bley}{{\mathbf \quley}}
\newcommand{\blee}{{\mathbf \qulee}}
\newcommand{\bles}{{\mathbf \qules}}
\newcommand{\bupey}{{\mathbf \quupey}}
\newcommand{\qulnnm}[1]{\|  #1\|  }
\newcommand{\qulntilde}[1]{\qreflink{quanttildop}{\tilde{#1}}}
\newcommand{\qulog}{\text{\rm log}\,}
\newcommand{\qultAsigmat}{A_{\qusigmat}^0}
\newcommand{\qultbulk}{\qubulk^0}
\newcommand{\qumathbb}[1]{\mathbb{#1}}
\newcommand{\qumddo}{{o_M}}
\newcommand{\qumetaz}{{{}_0\!\qumeta}}
\newcommand{\qumeta}{{a}}
\newcommand{\qumetb}{{b}}
\newcommand{\qumetc}{{c}}
\newcommand{\qumetell}{{\ell}}
\newcommand{\qumetg}{{g}}
\newcommand{\qumeth}{{h}}
\newcommand{\qumff}{{J}}
\newcommand{\qumubb}{{\overline{\qumub}}}
\newcommand{\qumub}{{\overline{\mu}}}
\newcommand{\qunL}{{L}}
\newcommand{\qunabla}{{{\mathchar"0272}}}
\newcommand{\qunatchar}{\chi}
\newcommand{\qunullxi}{{\quxb}}
\newcommand{\qunullzeta}{{\mathchar"0110}}
\newcommand{\qunulnorS}{\qunulnor}
\newcommand{\qunulnorU}{{\underline{\qunulnor}}}
\newcommand{\qunulnor}{{\bf N}}
\newcommand{\qunut}{\quwlntilde{\qunu}}
\newcommand{\qunu}{{\mathchar"0117}}
\newcommand{\qunwUs}[3]{\|  #1\|  _{\qusbW_\quxb^{#2}\qusbW_\quvb^{#3}(\quUz\vert \quXXbl_\qutxb)}}
\newcommand{\qunwU}[4]{\|  #1\|  _{\qusbW_\qusb^{#2}\qusbW_\quxb^{#3}\qusbW_\quvb^{#4}(\quUz\vert \quXXbl_\qutxb)}}
\newcommand{\qunws}[3]{\|  #1\|  _{\qusbW_\quxb^{#2}\qusbW_\quvb^{#3}(\quSigz\vert \quXbl_\qutxb)}}
\newcommand{\qunw}[4]{\|  #1\|  _{\qusbW_\qusb^{#2}\qusbW_\quxb^{#3}\qusbW_\quvb^{#4}(\quSigz\vert \quXbl_\qutxb)}}
\newcommand{\quoCS}{(4/3)^{1/2}}
\newcommand{\quoC}{{\overline{C}}}
\newcommand{\quoH}{{H_\circ\!\!\!}}
\newcommand{\quoN}{{\mathcal N}}
\newcommand{\quoeps}{{\overline{\epsilon}}}
\newcommand{\quomegbt}{\quwlntilde{\quomegb}}
\newcommand{\quomegb}{{\overline{\omega}}}
\newcommand{\quopsset}{{\mathcal \quops}}
\newcommand{\quops}{Y}
\newcommand{\quosize}{{\alpha}}
\newcommand{\qupark}{{k}}
\newcommand{\quparr}{\rho}
\newcommand{\qupass}{{\partial^2_\qucds \qumeta}}
\newcommand{\qupasv}{{\partial_v \partial_\qucds \qumeta}}
\newcommand{\qupasx}{{\partial_\qucdx \partial_\qucds \qumeta}}
\newcommand{\qupas}{{\partial_\qucds \qumeta}}
\newcommand{\qupavs}{{\partial_\qucdv \partial_\qucds \qumeta}}
\newcommand{\qupavv}{{\partial^2_\qucdv \qumeta}}
\newcommand{\qupavx}{{\partial_\qucdv \partial_\qucdx \qumeta}}
\newcommand{\qupav}{{\partial_\qucdv \qumeta}}
\newcommand{\qupaxs}{{\partial_\qucdx \partial_\qucds \qumeta}}
\newcommand{\qupaxx}{{\partial^2_\qucdx \qumeta}}
\newcommand{\qupax}{{\partial_\qucdx \qumeta}}
\newcommand{\qupbH}{{H}_{{\qudifset}}}
\newcommand{\qupbar}{{{\underline{\partial}}}}
\newcommand{\qupbass}{{\partial_{\qusb} \partial_{\qusb} \quab}}
\newcommand{\qupbasv}{{\partial_{\quvb} \partial_{\qusb} \quab}}
\newcommand{\qupbasx}{{\partial_{\quxb} \partial_{\qusb} \quab}}
\newcommand{\qupbas}{{\partial_\qusb \quab}}
\newcommand{\qupbavs}{{\partial_{\quvb} \partial_{\qusb} \quab}}
\newcommand{\qupbavv}{{\partial_{\quvb} \partial_{\quvb} \quab}}
\newcommand{\qupbavx}{{\partial_{\quvb} \partial_{\quxb} \quab}}
\newcommand{\qupbav}{{\partial_\quvb \quab}}
\newcommand{\qupbaxs}{{\partial_{\quxb} \partial_{\qusb} \quab}}
\newcommand{\qupbaxx}{{\partial_{\quxb} \partial_{\quxb} \quab}}
\newcommand{\qupbax}{{\partial_\quxb \quab}}
\newcommand{\qupbbsv}{{\partial_{\quvb} \partial_{\qusb} \qubb}}
\newcommand{\qupbbsx}{{\partial_{\quxb} \partial_{\qusb} \qubb}}
\newcommand{\qupbbs}{{\partial_{\qusb} \qubb}}
\newcommand{\qupbbvs}{{\partial_{\quvb} \partial_{\qusb} \qubb}}
\newcommand{\qupbbv}{{\partial_{\quvb} \qubb}}
\newcommand{\qupbbxs}{{\partial_{\quxb} \partial_{\qusb} \qubb}}
\newcommand{\qupbbx}{{\partial_{\quxb} \qubb}}
\newcommand{\qupbb}{{\overline{p}}}
\newcommand{\qupbcsx}{{\partial_{\quxb} \partial_{\qusb} \qucb}}
\newcommand{\qupbcs}{{\partial_{\qusb} \qucb}}
\newcommand{\qupbcv}{{\partial_{\quvb} \qucb}}
\newcommand{\qupbcx}{{\partial_{\quxb} \qucb}}
\newcommand{\qupbdlsv}{{\partial_\quvb \partial_\qusb \qudelb}}
\newcommand{\qupbdls}{{\partial_{\qusb} \qudelb}}
\newcommand{\qupbdlvv}{{\partial_\quvb \partial_\quvb \qudelb}}
\newcommand{\qupbdlv}{{\partial_{\quvb} \qudelb}}
\newcommand{\qupbgs}{{\partial_{\qusb} \qugamb}}
\newcommand{\qupbgvnl}{\left\| \qupbgv\right\| _{L^\infty(\quhSigmaz)}}
\newcommand{\qupbgvnp}{\left\| \qupbgv\right\| _{W^{m + 1, \infty}(\quhSigmaz)}}
\newcommand{\qupbgvn}{\left\| \qupbgv\right\| _{W^{m, \infty}(\quhSigmaz)}}
\newcommand{\qupbgv}{{\partial_{\quvb} \qugamb}}
\newcommand{\qupbgxnl}{\left\| \qupbgx\right\| _{L^\infty(\quhSigmaz)}}
\newcommand{\qupbgxnp}{\left\| \qupbgx\right\| _{W^{m + 1, \infty}(\quhSigmaz)}}
\newcommand{\qupbgxn}{\left\| \qupbgx\right\| _{W^{m, \infty}(\quhSigmaz)}}
\newcommand{\qupbgx}{{\partial_{\quxb} \qugamb}}
\newcommand{\qupblss}{{\partial_{\qusb} \partial_{\qusb} \qulb}}
\newcommand{\qupblsv}{{\partial_{\quvb} \partial_{\qusb} \qulb}}
\newcommand{\qupblsx}{{\partial_{\quxb} \partial_{\qusb} \qulb}}
\newcommand{\qupbls}{{\partial_\qusb \qulb}}
\newcommand{\qupblvs}{{\partial_{\quvb} \partial_{\qusb} \qulb}}
\newcommand{\qupblvv}{{\partial_{\quvb} \partial_{\quvb} \qulb}}
\newcommand{\qupblvx}{{\partial_{\quvb} \partial_{\quxb} \qulb}}
\newcommand{\qupblv}{{\partial_{\quvb} \qulb}}
\newcommand{\qupblxs}{{\partial_{\quxb} \partial_{\qusb} \qulb}}
\newcommand{\qupblxx}{{\partial_{\quxb} \partial_{\quxb} \qulb}}
\newcommand{\qupblx}{{\partial_{\quxb} \qulb}}
\newcommand{\qupbss}{{\partial^2_\qucds \qumetb}}
\newcommand{\qupbsv}{{\partial_v \partial_\qucds \qumetb}}
\newcommand{\qupbsx}{{\partial_\qucdx \partial_\qucds \qumetb}}
\newcommand{\qupbs}{{\partial_{\qucds} \qumetb}}
\newcommand{\qupbvs}{{\partial_\qucdv \partial_\qucds \qumetb}}
\newcommand{\qupbvx}{{\partial_\qucdv \partial_\qucdx \qumetb}}
\newcommand{\qupbv}{{\partial_{\qucdv} \qumetb}}
\newcommand{\qupbxs}{{\partial_\qucdx \partial_\qucds \qumetb}}
\newcommand{\qupbx}{{\partial_{\qucdx} \qumetb}}
\newcommand{\qupcss}{{\partial^2_\qucds \qumetc}}
\newcommand{\qupcsv}{{\partial_v \partial_\qucds \qumetc}}
\newcommand{\qupcsx}{{\partial_\qucdx \partial_\qucds \qumetc}}
\newcommand{\qupcs}{{\partial_{\qucds} \qumetc}}
\newcommand{\qupcvs}{{\partial_\qucdv \partial_\qucds \qumetc}}
\newcommand{\qupcv}{{\partial_{\qucdv} \qumetc}}
\newcommand{\qupcxs}{{\partial_\qucdx \partial_\qucds \qumetc}}
\newcommand{\qupcxx}{{\partial^2_\qucdx \qumetc}}
\newcommand{\qupcx}{{\partial_{\qucdx} \qumetc}}
\newcommand{\qupdiv}{\,\vert \,}
\newcommand{\qupgam}{\partial\qugamma}
\newcommand{\qupgi}{{\partial_{i} \qugamma}}
\newcommand{\qupgj}{{\partial_{j} \qugamma}}
\newcommand{\qupgss}{{\partial^2_\qucds \qugamma}}
\newcommand{\qupgsv}{{\partial_\qucds \partial_\qucdv \qugamma}}
\newcommand{\qupgsx}{{\partial_\qucds \partial_\qucdx \qugamma}}
\newcommand{\qupgs}{{\partial_{\qucds} \qugamma}}
\newcommand{\qupgv}{{\partial_{\qucdv} \qugamma}}
\newcommand{\qupgxx}{{\partial^2_\qucdx \qugamma}}
\newcommand{\qupgx}{{\partial_{\qucdx} \qugamma}}
\newcommand{\qupspig}{\partial_s^\ell \partial^I \left(\qugamma - \qugamma_M\right)}
\newcommand{\quqexpnum}[2]{\noexpand\noexpand\noexpand\quEEE#2}
\newcommand{\ququad}{\kern 4pt}
\newcommand{\qursc}{\hat{r}}
\newcommand{\qurz}{r_0}
\newcommand{\qusbH}{{H}}
\newcommand{\qusbW}{{W}}
\newcommand{\qusbn}{{n}}
\newcommand{\qusbt}{{\quwlntilde{\qusb}}}
\newcommand{\qusb}{{\overline{s}}}
\newcommand{\quscalnt}{\mu}
\newcommand{\quscaln}{{\overline{n}}}
\newcommand{\qusech}{\text{\rm sech\,}}
\newcommand{\qusighat}[1]{\widehat{#1}}
\newcommand{\qusigmat}{\sigma}
\newcommand{\qusigtu}{\qusighat{\qusigt}}
\newcommand{\qusigt}{{\qulntilde{\qusigmat}}}
\newcommand{\qusigu}{{\qusighat{\sigma}}}
\newcommand{\qusin}{\text{\rm sin}\,}
\newcommand{\quspX}{\quR^1 \times [\quKSk, \quKSk + \qudk]}
\newcommand{\qusupp}{\text{\rm supp}\,}
\newcommand{\qusup}{\mathop{\text{\rm sup}\,}\limits}
\newcommand{\qutAsigmat}{\qutA_\qusigmat}
\newcommand{\qutA}{{\bf A}}
\newcommand{\qutT}{\tilde{t}}
\newcommand{\qutaut}{{\quwlntilde{\qutau}}}
\newcommand{\qutauu}{{\qusighat{\mathchar"011C}}}
\newcommand{\qutau}{{\mathchar"011C}}
\newcommand{\qutcds}{{\omathbf{s}}}
\newcommand{\qutcdv}{{\omathbf{v}}}
\newcommand{\qutdhi}{{\overline{\delta h^{-1}}}}
\newcommand{\qutdh}{{\overline{\delta h}}}
\newcommand{\quthetab}{{\overline\theta}}
\newcommand{\quthetaz}{\theta_0}
\newcommand{\quthetsc}{\hat{\theta}}
\newcommand{\quthi}{{\quth^{-1}}}
\newcommand{\quth}{\quhbar}
\newcommand{\qutildop}{{\widetilde{\cdot}}}
\newcommand{\qutimvec}{{\bf T}}
\newcommand{\qutpi}{\overline{\pi}}
\newcommand{\qutri}{{\rm T}}
\newcommand{\qutr}{\text{\rm tr}\,}
\newcommand{\qutsbs}{{\qutsb^*}}
\newcommand{\qutsbt}{{\quwlntilde{\qutsb}}}
\newcommand{\qutsbzp}{\omathbf{s}_0}
\newcommand{\qutsbz}{{\qutsbzp}}
\newcommand{\qutsb}{{{\omathbf{s}}}}
\newcommand{\quttauu}{{\qusighat{\quttau}}}
\newcommand{\quttau}{{{\hbox{\mbf\char'034}}}}
\newcommand{\quttg}{\tilde{\tilde{\qugamma}}}
\newcommand{\qutvbss}{{\qutvb^{**}}}
\newcommand{\qutvbs}{{\qutvb^*}}
\newcommand{\qutvbu}{{\qusighat{\qutvb}}}
\newcommand{\qutvbz}{{\qutvb_0}}
\newcommand{\qutvb}{{{\omathbf{v}}}}
\newcommand{\qutvbpu}{{\widehat{\qutvb'}}}
\newcommand{\qutxbs}{{\qutxb^*}}
\newcommand{\qutxb}{{{\omathbf{x}}}}
\newcommand{\qutx}{\tilde{x}}
\newcommand{\qutz}{\tilde{z}}
\newcommand{\quuA}{\uvarb{A}}
\newcommand{\quuCnmi}{{\overline{C}}^{{\rm M}}}
\newcommand{\quuC}{{\underline{C}}}
\newcommand{\quuDelta}{\uvarb\Delta}
\newcommand{\quuGamma}{\uvarb\Gamma}
\newcommand{\quuJ}{{\overline{J}}}
\newcommand{\quuK}{{\overline{K}}}
\newcommand{\quuL}{{\overline{L}}}
\newcommand{\quuSigma}{\uvarb\Sigma}
\newcommand{\quuT}{{\underline{T}}}
\newcommand{\quuU}{\uvarb{U}}
\newcommand{\quub}{\quuu}
\newcommand{\quudifseta}{{{\mathcal C}_2}}
\newcommand{\quudifsetr}{{{\mathcal C}_1}}
\newcommand{\quudifsetz}{\quudifset^0}
\newcommand{\quudifset}{{{\mathcal C}}}
\newcommand{\quulee}{{{\overline{\qulee}}}}
\newcommand{\quules}{{{\overline{\qules}}}}
\newcommand{\quuley}{{{\overline{\quley}}}}
\newcommand{\quumu}{\uvarb\qumub}
\newcommand{\quupM}{{M}}
\newcommand{\quupN}{{N}}
\newcommand{\quupP}{{{\overline M}}}
\newcommand{\quupee}{{Q}}
\newcommand{\quupes}{{R}}
\newcommand{\quupey}{{P}}
\newcommand{\quupeep}{{\quupee + 3\quupey - 2}}
\newcommand{\quupeepppnp}{{\quupee + 3\quupey + 2\quupes - 2}}
\newcommand{\quupeepp}{{\quupee' + 3\quupey - 2}}
\newcommand{\quupeeppp}{{\quupee' + 3\quupey + 2\quupes - 2}}
\newcommand{\quupesp}{{\quupes}}
\newcommand{\quuprtl}{{D}}
\newcommand{\quusig}{{\underline{\sigma}}}
\newcommand{\quupsb}{{\underline{\mathchar"011D}}}
\newcommand{\quupsilon}{{\mathchar"011D}}
\newcommand{\quupsu}{{\qusighat{\mathchar"011D}}}
\newcommand{\quur}{\underline{r}}
\newcommand{\quuu}{{\underline{u}}}
\newcommand{\quu}{{\bf u}}
\newcommand{\quvbt}{{\quwlntilde{\quvb}}}
\newcommand{\quvbz}{{\quvb_0}}
\newcommand{\quvb}{{\overline{v}}}
\newcommand{\quvf}{{\mu}}
\newcommand{\quvolnorm}{\frac{\sqrt{\vert \widehat{\quth}\vert }}{\left(1 - \frac{\qucb}{2k}\right)^{1/2}}}
\newcommand{\quwab}{\quab}
\newcommand{\quwbb}{\qubb}
\newcommand{\quwbox}{{{\mathchar"0F03}}}
\newcommand{\quwcb}{\qucb}
\newcommand{\quwdab}{\qudab}
\newcommand{\quwdaib}{\qudaib}
\newcommand{\quwdlb}{\qudlb}
\newcommand{\quwdub}{{\widetilde{\overline{\delta\quupsilon}}}}
\newcommand{\quwlnsubtilde}[1]{\qreflink{quanttildop}{\widetilde{\csname kernquant#1\endcsname}}_{\csname indquant#1\endcsname}}
\newcommand{\quwlntilde}[1]{\qreflink{quanttildop}{\widetilde{#1}}}
\newcommand{\quwomegb}{{\widetilde{\quomegb}}}
\newcommand{\quwsqueezed}{concentrated}
\newcommand{\quwtbulk}{\quwlntilde{\qubulk}}
\newcommand{\quwups}{{\widetilde{\quupsilon}}}
\newcommand{\quxbt}{{\quwlntilde{\quxb}}}
\newcommand{\quxb}{{\overline{x}}}
\newcommand{\quxi}{\quxb}
\newcommand{\qux}{{{\bf x}}}
\newcommand{\quzEb}{{{}^0\!\quEb}}
\newcommand{\quzE}{{{}^0\!E}}
\newcommand{\quzIb}{{{}^0\!\overline{I}}}
\newcommand{\quzbI}{{{}^0\!\underline{I}}}
\newcommand{\quzbcI}{{{}^0\!\underline{I}^\circ}}
\newcommand{\quzbiI}{{{}^0\!\underline{I}^1}}
\newcommand{\quzb}{{\overline{z}}}
\newcommand{\quzcI}{{{}^0\!I^\circ}}
\newcommand{\quzeps}{{{}^0\!\epsilon}}
\newcommand{\quzeta}{{\mathchar"0110}}
\newcommand{\quzg}{{}^0 \qugamma}
\newcommand{\quzh}{{}^0 h}
\newcommand{\quzib}{{{}^0\!\overline{\quiota}}}
\newcommand{\quziob}{{{}^0\!\underline{\quiota}}}
\newcommand{\quziota}{{{}^0\!\quiota}}
\newcommand{\tphi}{{\widetilde{\phi}}}
\newcommand{\plC}{{\overline{C}}}
\newcommand{\lipschjc}{{\frak j}}
\newcommand{\vecx}{{\hbox{\bf x}}}
\newcommand{\vecxp}{{\vecx'}}
\newcommand{\vecphi}{{\phi_i}}
\newcommand{\vectphi}{{\tphi_i}}
\newcommand{\FlCb}{{\overline{C}}}
\newcommand{\FlCbb}{{\FlCb'}}
\newcommand{\FlFdom}{{X}}
\newcommand{\FlphiC}{{C_\phi}}
\newcommand{\FlphiCp}{{C'_\phi}}
\newcommand{\Flnumphi}{N}
\newcommand{\FlFexp}{p}
\newcommand{\usigf}{\lambda}
\newcommand{\Isn}{I}
\newcommand{\Ldeg}{r}
\newcommand{\qOmegbC}{C}
\newcommand{\hbmdC}{C_{\rm h}}
\newcommand{\qubmu}{\quwbox_{\quhbar_\mu}}
\newcommand{\qubmup}{\quwbox_{\quhbar_{\mu + 1}}}
\newcommand{\qubeta}{\quwbox_\eta}
\newcommand{\gambM}{M}
\newcommand{\gambMdiff}{\overline{\gambM}}
\newcommand{\nullf}{Y}
\newcommand{\nullfp}{\upsilon}
\newcommand{\nullfpm}{\upsilon_0}
\newcommand{\qunulfS}{\omega}
\newcommand{\qusbm}{m}
\newcommand{\quappfri}{\quappfr^0}
\newcommand{\quIi}{\quI_0}
\newcommand{\gord}[3]{{\left\{ \begin{matrix}#1\\#2\\#3\end{matrix}\right\}}}
\newcommand{\bne}{{\bf -1}}
\newcommand{\dxp}[1]{{\delta(#1)}}
\newcommand{\mxp}[1]{{N(#1)}}
\newcommand{\nLC}{{C^{\mathrm L}}}
\newcommand{\qudlbp}{{\qudlb'}}
\newcommand{\qubbp}{{\qubb'}}
\newcommand{\qucbp}{{\qucb'}}
\newcommand{\qugambp}{{\qugamb'}}
\newcommand{\quomegbp}{{\quomegb'}}
\newcommand{\qunup}{{\qunu'}}
\newcommand{\qusigmatp}{{\sigma'}}
\newcommand{\qusigmatz}{{\sigma_0}}
\newcommand{\enerCp}{C'}
\newbox\hattmpbox
\par\noindent
\section{INTRODUCTION}\label{introduction}
\par
\subsection{Overview}\label{iover} This paper addresses the challenge of constructing solutions to the vacuum Einstein equations which possess a fixed amount of gravitational energy concentrated on an arbitrarily small region. We consider this question for spacetimes possessing polarized $U(1)$ symmetry.
\par
Specifically, we consider $3 + 1$-dimensional spacetimes $(\quinM^{3 + 1}, \quing^{3 + 1})$ which admit an orthogonal splitting: topologically, $\quinM^{3 + 1} = \quinN^{2 + 1} \times \qubS^1$, while for $\quiny$ and $\quirad$ coordinates on $\quinN^{2 + 1}$ and $\qubS^1$, respectively, the metric $\quing^{3 + 1}$ is of the form
\begin{equation}\label{ingexp}
\quing^{3 + 1}(\quiny, \quirad) = e^{-2\qugamma(\quiny)} \qumeth^{2 + 1}_{ij}(\quiny)\,d\quiny^i \otimes d\quiny^j + e^{2\qugamma(\quiny)} d\quirad \otimes d\quirad
\end{equation}
for a Lorentzian metric $\qumeth^{2 + 1}$ on $\quinN^{2 + 1}$ and scalar function $\qugamma : \quinN^{2 + 1} \rightarrow \quR^1$. We abbreviate $\quinN^{2 + 1}$ to $\quinN$. As is well known (see \cite{choquetbruhat}, 13.11.1, \cite{huneauluk}, 2.1), for metrics of this form, the Einstein vacuum equations reduce to the system
\begin{equation}\label{intRicwav}
\quRic(\qumeth) = 2\qunabla\qugamma\otimes\qunabla\qugamma,\qquad \quwbox_\qumeth \qugamma = 0
\end{equation}
(note that these are the Einstein equations coupled to a massless free scalar field in $2 + 1$ dimensions), where $\quwbox_\qumeth$ is the wave (Laplace-Beltrami) operator of the $2 + 1$ metric $\qumeth$. We shall solve this system given characteristic initial data for $\qugamma$.
\par
We construct our spacetimes in a null geodesic gauge, foliating the $2 + 1$ spacetime region $\quinN$ by null geodesics determined by two parameters $\qucdx$, $\qucdv$ and equipped with an affine null parameter $\qucds$ in such a way that $\qucds$, $\qucdx$, $\qucdv$ is a coordinate system on $\quinN$. $\qucdx$ is chosen to be spacelike, while $\qucdv$ is null on the initial outgoing\footnote{Because we work with rectangular, rather than spherical, spatial sections, there is no natural notion of `incoming' or `outgoing' null hypersurfaces. For consistency with the terminology used in \cite{christodoulou}, \cite{klainrod}, etc., we use `outgoing' to refer to the null hypersurface on which the highly-concentrated initial data is supported, and `incoming' for the transverse null hypersurface. In $\quinN^{2 + 1}$ these will be $\quinin = \{ \qucds = 0 \}$ and $\quinout = \{ \qucdv = 0 \}$, respectively.} null hypersurface, $\quinin = \{ \qucds = 0 \}$. We will utilize the fact that, with respect to this coordinate system and assuming the wave equation $\quwbox_\qumeth \qugamma = 0$ holds, the system $\quRic(\qumeth) = 2\qunabla\qugamma\otimes\qunabla\qugamma$ is equivalent to a system of three {\it evolution equations\/} on $\quinN$, which are ODEs in $\qucds$, together with three {\it constraint equations\/} on $\quinin = \{ \qucds = 0 \}$, which are ODEs in $\qucdv$ and constitute constraints on the initial data for the evolution equations. The derivatives of $\qugamma$ appear in all six of these equations as source terms or coefficients. The resultant implicit self-coupling of $\qugamma$ in the wave equation $\quwbox_\qumeth \qugamma = 0$ is found to be effectively {\it cubic}.
\par 
Our work in this paper is connected to three interrelated challenges. The first is to construct a 1-parameter family of solutions to the Einstein vacuum equations over a fixed ``box'' ($\Gamma(\queT, \queTp)$ below), where the initial data has a fixed amount of gravitational energy yet is supported over a region $\quinR_\qupark\subset \quinN^{2+1}$ whose area approaches zero, as the parameter $\qupark$ tends to infinity. The ``squeezed'' support property of the solutions can be seen as a confinement of the gravitational waves in physical space. This confinement occurs in \emph{two} directions in the (3+1)-spacetime, but not in the third one. The second is to construct gravitational waves (solutions to the Einstein vacuum equations) with as high an amplitude as possible; in fact we seek to relate the amplitude of the waves to the size of their support. The third challenge is related to Thorne's hoop conjecture \cite{thornei}; this posits, roughly, ``Horizons form when and only when a mass $m$ gets compacted into a region whose circumference in {\it every\/} direction is $C \le 4\pi m$''. Viewed from one angle, the result here can be seen as evidence in favour of the ``only when'' direction of the hoop conjecture, in that we expect that for arbitrarily large values of the parameter $\qupark$, the spacetimes we construct contain no horizons (trapped surfaces), irrespective of the amount of squeezing. 
\par
In order to compare our results to those in the literature, we note the following characterization of the size of our initial data. The initial data for the wave $\qugamma$ on the characteristic surface $\quinin$ will be of the form
\begin{equation}\label{intgammaform}
\qugamma(0, \qucdx, \qucdv) = \qupark^{-\quiota} \quginit(\qupark^{1/2} \qucdx, \qupark \qucdv),
\end{equation}
where $\quginit$ is bounded (in $L^\infty$ and various relevant Sobolev spaces) independently of $\qupark$. Consider the full $3 + 1$ spacetime $\quinM^{3 + 1}$. As is common in general relativity, the size of our initial data can be probed via the shear $\quchihat$ of the initial null surface $\{ \qucds = 0 \} \times \qubS^1$ in $\quinM^{3 + 1}$, on which our incoming data is supported.\footnote{Had the initial data null hypersurface been located at past null infinity, the quantity $\int \vert \quchihat\vert ^2$ -- in a suitable limit -- would precisely correspond to the incoming gravitational energy.} On $\quinin = \{ \qucds = 0 \}$, the support of our initial data (the rectangle $\quinR^\qupark$ introduced above) is of dimensions $[0, \qupark^{-1/2}] \times [0, \qupark^{-1}]$. Moreover, for values of the scaling exponent $\iota \in [1/4, 1)$, the shear of $\quinin \times \qubS^1$ is dominated by $\partial_\qucdv \qugamma$ on $\quinR^\qupark \times \qubS^1$, while off this set, it vanishes.
By \eqref{intgammaform}, we obtain on $\quinR^\qupark \times \qubS^1$, pointwisely (recall $\iota \in [1/4, 1)$),
\begin{equation}\label{intchihatinf}
\vert \quchihat\vert  \sim \qupark^{1 - \quiota}
\end{equation}
and in $L^2$ (again, $\iota \in [1/4, 1)$),
\begin{equation}\label{iichihatbound}
\int \vert \quchihat\vert ^2 \sim \qupark^{\frac{1}{2} - 2\quiota}.
\end{equation}
In particular, it can be shown that this `total incoming gravitational energy' is uniformly positive (independently of $\qupark$) for the key value $\iota = 1/4$ (though not in general for any larger value of $\iota$). This is the sense in which our solutions represent uniformly positive incoming gravitational energy supported on an arbitrarily small region.
\par
(In terms of more conventional Sobolev bounds, we note that, for $\iota = 1/4$ and any $\epsilon > 0$, the initial data $\qugamma|_{\quinin}$ is, in $\qusbH^{1 + \epsilon}$, of size $\qupark^\epsilon$, and hence {\it unbounded\/} in the limit $\qupark \rightarrow \infty$.)
\par
\subsection{Related works}\label{relworks}
We now wish to situate this work in the context of the literature, especially that on solutions to the Einstein equations that display a {\it concentration\/} of the incoming data in at least one direction.
\par
The starting point of the many studies in this circle of questions is the landmark work of Christodoulou \cite{christodoulou}. The initial data used in \cite{christodoulou} can be described as follows.\footnote{It is also possible, following \cite{klainrod}, to describe the initial data as being given on a pair of intersecting null cones. Such a description would suggest analogies to the present work which are misleading.} Consider a single null cone originating from a point $o$. The initial data on this cone are Minkowskian up to some null parameter value $\qucduuz$, followed by large data imposed over the short interval $[\qucduuz, \qucduuz + \delta]$, where $\delta$ corresponds in our setting to $\qupark^{-1}$. In this context, `large' means in particular that the shear satisfies, on this `short-pulse' interval,
\begin{equation}\label{chiichihatbound}
\int_{\qucduuz}^{\qucduuz + \delta} \vert \quchihat\vert ^2(\qucduu, \vartheta)\,d\qucduu \geq M > 0
\end{equation}
{\it uniformly in\/} $\vartheta \in \qubS^2$, for some fixed $M$, independent of $\delta$.
In particular, this can be interpreted as saying that a fixed amount of gravitational energy enters the spacetime over an arbitrarily short time interval. The landmark work in \cite{christodoulou} shows that solutions corresponding to such initial data exist for a time independent of $\delta$, and develop trapped surfaces. In particular, \cite{christodoulou} was able to demonstrate (outside any special symmetry setting) the formation of trapped surfaces {\it in evolution\/} for the vacuum Einstein equations, from initial data where no such trapped surfaces are present.
\par
While \eqref{chiichihatbound} is similar to \eqref{iichihatbound} with $\iota = 1/4$, a closer inspection of the initial data in \cite{christodoulou} shows that it is more closely related to our initial data with $\iota = 1/2$. Indeed, \cite{christodoulou} imposes a scaling ansatz on an initial data quantity $\psi$ (which plays a role in \cite{christodoulou} similar to that played by $\qugamma$ for us) essentially of the form
\begin{equation}\label{chrinitpsidef}
\psi(\qucduu, \vartheta) = \qupark^{-1/2} \psi_0(\qupark \qucduu, \vartheta),
\end{equation}
where $\psi_0$ is independent of $\qupark$; this is analogous to \eqref{intgammaform} on $\qucds = 0$ with $\quiota = 1/2$, $\qucdv = \qucduu$, and {\it no\/} spatial scaling. (See \cite{christodoulou}, 2.1; \cite{thesis}, 0.1.) Furthermore, as exemplified by the ansatz in \eqref{chrinitpsidef}, \cite{christodoulou} requires that the lower bound on the incoming gravitational energy in \eqref{chiichihatbound} be satisfied {\it uniformly\/} in the angular variable $\vartheta \in \qubS^2$; \cite{christodoulou} also requires that the spherical derivatives $\qunabla_\vartheta \quchihat$ satisfy uniform upper bounds (independently of $\delta$). Both of these indicate that {\it spatial\/} concentration of the kind achieved in this paper is not allowed in \cite{christodoulou}. It is thus inappropriate to compare $L^2$ bounds (i.e., \eqref{chiichihatbound} with \eqref{iichihatbound}). \eqref{chiichihatbound} and \eqref{chrinitpsidef} suggest the $L^\infty$ estimate (in terms of $\qupark = \delta^{-1}$)
\begin{equation}\label{chintchihatinf}
\vert \quchihat\vert (\qucduu, \vartheta) \sim \qupark^{1/2}\hbox{ for } \qucduu \in [\qucduuz, \qucduuz + \qupark^{-1}],
\end{equation}
which is, again, clearly analogous to \eqref{intchihatinf} with $\quiota = 1/2$. 
\par
This landmark result of Christodoulou was extended and also simplified by the work of Klainerman and Rodnianski in \cite{klainrod}. Beyond the main result in their paper, they further note that they can prove existence to the Einstein equations with initial data satisfying an ansatz on the shear $\quchihat$ of the form
\begin{equation}\label{chihatkr}
\quchihat (\qucduu, \vartheta) = \qupark^{1/2} f_0(\qupark \qucduu, \qupark^{1/2} \vartheta),
\end{equation}
for $f_0$ independent of $\qupark$ but of suitably small support on $\qubS^2$ (compare \eqref{chrinitpsidef} and \eqref{chintchihatinf}, and \eqref{intgammaform} with $\qucds = 0$, recalling that for us $\quchihat \sim \partial_\qucdv\qugamma$). Such initial data would be supported on a region of dimensions $\sim \qupark^{-1/2} \times \qupark^{-1/2} \times \qupark^{-1}$, corresponding to concentration in {\it three\/} dimensions (in the $3 + 1$-dimensional spacetime), one more than in the present work. \eqref{chihatkr} gives the $L^\infty$ bound
\begin{equation*}
\vert \quchihat\vert  \sim \qupark^{1/2},
\end{equation*}
the same as in \cite{christodoulou}, corresponding to \eqref{intchihatinf} with $\quiota = 1/2$, rather than $\quiota = 1/4$ as obtained in the present work. Since the (pointwise) amplitude of the incoming gravitational energy is the same as in \cite{christodoulou}, while the support is much smaller, the total incoming energy must be much smaller, and in fact it is clear from \eqref{chihatkr} that it is of size $\qupark^{-1/2}$.
\par
The scaling relation \eqref{chihatkr} is the closest point of contact between the literature and the scaling used in the present paper. We emphasize, though, that our work is quite distinct from \cite{klainrod} because (a) we obtain solutions of much larger pointwise amplitude ($\vert \quchihat\vert  \sim \qupark^{3/4}$ instead of $\qupark^{1/2}$); (b) we work in polarized $U(1)$ symmetry and use noncompact (rectangular) spatial sections, whereas \cite{klainrod} works with no symmetry assumptions but with compact (spherical) spatial sections; (c) the main results in \cite{klainrod} use a different scaling which is imposed on norms rather than functions, in order to guarantee dynamical formation of a trapped surface, and this scaling is consistent with \eqref{chihatkr} only on the initial data. Our scaling \eqref{intgammaform}, on the other hand, is valid throughout the domain of existence of the solution.\footnote{It should be emphasized though that the function $\qugamb$ in \eqref{intgammaform} includes implicit lower-order dependence on the scaling parameter $\qupark$ off of the initial outgoing null hypersurface.} Thus, despite the similarities, the results in this paper are complementary to, and extend, those in \cite{klainrod}, for metrics enjoying polarized $U(1)$ symmetry.
\par
The ultimate aim of all the foregoing works was to prove the existence of dynamically evolved trapped surfaces, rather than to obtain concentrated solutions as here. The work of Luk and Rodnianski on impulsive gravitational waves, in particular \cite{lukrodiii} -- \cite{lukrod}, is more closely related to what we do here. In \cite{lukrodii}, in particular, the authors construct examples of {\it interacting impulsive gravitational wave\/} solutions to the Einstein equations. These follow from general existence theorems which allow {\it simultaneous\/} short-pulse data analogous to that in \cite{christodoulou} on each of two intersecting null cones. On the other hand, the size of the initial data is comparable to that in \cite{christodoulou} and \cite{klainrod}, i.e., it corresponds to $\quiota = 1/2$. Further, as in \cite{christodoulou}, the `largeness' of the data is exclusively in the null directions, while in the spherical directions much higher regularity is required -- excluding the possibility of spatial concentration.
\par
In \cite{lukrodiii} the authors were able to pass to the limit $\qupark\rightarrow\infty$ and derive weak convergence to a weak solution with a curvature singularity supported on (codimension-1) null hypersurfaces. Again this is for data enjoying higher regularity in the spatial directions, and the $L^2$ bounds on $\quchihat$ are consistent with a $\quchihat$ of size $\qupark^{1/2}$ in $L^\infty$ on an interval of length $\qupark^{-1}$ (hence corresponding still to $\quiota = 1/2$ in our setting). The authors have also extended their results to resolve many aspects of Burnett's conjecture; in particular the curvature singularity that appears in the limit can be interpreted as a (measure-valued) matter field, showing that {\it non-vacuum\/} spacetimes can appear as (weak) limits of vacuum spacetimes. 
More recently, Burnett's conjecture (and its reverse) have been considered in a generalized wave gauge, which allows for more general results -- see \cite{huneau2024burnetts}, \cite{touati2024reverse}. We note however that the solutions we construct here are more singular than those constructed in any of the above treatments of the Burnett conjecture (in the sense that all  Sobolev norms $H^s$ for $s > 1$ blow up for our solutions as $\qupark \rightarrow \infty$).
\par
Unlike the results of Christodoulou \cite{christodoulou} and Klainerman and Rodnianski \cite{klainrod}, the (large portion of the) initial data in \cite{lukrodiii} -- \cite{lukrod} can be supported on the {\it center\/} of the initial null hypersurfaces. Additionally, \cite{lukrodiii} -- \cite{lukrod} obtain solutions to the Einstein equations in which (curvature) singularities present in the initial data {\it persist\/} throughout the domain of existence of the solutions.
\par
We stress that all of the above results have no symmetry requirements of any kind.
\par
The foregoing comparisons can be summarized as follows. First, the initial data we allow display a `concentration' in a {\it spatial\/} direction as well as one null direction; this is not allowed in any of the prior works that we discussed, {\it except\/} (as explained) in \cite{klainrod}. Next, the (pointwise) amplitude of the initial data we allow is {\it strictly larger\/} than that constructed in any of the aforementioned papers: specifically, all of the aforementioned papers allow only initial data for which the key scaling exponent satisfies $\quiota \geq 1/2$, while our results allow for any $\quiota \geq 1/4$. This allows us to use initial data which, {\it in spite\/} of the additional spatial concentration, has total incoming gravitational energy which is uniformly positive. We note, however, that our methods only work in $U(1)$ symmetry.
\par
In addition to the foregoing, our second result proves existence of solutions to the Einstein vacuum equations which remain concentrated (in the sense described above) near an $\qubS^1$ family of null geodesics in the overall $3 + 1$-dimensional spacetime $\quinM$. We believe that this gives the first known example of solutions to the Einstein vacuum equations concentrated in {\it two\/} dimensions (in a $3 + 1$-dimensional spacetime) while possessing a certain amount of genericity (subject to $U(1)$ symmetry).
\par
All of the foregoing works, as ours, use a null foliation of the initial outgoing null hypersurface. Since the (optical) Raychaudhuri equation includes a source term proportional to $-|\quchihat|^2$, a `short-pulse' initial data with support of length $\qupark^{-1}$ and $L^\infty$ amplitude $|\quchihat| \sim \qupark^\alpha$ ($\alpha \geq 1/2$, say) will, generically, cause these null foliations to break down due to caustic formation by affine parameter $\sim \qupark^{1 - 2\alpha}$. For the previous works just reviewed, $\alpha = 1/2$, meaning that we do not expect caustic formation until affine parameter of order 1. For us, though, when $\iota = 1/4$, \eqref{intchihatinf} implies that $\alpha = 3/4$, and generically we expect caustic formation by affine parameter $\sim \qupark^{-1/2}$ -- in other words, not even our initial coordinate system would have a uniform existence time. We avoid such problems by, essentially, ruling them out: we require the initial data for the metric $\qumeth$ -- which is the geometric object via which caustic formation manifests itself -- to be Minkowskian to the entire future of the support of $\qugamma$, and require that support itself to be located at the {\it past\/} end of the initial outgoing null hypersurface. (Conceptually, we force the hypothetical caustics to occur to the {\it past\/} of the region on which we solve the equations.)\footnote{We thank Jonathan Luk for bringing these points to our attention.}
\par
Finally, we wish to expand on the connection of the result obtained here with the hoop conjecture (see \cite{thornei}). The hoop conjecture (purposely imprecisely formulated) stipulates that asymptotically flat data $(\Sigma^3, {\rm g}^3, {\rm k}^3)$, containing a fixed amount $M$ of mass (and thus energy) concentrated to a scale smaller than the Schwarzschild radius $r^{\rm Schwarz}(M)$ of $M$, must necessarily contain a trapped surface.
The term `concentrated' here is to be understood in the sense
that a (suitable quasi-local notion of) mass $M$ is  
contained in a domain 
$X\subset  \Sigma^3$, where $\partial X$ is bounded in circumference from \quemph{every} direction by  
$2\pi r^{\rm Schwarz}(M)$. (One can visualize this requirement as a hoop 
rotating around $\partial X$ in all possible directions and maintaining length no greater than $2\pi r^{\rm Schwarz}(M)$.) 
The conjecture is of course loosely stated; in particular  it relies on an appropriate quasi-local notion of gravitational energy contained in 
$X$.  
The conjecture also asserts that this 
requirement on energy concentration in \quemph{every} direction is 
\quemph{necessary} for the existence of a trapped surface.   
In particular, if $X$ is bounded by small circles in some directions but not others (like a pancake, or a cylinder), then trapped surfaces should not be present (see \cite{thornei}).
Our construction for $\quiota=1/4$ provides examples where (some notion of) gravitational energy is squeezed to arbitrarily small scales in the $2+1$-picture but is \quemph{not} squeezed in the remaining third direction $\quirad$.  We conjecture (but have not proved) that these spacetimes can be extended and `cut-off' in the $3+1$-picture
to obtain asymptotically flat data that does not contain trapped surfaces. If this is true, then morally our 
constructed spacetimes can be seen as evidence in favor of the necessity direction of the hoop
conjecture for {\it vacuum\/} solutions to the $3 + 1$-dimensional Einstein equations, with a choice of mass appropriate for our solutions. (See also \cite{schoenyau} and \cite{hirsch2023spectral} for older and newer results on the hoop conjecture.)
\par
\subsection{Statement of results}\label{statres}
We now present summary versions of our two results. See \quEEEinitdatbounds, \quEEEseqbound, and \quEEErbthm\ below for the detailed formulations.
\par
We recall that the surfaces $\quinin = \{ \qucds = 0 \}$ and $\quinout = \{ \qucdv = 0 \}$ in $\quinN^{2 + 1}$ (see \quEEEspacetimephyss) are null for the metric $\qumeth$, and that the constraint equations on the metric $\qumeth$ are ODEs in $\qucdv$ along $\quinin$. We combine our results on the constraint and evolutionary portions of \eqref{intRicwav} as follows.
\begin{figure}[h]\centering
\includegraphics[keepaspectratio]{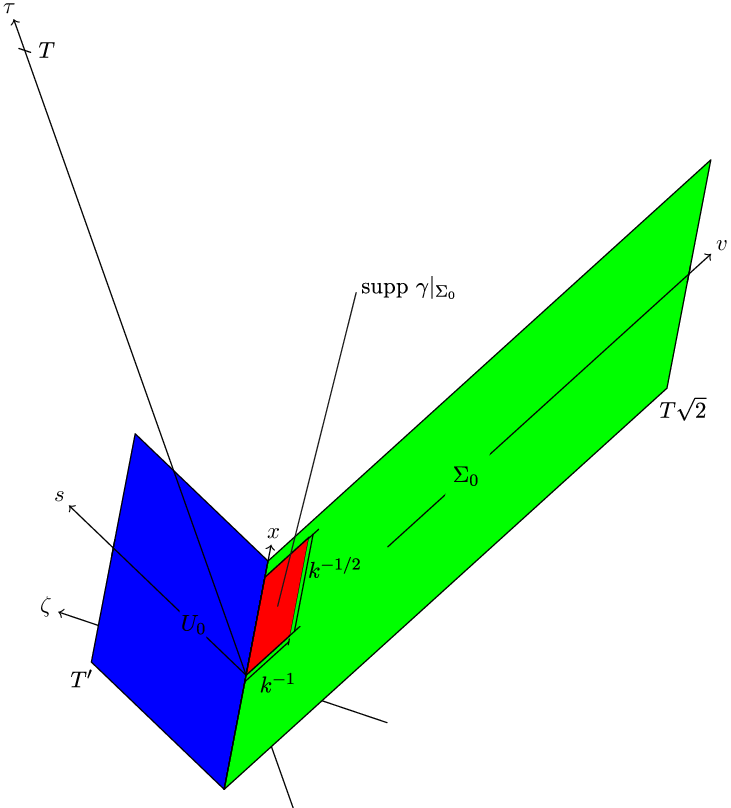}
\caption{Initial null hypersurfaces, and support of $\qugamma\vert _{\quSigz}$}\label{spacetimephyss}
\end{figure}
\par
\begin{theorem}\label{iseqbound} Let $\varpi \in C^\infty(\quR^2)$ be a fixed function supported on $(0, 1) \times (0, 1)$ and satisfying, for some $\qusbn \geq \qusbnmin$ and some $\qunu < 1$ sufficiently small (independently of $\qupark$),
\begin{equation*}
\sum_{{\scriptstyle I \leq 6\qusbn + 2}\atop{\scriptstyle J \leq \qusbn}} \| \partial_1^I \partial_2^J \varpi\| _{L^\infty((0, 1) \times (0, 1))} \leq \qunu.
\end{equation*}
Let $\qupark \geq 1$ be sufficiently large, and define $\qugamma^\qupark$ on $\quSigz$ by
\begin{equation}\label{gammainitdat}
\qugamma^\qupark(0, \qucdx, \qucdv) = \qupark^{-\quiota} \varpi(\qupark^{1/2} \qucdx, \qupark \qucdv).
\end{equation}
Then initial data for $\qumeth^\qupark$ on $\quSigz$, and $\qugamma^\qupark$ on $\quSigz \cup \quUz$, can be obtained such that
\begin{description}
\item{(i)} the constraint equations implied by \eqref{intRicwav} are satisfied on $\quSigz$;
\item{(ii)} the initial data for $\qugamma^\qupark$ are supported entirely on the region
\begin{equation*}
\{ (0, \qucdx, \qucdv) \in \quSigz\,\vert \,\qucdx \in [0, \qupark^{-1/2}],\,\qucdv \in [0, \qupark^{-1}] \} \cup \{ (\qucds, \qucdx, 0) \in \quUz\,\vert \,\qucdx \in [0, \qupark^{-1/2}] \};
\end{equation*}
\item{(iii)} the initial data for $\qugamma^\qupark$ satisfy the bounds
\begin{align}
\left[\sum_{J + K \leq \qusbn} \int_{\quSigz} \left(\partial_\qucdx^J \partial_\qucdv^K \qugamma^k\right)^2\,d\quxb\,d\quvb\right]^{1/2} &\leq C \qunu \qupark^{-\frac{3}{4} - \quiota + \frac{1}{2} J + K},\label{ifluxboundsone}\\
\left[\sum_{I + J \leq \qusbn} \int_{\quUz} \left(\partial_\qucds^I \partial_\qucdx^J \qugamma^k\right)^2\,d\qusb\,d\quxb\right]^{1/2} &\leq C \qunu \qupark^{-\frac{1}{4} - \quiota + \frac{1}{2} J},\label{ifluxboundstwo}
\end{align}
where $C > 0$ is a universal constant.
\end{description}
Furthermore, there is a polarized $U(1)$ symmetric Lorentzian metric $\qumetg^\qupark$ of the form \eqref{ingexp},
\begin{equation}\label{ningexp}
\qumetg^\qupark(\quiny, \quirad) = e^{-2\qugamma^\qupark(\quiny)} \qumeth^\qupark_{ij}(\quiny)\,d\quiny^i \otimes d\quiny^j + e^{2\qugamma^\qupark(\quiny)} d\quirad \otimes d\quirad,
\end{equation}
such that
\par
\begin{description}
\item{(i)} $\qumeth^\qupark$ and $\qugamma^\qupark$ agree with the constructed initial data on $\quSigz$ and $\quSigz \cup \quUz$, respectively;
\item{(ii)} $\qumetg^\qupark$ satisfies the vacuum Einstein equations $\quRic(\qumetg^\qupark) = 0$ on a domain $\qubulk(\queT, \queTp) \times \qubS^1$, where
\begin{equation*}
\qubulk(\queT, \queTp) = \{ (\qucds, \qucdx, \qucdv) \in \quR^3\,\vert \,\qucds \in [0, \queTp],\,\qucdx \in \quR^1,\,\qucdv \in [0, \queT\sqrt{2}]\},
\end{equation*}
and $\queT$, $\queTp$ are sufficiently small, depending on $\qunu$ but {\it independent\/} of $\qupark$.
\end{description}
\par\noindent
Moreover the metric $\qumeth^\qupark$ admits two transverse families of null hypersurfaces $C^1$ embedded in $\Gamma(\queT, \queTp)$, a family $\Sigma_{\qutcds}$ passing through $\{ \qucds = \qutcds, \qucdv = \queT\sqrt{2} - \qupark^{-1} \qutcds \}$ and a family $U_{\qutcdv}$ passing through $\{ \qucds = 0, \qucdv = \qutcdv \}$ (see \quEEEphysnullsurff), and the solution $\qugamma^\qupark$ satisfies the flux bounds
\begin{align}
\left[\int_{\Sigma_{\qutcds}} \qunulnorS^i \qutimvec^j Q_{ij}[\partial_\qucds^I \partial_\qucdx^J \partial_\qucdv^K \qugamma^\qupark]\,dV_{\Sigma_{\qutcds}}\right]^{1/2} &\leq C\qunu\qupark^{-\frac{3}{4} - \quiota + \frac{1}{2} J + K},\label{fluxboundsone}\\
\left[\int_{U_{\qutcdv}} \qunulnorU^i \qutimvec^j Q_{ij}[\partial_\qucds^I \partial_\qucdx^J \partial_\qucdv^K \qugamma^\qupark]\,dV_{U_{\qutcdv}}\right]^{1/2}&\le C\qunu \qupark^{-\frac{1}{4}-\quiota+\frac{1}{2}J + K},\label{fluxboundstwo}
\end{align}
where $I + J + K \leq \qusbn - 5$, $\qunulnorS$ and $\qunulnorU$ are (null) normal vectors to $\Sigma_{\qutcds}$ and $U_{\qutcdv}$, respectively, $\qutimvec$ is the family of future-directed normal vectors to the (spacelike) surfaces $\{ \qucds + \qupark \qucdv = \hbox{const} \}$, $Q_{ij}[f]$ is the stress-energy tensor of $f$ with respect to $\qumeth^\qupark$,
\begin{equation*}
Q_{ij}[f] = \partial_i f \partial_j f - \frac{1}{2} \qumeth^\qupark_{ij} ({\qumeth^\qupark}){}^{\ell m}\partial_\ell f \partial_m f,
\end{equation*}
and $C > 0$ is a universal constant. The area elements in \eqref{fluxboundsone} -- \eqref{fluxboundstwo} are the area elements on the null hypersurfaces $\Sigma_{\qutcds}$ and $U_{\qutcdv}$ induced by $\qumeth^\qupark$ and the null normal vectors $\qunulnorS$ and $\qunulnorU$.
\end{theorem}
\begin{figure}[h]\centering
\includegraphics[keepaspectratio]{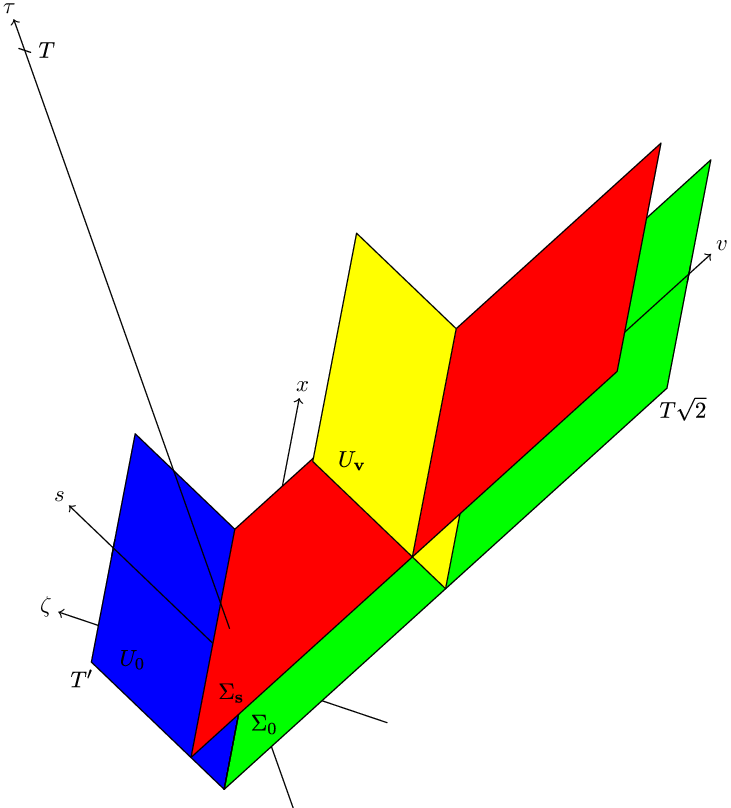}
\caption{Schematic of the two null foliations}\label{physnullsurff}
\end{figure}
\par\noindent
\begin{remark} We note that Theorem \ref{iseqbound} can be extended to include $\varpi$ supported on all of $\quR^1 \times (0, 1)$.\end{remark}
\begin{remark} The bounds in \eqref{ifluxboundsone} -- \eqref{ifluxboundstwo} are only necessary, not sufficient. See Theorem \ref{seqbound} for the full bounds. The solution we obtain is unique among a suitable class of functions, see Theorem \ref{seqbound} for the detailed formulation.\end{remark}
\begin{remark}
The flux bounds in \eqref{fluxboundsone} are not entirely natural since the higher-order derivatives are formed from the coordinate derivatives $\partial_\qucds$, $\partial_\qucdx$, $\partial_\qucdv$, instead of the tangent vectors to the surfaces $\Sigma_{\qutcds}$. (The tangent plane to $U_{\qutcdv}$ is spanned by $\partial_\qucdx$, $\partial_\qucdv$ so similar concerns do not apply to \eqref{fluxboundstwo}.) We formulate the result as here since we obtain only $C^1$ control on the surfaces $\Sigma_{\qutcds}$, and obtaining higher-order control appears nontrivial.
\end{remark}
\par
By applying Theorem \ref{iseqbound}\ to the specific (geometric-optics ansatz) choice
\begin{equation}\label{geooptvarpi}
\varpi(y, w) = A(y, w) \cos (\quparr w)
\end{equation}
for some smooth $A$ and fixed ($\qupark$-independent) $\quparr \gg 1$, we obtain solutions for which the `concentration' property of $\qugamma$ and its derivatives {\it persists\/} on the hypersurfaces $\Sigma_{\qutcds}$ in $\quinN^{2 + 1}$ (hence on $\Sigma_{\qutcds} \times \qubS^1$ in $\quinM^{3 + 1}$).
\par
\begin{theorem}\label{irbthm} Choose any desired fraction $1-\quconcrat, \quconcrat\in (0,1)$, and let $M > 4(\qusbn + 4)$. Then there is a quantity $\quparr = \quparr(\quconcrat)$, $\quparr \sim \quconcrat^{-\alpha(M)/M}$ for some $\alpha(M) \in (2, 4)$, and a constant $c > 0$ such that the following holds for $\quconcrat$ sufficiently small. Let $\qupark$ be sufficiently large (depending on $\quconcrat$). Let $A \in C^\infty(\quR^2)$ be an arbitrary fixed function (in particular, independent of $\quconcrat$, $\quparr$, and $\qupark$) supported on $(0, 1) \times (0, 1)$ and which is, together with sufficiently many of its derivatives, sufficiently small in $L^\infty(\quR^2)$. Define $\qugamma^\qupark$ on $\quSigz$ by (see \eqref{gammainitdat}, \eqref{geooptvarpi})
\begin{equation}\label{irbthmgammastip}
\qugamma^{\qupark, \quparr}(0, \qucdx, \qucdv) = c \qupark^{-\quiota} \quparr^{-M - 2\qusbn - 2} \queTp^{3/4} A(\qupark^{1/2} \qucdx, \qupark \qucdv) \cos (\qupark \quparr \qucdv).
\end{equation}
Then the conclusions of Theorem \ref{iseqbound}\ hold for this choice of $\qugamma^{\qupark, \quparr}\vert _\quSigz$, with $\qunu = \qunu(\quconcrat) \sim \quconcrat^{\alpha(M)}$. Moreover, the solution for $\qugamma^{\qupark, \quparr}$ on $\qubulk(\queT, \queTp)$ satisfies the following `persistence of concentration' property, with respect to the $\qumeth$-null foliation $\Sigma_{\qutcds}$: let $\quinR^\qupark(\Sigma_{\qutcds}) = \{ (\qucds, \qucdx, \qucdv) \in \Sigma_{\qutcds}\,\vert \,\qucdx \in [0, \qupark^{-1/2}],\,\qucdv \in [0, \qupark^{-1}] \}$; then
\begin{equation}\label{irbthmscalfrac}
  \frac{\| \partial_v \qugamma^{\qupark,\quparr} \| ^2_{L^2(R^\qupark({\Sigma}_{\qutcds}))}}{\|  \partial_v \qugamma^{\qupark,\quparr}\| ^2_{L^2({\Sigma}_{\qutcds})}+\|  \partial_x \qugamma^{\qupark,\quparr}\| ^2_{L^2({\Sigma}_{\qutcds})}}\ge 1-\quconcrat.
\end{equation}
In particular, the fraction of the total energy flux of $\qugamma$ that remains on $\quinR^\qupark({\Sigma}_{\qutcds})\subset {\Sigma}_{\qutcds}$ can be made arbitrarily close to 1. 
\end{theorem}
\par 
\subsection{Outline of the construction}\label{outconstr} The key steps in the proofs of the above results are as follows.
\par
\begin{figure}[h]\centering
\includegraphics[keepaspectratio]{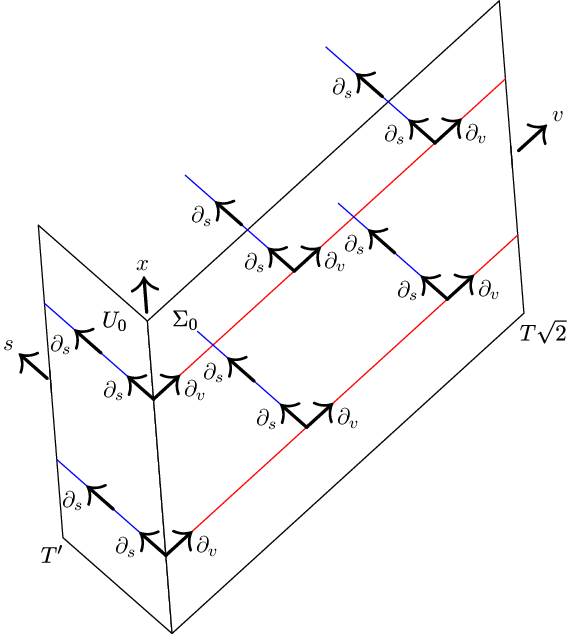}
\caption{Coordinate system}\label{scoordsys}
\end{figure}
We work with respect to a null geodesic coordinate system obtained by taking a null hypersurface $\quSigz$ for $\qumeth$ parameterized by a spacelike coordinate $\qucdx$ and a null coordinate $\qucdv$, and developing null geodesics transverse to $\quSigz$ with affine null parameter $\qucds$. (See Figure \ref{scoordsys} for a sketch, and \quEEEgaugechoice\ for the details.) We then introduce scaled coordinates
\begin{equation}
\qusb = \qucds,\qquad \quxb = \qupark^{1/2} \qucdx,\qquad \quvb = \qupark \qucdv,\label{iscalcdef}
\end{equation}
and define scaled functions by the relations
\begin{gather}
\qugamma(\qucds, \qucdx, \qucdv) = \qupark^{-\quiota} \qugamb(\qucds, \qupark^{1/2} \qucdx, \qupark \qucdv),\label{nintgammaform}\\
\quhbar_{ij} = \begin{pmatrix}
0&0&-1\\0&1&0\\-1&0&0
\end{pmatrix} + \qupark^{-2\quiota} \begin{pmatrix}
0&0&0\\0&2\qudlb + \qupark^{-2\quiota}\qudlb^2&\qubb\\0&\qubb&\qucb
\end{pmatrix}\label{ihbarexp}
\end{gather}
where $\quhbar = \qupark\qumeth$ and the above expression is with respect to the scaled coordinates $\qusb$, $\quxb$, $\quvb$. One finds that the gauge choice imposes the conditions 
\begin{equation}\label{intgaugecond}
\qubb = \qucb = \partial_\qusb \qucb = 0\hbox{ on }\quSigz.
\end{equation}
\par
Since this gauge has not appeared before in the literature, we must obtain the evolution and constraint equations corresponding to the reduced Einstein equations \eqref{intRicwav} from first principles. Standard geometric reasoning allows us to obtain the evolutionary part of $\quRic(\qumeth) = 2\qunabla\qugamma\otimes\qunabla\qugamma$
(here for convenience we define $\quellb = 1 + \qupark^{-2\quiota} \qudlb$, $\quab = \quellb^2$, $\qudaib = \qupark^{2\quiota} (\quab^{-1} - 1)$),
\begin{align}
\partial_\qusb^2 \qudlb &= -2\quellb (\partial_\qusb\qugamb)^2,\label{insricone}\\
\partial_\qusb^2 \qubb &= \frac{1}{\quellb} \qupark^{-2\quiota} (\partial_\qusb \qudlb) (\partial_\qusb\qubb) - 4 \partial_\qusb\qugamb (\partial_\quxb \qugamb + \qupark^{-2\quiota} \qubb \partial_\qusb\qugamb),\kern 12pt\label{insrictwo}\\
\partial_\qusb^2 \qucb &= -4 \partial_\qusb \qugamb \partial_\quvb \qugamb - \frac{2}{\quab} (\partial_\quxb\qugamb)^2\label{insricthree}\\
&\kern -0.125in\notag + \qupark^{-2\quiota} \left( \frac{1}{2\quab} (\partial_\qusb \qubb)^2 - 2\qucb(\partial_\qusb\qugamb)^2 - \frac{4}{\quab} \qubb \partial_\quxb \qugamb \partial_\qusb \qugamb \right) - \qupark^{-4\quiota} \frac{2}{\quab} \qubb^2 (\partial_\qusb\qugamb)^2,
\end{align}
while a direct calculation shows that the wave equation $\quwbox_\qumeth \qugamma = 0$ can be written as
\begin{multline}
\Biggl\{\left(-2\partial_\qusb\partial_\quvb + \partial_\quxb^2\right)\hfill\\
\shoveleft{\qupark^{-2\quiota} \Biggl[\qudaib \partial_\quxb^2 - \qucb \partial_\qusb^2 + 2\frac{\qubb}{\quab} \partial_\qusb \partial_\quxb}\\
\shoveright{- \left(\partial_\qusb \qucb - \frac{1}{\quab} \partial_\quxb \qubb + \frac{\partial_\quvb\qudlb}{\qulb}\right)\partial_\qusb + \left(\frac{1}{\quab} \partial_\qusb \qubb - \frac{\qulb\partial_\quxb\qudlb}{\quab^2}\right)\partial_\quxb - \frac{\partial_\qusb \qudlb}{\qulb}\partial_\quvb\Biggr]}\\
\shoveleft{\qupark^{-4\quiota} \left(\frac{\qubb^2}{\quab} \partial_\qusb^2 - \left(\qucb \frac{\partial_\qusb \qudlb}{\qulb} - 2\frac{\qubb\partial_\qusb \qubb}{\quab} + \frac{\qubb\,\quellb \partial_\quxb \qudlb}{\quab^2}\right) \partial_\qusb - \frac{\qubb\,\qulb\partial_\qusb\qudlb}{\quab^2}\partial_\quxb\right)}\\
 - \qupark^{-6\quiota} \frac{\qubb^2\qulb\partial_\qusb\qudlb}{\quab^2}\partial_\qusb\Biggr\}\qugamb = 0.\label{iswave}
\end{multline}
We obtain the constraint equations from a direct calculation of the Ricci tensor $\quRic(\qumeth)$, combined with the evolution equations \eqref{insricone} -- \eqref{insricthree}. The full form of these equations is quite complicated, but on $\quSigz$ we can utilize the gauge conditions \eqref{intgaugecond} to write them as 
\begin{align}
\partial_\quvb^2 \qudlb &= -2\quellb(\partial_\quvb\qugamb)^2,\label{iscalconsts}\\
\partial_\quvb (\quellb \partial_\qusb \qubb) &= 4\quellb \partial_\quxb \qugamb \partial_\quvb \qugamb,\label{iscalconste}\\
\quellb\cdot\qupbdlsv &= (\qupbgx)^2 + \frac{1}{2} \partial_\quxb \partial_\qusb \qubb + \frac{1}{4} \qupark^{-2\quiota} \left( (\partial_\qusb \qubb)^2 - \frac{2}{\quellb^3} \partial_\qusb \qubb \partial_\quxb \qudlb\right).\label{iscalconsty}
\end{align}
One must prove that the constraint equations are preserved by the evolution; as usual, this involves applying the contracted Bianchi identities, see \quEEEconstpres.
\par
Initial data for the system \eqref{insricone} -- \eqref{iswave} consist of
\begin{gather}
\partial_\qusb^\ell \qudlb,\,\partial_\qusb^\ell\qubb,\,\partial_\qusb^\ell\qucb\hbox{ on $\quSigz$, $\ell = 0$, $1$,}\label{qumetcompid}\\
\qugamb\hbox{ on }\quSigz\cup\quUz.\label{qugambcompid}
\end{gather}
\par
We may specify half of \eqref{qugambcompid} by fixing some $\varpi \in C^\infty(\quR^2)$ which is supported on $(0, 1) \times (0, 1)$ and stipulating
\begin{equation}
\qugamb|_\quSigz(0, \quxb, \quvb) = \varpi(\quxb, \quvb).\label{qugambquSigzvarpi}
\end{equation}
By the gauge condition \eqref{intgaugecond}, the only nonvanishing quantities in \eqref{qumetcompid} are
\begin{equation}
\qudlb,\,\partial_\qusb\qudlb,\,\partial_\qusb\qubb\hbox{ on $\quSigz$}.\notag
\end{equation}
\eqref{iscalconsts} -- \eqref{iscalconsty} shows that, given \eqref{qugambquSigzvarpi}, these quantities are entirely determined once they, together with $\partial_\quvb\qudlb$, are given on any line $\{ \quxb = \hbox{const} \}$ in $\quSigz$. In order to avoid caustics on $\quSigz$,\footnote{It is clear from \eqref{ihbarexp} and \eqref{iscalconsts} -- which is closely related to the Raychaudhuri equation -- that, generically, caustics can be expected to occur on at least one side of $\qusupp \qugamb|_\quSigz$ within a change in $\quvb$ of size $\sim \qupark^{2\quiota}$, which for $\iota = 1/4$ is $\qupark^{1/2} \ll \qupark$. Thus it is impossible for $\quSigz$ to remain smooth for parameter $\sim \qupark$ on both past and future sides of $\qusupp \qugamb|_\quSigz$. This is a concrete instance of the general considerations described towards the end of Subsection \ref{relworks} above.} we specify
\begin{equation}
\qudlb|_{\quvb = 1} = \partial_\quvb\qudlb|_{\quvb = 1} = \partial_\qusb \qudlb|_{\quvb = 1} = \partial_\qusb \qubb|_{\quvb = 1} = 0.
\end{equation}
This choice forces the metric $\quhbar$ to equal the Minkowski metric on $\{ (0, \quxb, \quvb)\in \quSigz\,\vert\,\quvb \geq 1 \}$.
\par
Finally, we then choose $\qugamb|_\quUz$ so that the a priori values of $\partial_\qusb^\ell \qugamb|_\quSigz$ for a suitable range of $\ell$ values (at least $\ell = 0, \ldots, \qusbn + 1$) vanish also on $\{ (0, \quxb, \quvb) \in \quSigz\,\vert\,\quvb = 1 \}$ (and hence on $\{ (0, \quxb, \quvb)\in\quSigz\,\vert\,\quvb \geq 1 \}$). These a priori values are determined by integrating $\qusb$ derivatives of the wave equation \eqref{iswave} after reducing higher $\qusb$ derivatives of $\qudlb$, $\qubb$, $\qucb$ via \eqref{insricone} -- \eqref{insricthree}. A priori bounds on the transverse derivatives of $\qudlb$, $\qubb$, $\qucb$, and $\qugamb$ on both $\quSigz$ and $\quUz$ can then be obtained from (differentiated forms of) \eqref{insricone} -- \eqref{iscalconsty}. The details are given in \quEEEinitdatbounds.
\par
We must then integrate the evolution equations \eqref{insricone} -- \eqref{iswave}. We prove existence and boundedness of solutions to the system \eqref{insricone} -- \eqref{iswave} from first principles, using a contraction-style argument like that in \cite{ringstrom}, 9.3, to produce Cauchy sequences which converge to solutions to \eqref{insricone} -- \eqref{iswave}.\footnote{Another approach would be to apply one of the general existence theorems for the Einstein equations to our situation to obtain local existence and continuation results and then use a bootstrap argument. The approach we use here is probably only marginally longer and also much cleaner.} Our approach involves obtaining energy estimates from the ODE-wave system \eqref{insricone} -- \eqref{iswave}, and is mostly standard. We point out two key steps. One involves a trick for avoiding loss of derivative: the coefficients in \eqref{iswave} depend on $\partial_\quxb \qubb$, which by \eqref{insrictwo} depends on $\partial_\quxb^2 \qugamb$. We get around this apparent loss of derivative by solving $\eqref{iswave}$ for $\partial_\quxb^2\qugamb$ and carrying extra $\qusb$ derivatives in our energies (see Definition \ref{eenerdefi}.) Additionally, the new Klainerman-Sobolev estimate we obtain allows us to conclude that a nonlinearity of order $q$ in $\qudlb$, $\qubb$, $\qucb$, $\qugamb$, and their derivatives will have decay $\sim \tau^{-(q - 1)/4}$, where $\tau$ is the timelike coordinate indicated in Figure \ref{spacetimephyss}. Roughly, then, the nonlinear terms in $\quwbox_\qumeth \qugamma = 0$ thus have decay $\sim \tau^{-1/2}$ and overall size no greater than $\sim \qupark^{-2\iota} \tau^{-1/2}$. From this, and similar estimates on the deformation tensor of $\qumeth$, we readily deduce existence up to scaled time $\sim \qupark^{4\iota}$, or unscaled time $\sim \qupark^{4\iota - 1}$, which has a uniform (in $\qupark$) lower bound precisely when $\iota \geq 1/4$.\footnote{The decay we obtain is much less than the $\sim \tau^{-(q - 1)/2}$ decay one would expect to obtain in two dimensions (see \cite{klainerman}). This is because we work on rectangular strips rather than regions with circular symmetry. See the proof of Proposition \ref{klainsobolv}, especially the discussion of Figure \ref{klainsobolevfigy} therein.}
\par
The idea behind the proof of Theorem \ref{irbthm} is to construct an explicit approximate solution $\qugambGNf$ (depending on a large parameter $\quparr$) to the wave equation \eqref{iswave} which is supported on the region $\quQS = \{ (\qusb, \quxb, \quvb) \in \Gamma(\queT, \queTp)\,\vert\,\quxb, \quvb \in (0, 1) \}$, construct the exact solution to the system \eqref{insricone} -- \eqref{iswave} agreeing with $\qugambGNf$ on $\quSigz$, and then show that the energy flux of the difference $\qugamb - \qugambGNf$ must be much smaller than the energy flux of $\qugambGNf$. Since $\qugambGNf$ is supported on $\quQS$, \eqref{irbthmscalfrac} follows. The argument is mostly a routine exercise in elementary geometric optics; see \quEEEboundsinitdatgamb\ and \quEEEremainbound\ for the details. Our main innovation is the following: since the non-Minkowski terms in the wave equation \eqref{iswave} are of size $\sim \qupark^{-2\iota}$, by taking $\qupark$ sufficiently large compared to the parameter $\quparr$ we are able to take $\qugambGNf$ to be a geometric optics approximate solution for the {\it Minkowski\/} wave equation.
\par
\subsection{Summary of the paper}\label{isummary} In \quEEEgaugechoicech\ we construct the null geodesic gauge and derive the unscaled versions of equations \eqref{insricone} -- \eqref{iswave}.
We then construct the unscaled versions of the constraint equations and show that they are preserved by the evolution implied by (the unscaled versions of) \eqref{insricone} -- \eqref{iswave},
and finally introduce the scaling and obtain \eqref{insricone} -- \eqref{iscalconsty}. In \quEEEinterlude\ we note some background algebraic and analytic results and prove our modified Klainerman-Sobolev estimate (see \quEEEklainsobolv). In \quEEEinitdat\ we construct and bound initial data satisfying the constraints \eqref{iscalconsts} -- \eqref{iscalconsty}. \quEEEenerineq\ is the core of the paper and presents the proof of the existence result in Theorem \ref{iseqbound}. In \quEEEfocsol\ we obtain the concentrated solutions which are our final goal.
\par
\subsection{Notations and conventions}\label{notations} The natural numbers $\quN$ are taken to include $0$.
We fix a particular Sobolev exponent $\qusbn \geq \qusbnmin$. We define
\begin{equation*}
\quRpos = \{ x \in \quR\,\vert \,x > 0\}.
\end{equation*}
\par
We always work with tensors in terms of their components with respect to a coordinate basis, and employ the Einstein summation convention throughout.
Indices $0$, $1$, and $2$ refer to the coordinates $\qucds$, $\qucdx$, $\qucdv$ in the unscaled picture, $\qusb$, $\quxb$, $\quvb$ or $\qutau$, $\quxi$, $\quzeta$ in the scaled picture, or (when working with the auxiliary coordinate system introduced in \quEEEfocsol) $\qutsb$, $\qutxb$, $\qutvb$. Index $3$, in the (very) few places where it is used, always refers to the coordinate $\qucdy$ in the unscaled $3 + 1$ picture.
\par
All $L^p$ and Sobolev spaces are taken with respect to coordinate measure and coordinate derivatives, except where noted otherwise. By a slight abuse of notation, we shall sometimes write things like $f \in L^2(\partial X)$ to mean that $f$ is defined on a set containing $\partial X$, and $f\vert _{\partial X} \in L^2(\partial X)$.
\par
We shall write $C(A, B, \ldots)$, etc., to indicate constants which depend only on upper bounds on the quantities $A$, $B$, $\ldots$, in the following sense: $(A, B, \ldots) \mapsto C(A, B, \ldots)$ defines a function, and if we have an inequality
\begin{equation}\label{constbaseineq}
f \leq C(A, B, \ldots) g,
\end{equation}
then whenever $A_0 \geq A$, $B_0 \geq B$, $\ldots$, also
\begin{equation}
f \leq C(A_0, B_0, \ldots) g.
\end{equation}
A sufficient condition is that $C(A, B, \ldots)$ be increasing in each variable independently. (This concept is more typically expressed by fixing an auxiliary variable which is an upper bound for the quantities $A$, $B$, $\ldots$, and writing $C$ in terms of that auxiliary upper bound. Such a method would be rather clumsy for us.) We note that any constant $C'(A, B, \ldots)$ which is bounded on compact sets can be replaced with one depending only on upper bounds on $A$, $B$, $\ldots$ in this sense. We shall tacitly assume that all constants are of this kind, even when not stated explicitly.
\par
By {\it numerical constant\/} we mean a quantity which can be evaluated to a definite number (like 197).
\par
\subsection{Acknowledgements}\label{acknowledgements} We thank J.\ Luk for helpful conversations, and J.\ Sbierski for suggesting we use a geometric optics ansatz in \quEEEfocsol. N.\ C.\ would like to acknowledge the influence of Ringstr\"om's text \cite{ringstrom} on various aspects of this paper. S.\ A.\ was partially supported by NSERC Grant RGPIN-2020-01113. N.\ C.\ was partially supported by an International Postdoctoral Exchange Talent-Introduction Program grant (YJ20210111).
\par
\section{GAUGE CHOICE, CONSTRAINT EQUATIONS, AND SCALING}\label{gaugechoicech}
\par
\subsection{Gauge choice}\label{gaugechoice}
\begin{figure}[h]\centering
\includegraphics[keepaspectratio]{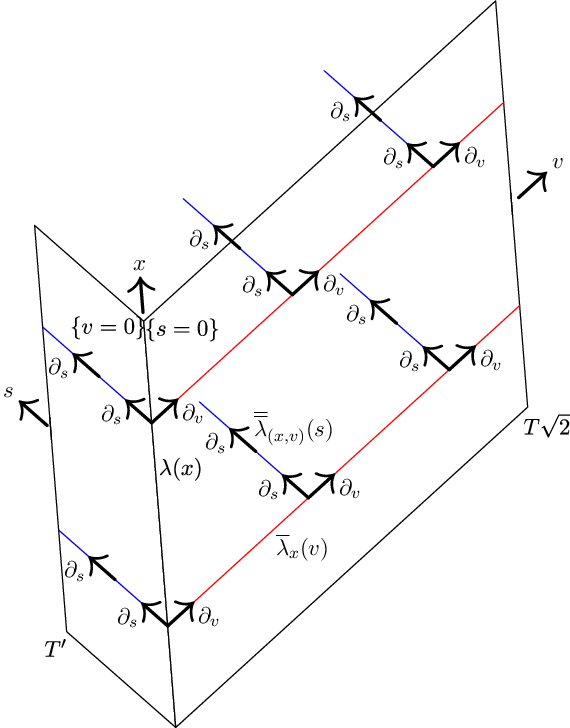}
\caption{Coordinate system}\label{fgaugechoice}
\end{figure}
We shall assume throughout this paper that our metric $\qumeth$ is such that
\begin{description}
\item{(i)} there exists a coordinate system $\qucds$, $\qucdx$, $\qucdv$ on some portion of the domain of $\qumeth$ whose image contains the set
\begin{equation}
\Gamma'_0 = \{ (\qucds, \qucdx, \qucdv) \in \quR^3\,\vert\,\qucds \in [0, S],\,\qucdv \in [0, V] \}
\end{equation}
for some $S$, $V > 0$, $S < \queTpupp$;
\item{(ii)} with respect to this coordinate system, the metric $\qumeth$ has the representation
\begin{equation}\label{gaugemetform}
\qumeth_{ij} = \begin{pmatrix}
0&0&-1\\0&\qumeta&\qumetb\\-1&\qumetb&\qumetc
\end{pmatrix};
\end{equation}
\item{(iii)} the quantities $\qumeta$, $\qumetb$, $\qumetc : \Gamma'_0 \rightarrow \quR^1$ are smooth and satisfy
\begin{equation}\label{gaugemetcond}
\qumeta > 0\hbox{ on }\Gamma'_0,\qquad \qumetb = \qumetc = \partial_\qucds \qumetc = 0\hbox{ on }\Gamma'_0 \cap \{ \qucds = 0 \}.
\end{equation}
\end{description}
\par
We note that, locally, any $2 + 1$ Lorentzian metric can be put in the form \eqref{gaugemetform} (see, e.g., \cite{thesis}, Section 1.2, for a sketch).
\par
The coordinate system $\qucds$, $\qucdx$, $\qucdv$ has an intrinsic geometric description. To give it we require the Christoffel symbols of $\qumeth$; see \cite{thesis}, Section 2.2 for the proof:
\par
\begin{lemma}\label{methChristoffel} The Christoffel symbols for a metric of the form \eqref{gaugemetform} are given by
\begin{align*}
\Gamma^0_{ij} &= \frac{1}{2a} \begin{pmatrix}
0 & b\qupas - a\qupbs & b\qupbs - a \qupcs\\
\intertext{}
\bullet & \begin{matrix}b\qupax\\-(b^2 - ac)\qupas\\ - a(2\qupbx - \qupav)\end{matrix} & \begin{matrix}-(b^2 - ac) \qupbs\\ + b\qupav - a \qupcx\end{matrix}\\
\intertext{}
\bullet & \bullet & \begin{matrix}-(b^2 - ac) \qupcs\\ + b(2\qupbv - \qupcx)\\ - a\qupcv\end{matrix}
\end{pmatrix}\\
\Gamma^1_{ij} &= \frac{1}{2a} \begin{pmatrix}
0&\qupas&\qupbs\\\bullet&\qupax - b\qupas&\qupav - b\qupbs\\\bullet&\bullet& 2\qupbv - \qupcx - b\qupcs
\end{pmatrix}\\
\Gamma^2_{ij} &= \frac{1}{2} \partial_\qucds \qumeth = \frac{1}{2} \begin{pmatrix}
0&0&0\\0&\qupas&\qupbs\\0&\qupbs&\qupcs
\end{pmatrix}.
\end{align*}
\end{lemma}
\par
\par
\begin{proposition}\label{gaugemetii} Let $\quaph$ be any Lorentzian metric satisfying conditions (i) -- (iii) above. Then on $\Gamma'_0$
\begin{description}
\item{(i)} Curves $\qucds = 0$, $\qucdx = \hbox{const}$ are null geodesics for $\quaph$ with affine parameter $\qucdv$;
\item{(ii)} Curves $\qucdx = \hbox{const}$, $\qucdv = \hbox{const}$ are null geodesics for $\quaph$ with affine parameter $\qucds$.
\end{description}
\end{proposition}
\par
\begin{proof} Note that (i) is equivalent to the requirement $\Gamma^i_{22} = 0$ for $i = 0, 1, 2$ when $\qucds = 0$, and this easily follows from Lemma \ref{methChristoffel} and the condition \eqref{gaugemetcond}. Similarly, (ii) is equivalent to the requirement $\Gamma^i_{00} = 0$ on $\Gamma'_0$, and this follows from Lemma \ref{methChristoffel}.\end{proof}
\par
See \quEEEfgaugechoice\ for a graphical depiction of the situation. Here $\lambda$ is the bifurcation curve $\qucds = \qucdv = 0$, while $\qulambdab$ and $\qulambdabb$ denote the geodesics in (i) and (ii) above.
\par
We note that curves $\qucds = \hbox{const} \neq 0$, $\qucdx = \hbox{const}$ are in general {\it not\/} geodesics for a metric satisfying (i) -- (iii). Thus, while sets of constant $\qucdv$ are always null for $\qumeth$, sets of constant nonzero $\qucds$ are not. See our discussion on the inappropriateness of the double null gauge in our setting in Section \ref{outconstr}.
\par
Let $\qupark \geq 1$ be the scaling parameter introduced in Section \ref{introduction}, and let $\queT$, $\queTp > 0$, $\queTp \leq \queTpupp$. In \quEEEenerineq\ we shall show that, under appropriate conditions, our solution for the metric $\qumeth$ satisfies, in addition to conditions (i) -- (iii) at the start of this subsection,
\begin{description}
\item{(iv)} There is a function
\begin{equation}\label{Bsigffdefeq}
\quBsigff : [0, \queT\sqrt{2}] \rightarrow [\queTp/2, \queTp],\hbox{ satisfying }\quBsigff(0) = \queTp,
\end{equation}
such that
\begin{multline}\label{Bsigffdef}
\hbox{$\quBsigf = \{ (\qucds, \qucdx, \qucdv) \in \hbox{dom}\,\qumeth\,\vert \,\qucds = \quBsigff(\qucdv) \}$ is smooth, and spacelike for $\qumeth$.}
\end{multline}
\end{description}
For now we assume that (iv) is satisfied and define
\begin{multline}\label{bulkdef}
\qubulkz = \qubulkz(\queT, \queTp) =\\
\{ (\qucds, \qucdx, \qucdv) \in \quR^3\,\vert \,\qucdx \in \quR^1, \qucdv \in [0, \queT\sqrt{2}], \qucds \in [0, \quBsigff(\qucdv)], \qupark^{-1} \qucds + \qucdv \leq \queT\sqrt{2} \},
\end{multline}
\begin{gather}
\quSigzz = \{ (\qucds, \qucdx, \qucdv) \in \qubulkz\,\vert \,\qucds = 0 \},\ququad \quUzz = \{ (\qucds, \qucdx, \qucdv) \in \qubulkz\,\vert \,\qucdv = 0 \}.\label{Sigzzdef}
\end{gather}
\par
Note for future reference that $\qumeth$ has determinant $-\qumeta$ and inverse with representation
\begin{equation}\label{hinvexp}
\qumeth^{ij} = \frac{1}{\qumeta} \begin{pmatrix}
 \qumetb^2 - \qumeta\qumetc & \qumetb & -\qumeta\\\qumetb&1&0\\ -\qumeta&0&0
\end{pmatrix} = \begin{pmatrix}
\frac{\qumetb^2}{\qumeta} - \qumetc & \frac{\qumetb}{\qumeta} & -1\\\frac{\qumetb}{\qumeta}&\frac{1}{\qumeta}&0\\ -1&0&0
\end{pmatrix}
\end{equation}
with respect to the $\qucds$, $\qucdx$, $\qucdv$ coordinate system.
\par
\subsection{Einstein equations in the null geodesic gauge}\label{Eeqgauge} The metric quantities $\qumeta$, $\qumetb$, and $\qumetc$ satisfy evolution equations which can be derived geometrically as follows. (In the next section, we shall show how these equations can also be obtained directly from the Ricci tensor $\quRic(\qumeth)$.) We work with respect to the $\qucds$, $\qucdx$, $\qucdv$ coordinate system throughout.
\par
Let $\quNb = \partial_\qucds$ be the tangent vector to the null geodesic congruence foliating $\qubulkz$. Define the tensor
\begin{equation*}
K_{ij} = \qunabla_i \quNb_j,
\end{equation*}
where $\qunabla$ is the covariant derivative operator arising from $\qumeth$. ($K_{ij}$ is effectively the second fundamental form corresponding to the foliation of $\qubulkz$ by the null hypersurfaces $\{ \qucdv = \hbox{const} \}$, except that we allow the indices $i$ and $j$ to run from $0$ to $2$.) We have
\begin{equation}\label{abcK}
\partial_\qucds \qumeth_{ij} = 2K_{ij},
\end{equation}
and $K_{ij}$ satisfies a Riccati evolution ODE in $\qucds$:
\par
\begin{proposition}\label{Keq} Assume that $\quRic(\qumeth) = 2\qunabla\qugamma\otimes\qunabla\qugamma$ on $\qubulkz$. Then, on $\qubulkz$, $K_{ij}$ satisfies
\begin{equation*}
\partial_\qucds K_{ij} = \qumeth^{kl} K_{ik} K_{lj} - 2\qumeth_{ij} \left(\partial_\qucds \qugamma\right)^2 + 2\qumeth_{i0} \partial_\qucds\qugamma \partial_j \qugamma + 2\qumeth_{0j} \partial_\qucds\qugamma \partial_i \qugamma - \qumeth^{kl} \partial_k\qugamma \partial_l\qugamma \qumeth_{j0} \qumeth_{i0},
\end{equation*}
or alternatively
\begin{equation}\label{Kriceq}
\partial_\qucds K_{ij} = \qumeth^{kl} K_{ik} K_{lj} - 2\qumeth_{ij} \left(\partial_\qucds\qugamma\right)^2 - 2\delta_{i2} \partial_\qucds \qugamma \partial_j \qugamma - 2 \delta_{j2} \partial_\qucds \qugamma \partial_i \qugamma - \delta_{i2} \delta_{j2} \qumeth^{kl} \partial_k \qugamma \partial_l\qugamma,
\end{equation}
where $\delta_{ij}$ is the Kronecker delta.
\end{proposition}
\par
\begin{proof} Let $K_i^j = \qumeth^{jk} K_{ik} = \qunabla_i \quNb^j$. Straightforward calculations then give
\begin{equation*}
\partial_\qucds K_i^j = (\qunabla_\quNb K)_i^j = -K_i^k K_k^j - R^j_{0i0},
\end{equation*}
where $R^i_{jkl}$ is the Riemann curvature tensor of $\qumeth$. From this and \eqref{abcK} we obtain
\begin{equation*}
\partial_\qucds K_{ij} = \qumeth^{kl} K_{ik} K_{lj} - R_{j0i0}.
\end{equation*}
Using the expression (\cite{wald}, (3.2.28)) 
\begin{equation*}
R_{ijkl} = C_{ijkl} + 2 (\qumeth_{i[k} R_{l]j} - \qumeth_{j[k} R_{l]i}) - R \qumeth_{i[k} \qumeth_{l]j}
\end{equation*}
and the fact that the Weyl tensor $C_{ijkl}$ vanishes in 3 dimensions, we obtain the first of the two equations above. The second follows by noting that $\qumeth_{i0} = \qumeth_{0i} = -\delta_{i2}$.\end{proof}
\par
A routine calculation then shows that $\qumeta$, $\qumetb$, and $\qumetc$ obey the following ODEs on $\qubulkz$:
\begin{align}
\partial_\qucds^2 \qumeta &= \frac{\left(\partial_\qucds \qumeta\right)^2}{2a} - 4\qumeta\left(\partial_\qucds\qugamma\right)^2\label{oricone}\\
\partial_\qucds^2 \qumetb &= \frac{1}{2\qumeta} \left(\partial_\qucds \qumeta\right)\left(\partial_\qucds \qumetb\right) - 4\partial_\qucds\qugamma \left(\qumetb\partial_\qucds\qugamma + \partial_\qucdx\qugamma\right)\label{orictwo}\\
\partial_\qucds^2 \qumetc &= \frac{(\qupbs)^2}{2\qumeta} - 2\qupgs\left(2\qupgv + \left(\frac{\qumetb^2}{\qumeta} + \qumetc\right) \qupgs + 2\frac{\qumetb}{\qumeta} \qupgx\right) - \frac{2}{\qumeta} (\qupgx)^2\label{oricthree}
\end{align}
Let $\qumetell = \sqrt{\qumeta}$; then $\qumetell$ satisfies
\begin{equation}\label{oriconep}
\partial_\qucds^2\qumetell = -2\qumetell(\partial_\qucds\qugamma)^2.
\end{equation}
With this change of variables, the three equations \eqref{oriconep}, \eqref{orictwo}, and \eqref{oricthree} are linear for $\qumetell$, $\qumetb$, and $\qumetc$ when solved in that order. We shall refer to these equations (or the equivalent system obtained by replacing \eqref{oriconep} by \eqref{oricone}) as the Riccati equations for the metric components. Recalling that in our gauge $\qumetb = \qumetc = \partial_\qucds \qumetc = 0$ on $\quSigzz$ (see \eqref{gaugemetcond}), complete initial data for this system is obtained by specifying
\begin{equation*}
\qumetell,\ququad \partial_\qucds \qumetell, \ququad\partial_\qucds \qumetb\hbox{ on $\quSigzz$}.
\end{equation*}
Note that \eqref{gaugemetcond} implies that $\qumetell|_{\quSigzz} > 0$.
\par
By another routine, if lengthy, calculation, we find that the wave equation $\quwbox_\qumeth \qugamma = 0$ becomes
\begin{multline}
\Biggl[\left(\frac{\qumetb^2}{\qumeta} - \qumetc\right)\partial_\qucds^2 + 2\frac{\qumetb}{\qumeta} \partial_\qucds\partial_\qucdx - 2\partial_\qucds\partial_\qucdv + \frac{1}{\qumeta}\partial_\qucdx^2\\
- \frac{1}{2}\left(\left(\frac{\qumetb^2}{\qumeta} + \qumetc\right)\frac{\partial_\qucds \qumeta}{\qumeta} - 4\frac{\qumetb}{\qumeta}\partial_\qucds \qumetb + \frac{\qumetb}{\qumeta^2} \partial_\qucdx \qumeta + 2\partial_\qucds \qumetc - \frac{2}{\qumeta} \partial_\qucdx \qumetb + \frac{\partial_\qucdv \qumeta}{\qumeta}\right)\partial_\qucds\\
\qquad- \frac{1}{2} \left(\frac{\qumetb}{\qumeta^2} \partial_\qucds \qumeta - \frac{2}{\qumeta} \partial_\qucds \qumetb + \frac{\partial_\qucdx \qumeta}{\qumeta^2}\right) \partial_\qucdx - \frac{1}{2} \frac{\partial_\qucds \qumeta}{\qumeta} \partial_\qucdv\Biggr] \qugamma = 0.\label{oowave}
\end{multline}
As initial data we specify $\qugamma$ on the two null hypersurfaces $\quSigzz = \qubulkz \cap \{ \qucds = 0 \}$ and $\quUzz = \qubulkz \cap \{ \qucdv = 0 \}$. These are subject to the consistency conditions (see \cite{rendall}, Section 4)
\begin{align}\label{consistencycond}
\begin{split}
\lim_{\qucdv \rightarrow 0^+} \left.\frac{\partial^\ell \qugamma}{\partial \qucds^\ell}\right\vert _{s = 0} &= \lim_{s \rightarrow 0^+} \left.\frac{\partial^\ell \qugamma}{\partial \qucds^\ell}\right\vert _{\qucdv = 0},\\
\lim_{\qucdv \rightarrow 0^+} \left.\frac{\partial^\ell \qugamma}{\partial \qucdv^\ell}\right\vert _{s = 0} &= \lim_{s \rightarrow 0^+} \left.\frac{\partial^\ell \qugamma}{\partial \qucdv^\ell}\right\vert _{\qucdv = 0},
\end{split}
\end{align}
where the degree $\ell$ takes values up to some $N$ depending on the Sobolev space in which we solve our equations, and the transverse derivatives in \eqref{consistencycond} are a priori values computed from the wave and Riccati equations, see \quEEEinitdat. We leave a treatment of the conditions \eqref{consistencycond} for \quEEEinitdat.
\par
\subsection{Constraint equations}\label{motconst} Let $R_{ij}$ denote the Ricci tensor for $\qumeth$ with respect to the $\qucds$, $\qucdx$, $\qucdv$ coordinate system. By writing out $R_{ij}$ explicitly, it is possible to identify three linear combinations of the equations $R_{ij} = 2\partial_i\qugamma\partial_j\qugamma$ which, taken together, are equivalent to the Riccati equations \eqref{oricone} -- \eqref{oricthree}. When \eqref{oricone} -- \eqref{oricthree} hold, all complementary linear combinations of the equations $R_{ij} = 2\partial_i\qugamma\partial_j\qugamma$ are equivalent, and any such combination can be taken as the set of constraint equations. We shall show that any such set of equations must hold on $\qubulkz$ whenever it holds on $\quSigzz$ and \eqref{oricone} -- \eqref{oricthree}, \eqref{oowave} hold on $\qubulkz$ and fix a particularly convenient set of constraint equations.
\par
In detail, a few pages of algebra (see \cite{thesis}, 2.2) shows that the Ricci tensor of a metric of the form \eqref{gaugemetform} is given by (recall that indices $0$, $1$, $2$ refer to $\qucds$, $\qucdx$, $\qucdv$)
\begin{gather}\label{Ricexp}
R_{00} = \frac{1}{2\qumeta}\left(-\qupass + \frac{(\qupas)^2}{2\qumeta}\right),\\
R_{01} = \frac{1}{2\qumeta}\left(\qumetb\qupass - \qumeta\qupbss - \frac{1}{2} \left(\qumetb\frac{(\qupas)^2}{\qumeta} - \qupbs\qupas\right)\right),\notag
\end{gather}
\begin{multline*}
R_{02} = \frac{1}{2\qumeta}\Biggl( \qumetb\qupbss - \qumeta\qupcss + \qupbsx - \qupasv + (\qupbs)^2\\
- \frac{\qumetb\qupbs\qupas}{2\qumeta} - \frac{\qupbs\qupax}{2\qumeta} + \frac{\qupas\qupav}{2\qumeta} - \frac{1}{2} \qupas\qupcs \Biggr),
\end{multline*}
\begin{multline*}
R_{11} = \frac{1}{2\qumeta}\Biggl( \qumeta\qupcs\qupas - (\qumetb^2 - \qumeta\qumetc) \qupass + \qupbs\qupax - \qumeta(2\qupbxs - 2\qupasv)\\
+ \frac{\qumetb^2(\qupas)^2}{2\qumeta} - \frac{1}{2} \qumetc(\qupas)^2 - \qupas\qupav - \qumeta(\qupbs)^2 \Biggr),
\end{multline*}
\begin{multline*}
 R_{12} =\frac{1}{2\qumeta}\Biggl( -\qumetb(\qupbs)^2 - (\qumetb^2 - \qumeta\qumetc)\qupbss + \qumetb(\qupasv - \qupbsx)\\
 + \frac{\qumetb^2 \qupas\qupbs}{2\qumeta} - \frac{\qumetb\qupav\qupas}{2\qumeta} + \frac{\qumetb\qupbs\qupax}{2\qumeta} + \qumeta(\qupbsv - \qupcxs) + \qumetb\qupav\qupcs\\
- \frac{1}{2} \qumetc\qupas\qupbs + \frac{1}{2} \qupas\qupcx + \frac{1}{2} \qupav\qupbs \Biggr),
\end{multline*}
\begin{multline*}
R_{22} = \frac{1}{2\qumeta}\Biggl( -(\qumetb^2 - \qumeta\qumetc) \qupcss + \qupbs\qupcx + 2\qumetb (\qupbvs - \qupcxs) + \frac{\qumetb^2 \qupas\qupcs}{2\qumeta}\\
- \frac{\qumetb\qupas}{2\qumeta} (2\qupbv - \qupcx) + 2\qupbvx - \qupcxx - \qupavv - \qupbx\qupcs - \frac{\qupax}{2\qumeta} (2\qupbv - \qupcx + \qumetb\qupcx)\\
+ \frac{(\qupav)^2}{2\qumeta} - \qumetc(\qupbs)^2 + \frac{1}{2} \qupas \qumetc \qupcs - \frac{1}{2} \qupas \qupcv + \frac{1}{2} \qupav\qupcs \Biggr).
\end{multline*}
Thus the equations $R_{00} = 2(\qupgs)^2$, $R_{01} = 2\qupgs\qupgx$, $R_{02} = 2\qupgs\qupgv$, and $R_{11} = 2(\qupgx)^2$ give, respectively,
\begin{gather*}
-\frac{1}{2\qumeta}\qupass + \frac{(\qupas)^2}{4\qumeta^2} = 2(\qupgs)^2,\\
-\frac{1}{2} \qupbss + \frac{b}{2\qumeta} \qupass - \frac{1}{4\qumeta}\left(\qumetb\frac{(\qupas)^2}{\qumeta} - \qupbs\qupas\right) = 2\qupgs\qupgx,
\end{gather*}
\begin{multline*}
-\frac{1}{2}\qupcss + \frac{1}{2\qumeta} \Biggl(\qumetb\qupbss + \qupbsx - \qupasv + (\qupbs)^2 - \frac{\qumetb\qupbs\qupas}{2\qumeta}\\
- \frac{\qupbs\qupax}{2\qumeta} + \frac{\qupas\qupav}{2\qumeta} - \frac{1}{2} \qupas\qupcs\Biggr) = 2\qupgs\qupgv
\end{multline*}
\begin{multline*}
-(\qupbxs - \qupavs) + \frac{\qupbs\qupax}{2\qumeta} - \frac{\qupas\qupav}{2\qumeta} + \frac{1}{2} \qupcs\qupas - \frac{1}{2a} (\qumetb^2 - \qumeta\qumetc)\qupass\\
 + \frac{\qumetb^2 (\qupas)^2}{4\qumeta^2} - \frac{1}{4\qumeta} \qumetc (\qupas)^2 - \frac{1}{2} (\qupbs)^2 = 2(\qupgx)^2.
\end{multline*}
As noted above, the system \eqref{oricone} -- \eqref{oricthree} is equivalent to three linear combinations of these four equations:
\par
\begin{proposition}\label{ricric} We have the following equivalences:
\begin{description}
\item{(i)} Equation \eqref{oricone} holds if and only if $R_{00} = 2(\qupgs)^2$.
\item{(ii)} Equations \eqref{oricone} and \eqref{orictwo} hold if and only if $R_{00} = 2(\qupgs)^2$ and $R_{01} = 2\qupgs\qupgx$.
\item{(iii)} Equations \eqref{oricone} -- \eqref{oricthree} hold if and only if $R_{00} = 2(\qupgs)^2$, $R_{01} = 2\qupgs\qupgx$, and $2R_{02} + \frac{1}{\qumeta} R_{11} = 2(2\qupgs\qupgv) + \frac{2}{\qumeta} (\qupgx)^2$.
\end{description}
\end{proposition}
\par
\begin{proof} (i) is clear; (ii) follows by writing out the equation for $2 \qumetb R_{00} + 2 R_{01}$; and (iii) follows by writing out the equation for
\begin{gather}
2\left(R_{02} + \qumetb \left[\frac{\qumetb}{\qumeta} R_{00} + \frac{1}{\qumeta} R_{01}\right]\right) + \frac{1}{\qumeta} \left( R_{11} - (\qumetb^2 - \qumeta\qumetc) R_{00}\right).
\end{gather}
\end{proof}
\par
Define the tensors
\begin{gather}\label{TSSbdef}
T_{ij} = 2\qupgi\qupgj - \qumeth_{ij} \qumeth^{k\ell} \partial_k \qugamma \partial_\ell \qugamma,\qquad S_{ij} = R_{ij} - \left(T_{ij} - \qumeth_{ij} T\right).
\end{gather}
Here $T_{ij}$ is the stress-energy tensor of the scalar field $\qugamma$, and $T = \qumeth^{ij} T_{ij}$ is its trace.
Note that for any $i$, $j$, $S_{ij} = 0$ if and only if $R_{ij} = 2\qupgi\qupgj$. Proposition \ref{ricric}\ implies that, at any point of $\qubulkz$, the system \eqref{oricone} -- \eqref{oricthree} is equivalent to the system
\begin{equation}\label{Ssyseq}
S_{00} = S_{01} = 2S_{02} + \frac{1}{\qumeta} S_{11} = 0.
\end{equation}
Thus, at any point of $\qubulkz$, $R_{ij} = 2\qupgi\qupgj$ is equivalent to \eqref{Ssyseq} together with either of the systems
\begin{equation}\label{cnstEqchoice}
S_{12} = S_{22} = S_{02} = 0,\qquad S_{12} = S_{22} = S_{11} = 0.
\end{equation}
The next proposition shows that, given the Riccati equations \eqref{oricone} -- \eqref{oricthree}, together with the wave equation \eqref{oowave} (to ensure $\qunabla^j T_{ij} = 0$), it is sufficient to impose either of the systems in \eqref{cnstEqchoice} on the initial outgoing null hypersurface $\quSigzz$ alone. In other words, this proposition gives propagation of constraints in our gauge.
\par
\begin{proposition}\label{constpres} Let $\quSu$ be an open subset of $\quSigzz$, let $\quTu = \{ (\qucds', \qucdx', \qucdv') \in \qubulkz\,\vert \,(\qucdx', \qucdv') \in \quSu \}$, and suppose that the system \eqref{oricone} -- \eqref{oricthree} and the wave equation \eqref{oowave} hold on $\quTu$. Suppose also that on $\quSu$ 
\begin{equation}\label{partconstr}
S_{12} = S_{22} = S_{02} = 0.
\end{equation}
Then \eqref{partconstr} holds on $\quTu$.
\end{proposition}
\par
\begin{proof} Since \eqref{oricone} -- \eqref{oricthree} hold on $\quTu$, \eqref{Ssyseq} must hold on $\quTu$, whence on $\quTu$ the tensor $S_{ij}$ has trace (with respect to $\qucds,\,\qucdx,\,\qucdv$)
\begin{equation}\label{Scalc}
S = \qumeth^{ij} S_{ij} = \left(\frac{\qumetb^2}{\qumeta} - \qumetc\right) S_{00} + 2\frac{\qumetb}{\qumeta} S_{01} - 2 S_{02} + \frac{1}{\qumeta} S_{11} = -4S_{02}.
\end{equation}
Further, the wave equation on $\quTu$ implies that $\qunabla^j T_{ij} = 0$ on $\quTu$, and since the Einstein tensor has identically vanishing divergence by the Bianchi identity, we have on $\quTu$
\begin{equation*}
\qunabla^j \left( S_{ij} - \frac{1}{2}\qumeth_{ij} S\right) = \qunabla^j \left( R_{ij} - \frac{1}{2} \qumeth_{ij} R - T_{ij}\right) = 0,
\end{equation*}
which together with \eqref{Scalc} gives
\begin{equation}\label{SijSeq}
\qunabla^j S_{ij} + 2\partial_i S_{02} = 0.
\end{equation}
By Lemma \ref{methChristoffel}\ and \eqref{hinvexp},
\begin{gather}
\Gamma^{\ell}_{00} = 0,\qquad \Gamma^2_{0i} = 0,\qquad \Gamma^1_{01} = \frac{\qupas}{2\qumeta},\qquad \qumeth^{jk} \Gamma^2_{jk} = \frac{\qupas}{2\qumeta},\label{Gammaeqns}\\
\qumeth^{ij} = \begin{pmatrix}
\frac{\qumetb^2}{\qumeta} - \qumetc & \frac{\qumetb}{\qumeta} & -1\\\frac{\qumetb}{\qumeta}&\frac{1}{\qumeta}&0\\ -1&0&0
\end{pmatrix},\label{consthinvexp}
\end{gather}
from which \eqref{SijSeq} with $i = 0$ gives, after some work,
\begin{align}\label{Sztp}
\begin{split}
0 &= \qumeth^{jk} \qunabla_k S_{0j} + 2\partial_\qucds S_{02}\\
&= \qumeth^{jk} \left( \partial_k S_{0j} - \Gamma^\ell_{k0} S_{\ell j} - \Gamma^\ell_{kj} S_{0\ell}\right) + 2\partial_\qucds S_{02}\\
&= -\partial_\qucds S_{02} - \frac{\qupas}{2\qumeta^2} S_{11} - \frac{\qupas}{2\qumeta} S_{02} + 2\partial_\qucds S_{02} = \partial_\qucds S_{02} + \frac{\qupas}{2\qumeta} S_{02}.
\end{split}
\end{align}
Since, by \eqref{partconstr}, $S_{02} = 0$ on $\quSu$, \eqref{Sztp} shows that $S_{02}$ must vanish on $\quTu$. By \eqref{Ssyseq}, $S_{11}$ must vanish on $\quTu$ as well.
\par
Equation \eqref{SijSeq} now becomes
\begin{equation*}
\qunabla^j S_{ij} = 0.
\end{equation*}
For $i = 1, 2$, using $S_{00} = S_{01} = S_{02} = S_{11} = 0$ on $\quTu$ and applying \eqref{Gammaeqns} and \eqref{consthinvexp}, we obtain after dropping terms which vanish on $\quTu$
\begin{equation}\label{Sitwoeq}
-\partial_\qucds S_{i2} - \frac{\partial_\qucds \qumeta}{2\qumeta} S_{i2} - \left(\Gamma^1_{0i} + \qumeth^{1k} \Gamma^2_{ik}\right) S_{12} = 0;
\end{equation}
\eqref{Sitwoeq} with $i = 1$, together with $S_{12}\vert _{\quSu} = 0$, shows that $S_{12}$ must vanish on $\quTu$, after which \eqref{Sitwoeq} with $i = 2$ together with $S_{22}\vert _{\quSu} = 0$ shows that $S_{22}$ must vanish on $\quTu$. This completes the proof.\end{proof}
\par
Applying the condition \eqref{gaugemetcond} to the expressions in \eqref{Ricexp}, we thus obtain the following corollary.
\par
\begin{corollary}\label{Ricconst} Suppose that on $\qubulkz$ the system \eqref{oricone} -- \eqref{oricthree}, \eqref{oowave} holds, 
and that on $\quSigzz$ the system
\begin{align}
-\qupavv + \frac{(\qupav)^2}{2\qumeta} &= 4\qumeta(\qupgv)^2\label{constrthree}\\
\qumeta\qupbvs + \frac{1}{2}\qupav\qupbs &= 4\qumeta\qupgx\qupgv\label{constrtwo}\\
\qupbs\qupax - 2\qumeta(\qupbxs - \qupavs) - \qupas\qupav - \qumeta(\qupbs)^2 &= 4\qumeta(\qupgx)^2\label{constrone}
\end{align}
holds. Then the equations \eqref{intRicwav}
\begin{equation*}
\quRic(\qumeth) = 2\qunabla\qugamma\otimes\qunabla\qugamma,\qquad \quwbox_\qumeth \qugamma = 0
\end{equation*}
hold on $\qubulkz$, and hence the metric $\qumetg$ in \eqref{ingexp} solves the vacuum Einstein equations $\quRic(\qumetg) = 0$ on $\qubulkz \times \qubS^1$.
\end{corollary}
\par
\begin{proof} By \eqref{gaugemetcond} and \eqref{Ricexp}, the system \eqref{constrthree} -- \eqref{constrone} is equivalent on $\quSigzz$ to the set of constraint equations $S_{22} = S_{12} = S_{11} = 0$. The result then follows from Proposition \ref{constpres}.\end{proof}
\par
\subsection{Scaling}\label{scalcut} Note that the Minkowski wave operator in $\qucds$, $\qucdx$, $\qucdv$ coordinates is given by
\begin{equation}
\quwbox_\eta = -2\partial_\qucds\partial_\qucdv + \partial_\qucdx^2.
\end{equation}
It is evident that the only scaling which preserves $\quwbox_\eta$ up to a multiplicative constant, and does not scale $\qucds$, must be of the form
\begin{equation}
\qucds \mapsto \qucds,\qquad\qucdx\mapsto\qupark^{1/2} \qucdx,\qquad\qucdv\mapsto\qupark \qucdv.
\end{equation}
It can be shown that the system \eqref{oricone} -- \eqref{oricthree}, \eqref{oowave} has an exact scaling symmetry given by the above coordinate scaling together with
\begin{equation}
\qumeta \mapsto \qumeta,\quad\qumetb\mapsto\qupark^{1/2}\qumetb,\quad\qumetc\mapsto\qupark\qumetc,\quad\qugamma\mapsto\qugamma.
\end{equation}
If we require instead, by analogy with \cite{christodoulou}, \cite{klainrod}, \cite{lukrod}, etc., that $\qugamma$ satisfy a scaling of the form
\begin{equation}
\qugamma\mapsto\qupark^{-\quiota}\qugamma,
\end{equation}
then the detailed structure of the system \eqref{oricone} -- \eqref{oricthree}, \eqref{oowave} suggests scaling behaviour for $\qumeta$, $\qumetb$, $\qumetc$ as in the following definition. Here and throughout the rest of the paper we fix some particular choice of exponent $\iota \geq 1/4$.
\par
\begin{definition}\label{scalquantdef} Let $\qupark \geq 1$. Define {\it scaled coordinates\/} $\qusb,\,\quxb,\,\quvb$ by
\begin{equation}\label{fscaldef}
\qusb = \qucds,\qquad \quxb = \qupark^{1/2} \qucdx,\qquad \quvb = \qupark \qucdv.
\end{equation}
Define {\it scaled quantities} $\quab$, $\qudab$, $\quellb$, $\qudlb$, $\qudaib$, $\qubb$, $\qucb$, and $\qugamb$ by the following equations:
\begin{gather}
\qumeta(\qucds, \qucdx, \qucdv) = \quab(\qusb, \quxb, \quvb) = 1 + \qupark^{-2\quiota} \qudab(\qusb, \quxb, \quvb),\label{fscalcdef}\\
\quellb = \sqrt{\quab} = 1 + \qupark^{-2\quiota} \qudlb(\qusb, \quxb, \quvb),\ququad 
\qudaib = \qupark^{2\quiota}\left(\quab^{-1} - 1\right) = -\qudab/\quab,\label{fscalcdefii}\\
\qumetb(\qucds, \qucdx, \qucdv) = \qupark^{1/2 - 2\quiota} \qubb(\qusb, \quxb, \quvb),\ququad \qumetc(\qucds, \qucdx, \qucdv) = \qupark^{1 - 2\quiota}\qucb(\qusb, \quxb, \quvb),\label{fscalcdefiii}\\
\qugamma(\qucds, \qucdx, \qucdv) = \qupark^{-\quiota} \qugamb(\qusb, \quxb, \quvb).\label{fscalgamdef}
\end{gather}
\end{definition}
\par
We will refer to this as the `blown-up picture', or the `scaled picture', for the remainder of this paper. We will ultimately show that, in the scaled picture, $\qudlb$, $\qubb$, $\qucb$, and $\qugamb$ have bounds independent of $\qupark$ (see \quEEEenerineq, especially \quEEEseqbound).
\par
The scaling in Definition \ref{scalquantdef} involves both a geometric scaling of the coordinates and a subsequent nongeometric scaling of $\qugamma$, $\qumeta$, $\qumetb$, and $\qumetc$. We define for convenience the {\it scaled metric\/} $\quhbar = \qupark\qumeth$, which together with its inverse is given with respect to $\qusb$, $\quxb$, $\quvb$ by
\begin{align}\label{hbardef}
\begin{split}
\quhbar_{ij} &= \begin{pmatrix}
0&0&-1\\0&1&0\\-1&0&0
\end{pmatrix} + \qupark^{-2\quiota} \begin{pmatrix}
0&0&0\\0&\qudab&\qubb\\0&\qubb&\qucb
\end{pmatrix}\\
&= \queta_{ij} + \qupark^{-2\quiota} \qutdh,\\
(\quhbar^{-1})^{ij} &= \begin{pmatrix}
0&0&-1\\0&1&0\\-1&0&0
\end{pmatrix} + \frac{\qupark^{-2\quiota}}{\quab} \begin{pmatrix}
\qupark^{-2\quiota} \qubb^2 - \quab\qucb & \qubb & 0\\ \qubb & -\qudab & 0 \\ 0 & 0 & 0 
\end{pmatrix}\\
&= \eta^{ij} + \qupark^{-2\quiota} \qutdhi,\\
\end{split}
\end{align}
where $\eta$ is the Minkowski metric and the second and fourth lines define $\qutdh$ and $\qutdhi$.
\par
Finally, we let $\quSigz$ and $\quUz$ denote the representations of the sets $\quSigzz$ and $\quUzz$, respectively, with respect to the scaled coordinates; thus
\begin{gather*}
\quSigz = \{ (0, \quxb, \quvb)\in\quR^3\,\vert \,\quxb\in\quR^1,\,\quvb \in [0, \qupark\queT\sqrt{2}]\},\\
\quUz = \{ (\qusb, \quxb, 0)\in\quR^3\,\vert \,\qusb \in [0, \queTp],\,\quxb \in \quR^1 \}.
\end{gather*}
We leave a detailed description of the bulk region in scaled coordinates, $\qubulk$, for \quEEEenerineq\ (see Definition \ref{longbulkdefin}).
\par
\subsection{Scaled equations}\label{scaledeqintro} In terms of the above variables, it is straightforward (if slightly lengthy) to show that the Riccati and wave equations \eqref{oricone} -- \eqref{oricthree}, \eqref{oowave} become (cf.\ \cite{thesis}, Sections 3.2, 3.3, for the case $\quiota = 1/2$)
\begin{align}
\partial_\qusb^2 \quab &= \frac{(\partial_\qusb\quab)^2}{2\quab} - 4\quab \qupark^{-2\quiota} (\partial_\qusb \qugamb)^2,\quad \partial_\qusb^2 \qudlb = -2\quellb (\partial_\qusb\qugamb)^2,\label{nsricone}\\
\partial_\qusb^2 \qubb &= \frac{1}{\quellb} \qupark^{-2\quiota} (\partial_\qusb \qudlb) (\partial_\qusb\qubb) - 4 \partial_\qusb\qugamb (\partial_\quxb \qugamb + \qupark^{-2\quiota} \qubb \partial_\qusb\qugamb),\kern 12pt\label{nsrictwo}\\
\partial_\qusb^2 \qucb &= -4 \partial_\qusb \qugamb \partial_\quvb \qugamb - \frac{2}{\quab} (\partial_\quxb\qugamb)^2\label{nsricthree}\\
&\kern -0.125in\notag + \qupark^{-2\quiota} \left( \frac{1}{2\quab} (\partial_\qusb \qubb)^2 - 2\qucb(\partial_\qusb\qugamb)^2 - \frac{4}{\quab} \qubb \partial_\quxb \qugamb \partial_\qusb \qugamb \right) - \qupark^{-4\quiota} \frac{2}{\quab} \qubb^2 (\partial_\qusb\qugamb)^2,
\end{align}
\begin{multline}
\Biggl\{\left(-2\partial_\qusb\partial_\quvb + \partial_\quxb^2\right) + \qupark^{-2\quiota} \Biggl[\qudaib \partial_\quxb^2 - \qucb \partial_\qusb^2 + 2\frac{\qubb}{\quab} \partial_\qusb \partial_\quxb\\
\shoveright{- \left(\partial_\qusb \qucb - \frac{1}{\quab} \partial_\quxb \qubb + \frac{\partial_\quvb\qudlb}{\qulb}\right)\partial_\qusb + \left(\frac{1}{\quab} \partial_\qusb \qubb - \frac{\qulb\partial_\quxb\qudlb}{\quab^2}\right)\partial_\quxb - \frac{\partial_\qusb \qudlb}{\qulb}\partial_\quvb\Biggr]}\\
\shoveleft{\qupark^{-4\quiota} \left(\frac{\qubb^2}{\quab} \partial_\qusb^2 - \left(\qucb \frac{\partial_\qusb \qudlb}{\qulb} - 2\frac{\qubb\partial_\qusb \qubb}{\quab} + \frac{\qubb\,\quellb \partial_\quxb \qudlb}{\quab^2}\right) \partial_\qusb - \frac{\qubb\,\qulb\partial_\qusb\qudlb}{\quab^2}\partial_\quxb\right)}\\
 - \qupark^{-6\quiota} \frac{\qubb^2\qulb\partial_\qusb\qudlb}{\quab^2}\partial_\qusb\Biggr\}\qugamb = 0.\label{swave}
\end{multline}
The constraint equations \eqref{constrthree} -- \eqref{constrone} become (cf.\ \cite{thesis}, Section 3.4)
\begin{align}
\partial_\quvb^2 \qudlb &= -2\quellb(\partial_\quvb\qugamb)^2,\label{scalconsts}\\
\partial_\quvb (\quellb \partial_\qusb \qubb) &= 4\quellb \partial_\quxb \qugamb \partial_\quvb \qugamb,\label{scalconste}\\
\quellb\cdot\qupbdlsv &= (\qupbgx)^2 + \frac{1}{2} \partial_\quxb \partial_\qusb \qubb + \frac{1}{4} \qupark^{-2\quiota} \left( (\partial_\qusb \qubb)^2 - \frac{2}{\quellb^3} \partial_\qusb \qubb \partial_\quxb \qudlb\right).\label{scalconsty}
\end{align}
These are precisely the evolution and constraint equations given in \eqref{insricone} -- \eqref{iscalconsty}.
\par
In \quEEEinitdat\ and \quEEEenerineq\ we shall show how to solve the constraint and evolution equations, respectively.
\par
\section{INTERLUDE: ANALYTIC PRELIMINARIES}\label{interlude}
\par
\subsection{Algebraic and calculus results}\label{algres} We have the following lemmata.
\par
\begin{lemma}\label{quadineq} Let $z$, $d$, $e \in \quRpos$ be positive real numbers satisfying
\begin{equation*}
z^2 \leq dz + e.
\end{equation*}
Then 
\begin{equation*}
z^2 \leq 2\left(d^2 + e\right).
\end{equation*}
\end{lemma}
\par
\begin{proof} The quadratic equation $z^2 - dz - e = 0$ has roots $\frac{1}{2} d \pm \frac{1}{2} \left(d^2 + 4e\right)^{1/2}$, one of which is positive and the other negative, and the positive root is bounded above by $d + e^{1/2}$.\end{proof}
\par
We define the Hilbert-Schmidt norm of any matrix $M$ (not necessarily square) by
\begin{equation*}
\quHS{M} = \left(\sum_{cd} \vert M_{cd}\vert ^2\right)^{1/2} = \left({\rm Tr\,}M^T M\right)^{1/2}
\end{equation*}
and recall that it possesses the following properties.
\par
\begin{lemma}\label{HSinequal} The Hilbert-Schmidt norm has the following properties:
\begin{description}
\item{(i)} If $V$ is a covector in some Euclidean space, $\vert V\vert $ denotes the Euclidean norm, and $(V_i V_j)$ denotes the matrix with elements $V_i V_j$, then $\quHS{V_i V_j} = \vert V\vert ^2$.
\item{(ii)} If $A$ and $B$ are two square matrices, then $\left\vert \hbox{Tr}\,AB\right\vert  \leq \quHS{A} \quHS{B}$.
\end{description}
\end{lemma}
\par
We shall need the following reordering lemma multiple times below.
\par
\begin{lemma}\label{ishuffleops} Let
\begin{equation*}
\quopsset = (\quops_1, \cdots, \quops_q),
\end{equation*}
$q \geq 2$, be a collection of (not necessarily first-order) differential operators satisfying
\begin{equation}\label{opssetbrackid}
[\quops_i, \quops_j] = \sum_{\ell = 1}^{q'} C^\ell_{ij} \quops_\ell,
\end{equation}
where the $C^\ell_{ij}$ are real numbers satisfying $\left\vert C^\ell_{ij}\right\vert  \leq \quCopss$ for some constant $\quCopss \geq 1$, and $q' \leq q$ (i.e., the derived algebra $[\quopsset, \quopsset]$ is spanned by $(\quops_1, \cdots, \quops_{q'})$). Let $\quindset = \{ 1, \cdots, n \}$, and let $\pi : \quindset \rightarrow \{ 1, \cdots, q \}$ be any map. Then for all $F$ and $\qux_0$ for which the necessary derivatives are defined, 
\begin{equation}\label{lishopresest}
\left\vert \left[\prod_{i = 1}^n \quops_{\pi(i)}\right] F(\qux_0)\right\vert  \leq \left\vert \quopsset^{I_0} F(\qux_0)\right\vert  + n! (q'\quCopss)^n \sum_{0 < \vert I\vert  < n} \left\vert \quopsset^I F(\qux_0)\right\vert ,
\end{equation}
where $I = (I_1, \cdots, I_q)$ in the sum is a $q$-multiindex, $\quopsset^I = \prod_{j = 1}^q \quops_j^{I_j}$, and the multiindex $I_0$ is such that $\quopsset^{I_0}$ is a reordering of $\prod_i \quops_{\pi(i)}$ (more precisely, $I_{0, j} = \vert \pi^{-1} \{ j \}\vert $ for $j = 1, \cdots, q$). Moreover, each multiindex $I$ in the sum can be taken to satisfy
\begin{equation}\label{lishopIcond}
I_p \leq I_{0, p}\ququad\hbox{for all}\ququad p > q',
\end{equation}
and if the $\quops_{q' + 1}, \cdots, \quops_q$ commute with each other, then the sum may be restricted to $I$ satisfying
\begin{equation}\label{lishopIconder}
\sum_{i = 1}^{q'} I_i \leq \sum_{i = 1}^{q'} I_{0, i}.
\end{equation}
\end{lemma}
\par
\begin{proof} This follows by a simple induction on $n$. For $n = 1$ the result is obvious (taking the empty sum to equal 0). Now suppose that the result holds for $n \leq N$, where $N \geq 1$. By \eqref{opssetbrackid}, if $1 \leq m \leq N$,
\begin{multline}\label{ishopbaseest}
\left\vert \quops_{\pi(1)} \cdots \quops_{\pi(N + 1)} F(\qux_0)\right\vert  \leq \left\vert \quops_{\pi(1)} \cdots \quops_{\pi(m + 1)} \quops_{\pi(m)} \cdots \quops_{\pi(N + 1)} F(\qux_0)\right\vert\\
+ \quCopss\sum_{j = 1}^{q'} \left\vert \quops_{\pi(1)} \cdots \quops_j \cdots \quops_{\pi(N + 1)} F(\qux_0)\right\vert,
\end{multline}
where $\quops_j$ in the last product stands where $\quops_{\pi(m)} \quops_{\pi(m + 1)}$ stood in the first, and we note that the operator in the last product is a product of at most $N$ elements of $\quopsset$. Let $p = \max \pi(\quindset)$. Applying \eqref{ishopbaseest} $N' \leq N$ times, we may rearrange $\prod_i \quops_{\pi(i)}$ to make $\quops_p$ appear as the rightmost operator. More precisely, there is a (not necessarily unique) collection of index maps $\qutpi_\ell : \{ 1, \cdots, N\} \rightarrow \{ 1, \cdots, q\}$, $\ell = 0, \cdots, M \leq q'N' \leq q'N$ (one for each term in the sum in \eqref{ishopbaseest}, for each of the $N'$ applications of \eqref{ishopbaseest}), with $(\qutpi_0(1), \cdots, \qutpi_0(N), p)$ a rearrangement of $(\pi(1), \cdots, \pi(N + 1))$, such that
\begin{multline}\label{ishopbaseestyi}
\left\vert \quops_{\pi(1)} \cdots \quops_{\pi(N + 1)} F(\qux_0)\right\vert  \leq \left\vert \prod_{i = 1}^N \quops_{\qutpi_0(i)} \quops_p F(\qux_0)\right\vert\\
 + \quCopss\sum_{\ell = 1}^M \left\vert \quops_{\qutpi_\ell(1)} \cdots \quops_{\qutpi_\ell(N)} F(\qux_0)\right\vert.
\end{multline}
Applying the induction hypothesis to each term on the right-hand side of \eqref{ishopbaseestyi}, we obtain
\begin{multline}
\left\vert \quops_{\pi(1)} \cdots \quops_{\pi(N + 1)} F(\qux_0)\right\vert\\
\leq \left\vert \quopsset^{I_0} F(\qux_0)\right\vert  + N! (q'\quCopss)^N \left[\sum_{0 < \vert I\vert  < N}\!\!\!\! \left\vert \quopsset^I \quops_p F(\qux_0)\right\vert  + \quCopss q'N\!\!\!\! \sum_{0 < \vert I\vert  \leq N}\!\!\!\! \left\vert \quopsset^I F(\qux_0)\right\vert \right]\\
\leq \left\vert \quopsset^{I_0} F(\qux_0)\right\vert  + (N + 1)! (q'\quCopss)^{N + 1} \sum_{0 < \vert I\vert  \leq N} \left\vert \quopsset^I F(\qux_0)\right\vert ,\label{ishopbaseester}
\end{multline}
where $\quopsset^{I_0}$ is a reordering of $\prod_i \quops_{\pi(i)}$. This completes the induction step and establishes \eqref{lishopresest}. \eqref{lishopIcond} follows by observing that \eqref{ishopbaseest} and \eqref{opssetbrackid} imply that each $\qutpi_\ell$ in \eqref{ishopbaseestyi} must satisfy, for $r > q'$,
\begin{equation}\label{lishoptpicond}
\vert \qutpi_\ell^{-1} \{ r\}\vert  \leq \vert \pi^{-1} \{ r \}\vert ,
\end{equation}
so by induction the multiindices $I$ in \eqref{ishopbaseester} can all be taken to satisfy \eqref{lishopIcond}. When $\quops_{q' + 1}, \cdots, \quops_q$ commute with each other, \eqref{lishopIconder} follows similarly by observing that in this case \eqref{ishopbaseest} implies that each $\qutpi_\ell$ in \eqref{ishopbaseestyi} must satisfy \eqref{lishoptpicond} for $r \leq q'$ as well.\end{proof}
\par
The following $L^\infty$ Sobolev spaces are useful when dealing with transverse derivatives, and will be used extensively in our treatment of the constraint equations in \quEEEinitdat.
\par
\begin{definition}\label{sbWspace} Let $\quElinset$ be any open set in some linear submanifold of $\quR^p$, and let $J$ denote a multiindex in derivatives tangent to $\quElinset$. For any subset $\quElinsubset \subset \quElinset$ and any $m \geq 0$, we define Sobolev norms on real-, $\quR^q$-, and matrix-valued functions on $\quElinsubset$ by
\begin{equation}\label{Winftydef}
\| F\| _{\qusbW^{m, \infty}(\quElinset\vert \quElinsubset)} = \sum_{\vert J\vert  \leq m} \left\|\left\vert \partial^J F\right\vert\right\| _{L^\infty(\quElinsubset)},
\end{equation}
where $\left\vert\partial^J F\right\vert$ denotes, respectively, the absolute value, $\quR^q$-norm, and Hilbert-Schmidt norm of $F$, respectively. We let $\qusbW^{m, \infty}(\quElinset\vert \quElinsubset)$ denote the set of all real-, $\quR^q$-, or matrix-valued functions on $\quElinsubset$ for which the corresponding quantity in \eqref{Winftydef} is defined and finite. For $\phi$ any (real-, $R^q$-, or matrix-valued) function on $[0, T] \times \quElinset \rightarrow \quR^1$, $t \in [0, T]$, and $n \geq 0$, we write as convenient shorthands
\begin{gather*}
\| \phi(t, \cdot)\| _{\qusbW^{m, \infty}(\quElinset\vert \quElinsubset)} = \left\| \phi\vert _{\{t\}\times\quElinset}\right\| _{\qusbW^{m, \infty}(\quElinset\vert \quElinsubset)},\\
\| \phi\| _{W^{n,\infty}W^{m,\infty}([0, T]\times \quElinset\vert \quElinsubset)} = \qusup_{t \in [0, T]} \sum_{i = 0}^n \| \partial_t^i \phi(t, \cdot)\| _{\qusbW^{m, \infty}(\quElinset\vert \quElinsubset)},\\
\| \phi\| _{W^{n,1}W^{m,\infty}([0, T]\times \quElinset\vert \quElinsubset)} = \int_0^T \sum_{i = 0}^n \| \partial_t^i \phi(t, \cdot)\| _{\qusbW^{m, \infty}(\quElinset\vert \quElinsubset)}.
\end{gather*}
As usual, we write $L^\infty$ instead of $W^{0, \infty}$.
\end{definition}
\par
We note the following elementary lemmata.
\par
\begin{lemma}\label{inverse} Let $\qux_0 \in \quR^n$, and let $g$ be a function on a neighbourhood of $\qux_0$, $g(\qux_0) \neq 0$. Let $p \geq 1$ and suppose that $I$ is any multiindex for which $\partial^I g(\qux_0)$ exists. Let $\mathcal{K}$ denote the set of all collections of multiindices $\{ K_k \}$ whose sum equals $I$. There is a collection of combinatorial constants $\{ C^p_{\{ K_k \}}\,\vert \, \{K_k\}\in\mathcal{K}\}$ such that at $\qux_0$
\begin{equation*}
\partial^I \frac{1}{g^p} = \sum_{\{K_k\} \in \mathcal{K}} C^p_{\{K_k\}} \frac{\prod\limits_{K \in \{ K_k \}} \partial^K g}{g^{p + \vert \{K_k\}\vert }}.
\end{equation*}
\end{lemma}
\par
\par
\begin{lemma}\label{Winftyalg} For any $m \geq 0$ there exist constants $C_m$ and $C^{\rm I}_m$ such that the following hold. If $f, g \in \qusbW^{m, \infty}(\quElinset\vert \quElinsubset)$, then
\begin{equation}\label{prodrulebound}
\| fg\| _{\qusbW^{m, \infty}(\quElinset\vert \quElinsubset)} \leq C_m \| f\| _{\qusbW^{m, \infty}(\quElinset\vert \quElinsubset)}\| g\| _{\qusbW^{m, \infty}(\quElinset\vert \quElinsubset)},
\end{equation}
for $f$, $g$ any real-, $\quR^q$-, or matrix-valued functions for which the product is defined. For any {\it real-valued\/} $f \in \qusbW^{m, \infty}(\quElinset\vert \quElinsubset)$ for which $\| 1/f\| _{L^\infty(\quElinsubset)}$ is finite,
\begin{equation*}
\left\| \frac{1}{f}\right\| _{\qusbW^{m, \infty}(\quElinset\vert \quElinsubset)} \leq C^{\rm I}_m \left(1 + \| f\| _{\qusbW^{m, \infty}(\quElinset\vert \quElinsubset)}\right)^m \left\| \frac{1}{f}\right\| _{L^\infty(\quElinsubset)}^{m + 1}.
\end{equation*}
\end{lemma}
\par
Let $M_{q\times q}(\quR)$, where $q$ is a positive integer, denote the set of $q \times q$ matrices over $\quR$. The following estimate follows from a Gr\"onwall estimate and an induction on the order of the Sobolev space.
\par
\begin{proposition}\label{refgronwallcont} Let $[0, T]$ be some compact interval in $\quR^1$ and $\quElinset \subset \quR^p$ be open, and suppose that $\qub, \quu : [0, T] \times \quElinset \rightarrow \quR^q$ and $\quM : [0, T] \times \quElinset \rightarrow M_{q\times q}(\quR)$ are $C^\infty$ functions on $[0, T]$ which satisfy
\begin{equation}\label{refgronwallde}
\qudu = \quM\quu + \qub
\end{equation}
on $(0, T) \times \quElinset$. Let $U \subset \quElinset$, $n \geq 1$, $m \geq 0$. Then there is a constant
\begin{equation}
C(n, m, T, \| \quM\| _{\qusbW^{n - 1, \infty} \qusbW^{m, \infty}([0, T] \times \quElinset\vert U)})
\end{equation}
depending only on $n$, $m$, $T$, and (an upper bound on)
\begin{equation}
\| \quM\| _{\qusbW^{n - 1, \infty} \qusbW^{m, \infty}([0, T] \times \quElinset\vert U)}
\end{equation}
such that
\begin{multline}\label{refgronwalleqyi}
\| \quu\| _{\qusbW^{n, \infty} \qusbW^{m, \infty} ([0, T] \times \quElinset\vert U)} \leq C(n, m, T, \| \quM\| _{\qusbW^{n - 1, \infty} \qusbW^{m, \infty}([0, T] \times \quElinset\vert U)})\\
\cdot \biggl\{ \| \quu(0, \cdot)\| _{\qusbW^{m, \infty}(\quElinset\vert U)} + \|\qub(s, \cdot)\| _{\qusbW^{n - 1, 1} \qusbW^{m, \infty}([0, T] \times \quElinset\vert U)} \biggr\}.
\end{multline}
\end{proposition}
\par
\subsection{Sobolev-type inequalities}\label{PSti} The spatial sections $\quAsigmat$ on which we define energies (see Definitions \ref{longbulkdefin}, \ref{difrerset}) have width that vanishes in the limit $\qusigmat \rightarrow 0$, and moreover depends on the existence time $\queTp$. Since an ordinary Sobolev inequality on such regions would have a constant diverging as $\qusigmat \rightarrow 0$ and depending on $\queTp$, we derive an alternate Sobolev inequality with bounded constant independent of $\qusigmat$ and $\queTp$ given auxiliary $L^2$ bounds on one of the boundaries (see \quEEEsobolevemb).
We then apply similar ideas to derive a Klainerman-Sobolev inequality for functions defined on a strip of the form $\quR^1 \times [\quKSk, \quKSk + \qudk]$ and which are bounded, together with various of their derivatives, in $L^\infty(\quR^1 \times \{ \quKSk + \qudk \})$ (see \quEEEklainsobolv).
\par
We first define the following Sobolev norms, cp.\ Definition \ref{sbWspace}.
\par
\begin{definition}\label{oHspace} Let $\quElinsetb$ be any open set in some linear submanifold of $\quR^p$ together with any portion (possibly empty) of its boundary, and let $J$ denote a multiindex in derivatives tangent to $\quElinsetb$. For $m \geq 1$ we define 
\begin{equation}\label{specsoboldefyi}
\| f\| _{\quoH^m(\quElinsetb)}^2 = \sum_{1 \leq \vert J\vert  \leq m} \| \partial^J f\| _{L^2(\quElinsetb)}^2,\qquad \| f\| _{\qusbH^m(\quElinsetb)}^2 = \| f\| _{\quoH^m(\quElinsetb)}^2 + \| f\| _{L^2(\quElinsetb)}^2.
\end{equation}
If $Z > 0$, we define $\Omega_Z = \quR^1 \times [-Z, Z]$, $\qupbar \Omega_Z = \quR^1 \times \{ Z \} \subset \partial\Omega_Z$, and obtain in particular
\begin{equation}\label{specsoboldefer}
\| f\| _{\qusbH^1(\qupbar\Omega_Z)}^2 = \int_{-\infty}^{+\infty} \vert f(x, Z)\vert ^2\,dx + \int_{-\infty}^{+\infty} \left\vert \partial_x f(x, Z)\right\vert ^2\,dx.
\end{equation}
\eqref{specsoboldefyi} -- \eqref{specsoboldefer} are understood to be for all $f$ for which the right-hand sides are defined.
\end{definition}
\par
\begin{remark} It should be noted that \quEEEsobolevemb\ is translation invariant, so we could just as well take $\Omega_Z = \quR^1 \times [C - Z, C + Z]$ for any $C \in \quR$. \end{remark}
\par
The proof of the next result is modified from \cite{adams}, Lemma 4.15. 
\par
\begin{lemma}\label{onedsobolev} Let $a, b \in \quR^1$, $a < b$, and let $f \in \qusbH^1([a, b])$. Then
\begin{equation}\label{onedintineq}
\| f\| _{L^\infty([a, b])} \leq \left[ \vert b - a\vert ^{-1} + \frac{4}{9} \vert b - a\vert \right]^{1/2} \| f\| _{\qusbH^1([a, b])}.
\end{equation}
Thus if $f \in \qusbH^1(\quR^1)$, then
\begin{equation}\label{onedlineineq}
\| f\| _{L^\infty(\quR^1)} \leq \quoCS \| f\| _{\qusbH^1(\quR^1)}.
\end{equation}
\end{lemma}
\par
\begin{proof} \eqref{onedlineineq} follows from \eqref{onedintineq} by restricting to an interval of length $3/2$. To prove \eqref{onedintineq}, let $x \in [a, b]$; then
\begin{multline*}
\vert b - a\vert  \vert f(x)\vert  \leq \int_a^b \left[ \vert f(y)\vert  + \int_y^x \vert f'(u)\vert \,du\right]\,dy \leq \vert b - a\vert ^{1/2} \| f\| _{L^2([a, b])}\\
+ \left[\int_a^b \vert y - x\vert ^{1/2}\,dy\right] \| f'\| _{L^2([a, b])};
\end{multline*}
since $\int_a^b \vert y - x\vert ^{1/2}\,dy = \frac{2}{3} \left[(b - x)^{3/2} + (x - a)^{3/2}\right] \leq \frac{2}{3} \vert b - a\vert ^{3/2}$, the result follows by applying the H\"older inequality.\end{proof}
\par
\begin{proposition}\label{sobolevemb} Let $Z > 0$, $Z \leq 1/2$. Then for all $f \in \quoH^2(\Omega_Z) \cap \qusbH^1(\qupbar \Omega_Z)$
\begin{equation}\label{poneres}
\| f\| _{L^\infty(\Omega_Z)} \leq \quoCS \left[\| f\| _{\quoH^2(\Omega_Z)} + \| f\| _{\qusbH^1(\qupbar\Omega_Z)}\right].
\end{equation}
\end{proposition}
\par
\begin{proof} It suffices to show this for $f \in C^\infty(\Omega_Z)$ having support which is compact in the first variable. Let $Z \leq 1/2$. For any $(x, y) \in \Omega_Z$, we have by Lemma \ref{onedsobolev}
\begin{align}\label{sobolevembeqi}
\vert f(x, y)\vert  &\leq \vert f(x, Z)\vert  + \vert f(x, y) - f(x, Z)\vert\\
&\leq \quoCS \left[\| f\| _{\qusbH^1(\qupbar \Omega_Z)} + \| f(\cdot, y) - f(\cdot, Z)\| _{\qusbH^1(\quR^1)}\right].\notag
\end{align}
Now if $g \in C^\infty([-Z, Z])$, then for any $y \in [-Z, Z]$
\begin{equation}\label{gdiffeq}
\vert g(y) - g(Z)\vert  \leq \int_{-Z}^Z \vert g'(y)\vert  dy \leq (2Z)^{1/2} \| g'(y)\| _{L^2([-Z, Z])},
\end{equation}
and using this in \eqref{sobolevembeqi} gives the desired result.\end{proof}
\par
Finally, we have the following Klainerman-Sobolev inequality (cf.\ \cite{klainerman}, Proposition 3). Let $I = (I_1, I_2, I_3)$ denote a three-variable multiindex, and let $\Omega = x\partial_y - y \partial_x$ be the generator of counterclockwise rotations around the origin in $\quR^2$.
\par
\begin{proposition}\label{klainsobolv} There is a universal constant $\quCrip \geq 1$ such that the following holds: if $\quKSk, \qudk \in \quR$, $\qudk > 0$, $\quKSk \geq 1$, $\qudk \leq 1$, then
\begin{multline}\label{kseqero}
\| f\| _{L^\infty(\quR^1 \times [\quKSk, \quKSk + \qudk])} \leq \quCrip \Biggl\{ \sum_{i = 0}^1 \left\| \Omega^i f\right\| _{L^\infty(\quR^1 \times \{ \quKSk + \qudk \})}\\
+ \qudk^{-3/4} (\quKSk + \qudk)^{-1/4} \Biggl[\sum_{\vert I\vert  \leq 2} \left\| \partial_x^{I_1} \partial_y^{I_2} \Omega^{I_3} f\right\| _{L^2(\quR^1\times [\quKSk, \quKSk + \qudk])}^2\Biggr]^{1/2}\Biggr\}
\end{multline}
for all $f : \quR^1 \times [\quKSk, \quKSk + \qudk] \rightarrow \quR^1$ for which the sum on the right-hand side exists and is finite. In particular, the constant $\quCrip$ is independent of $\quKSk$ and $\qudk$.
\end{proposition}
\par
\begin{proof} Note first that $\qudk \leq \quKSk$. This implies that
\begin{equation*}
(\quKSk + \qudk)^{-1/4} = \quKSk^{-1/4} (1 + \qudk/\quKSk)^{-1/4} \geq 2^{-1/4} \quKSk^{-1/4},
\end{equation*}
so it suffices to replace $(\quKSk + \qudk)^{-1/4}$ with $\quKSk^{-1/4}$ in \eqref{kseqero} and show instead the modified inequality
\begin{multline}\label{kseqer}
\| f\| _{L^\infty(\quR^1 \times [\quKSk, \quKSk + \qudk])} \leq \quCrip \Biggl\{ \sum_{i = 0}^1 \left\| \Omega^i f\right\| _{L^\infty(\quR^1 \times \{ \quKSk + \qudk \})}\\
+ \qudk^{-3/4} \quKSk^{-1/4} \Biggl[\sum_{\vert I\vert  \leq 2} \left\| \partial_x^{I_1} \partial_y^{I_2} \Omega^{I_3} f\right\| _{L^2(\quR^1\times [\quKSk, \quKSk + \qudk])}^2\Biggr]^{1/2}\Biggr\}.
\end{multline}
We shall show this by using techniques similar to those in Proposition \ref{sobolevemb}, but applied in {\it polar\/} coordinates. Let $f : \quspX \rightarrow \quR^1$ be bounded and satisfy
\begin{equation}\label{kseqsan}
\sum_{i = 0}^1 \left\| \Omega^i f\right\| _{L^\infty(\quR^1 \times \{ \quKSk + \qudk \})} < \infty,\qquad \sum_{\vert I\vert  \leq 2} \left\| \partial_x^{I_1} \partial_y^{I_2} \Omega^{I_3} f\right\| _{L^2(\quspX)}^2 < \infty.
\end{equation}
It is worth noting that the second inequality in \eqref{kseqsan} implies, by ordinary Sobolev embedding (see, e.g., \cite{adams}, Theorem 4.12), that $f$ is continuous on $\quspX$. Define polar coordinates $(r, \theta)$, $r > 0$, $\theta \in (0, 2\pi)$ on the cut plane $\quR^2_* = \quR^2 \backslash \{ (x, y) \in \quR^2\,\vert \,x \geq 0,\,y = 0 \} \supseteq \quspX$ by $x = r\cos\theta$, $y = r\qusin\theta$. The region $\quspX$ in rectangular coordinates corresponds to the polar coordinate region
\begin{equation}\label{ksUsdef}
U = \{ (r, \theta)\in\quR^2\,\vert \,\theta \in (0, \pi),\, \quKSk\csc\theta \leq r \leq (\quKSk + \qudk)\csc\theta \}
\end{equation}
while the set $\quR^1 \times \{ \quKSk + \qudk \}$ corresponds to (see \quEEEklainsobolevfigy)
\begin{equation}\label{kspbUdef}
\qupbar U = \{ (r, \theta) \in \quR^2\,\vert \, r = (\quKSk + \qudk)\csc\theta \}.
\end{equation}
By abuse of notation, we shall use $f$ to denote also the function $U \rightarrow \quR^1$, $(r, \theta) \mapsto f(r\cos\theta, r\qusin\theta)$. We claim that there is a constant $C'$, independent of $\quKSk$, such that
\begin{equation}\label{polSineq}
\| f\| _{L^\infty(U)} \leq C' \left[\sum_{i = 0}^1 \left\| \partial_\theta^i f\right\| _{L^\infty(\qupbar U)} + \delta^{-3/4} \quKSk^{1/4} \| f\| _{H^2(U)}\right],
\end{equation}
where $H^2(U)$ denotes the Sobolev space on $U$ with respect to the {\it coordinate\/} (not area!) measure $dr\,d\theta$ and the (polar) coordinate derivatives $\partial_r$, $\partial_\theta$. Granting \eqref{polSineq}, we show how to obtain \eqref{kseqer}. Clearly $\| f\| _{L^\infty(U)} = \| f\| _{L^\infty(\quspX)}$, while $\left\| \partial_\theta^i f\right\| _{L^\infty(\qupbar U)} = \left\| \Omega^i f\right\| _{L^\infty(\quR^1 \times \{ \quKSk + \qudk \})}$. Further, since for all $(x, y) \in \quspX$, $\sqrt{x^2 + y^2} \geq \quKSk \geq 1$, we have
\begin{align}
\left\vert \partial_r f\right\vert  = \left\vert \frac{1}{\sqrt{x^2 + y^2}} \left(x\partial_x + y\partial_y\right) f\right\vert  &\leq \left\vert \partial_x f\right\vert  + \left\vert \partial_y f\right\vert ,\label{prfer}\\
\left\vert \partial_r^2 f\right\vert ^2 &\leq 18 \sum_{j_1, j_2 \leq 2} \left\vert \partial_x^{j_1} \partial_y^{j_2} f\right\vert ^2,\label{prferr}\\
\left\vert \partial_\theta^i f\right\vert  &= \left\vert \Omega^i f\right\vert .\label{pthetf}
\end{align}
Since $\quKSk^{-1} r \geq 1$ on $U$, we thus obtain 
\begin{align}\label{fHTU}
\begin{split}
\| f\| _{H^2(U)}^2 &= \sum_{i, j \leq 2} \int_0^{\pi} \int_{\quKSk \csc\theta}^{(\quKSk + \qudk) \csc\theta} \left\vert \partial_r^i \partial_\theta^j f\right\vert ^2\,dr\,d\theta\\
&\leq \sum_{i, j \leq 2} \int_0^{\pi} \int_{\quKSk \csc\theta}^{(\quKSk + \qudk) \csc\theta} \left\vert \partial_r^i \partial_\theta^j f\right\vert ^2\,\quKSk^{-1} r\,dr\,d\theta\\
&\kern -0.5pt\leq 20 \quKSk^{-1} \sum_{\vert I\vert  \leq 2} \int_{-\infty}^{\infty} \int_\quKSk^{\quKSk + \qudk} \left\vert \partial_x^{I_1}\partial_y^{I_2} \Omega^{I_3} f\right\vert ^2\,dy\,dx\\
&= 20 \quKSk^{-1} \sum_{\vert I\vert  \leq 2} \left\| \partial_x^{I_1} \partial_y^{I_2} \Omega^{I_3} f\right\| _{L^2(\quspX)}^2.
\end{split}
\end{align}
\eqref{fHTU} combined with \eqref{polSineq} then gives \eqref{kseqer}.
\begin{figure}[h]\centering
\includegraphics[keepaspectratio]{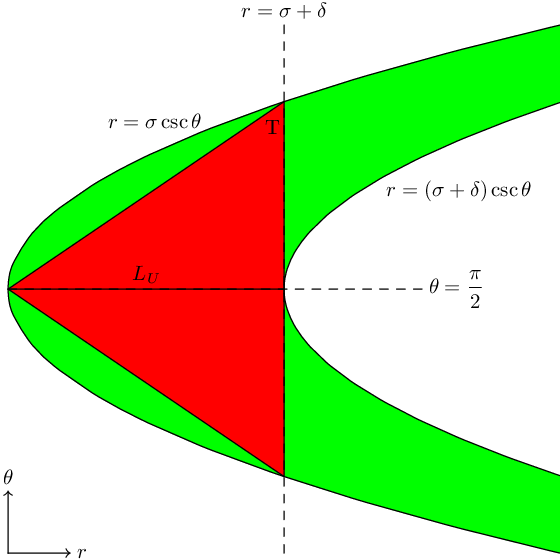}
\caption{Regions in the Klainerman-Sobolev inequality}\label{klainsobolevfigy}
\end{figure}
\par
For the rest of this proof we work entirely with respect to polar coordinates and take all $L^2$ and Sobolev spaces to be with respect to the coordinate measure $dr\,d\theta$ and coordinate derivatives $\partial_r$, $\partial_\theta$ (as applicable). We begin with some facts about $\csc$. We recall that $\csc \pi/2 = 1$, $\csc \pi/6 = \csc 5\pi/6 = 2$. Since the derivative of $\csc\theta$ vanishes at $\theta = \pi/2$, by considering the Taylor expansion of $\csc\theta$ about $\theta = \pi/2$ and restricting to the interval $[\pi/6, 5\pi/6] = [\pi/2 - \pi/3, \pi/2 + \pi/3]$ we see that there is some constant $C_C > 0$ such that for $\theta \in [\pi/6, 5\pi/6]$,
\begin{equation}\label{cscbound}
\csc \theta \leq 1 + C_C (\theta - \pi/2)^2.
\end{equation}
Taking $\quarccsc$ to be the branch mapping $[1, +\infty)$ into $(0, \pi/2]$ (which is decreasing on $[1, +\infty)$), we obtain from \eqref{cscbound} that for $x \in [0, 1]$
\begin{equation}\label{arccscbound}
\quarccsc (1 + x) \leq \pi/2 - \sqrt{\frac{x}{C_C}}.
\end{equation}
\par
We now bound $f$ on the line segment (see \quEEEklainsobolevfigy)
\begin{equation*}
\quUln = \{ (r, \theta) \in U\,\vert \,\theta = \pi/2 \}.
\end{equation*}
Let $\qutri \subset U$ be the triangle with vertices (see \quEEEklainsobolevfigy)
\begin{equation*}
(\quKSk, \pi/2),\ququad (\quKSk + \qudk, \quarccsc (1 + \qudk/\quKSk)), (\quKSk + \qudk, \pi - \quarccsc (1 + \qudk/\quKSk)),
\end{equation*}
where as above we use the branch of $\quarccsc$ mapping $[1, +\infty)$ into $(0, \pi/2]$. By \eqref{arccscbound}, for $\qudk \leq \quKSk$, $\qutri$ will be a triangle of height $\qudk$ and base bounded by a constant multiple of $\sqrt{\qudk/\quKSk}$, which suggests that the Sobolev constant on $\qutri$ will involve $\qudk$ and $\quKSk$. (It should also be noted that the apex angle of $\qutri$ is of size $\sim \qudk^{-1/2} \quKSk^{-1/2}$, so that $\qutri$ is very narrow for $\quKSk \gg 1/\qudk$.) We make this precise by using a scaling argument. Let $\qudthet = [\pi/2 - \quarccsc(1 + \qudk/\quKSk)]/2$ (so that $0 < \qudthet < 1$) and define scaled coordinates $\qursc$, $\quthetsc$ by
\begin{equation*}
\qursc = \qudk^{-1} r,\ququad \quthetsc = \qudthet^{-1} (\theta - \pi/2),
\end{equation*}
so that the vertices of $\qutri$ are given with respect to the $\qursc$, $\quthetsc$ system by
\begin{equation*}
(\qudk^{-1} \quKSk, 0),\ququad (\qudk^{-1} \quKSk + 1, -2),\ququad (\qudk^{-1} \quKSk + 1, 2),
\end{equation*}
i.e., in the $\qursc$, $\quthetsc$ system the triangle $\qutri$ is represented by an isosceles triangle with sidelengths $4$ and $\sqrt{5}$. By the standard Sobolev embedding theorem (see, e.g., \cite{adams}, Theorem 4.12 I(A)) applied in the $\qursc$, $\quthetsc$ coordinate system, and scaling properties of Sobolev spaces, there is a constant $C_T > 0$, independent of $\quKSk$ and $\qudk$, such that (since $\qudk, \qudthet \leq 1$)
\begin{equation*}
\| f\| _{L^\infty(\qutri)} \leq C_T \qudk^{-1/2} \qudthet^{-1/2} \| f\| _{H^2(\qutri)}.
\end{equation*}
By \eqref{arccscbound},
\begin{equation*}
\qudthet = \frac{1}{2} \left[ \frac{\pi}{2} - \quarccsc \left(1 + \frac{\qudk}{\quKSk}\right)\right] \geq \frac{1}{2\sqrt{C_C}} \sqrt{\frac{\qudk}{\quKSk}},
\end{equation*}
and we thus obtain for some constant $C_L > 0$
\begin{equation}\label{fLinfUln}
\| f\| _{L^\infty(\quUln)} \leq \| f\| _{L^\infty(\qutri)} \leq C_L \qudk^{-3/4} \quKSk^{1/4} \| f\| _{H^2(U)}.
\end{equation}
\begin{figure}[h]\centering
\includegraphics[keepaspectratio]{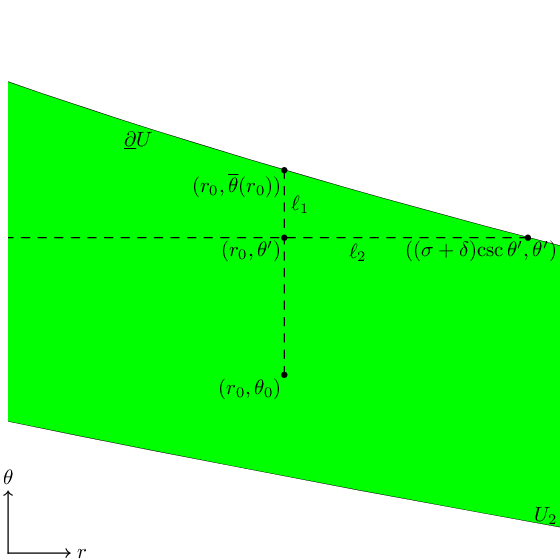}
\caption{Detailed view}\label{klainsobolevfige}
\end{figure}
\par
To complete the proof, we must apply two separate Sobolev embeddings on lines. First note that, by Lemma \ref{onedsobolev}, for any interval $[a, b] \subset \quR^1$ with $b - a \geq 1$, and any function $f_0 \in H^1([a, b])$,
\begin{equation}\label{Ctydef}
\| f_0\| _{L^\infty([a, b])} \leq \| f_0\| _{H^1([a, b])}.
\end{equation}
Second, as in \eqref{gdiffeq}, if $b - a \leq 1$ we have also
\begin{equation}\label{Ctysdef}
\| f_0\| _{L^\infty([a, b])} \leq \left\vert f_0(b)\right\vert  + \| f_0'\| _{L^2([a, b])}.
\end{equation}
Combining \eqref{Ctydef} and \eqref{Ctysdef}, we obtain
\begin{equation}\label{yidSemb}
\| f_0\| _{L^\infty([a, b])} \leq \left\vert f_0(b)\right\vert  + \| f_0\| _{H^1([a, b])}
\end{equation}
for any interval $[a, b]$. In \eqref{Ctysdef} and hence \eqref{yidSemb} $f_0(b)$ can be replaced by $f_0(a)$; we refer to the choice of $a$ or $b$ as the {\it endpoint\/} for which \eqref{yidSemb} is applied. The idea is to apply \eqref{yidSemb} twice, along perpendicular lines like $\ell_1$ and $\ell_2$ in \quEEEklainsobolevfige.
\par
Thus, fix some point $(\qurz, \quthetaz) \in U$, and assume without loss of generality that $\quthetaz \leq \pi/2$. Define the function $\quthetab$ on $[\quKSk, \infty)$ by
\begin{equation*}
\quthetab(r) = \begin{cases}
\quarcsin \displaystyle{\frac{\quKSk + \qudk}{r}},&\ququad r \geq \quKSk + \qudk,\\ \pi/2,&\ququad \quKSk \leq r < \quKSk + \qudk,
\end{cases}
\end{equation*}
where we use the branch of $\quarcsin$ mapping $[0, 1]$ into $[0, \pi/2]$ (consistent with our choice of branch for $\quarccsc$ above). Note that the graph of $\quthetab$ in the $r,\,\theta$ plane is that portion of $\qupbar U$ where $\theta \leq \pi/2$ together with the line $\quUln$, see \quEEEklainsobolevfigy, and that
\begin{equation*}
r \leq (\quKSk + \qudk) \csc \quthetab(r)\ququad\hbox{for all $r \geq \quKSk$}.
\end{equation*}
Applying \eqref{yidSemb} to the function $\theta \mapsto f(\qurz, \theta)$ on the segment $[\quthetab(\qurz), \quthetaz]$ with $\quthetab(\qurz)$ as endpoint gives
\begin{equation}\label{fineqyiver}
\left\vert f(\qurz, \quthetaz)\right\vert  \leq \left\| f\right\| _{L^\infty(\qupbar U \cup \quUln)} + \left\| f(\qurz, \cdot)\right\| _{H^1([\quthetab(\qurz), \quthetaz])}.
\end{equation}
Now fix a $\theta' \in [\quthetab(\qurz), \quthetaz]$. Note first that
\begin{equation*}
\quKSk \csc\theta' \leq \quKSk\csc\quthetaz \leq \qurz \leq (\quKSk + \qudk) \csc \quthetab(\qurz) \leq (\quKSk + \qudk) \csc \theta',
\end{equation*}
so that by applying \eqref{yidSemb} to $r \mapsto f(r, \theta')$ on $[\quKSk \csc\theta', (\quKSk + \qudk) \csc\theta']$ with $(\quKSk + \qudk) \csc\theta'$ as endpoint we obtain ($i = 0, 1$)
\begin{equation}\label{fineqerver}
\left\vert \partial_\theta^i f(\qurz, \theta')\right\vert  \leq \left\| \partial_\theta^i f\right\| _{L^\infty(\qupbar U)} + \left\| \partial_\theta^i f(\cdot, \theta')\right\| _{H^1([\quKSk \csc\theta', (\quKSk + \qudk)\csc\theta'])};
\end{equation}
note that $((\quKSk + \qudk)\csc\theta', \theta') \in \qupbar U$, so we do not need bounds on $\partial_\theta f$ on the line $\quUln$ (which we do not have). Combining this with \eqref{fineqyiver} and using $\vert \quthetaz - \quthetab(\qurz)\vert  \leq \pi$ to bound the length of $[\quthetab(\qurz), \quthetaz]$ gives that for some constant $C > 0$
\begin{equation*}
\left\vert f(\qurz, \quthetaz)\right\vert  \leq \| f\| _{L^\infty(\qupbar U \cup \quUln)} + C \left\| \partial_\theta f\right\| _{L^\infty(\qupbar U)} + C \| f\| _{H^2(U)}.
\end{equation*}
Using the bound on $\| f\| _{L^\infty(\quUln)}$ in \eqref{fLinfUln} together with $\qudk \leq 1$, $\quKSk \geq 1$, we thus obtain \eqref{polSineq} and hence \eqref{kseqer}.\end{proof}
\par
\begin{remark} Consideration of functions of the form $(\quKSk + \qudk - r)^n$ on $\{ (r, \theta) \in U\,\vert \,\quKSk \csc\theta \leq r \leq \quKSk + \qudk \}$ suggests that the factor of $\qudk^{-3/4} (\quKSk + \qudk)^{-1/4}$ is optimal given the hypotheses of the theorem. It should be noted -- as suggested by this example -- that this factor comes from bounding $f$ on the portion of the line $\quUln$ near $r = \quKSk$, or more roughly, from bounding $f$ on the `tip region' $r \sim \quKSk$ in $U$. The standard form of the Klainerman-Sobolev inequality obtains (in two dimensions) a factor of $(\quKSk + \qudk)^{-1/2}$ instead of $(\quKSk + \qudk)^{-1/4}$ since it applies to regions with rotational symmetry, such as the entire plane or an annular region, which do not have sharp `tips' when represented in polar coordinates.\end{remark}
\par
\section{CONSTRUCTION OF INITIAL DATA}\label{initdat}
\subsection{Statement of results}\label{initdatintro} We continue by obtaining initial data for the Riccati-wave system \eqref{nsricone} -- \eqref{swave} as described in Theorem \ref{iseqbound}. We refer the reader to Subsection \ref{outconstr} for an outline of the procedure.
\par
We shall use the following norms: if $Y$ is one of $\quSigz$, $\quUz$, and $\quSigz \cup \quUz$, $\quElinsubsetp \subset Y$, and $\quley$, $\qulee$, $\qules \in \quN$, $\quley$, $\qulee$, $\qules \geq 0$, we define
\begin{equation}\label{sbWsbxbvbdef}
\| f\| _{\qusbW_\qusb^\quley \qusbW_\quxb^\qulee \qusbW_\quvb^\qules(Y\vert \quElinsubsetp)} = \sum_{{\quley'} = 0}^\quley \sum_{{\qulee'} = 0}^\qulee \sum_{{\qules'} = 0}^\qules \| \partial_\qusb^{\quley'} \partial_\quxb^{\qulee'} \partial_\quvb^{\qules'} f\| _{L^\infty(\quElinsubsetp)}
\end{equation}
for all $f$ for which the right-hand side exists and is finite. If any of $\quley$, $\qulee$, or $\qules$ is equal to zero, we drop the corresponding $\qusbW$ from the notation; we replace $Y\vert Y$ with $Y$.
\par
We shall prove the following result. Note that quantities such as $\partial_\qusb^\quley \qudlb$ on $\quSigz$, which appear in the proposition, are in general not (yet) derivatives of the function $\qudlb$. We nevertheless identify $\partial_\qusb^0 \qudlb$ with $\qudlb$.
\par
\begin{proposition}\label{initdatbounds} Let $\quupey$, $\quupee$, $\quupes \in \quN$, $\quupey \geq 3$, $\quupee$, $\quupes \geq 2$. Let $\quginit \in C^\infty(\quR^2)$ be supported on $(0, 1) \times (0, 1)$. Then there is a unique collection of functions, which we denote $\partial_\qusb^\quley \qudlb$, $\partial_\qusb^\quley \qubb$, $\partial_\qusb^\quley \qucb$, $\partial_\qusb^\quley \qugamb$, $0 \leq \quley \leq \quupey$, on $\quSigz$, and $\partial_\quvb^\qules \qudlb$, $\partial_\quvb^\qules \qubb$, $\partial_\quvb^\qules \qucb$, $\partial_\quvb^\qules \qugamb$, $0 \leq \qules \leq \quupes$, on $\quUz$, which satisfy the following conditions:
\begin{description}
\item{(i)} $\qugamb|_{\quSigz} = \quginit$;
\item{(ii)} $\qugamb|_{\quUz} = \sum_{\quley = 0}^{\quupey} \frac{1}{\quley!} \qusb^\quley \partial_\qusb^\quley \qugamb(0, \quxb, 0)$;
\item{(iii)}\label{idsolnsupp} $\partial_\qusb^\quley \qudlb$, $\partial_\qusb^\quley \qubb$, $\partial_\qusb^\quley \qucb$, $\partial_\qusb^\quley \qugamb$, $0 \leq \quley \leq \quupey$,  vanish on $\{ (0, \quxb, \quvb) \in \quSigz\,|\,\quvb \geq 1 \}$;
\item{(iv)} $\qudlb$, $\partial_\qusb \qudlb$, and $\partial_\qusb \qubb$ satisfy the constraint equations \eqref{scalconsts} -- \eqref{scalconsty} on $\quSigz$, while $\qubb = \qucb = \partial_\qusb \qucb = 0$ on $\quSigz$;
\item{(v)} on $\quSigz$, $\partial_\qusb^\quley \partial_\quxb^\qulee \partial_\quvb^\qules \qudlb$, $\partial_\qusb^\quley \partial_\quxb^\qulee \partial_\quvb^\qules \qubb$, $\partial_\qusb^\quley \partial_\quxb^\qulee \partial_\quvb^\qules \qucb$, $2 \leq \quley \leq \quupey$, $0 \leq \qulee \leq \quupee$, $0 \leq \qules \leq \quupes$, and $\partial_\qusb^\quley \partial_\quxb^\qulee \partial_\quvb^\qules \qugamb$, $1 \leq \quley \leq \quupey$, $0 \leq \qulee \leq \quupee$, $0 \leq \qules \leq \quupes$, satisfy the $\partial_\qusb^{\quley - 2} \partial_\quxb^\qulee \partial_\quvb^\qules$-differentiated Riccati equations and $\partial_\qusb^{\quley - 1} \partial_\quxb^\qulee \partial_\quvb^\qules$-differentiated wave equation (see \eqref{nsricone} -- \eqref{nsricthree} and \eqref{swave}), respectively;
\item{(vi)} on $\quUz$, $\partial_\qusb^\quley \partial_\quxb^\qulee \partial_\quvb^\qules \qudlb$, $\partial_\qusb^\quley \partial_\quxb^\qulee \partial_\quvb^\qules \qubb$, $\partial_\qusb^\quley \partial_\quxb^\qulee \partial_\quvb^\qules \qucb$, $0 \leq \quley \leq \quupey$, $0 \leq \qulee \leq \quupee$, $0 \leq \qules \leq \quupes$, and $\partial_\qusb^\quley \partial_\quxb^\qulee \partial_\quvb^\qules \qugamb$, $0 \leq \quley \leq \quupey$, $0 \leq \qulee \leq \quupee$, $1 \leq \qules \leq \quupes$, satisfy the $\partial_\qusb^{\quley} \partial_\quxb^\qulee \partial_\quvb^\qules$-differentiated Riccati equations and $\partial_\qusb^\quley \partial_\quxb^\qulee \partial_\quvb^{\qules - 1}$-differentiated wave equation, respectively. 
\end{description}
Moreover, there is a constant $C = C(\quupey, \quupee, \quupes) > 1$ depending only on $\quupey$, $\quupee$, and $\quupes$, such that if $\quginit$ satisfies
\begin{equation}\label{basequginitsmall}
\|\quginit\|_{\qusbW_\quxb^\quupeepppnp \qusbW_\quvb^\quupesp((0, 1) \times (0, 1))} \leq 1/(4C),
\end{equation}
then the functions $\partial_\qusb^\quley \qudlb$, $\partial_\qusb^\quley \qubb$, $\partial_\qusb^\quley \qucb$, $\partial_\qusb^\quley \qugamb$, $0 \leq \quley \leq \quupey$ on $\quSigz$ and $\partial_\quvb^\qules \qugamb$, $0 \leq \qules \leq \quupes$ on $\quUz$ satisfy the following bounds: 
\begin{gather}\label{stuffdelbound}
\|\qudlb\|_{\qusbW_\qusb^\quupey \qusbW_\quxb^\quupee \qusbW_\quvb^\quupes(\quSigz)}, \|\qubb\|_{\qusbW_\qusb^\quupey \qusbW_\quxb^\quupee \qusbW_\quvb^\quupes(\quSigz)}, \|\qucb\|_{\qusbW_\qusb^\quupey \qusbW_\quxb^\quupee \qusbW_\quvb^\quupes(\quSigz)},\\
\|\qugamb\|_{\qusbW_\qusb^\quupey \qusbW_\quxb^\quupee \qusbW_\quvb^\quupes(\quSigz \cup \quUz)} \leq C(\quupey, \quupee, \quupes) \|\quginit\|_{\qusbW_\quxb^\quupeepppnp \qusbW_\quvb^\quupesp((0, 1) \times (0, 1))}.\notag
\end{gather}
\end{proposition}
\par
This section is devoted to a proof of Proposition \ref{initdatbounds}, which we complete in Subsection \ref{initdatbdproof}.
\par
We now fix a choice of $\quginit \in C^\infty(\quR^2)$ supported on $(0, 1) \times (0, 1)$, as well as a choice of exponents $\quupey$, $\quupee$, $\quupes \in \quN$, $\quupey \geq 3$, $\quupee$, $\quupes \geq 2$. Further, in accordance with (i) above, we define $\qugamb$ on $\quSigz$ by
\begin{equation}\label{qugambinitdatinitdef}
\qugamb|_{\quSigz}(0, \quxb, \quvb) = \quginit(\quxb, \quvb).
\end{equation}
Finally, we point out that the bounds we achieve can be sharpened to {\it linewise\/} bounds on lines of the form ($\qutxb \in \quR^1$)
\begin{equation*}
\quXbl_\qutxb = \{ (0, \quxb, \quvb) \in \quSigz\,\vert \,\quxb = \qutxb \},\qquad \quXXbl_\qutxb = \{ (\qusb, \quxb, 0) \in \quUz\,\vert \,\quxb = \qutxb \}.
\end{equation*}
\par
\subsection{Bounds on the initial data: the hypersurface $\quSigz$}\label{sinitdat} We begin by obtaining bounds on solutions to the constraint equations \eqref{scalconsts} -- \eqref{scalconsty}, rewritten as follows:
\begin{align}
\partial_\quvb^2 \qudlb &= -2\qupark^{-2\quiota} (\partial_\quvb \qugamb)^2 \qudlb - 2(\partial_\quvb \qugamb)^2\label{dlbeqone}\\
\partial_\quvb (\partial_\qusb \qubb) &= -\qupark^{-2\quiota} \frac{\partial_\quvb \qudlb}{\quellb} \partial_\qusb \qubb + 4\partial_\quxb \qugamb \partial_\quvb \qugamb\label{psbbeqone}\\
\partial_\quvb (\partial_\qusb \qudlb) &= \frac{1}{\quellb} (\partial_\quxb \qugamb)^2 + \frac{1}{2\quellb} \partial_\quxb \partial_\qusb \qubb + \frac{1}{4\quellb} \qupark^{-2\quiota} \left((\partial_\qusb \qubb)^2 - \frac{2}{\quellb^3} \partial_\qusb \qubb \partial_\quxb \qudlb\right),\label{psdlbeqone}
\end{align}
where $\qugamb$ is as defined in \eqref{qugambinitdatinitdef}.
We first bound $\qudlb$ so as to bound $1/\quellb$:
\par
\begin{proposition}\label{sclinitboundsi} Let $\qudlb : \quSigz \rightarrow \quR^1$ satisfy \eqref{dlbeqone} as well as the initial conditions
\begin{equation}\label{qudlbcinitdatcond}
\qudlb(0, \qutxb, 1) = \partial_\quvb \qudlb(0, \qutxb, 1) = 0.
\end{equation}
Then there is a constant $C(\| \quginit\| _{\qusbW^1_\quvb(\quSigz)})$, depending only on an upper bound on $\| \quginit\| _{\qusbW^1_\quvb(\quSigz)}$, such that
\begin{equation}\label{qudlbchulinf}
\| \qudlb\| _{L^\infty(\quSigz)} \leq C(\| \quginit\| _{\qusbW^1_\quvb(\quSigz)}) \| \quginit\| ^2_{\qusbW^1_\quvb(\quSigz)}.
\end{equation}
In particular, for $\| \quginit\| _{\qusbW^1_\quvb(\quSigz)} \leq \min \{ 1, 1/(4C(1)) \}$, we have
\begin{equation}\label{ellbinvlb}
\left\| \qudlb\right\| _{L^\infty(\quSigz)} < 1/4,\ququad \| \quellb^{-1}\| _{L^\infty(\quSigz)} < 4/3.
\end{equation}
\end{proposition}
\par
\begin{proof} \eqref{qudlbchulinf} follows easily from Proposition \ref{refgronwallcont} applied to \eqref{dlbeqone}, and \eqref{ellbinvlb} follows immediately from \eqref{qudlbchulinf} given the bound on $\quginit$ and using $\qupark \geq 1$.\end{proof}
\par
We now construct the remaining functions $\partial_\qusb^\quley \qudlb$, $\partial_\qusb^\quley \qubb$, $\partial_\qusb^\quley \qucb$, $\partial_\qusb^\quley \qugamb$, $1 \leq \quley \leq \quupey$, on $\quSigz$ and prove the bounds on these functions given in \eqref{stuffdelbound}. Proving the bounds is mostly an exercise in counting derivatives, and for convenience we introduce the following notation. We let boldface letters $\bley_1$, etc., denote triples of integers $(\quley_1, \qulee_1, \qules_1)$, order them by $(\quley_1, \qulee_1, \qules_1) \leq (\quley_2, \qulee_2, \qules_2)$ when $\quley_1 \leq \quley_2$, $\qulee_1 \leq \qulee_2$, and $\qules_1 \leq \qules_2$, and write, for $\bley = (\quley, \qulee, \qules)$, $Y$ one of $\quSigz$, $\quUz$, and $\quSigz \cup \quUz$,
$\|f\|_{\qusbW^\bley(Y)} = \|f\|_{\qusbW_\qusb^{\quley} \qusbW_\quxb^{\qulee} \qusbW_\quvb^{\qules} (Y)}$. We define $\bupey = (\quupey, \quupee, \quupes)$. Suppose that we have constructed $\partial_\qusb^\quley \qugamb$, $0 \leq \quley \leq \quupey$, on $\quSigz$. For $\bley_1$, $\bley_2$, $\bley_3 \in \quN^3$, $\bley_1 \leq \bupey - (1, 0, 0)$, $\bley_2$, $\bley_3 \leq \bupey$, we write $\gord{\bley_1}{\bley_2}{\bley_3}$ to denote any function $f$ on $\quSigz$ which satisfies
\begin{gather}\label{gordineqdef}
\|f\|_{L^\infty(\quSigz)} \leq C(\bley_1, \bley_2, \bley_3, \|\partial_\qusb \qugamb\|_{W^{\bley_1}(\quSigz)}, \|\partial_\quxb \qugamb\|_{W^{\bley_2}(\quSigz)}, \|\partial_\quvb \qugamb\|_{W^{\bley_3}(\quSigz)})\\
\cdot \bigl[ \|\partial_\qusb \qugamb\|_{W^{\bley_1}(\quSigz)}^2 + \|\partial_\quxb \qugamb\|_{W^{\bley_2}(\quSigz)}^2 + \|\partial_\quvb \qugamb\|_{W^{\bley_3}(\quSigz)}^2\bigr],\notag
\end{gather}
where $C$ is a constant depending only on the quantities
\begin{equation*}
\bley_1, \bley_2, \bley_3, \|\partial_\qusb \qugamb\|_{W^{\bley_1}(\quSigz)}, \|\partial_\quxb \qugamb\|_{W^{\bley_2}(\quSigz)}, \|\partial_\quvb \qugamb\|_{W^{\bley_3}(\quSigz)}.
\end{equation*}
We extend $\gord{\bley_1}{\bley_2}{\bley_3}$ to $\bley_1$, $\bley_2$, $\bley_3 \in \quZ \times \quN^2$ by dropping any term in \eqref{gordineqdef} which has a negative exponent, and we set for convenience $\bne = (-1, 0, 0)$. 
Finally, letting $\bley \vee \bley' = ( \max\{ \quley, \quley'\}, \max\{\qulee, \qulee'\}, \max\{\qules, \qules'\} )$, we see that
\begin{equation}\label{gordmultineq}
\gord{\bley_1}{\bley_2}{\bley_3} + \gord{\bley'_1}{\bley'_2}{\bley'_3},\, \gord{\bley_1}{\bley_2}{\bley_3} \gord{\bley'_1}{\bley'_2}{\bley'_3} \leq \gord{\bley_1\vee\bley'_1}{\bley_2\vee\bley'_2}{\bley_3\vee\bley'_3}.
\end{equation}
\par
The wave equation $\quwbox_\quhbar \qugamb = 0$ can be rewritten as (see \eqref{swave})
\begin{multline}
\partial_\qusb\partial_\quvb\qugamb = \qupark^{-2\iota} \frac{\partial_\quvb\qudlb}{2\qulb}\partial_\qusb\qugamb  - \qupark^{-2\iota} \frac{\partial_\qusb \qudlb}{2\qulb}\partial_\quvb\qugamb + \frac{1}{2} \Biggl\{\partial_\quxb^2\\
+ \qupark^{-2\quiota} \Biggl[\qudaib \partial_\quxb^2 - \qucb \partial_\qusb^2 + 2\frac{\qubb}{\quab} \partial_\qusb \partial_\quxb - \left(\partial_\qusb \qucb - \frac{1}{\quab} \partial_\quxb \qubb\right)\partial_\qusb + \left(\frac{1}{\quab} \partial_\qusb \qubb - \frac{\qulb\partial_\quxb\qudlb}{\quab^2}\right)\partial_\quxb\Biggr]\\
\shoveleft{\qupark^{-4\quiota} \left(\frac{\qubb^2}{\quab} \partial_\qusb^2 - \left(\qucb \frac{\partial_\qusb \qudlb}{\qulb} - 2\frac{\qubb\partial_\qusb \qubb}{\quab} + \frac{\qubb\,\quellb \partial_\quxb \qudlb}{\quab^2}\right) \partial_\qusb - \frac{\qubb\,\qulb\partial_\qusb\qudlb}{\quab^2}\partial_\quxb\right)}\\
 - \qupark^{-6\quiota} \frac{\qubb^2\qulb\partial_\qusb\qudlb}{\quab^2}\partial_\qusb\Biggr\}\qugamb.\label{constswave}
\end{multline}
\par
Now let $\qudlb : \quSigz \rightarrow \quR^1$ be as constructed in Proposition \ref{sclinitboundsi}, and let $\partial_\qusb \qubb, \partial_\qusb \qudlb : \quSigz \rightarrow \quR^1$ be the unique solutions to \eqref{psbbeqone} -- \eqref{psdlbeqone}, respectively, on $\quSigz$ satisfying
\begin{equation}
\partial_\qusb \qubb(0, \qutxb, 1) = \partial_\qusb \qudlb(0, \qutxb, 1) = 0.
\end{equation}
Define $\qubb, \qucb, \partial_\qusb \qucb : \quSigz \rightarrow \quR^1$ by $\qubb = \qucb = \partial_\qusb \qucb = 0$, and let $\partial_\qusb \qugamb : \quSigz \rightarrow \quR^1$ be the unique solution on $\quSigz$ to \eqref{constswave} with all terms containing a factor of $\qubb$, $\qucb$, or $\partial_\qusb \qucb$ dropped, satisfying
\begin{equation}
\partial_\qusb \qugamb(0, \qutxb, 1) = 0.
\end{equation}
Finally, define $\partial_\qusb^\quley \qudlb$, $\partial_\qusb^\quley \qubb$, $\partial_\qusb^\quley \qucb$, $\partial_\qusb^\quley \qugamb : \quSigz \rightarrow \quR^1$, $2 \leq \quley \leq \quupey$, inductively, by first obtaining $\partial_\qusb^\quley \qudlb$, $\partial_\qusb^\quley \qubb$, $\partial_\qusb^\quley \qucb : \quSigz \rightarrow \quR^1$ from \eqref{nsricone} -- \eqref{nsricthree} differentiated $\quley - 2$ times with respect to $\qusb$ (note that this is a direct {\it definition}, there is nothing to solve), and then defining $\partial_\qusb^\quley \qugamb : \quSigz\rightarrow\quR^1$ to be the unique solution on $\quSigz$ to the equation obtained from \eqref{constswave} by differentiating $\quley - 1$ times with respect to $\qusb$ and then dropping all terms containing a factor of $\qubb$, $\qucb$, or $\partial_\qusb \qucb$, subject to the condition
\begin{equation}\label{gambsinitzed}
\partial_\qusb^\quley \qugamb(0, \qutxb, 1) = 0.
\end{equation}
The existence and uniqueness of the functions $\partial_\qusb^\quley \qudlb$, $\partial_\qusb^\quley \qubb$, $\partial_\qusb^\quley \qucb$, $\partial_\qusb^\quley \qugamb$, $0 \leq \quley \leq \quupey$, follows from standard ODE theory and Proposition \ref{sclinitboundsi}, noting that the equations \eqref{psbbeqone} -- \eqref{psdlbeqone} and \eqref{constswave} are {\it linear\/} for $\partial_\qusb \qubb$, $\partial_\qusb \qudlb$, and $\partial_\qusb \qugamb$, respectively.
\par
\begin{proposition}\label{sclinitboundsallgamb} Assume that \eqref{ellbinvlb} holds. Let $\quupee' \geq 2$, $\bupey' = (\quupey, \quupee', \quupes)$. The functions $\partial_\qusb^\quley \qudlb$, $\partial_\qusb^\quley \qubb$, $\partial_\qusb^\quley \qucb$, $\partial_\qusb^\quley \qugamb : \quSigz \rightarrow \quR^1$, $0 \leq \quley \leq \quupey$, constructed as above, satisfy the bounds, for some constant $C(\bupey', \|\quginit\|_{\qusbW^{\quupeepp}_\quxb \qusbW^{\quupesp}_\quvb(\quSigz)})$ depending only on $\bupey'$ and $\|\quginit\|_{\qusbW^{\quupeepp}_\quxb \qusbW^{\quupesp}_\quvb(\quSigz)}$
\begin{gather}\label{allSigzbounds}
\|\qudlb\|_{\qusbW^{\bupey'}(\quSigz)}, \|\qubb\|_{\qusbW^{\bupey'}(\quSigz)}, \|\qucb\|_{\qusbW^{\bupey'}(\quSigz)}, \|\qugamb\|^2_{\qusbW^{\bupey'}(\quSigz)}\\
\leq C({\bupey'}, \|\quginit\|_{\qusbW^{\quupeepp}_\quxb \qusbW^{\quupesp}_\quvb(\quSigz)}) \|\quginit\|^2_{\qusbW^{\quupeepp}_\quxb \qusbW^{\quupesp}_\quvb(\quSigz)}.\notag
\end{gather}
\end{proposition}
\par
\begin{proof} We first note the following fact: let functions $\partial_\qusb^\quley f_i$, $i = 1, \cdots, m$, $\quley = 1, \cdots, \quupey_i$ (for some set of $\quupey_i \in \quN$) be given on $\quSigz$, together with triples $\bley^i$, $\bley^i_j$, $i = 1, \cdots, m$, $j = 1, 2, 3$. Let $F(\partial_\qusb^\quley f_i)$ be any (suitably smooth) fixed function of (some subcollection of the) $\partial_\qusb^\quley f_i$, and let $I = (\quley_0, \qulee_0, \qules_0)$ be a multiindex for which $\partial^I [F(\partial_\qusb^\quley f_i)]$ is defined (where $\partial_\qusb^{\quley'}$ is defined in the obvious way: $\partial_\qusb^{\quley'} \partial_\qusb^\quley f_i = \partial_\qusb^{\quley + \quley'} f_i$). By the chain and product rules, $\partial^I [F(\partial_\qusb^\quley f_i)]$ is a sum of terms of the form
\begin{equation}\label{sclinitFlipschyi}
(\partial^J F)(\partial_\qusb^\quley f_i) \prod \partial^{I_j} \partial_\qusb^{\quley_j} f_{i_j}.
\end{equation}
Combining this with \eqref{gordmultineq}, we see that, for fixed $F$,
\begin{equation}\label{sclinitFlipschr}
\|\partial_\qusb^\quley f_i\|_{\qusbW^I(\quSigz)} \leq \gord{\bley^i_1}{\bley^i_2}{\bley^i_3} \hbox { implies } \|F(\partial_\qusb^\quley f_i)\|_{\qusbW^I (\quSigz)} \leq \gord{\max \{ \bley^i_1 \}}{\max \{ \bley^i_2 \}}{\max \{ \bley^i_3\}}.
\end{equation}
(Here we allow the constant implicit in the right-hand side to depend on the function $F$, since we consider $F$ to be fixed.) Given \eqref{sclinitFlipschr}, the derivation of the bounds below is straightforward, and we leave it to the reader. Applying Proposition \ref{refgronwallcont} to \eqref{dlbeqone} -- \eqref{psdlbeqone} in that order gives the bounds, for $\qules \geq 2$,
\begin{gather}
\|\qudlb\|_{\qusbW_\qusb^0\qusbW_\quxb^\qulee\qusbW_\quvb^\qules(\quSigz)} \leq \gord{\bne}{\bne}{(0, \qulee, \qules - 2)},\, \|\partial_\qusb\qubb\|_{\qusbW_\qusb^0 \qusbW_\quxb^\qulee\qusbW_\quvb^\qules(\quSigz)} \leq \gord{\bne}{(0, \qulee, \qules - 1)}{(0, \qulee, \qules - 1)},\label{dlbjib}\\
\|\partial_\qusb\qudlb\|_{\qusbW_\qusb^0 \qusbW_\quxb^\qulee\qusbW_\quvb^\qules(\quSigz)} \leq \gord{\bne}{(0, \qulee + 1, \qules - 1)}{(0, \qulee + 1, \qules - 2)}.\label{psdlbjib}
\end{gather}
For any $z \in \quR^1$, define $z^+ = \max \{ z, 0 \}$. We claim that $\qudlb$, $\qubb$, $\qucb$ satisfy the following bounds, for $\bley = (\quley, \qulee, \qules) \leq \bupey$, $\qules \geq 2$:
\begin{align}
\|\qudlb\|_{\qusbW^\bley(\quSigz)} &\leq \gord{(\quley - 2, \qulee, \qules)}{(0, \qulee + 1, \qules - 1)}{(0, \qulee + 1, \qules - 2)},\, \|\qubb\|_{\qusbW^\bley(\quSigz)} &\leq \gord{(\quley - 2, \qulee, \qules)}{((\quley - 2)^+, \qulee + 1, \qules)}{(0, \qulee + 1, \qules - 1)},\label{dlbbbmainbound}\\
\|\qucb\|_{\qusbW^\bley(\quSigz)} &\leq \gord{(\quley - 2, \qulee, \qules)}{(\quley - 2, \qulee + 1, \qules)}{(\quley - 2, \qulee + 1, \qules)}.\label{cbmainbound}
\end{align}
For $\quley < 2$, these follow from \eqref{dlbjib} -- \eqref{psdlbjib} and $\qubb = \qucb = \partial_\qusb \qucb = 0$ on $\quSigz$. For $\quley \geq 2$, they follow by induction from the following bounds, which themselves follow from the Riccati equations \eqref{nsricone} -- \eqref{nsricthree}:
\begin{align}
\|\qudlb\|_{\qusbW^\bley(\quSigz)} &\leq \gord{(\quley - 2, \qulee, \qules)}{\bne}{\bne} + \left(\gord{(\quley - 2, \qulee, \qules)}{-1}{-1} + 1\right) \|\qudlb\|_{\qusbW^{\bley - (1, 0, 0)}(\quSigz)},\\
\|\qubb\|_{\qusbW^\bley(\quSigz)} &\leq \gord{(\quley - 2, \qulee, \qules)}{(\quley - 2, \qulee, \qules)}{\bne} + \left(\gord{(\quley - 2, \qulee, \qules)}{(0, \qulee + 1, \qules - 1)}{(0, \qulee + 1, \qules - 2)} + 1\right) \|\qubb\|_{\qusbW^{\bley - (1, 0, 0)}(\quSigz)},\\
\|\qucb\|_{\qusbW^\bley(\quSigz)} &\leq \gord{(\quley - 2, \qulee, \qules)}{(\quley - 2, \qulee + 1, \qules)}{(\quley - 2, \qulee + 1, \qules)} + \left(\gord{(\quley - 2, \qulee, \qules)}{\bne}{\bne} + 1\right) \|\qucb\|_{\qusbW^{\bley - (1, 0, 0)}(\quSigz)}.
\end{align}
Next, Proposition \ref{refgronwallcont} applied to \eqref{constswave}, combined with the bounds in \eqref{dlbbbmainbound} -- \eqref{cbmainbound}, gives next
\begin{equation}\label{qugambbasebound}
\|\partial_\qusb \qugamb\|^2_{\qusbW_\qusb^0 \qusbW_\quxb^{\qulee} \qusbW_\quvb^{\qules}(\quSigz)} \leq \gord{\bne}{(0, \qulee + 1, \qules - 1)}{(0, \qulee + 1, \qules - 1)}.
\end{equation}
Finally, consider the equation obtained from \eqref{constswave} by applying the differential operator $\partial_\qusb^\quley \partial_\quxb^\qulee \partial_\quvb^{\qules - 1}$ for some $\bley = (\quley, \qulee, \qules) \leq \bupey$. By inspection, we find that the forcing terms can be bounded in terms of the quantities
\begin{gather*}
\|\qudlb\|_{\qusbW^{(\quley + 1, \qulee + 1, \qules)}(\quSigz)} \leq \gord{(\quley - 1, \qulee + 1, \qules)}{(0, \qulee + 2, \qules - 1)}{(0, \qulee + 2, \qules - 2)},\\
\|\qubb\|_{\qusbW^{(\quley + 1, \qulee + 1, \qules - 1)}(\quSigz)} \leq \gord{(\quley - 1, \qulee + 1, \qules - 1)}{((\quley - 1)^+, \qulee + 2, \qules - 1)}{(0, \qulee + 2, \qules - 2)},\\
\|\qucb\|_{\qusbW^{(\quley + 1, \qulee, \qules - 1)}(\quSigz)} \leq \gord{(\quley - 1, \qulee, \qules - 1)}{(\quley - 1, \qulee + 1, \qules - 1)}{(\quley - 1, \qulee + 1, \qules - 1)}
\end{gather*}
and (fewer $\qusb$ derivatives of $\partial_\qusb\qugamb$, $\partial_\qusb^2 \qugamb$, and $\partial_\qusb\partial_\quxb\qugamb$ are needed because $\qubb = \qucb = \partial_\qusb \qucb$ on $\quSigz$)
\begin{gather*}
\|\partial_\qusb\qugamb\|^2_{\qusbW^{(\quley - 1, \qulee, \qules - 1)}(\quSigz)} \leq \gord{(\quley - 1, \qulee, \qules - 1)}{\bne}{\bne},\, \|\partial_\quxb\qugamb\|^2_{\qusbW^{(\quley, \qulee, \qules - 1)}(\quSigz)} \leq \gord{\bne}{(\quley, \qulee, \qules - 1)}{\bne},\\
\|\partial_\quvb \qugamb\|^2_{\qusbW^{(\quley, \qulee, \qules - 1)}(\quSigz)} \leq \gord{\bne}{\bne}{(\quley, \qulee, \qules - 1)},\,\|\partial_\qusb^2\qugamb\|^2_{\qusbW^{(\quley - 2, \qulee, \qules - 1)}(\quSigz)} \leq \gord{(\quley - 1, \qulee, \qules - 1)}{\bne}{\bne},\\
\|\partial_\qusb\partial_\quxb \qugamb\|^2_{\qusbW^{(\quley - 1, \qulee, \qules - 1)}(\quSigz)} \leq \gord{\bne}{(\quley, \qulee, \qules - 1)}{\bne},\, \|\partial_\quxb^2 \qugamb\|^2_{\qusbW^{(\quley, \qulee, \qules - 1)}(\quSigz)} \leq \gord{\bne}{(\quley, \qulee + 1, \qules - 1)}{\bne}.
\end{gather*}
By Proposition \ref{refgronwallcont}, then, we obtain
\begin{equation}
\|\partial_\qusb\qugamb\|^2_{\qusbW^{(\quley, \qulee, \qules)}(\quSigz)} \leq \gord{(\quley - 1, \qulee + 1, \qules)}{(\quley, \qulee + 2, \qules - 1)}{(\quley, \qulee + 2, \qules - 1)},
\end{equation}
from which we obtain by induction and \eqref{qugambbasebound} that
\begin{equation}
\|\partial_\qusb\qugamb\|^2_{\qusbW^{(\quley, \qulee, \qules)}(\quSigz)} \leq \gord{(0, \qulee + 3\quley, \qules)}{(0, \qulee + 3\quley - 1, \qules - 1)}{(0, \qulee + 3\quley - 1, \qules - 1)} \leq \gord{\bne}{(0, \qulee + 3\quley + 1, \qules - 1)}{(0, \qulee + 3\quley + 1, \qules - 1)}.
\end{equation}
This gives, finally, for $\bley = (\quley, \qulee, \qules) \leq \bupey$, $\quley \geq 2$, $\qules \geq 2$,
\begin{gather}
\|\qudlb\|_{\qusbW^\bley(\quSigz)} \leq \gord{\bne}{(0, \qulee + 3\quley - 5, \qules - 1)}{(0, \qulee + 3\quley - 5, \qules - 1)},\label{interdlbbound}\\
\|\qubb\|_{\qusbW^\bley(\quSigz)} \leq \gord{\bne}{(0, \qulee + 3\quley - 3, \qules)}{(0, \qulee + 3\quley - 3, \qules - 1)},\,\|\qucb\|_{\qusbW^\bley(\quSigz)} \leq \gord{\bne}{(0, \qulee + 3\quley - 3, \qules)}{(0, \qulee + 3\quley - 3, \qules)},\notag\\
\|\partial_\qusb \qugamb\|^2_{\qusbW^\bley(\quSigz)} \leq \gord{\bne}{(0, \qulee + 3\quley + 1, \qules - 1)}{(0, \qulee + 3\quley + 1, \qules - 1)},\, \|\partial_\quxb\qugamb\|^2_{\qusbW^\bley(\quSigz)} \leq \gord{\bne}{(0, \qulee + 3\quley - 1, \qules)}{(0, \qulee + 3\quley - 1, \qules - 1)},\notag\\
\|\partial_\quvb\qugamb\|^2_{\qusbW^\bley(\quSigz)} \leq \gord{\bne}{(0, \qulee + 3\quley - 2, \qules)}{(0, \qulee + 3\quley - 2, \qules)},\notag
\end{gather}
from which the bounds in \eqref{allSigzbounds} follow immediately.
\end{proof}
\par
\subsection{Bounds on the initial data: the hypersurface $\quUz$}\label{vinitdat} Now let $\partial_\qusb^\quley \qudlb$, $\partial_\qusb^\quley \qubb$, $\partial_\qusb^\quley \qucb$, $\partial_\qusb^\quley \qugamb : \quSigz \rightarrow \quR^1$, $0 \leq \quley \leq \quupey$ be as constructed in the previous subsection, so that the bounds in \eqref{allSigzbounds} hold.
We specify initial data for $\qugamb$ on $\quUz$ by
\begin{equation}\label{Ue}
\qugamb\vert_{\quUz} = \sum_{\quley = 0}^\quupey \frac{1}{\quley!} \qusb^{\quley} \partial_\qusb^\quley \qugamb(0, \quxb, 0).
\end{equation}
\begin{remark} Note that $\partial_\qusb^\quley \qugamb|_{\quUz}$ vanishes for all $\quley > \quupey$.\end{remark}
\par
\begin{lemma}\label{gambUzSigzsbz} Replacing \eqref{gambsinitzed} by \eqref{Ue} in the procedure in Subsection \ref{sinitdat} produces the same functions $\partial_\qusb^\quley \qudlb$, $\partial_\qusb^\quley \qubb$, $\partial_\qusb^\quley \qucb$, $\partial_\qusb^\quley \qugamb : \quSigz \rightarrow \quR^1$, $0 \leq \quley \leq \quupey$.
\end{lemma}
\par
\begin{proof} This follows by elementary ODE theory applied to \eqref{constswave}, combined with \eqref{nsricone} -- \eqref{nsricthree} and an induction argument.\end{proof}
\par
\begin{remark} Lemma \ref{gambUzSigzsbz} is only of conceptual, rather than computational, significance.\end{remark}
\par
\eqref{Ue} completes the specification of the initial data for \eqref{nsricone} -- \eqref{swave}. We immediately obtain the following bounds on $\qugamb|_{\quUz}$:
\par
\begin{proposition}\label{Ucont}
Let $\quley \geq 0$, $\quupee' \geq 2$. Then
\begin{equation}\label{Uconteqyi}
\| \qugamb\| _{\qusbW^\quley_\qusb \qusbW^{\quupee'}_\quxb(\quUz)} \leq C(\quley, \quupey, \quupee', \quupes, \|\quginit\|_{\qusbW^{\quupeepp}_\quxb \qusbW^{\quupesp}_\quvb(\quSigz)}) \|\quginit\|_{\qusbW^{\quupeepp}_\quxb \qusbW^{\quupesp}_\quvb(\quSigz)}
\end{equation}
for some constant $C > 0$ depending only on the indicated quantities.
\end{proposition}
\par
\begin{proof} This follows directly from Proposition \ref{sclinitboundsallgamb} applied to \eqref{Ue}.\end{proof}
\par
It remains only to bound the $\quvb$ derivatives of $\qudlb$, $\qubb$, $\qucb$, and $\qugamb$ on $\quUz$. We first bound $\quellb^{-1}$:
\par
\begin{proposition}\label{ellbinssmall} Let $\quley \geq 0$, $\quupee' \geq 2$. There is a constant
\begin{equation}
C = C(\quley, \quupey, \quupee', \quupes, \|\quginit\|_{\qusbW^{\quupeepp}_\quxb \qusbW^{\quupesp}_\quvb(\quSigz)}) > 1,
\end{equation}
depending only on $\quley$, $\quupey$, $\quupee'$, $\quupes$, and $\|\quginit\|_{\qusbW^{\quupeepp}_\quxb \qusbW^{\quupesp}_\quvb(\quSigz)}$, such that
\begin{equation}\label{qudlbchulvb}
\|\qudlb\|_{\qusbW^{\quley}_\qusb \qusbW^{\quupee'}_\quxb (\quUz)} \leq C(\quley, \quupey, \quupee', \quupes, \|\quginit\|_{\qusbW^{\quupeepp}_\quxb \qusbW^{\quupesp}_\quvb(\quSigz)}) \|\quginit\|^2_{\qusbW^{\quupeepp}_\quxb \qusbW^{\quupesp}_\quvb(\quSigz)}.
\end{equation}
In particular, for $\|\quginit\|_{\qusbW^{\quupeepp}_\quxb \qusbW^{\quupesp}_\quvb(\quSigz)} \leq \min \{ 1, 1/(2C(1, \quupey, 2, \quupes, 1)) \}$, we have
\begin{equation}\label{ellbinvlbs}
\left\| \quellb^{-1}\right\| _{L^\infty(\quUz)} < 4/3.
\end{equation}
\end{proposition}
\par
\begin{proof} This follows from the Riccati equation \eqref{nsricone} and Propositions \ref{refgronwallcont}, \ref{sclinitboundsallgamb}, and \ref{Ucont}. We omit the details.\end{proof}
\par
We may now bound all derivatives of all four quantities $\qudlb$, $\qubb$, $\qucb$, and $\qugamb$ on $\quUz$. 
\par
\begin{proposition}\label{gambvbound} Assume that \eqref{ellbinvlbs} holds. Let $\quupee' \geq 2$, $\bupey' = (\quupey, \quupee', \quupes)$. There is a constant
\begin{equation}
C = C(\quupey, \quupee', \quupes, \|\quginit\|_{\qusbW^{\quupeeppp}_\quxb \qusbW^{\quupesp}_\quvb(\quSigz)}) > 0,
\end{equation}
depending only on $\quupey$, $\quupee'$, $\quupes$, and $\|\quginit\|_{\qusbW^{\quupeeppp}_\quxb \qusbW^{\quupesp}_\quvb(\quSigz)}$, such that
\begin{gather}\label{allUzbounds}
\|\qudlb\|_{\qusbW^{\bupey'}(\quUz)}, \|\qubb\|_{\qusbW^{\bupey'}(\quUz)}, \|\qucb\|_{\qusbW^{\bupey'}(\quUz)}, \|\qugamb\|_{\qusbW^{\bupey'}(\quUz)}\\
\leq C({\bupey'}, \|\quginit\|_{\qusbW^{\quupeeppp}_\quxb \qusbW^{\quupesp}_\quvb(\quSigz)}) \|\quginit\|_{\qusbW^{\quupeeppp}_\quxb \qusbW^{\quupesp}_\quvb(\quSigz)}.\notag
\end{gather}
\end{proposition}
\par
\begin{proof} 
We write $C(\ldots)$ to denote constants depending only on the quantities indicated. Fix $\qules \in \quN$, $1 \leq \qules \leq \quupes$, and define
\begin{gather}
\quu_{\qules} = (\partial_\quvb^{\qules}\qudlb, \partial_\quvb^{\qules - 1}\qubb, \partial_\quvb^{\qules - 1} \partial_\quxb \qubb, \partial_\quvb^{\qules - 1}\qucb, \partial_\quvb^{\qules} \qugamb, \partial_\qusb \partial_\quvb^{\qules}\qudlb, \partial_\qusb \partial_\quvb^{\qules - 1} \partial_\quxb \qubb, \partial_\qusb \partial_\quvb^{\qules - 1} \qucb).
\end{gather}
Moreover, define
\begin{equation}
\quu_0 = (\qudlb, \qugamb, \partial_\qusb \qudlb),
\end{equation}
and note that, by \eqref{interdlbbound} and Propositions \ref{Ucont} and \ref{ellbinssmall}, for any $\quley \geq 0$,
\begin{equation}\label{quuzedbounds}
\|\quu_0\|_{\qusbW^{\quley}_\qusb \qusbW^{\quupee'}_\quxb(\quUz)} \leq C(\quley, \quupey, \quupee', \quupes, \|\quginit\|_{\qusbW^{\quupeepp}_\quxb \qusbW^{\quupesp}_\quvb(\quSigz)}) \|\quginit\|_{\qusbW^{\quupeepp}_\quxb \qusbW^{\quupesp}_\quvb(\quSigz)}.
\end{equation}
Differentiating \eqref{nsricone} $\qules$ times with respect to $\quvb$ and \eqref{nsrictwo} -- \eqref{nsricthree} and \eqref{constswave} $\qules - 1$ times with respect to $\quvb$, we obtain a {\it linear\/} system 
\begin{equation}\label{psbusys}
\partial_\qusb \quu_{\qules} = \quM_{\qules} \quu_{\qules} + \qub_{\qules},
\end{equation}
where $\quM_{\qules}$ and $\qub_{\qules}$ are polynomials in $u_{\qules'}$, $\qules' \leq \qules - 1$, together with the quantities
\begin{equation}\label{veqpquant}
\quellb^{-1}, \partial_\quvb^{\qules'} \partial_\quxb \qudlb, \partial_\qusb \partial_\quvb^{\qules'} \partial_\quxb \qudlb, \partial_\quvb^{\qules'} \partial_\qusb^{\quley'} \partial_\quxb^{\qulee'} \qugamb,
\end{equation}
where $\qules' \leq \qules - 1$ and $\quley' + \qulee' \leq 2$.
Noting the bounds 
\begin{gather}
\|\partial_\quvb^{\qules'} \partial_\qusb^{\quley'} \partial_\quxb^{\qulee'} \qugamb\|_{\qusbW_\qusb^\quley \qusbW_\quxb^{\quupee'}} \leq \|\quu_{\qules'}\|_{\qusbW_\qusb^{\quley + 2} \qusbW_\quxb^{\quupee' + 2}},\\
\|\partial_\quvb^{\qules'} \partial_\quxb \qudlb\|_{\qusbW_\qusb^\quley \qusbW_\quxb^{\quupee'}} \leq \|\quu_{\qules'}\|_{\qusbW_\qusb^{\quley} \qusbW_\quxb^{\quupee' + 1}},
\end{gather}
and using Proposition \ref{sclinitboundsallgamb} to bound the initial data for $\quu_\qules$ on $\quSigz \cap \quUz$, we obtain from Proposition \ref{refgronwallcont}, Proposition \ref{Ucont}, and Proposition \ref{ellbinssmall} that 
\begin{gather*}
\|\quu_\qules\|_{\qusbW_\qusb^\quley \qusbW_\quxb^{\quupee'}(\quUz)} \leq C(\quley, \qules, \quupey, \quupee', \quupes, \sum_{\qules' \leq \qules - 1} \|\quu_{\qules'}\|_{\qusbW_\qusb^{\quley + 2} \qusbW_\quxb^{\quupee' + 2}(\quUz)})\\
\cdot \sum_{\qules' \leq \qules - 1} \|\quu_{\qules'}\|_{\qusbW_\qusb^{\quley + 2} \qusbW_\quxb^{\quupee' + 2}(\quUz)}.
\end{gather*}
An induction based on \eqref{quuzedbounds} thus clearly gives
\begin{equation}
\|\quu_\qules\|_{\qusbW_\qusb^\quley \qusbW_\quxb^{\quupee'}(\quUz)} \leq C(\quley, \qules, \quupey, \quupee', \quupes, \|\quginit\|_{\qusbW_\quxb^{\quupee' + 3\quupey + 2\quupes - 2}\qusbW_\quvb^\quupes(\quSigz)}) \|\quginit\|_{\qusbW_\quxb^{\quupee' + 3\quupey + 2\quupes - 2}\qusbW_\quvb^\quupes(\quSigz)},
\end{equation}
and from this the result follows.\end{proof}
\par
\begin{remark} The fundamentally different methods of proof used in Propositions \ref{sclinitboundsallgamb} and \ref{gambvbound} are needed because the transverse ($\qusb$) derivatives of the constraint equations in \eqref{scalconsts} -- \eqref{scalconsty} do not, in general, hold on $\quSigz$.\end{remark}
\par
\subsection{Proof of Proposition \ref{initdatbounds}}\label{initdatbdproof} By combining Propositions \ref{sclinitboundsallgamb}\ and \ref{gambvbound} we may prove Proposition \ref{initdatbounds}, as follows.
\par
\begin{proof}[Proof of Proposition \ref{initdatbounds}] Let $C_1$, $C_2$ denote the constants in Propositions \ref{sclinitboundsallgamb} and \ref{gambvbound}, respectively, and define
\begin{equation}
C(\quupey, \quupee, \quupes) = \max \{ C_1(\quupey, \quupee, \quupes, 1), C_2(\quupey, \quupee, \quupes, 1) \}.
\end{equation}
Then \eqref{ellbinvlb} and \eqref{ellbinvlbs} hold by Propositions \ref{sclinitboundsi} and \ref{ellbinssmall}, respectively. Proposition \ref{initdatbounds} then follows immediately from Propositions \ref{sclinitboundsallgamb} and \ref{gambvbound}.\end{proof}
\par
\section{GENERAL EXISTENCE}\label{enerineq}
\subsection{Statement of results}\label{existintro} We now proceed to prove existence to the evolutionary system \eqref{nsricone} -- \eqref{swave} for a class of initial data which includes (for $\quginit$ sufficiently small) that just constructed in the previous section. By Corollary \ref{Ricconst}, this will complete the proof of existence of solutions to the reduced Einstein equations \eqref{intRicwav}.
\par
Since there is no local existence result available in our gauge, we prove existence from first principles, using an iterative argument.
\par
\begin{figure}[h]\centering
\includegraphics[keepaspectratio]{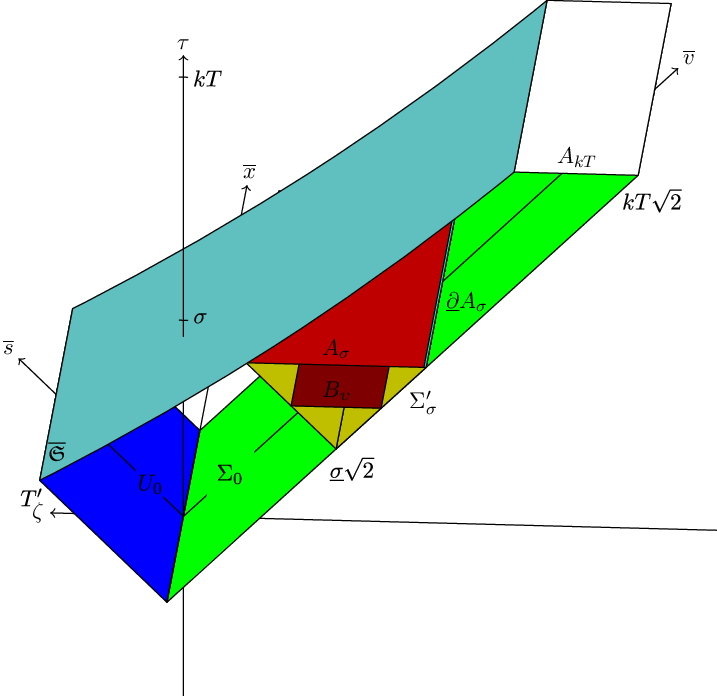}
\caption{Spacetime regions. $\qubulks$ is the triangular prism bounded by $\quSigs$, $\quAsigmat$, and $\quvb = \quusig\sqrt{2}$.}\label{scalbulk}
\end{figure}

We define the following Sobolev constant, which we note is a fixed numerical constant.
\begin{definition}\label{specconstdef} Define
\begin{equation}\label{CKSdefeq}
\quCKS = 2^{5/8} (4/3)^{1/2} \quCrip,\qquad\hbox{$\quCrip$ the constant in Proposition \ref{klainsobolv}}.
\end{equation}
\end{definition}
We shall make use of the following regions. See Figure \ref{scalbulk}.
\par
\begin{definition}\label{longbulkdefin} Define coordinates
\begin{equation}\label{tauzetadef}
\qutau = \frac{1}{\sqrt{2}} (\qusb + \quvb),\qquad \quzeta = \frac{1}{\sqrt{2}} (\qusb - \quvb).
\end{equation}
Define the {\it scaled bulk region\/} by (cf.\ \eqref{bulkdef})
\begin{equation}\label{sbulkdef}
\qubulk = \{ (\qusb, \quxb, \quvb)\in\quR^3\,\vert \,0 \leq \quvb \leq \qupark\queT\sqrt{2},\, 0 \leq \qusb \leq \quBsigffb(\quvb),\, 0 \leq \qutau \leq \qupark\queT \}
\end{equation}
where
\begin{gather}\label{Bsigffbdefeq}
\quBsigffb(\quvb) = \queTp - 4\OmegbC\quCKS\qunu^2\qupark^{-2\quiota} \left[(1 + \quvb)^{1/2} - 1\right].
\end{gather}
Further, let
\begin{gather}\label{tauSUSAdef}
\quSigz = \qubulk \cap \{ \qusb = 0 \},\ququad \quUz = \qubulk \cap \{ \quvb = 0 \},\ququad \quBsigfb = \{ (\qusb, \quxb, \quvb) \in \qubulk\,\vert \,\qusb = \quBsigffb(\quvb) \},\\
\qreflink{quantAsigmat}{A}_{\qupark\queT} = \qubulk \cap \{ \qutau = \qupark\queT \}\notag
\end{gather}
denote its boundary surfaces, and
\begin{equation}\label{pbarAsigmdef}
\quAsigmat = \qubulk \cap \{ \qutau = \qusigmat \},\ququad \qupbar \quAsigmat = \quSigz \cap \{ \qutau = \qusigmat \} = \quSigz \cap \quAsigmat
\end{equation}
be foliations on $\qubulk$ and $\quSigz$, respectively. Finally, define the subregions
\begin{gather}
\qubulks = \{ (\qusb, \quxb, \quvb) \in \qubulk\,\vert \, \quusig\sqrt{2} \leq \quvb \leq \qusigmat\sqrt{2},\,\qutau \leq \qusigmat \},\ququad\quSigs = \qubulks \cap \quSigz,\label{bulksSBdef}
\end{gather}
where ($\alpha = 4\OmegbC\quCKS\qunu^2\qupark^{-2\quiota}$)
\begin{equation}\label{quusigdef}
\quusig = \qusigmat - 2^{-1/2} \queTp + \frac{1}{2\sqrt{2}} \alpha(\alpha - 2) \left[ 1 - \left(1 - 4\frac{\queTp - \qusigmat\sqrt{2}}{(\alpha - 2)^2}\right)^{1/2}\right]
\end{equation}
is such that $\quAsigmat$ intersects $\quBsigfb$ at $\quvb = \quusig\sqrt{2}$.
\end{definition}
Note that $\qubulks = \quSigs = \emptyset$ for $\qusigmat < 0$. Note also that the geometric description of $\quusig$ implies that either $\quusig \geq \qusigmat - 2^{-1/2} \queTp$ or $\quusig \leq 0$.
\par
We shall use the following commutation vector fields (which involve rotations, boosts, and isotropic scaling), seminorms on initial data, and energies.
\par
\begin{definition}\label{difrerset} Define the (ordered) sets of vector fields\footnote{The inclusion of $\partial_\qutau$ in $\quudifsetr$ does not seem to be necessary but simplifies the argument.}
\begin{align}
\quudifsetr &= ( \partial_\quxi, \partial_\quzeta, \partial_\qutau, \quxi \partial_\quzeta - \quzeta \partial_\quxi ),\label{simpdifrset}\\
\quudifseta &= ( \partial_\quxi, \partial_\quzeta, \partial_\qutau, \quxi \partial_\quzeta - \quzeta \partial_\quxi, \quxi \partial_\qutau + \qutau \partial_\quxi, \quzeta \partial_\qutau + \qutau \partial_\quzeta, \qutau \partial_\qutau + \quxi \partial_\quxi + \quzeta \partial_\quzeta ),\label{simpdifaset}
\end{align}
and let $\qudifsetr$, $\qudifseta$ denote the Lie algebra generated by $\quudifsetr$, $\quudifseta$, respectively. Let $\quudifset$ denote a particular choice of $\quudifsetr$ and $\quudifseta$ and $\qudifset$ the corresponding Lie algebra, and let $(\qucvf_1, \cdots, \qucvf_{|\quudifset|})$ be the ordering of $\quudifset$ as given in \eqref{simpdifrset} -- \eqref{simpdifaset}.
For a multiindex $I = (I_1, \cdots, I_{|\quudifset|})$, define
\begin{equation}\label{uprtlIdef}
\quuprtl^I = \prod_{\qucvf_i \in \quudifset} \qucvf_i^{I_i} = Y_1^{I_1} \cdots Y_{|\quudifset|}^{I_{|\quudifset|}}.
\end{equation}
For $\quaph$ any smooth Lorentzian metric on $\qubulk$, and $f$ any sufficiently smooth function on $\qubulk$, define the stress-energy tensor of $f$ with respect to $\quaph$ by
\begin{equation}\label{QTaphdef}
\quQT_\quaph[f]_{ij} = \partial_i f\,\partial_j f - \frac{1}{2} \quaph_{ij} \quaph^{\ell m} \partial_\ell f\,\partial_m f
\end{equation}
and the energy current of $f$ across the surfaces $\quAsigmat$ by ($\partial_\qutau^j$ is the $j$th component of the vector field $\partial_\qutau$)
\begin{equation}\label{mffdef}
\qumff_\quaph[f]^i = -\quaph^{i\ell} \quQT_\quaph[f]_{j\ell} \partial_\qutau^j = -\quaph^{i\ell} \partial_\ell f\,\partial_\qutau f + \frac{1}{2} \quaph^{j\ell} \partial_j f\,\partial_\ell f\,\partial_\qutau^i.
\end{equation}
Finally, define (letting $X$ denote one of $\quSigz$ and $\quUz$)
\begin{gather}\label{ibounddef}
\quIb_{\quSigz, \quaph}[f] = \int_{\quSigz} d\qusb_i \qumff_\quaph[f]^i \vert \qudet\quaph\vert ^{1/2}\,d\quvb\,d\quxb,\ququad \quIb_{\quUz, \quaph}[f] = \int_{\quUz} d\quvb_i \qumff_\quaph[f]^i \vert \qudet\quaph\vert ^{1/2}\,d\qusb\,d\quxb,\\
\quib_{X, \quaph, \qusbn, \ell}[f] = \sum_{\vert I\vert  \leq \qusbn - 1} \quIb_{X, \quaph} [\quuprtl^I \partial_\qusb^\ell f],\ququad \quib_{X, \quaph, \qusbn}[f] = \sum_{\ell = 0}^1 \quib_{X, \quaph, \qusbn, \ell}[f],\notag
\end{gather}
\begin{gather}
\quepsn_\quaph[f](\qusigmat) = \int_{\quAsigmat} d\qutau_i \qumff_\quaph[f]^i \vert \qudet\quaph\vert ^{1/2} \,d\quxi\,d\quzeta,\label{epsdef}\\
\quEb_{\qusbn, \quaph}[\qugamb](\qusigmat) = \sum_{\vert I\vert  \leq \qusbn - 1} \sum_{\ell = 0}^1 \quepsn_\quaph[\quuprtl^I \partial_\qusb^\ell \qugamb](\qusigmat).\label{eenerdefi}
\end{gather}
\end{definition}
\begin{remark} These seminorms are standard except for the additional $\qusb$ derivative. We include this extra $\qusb$ derivative to ease the energy estimates for the Riccati equations \eqref{nsricone} -- \eqref{nsricthree}, though as mentioned in Subsection \ref{outconstr} it also seems to play a deeper (and as yet not fully understood) role in the structure of the full coupled Riccati-wave system \eqref{nsricone} -- \eqref{swave}.\end{remark}
\par
Our existence result can now be formulated as follows. Here and in many places below we use the convenient shorthand
\begin{equation}\label{hatfuncdef}
\qusighat{u} = \max \{ 1, u \},\quad\hbox{$u \in \quR^1$.}
\end{equation}
\par
\begin{theorem}\label{seqbound} Let $\qusbn \geq 7$. There are constants $C_1, C_2 > 0$, depending only on $\qusbn$, such that the following holds. Let $\queT$, $\queTp > 0$, $\queTp \leq \queTpupp$, $\qupark \geq 1$, and $\qunu \in (0, 1)$ be such that
\begin{gather}
\qunu \qupark^{-2\quiota} \leq 1/(12C_1\quCKS),\qquad \qunu \qupark^{1/2 - 2\iota} \leq 2^{13/4} C_1 \quCKS\queTp\queT^{-1/2},\label{mainprkestc}\\
\qunu \queT^{1/2} \leq \frac{1}{C_2}.\label{qunuqueTcondc}
\end{gather}
Let compactly supported initial data
\begin{gather}
\partial_\qusb^\ell \qudlb|_{\quSigz},\,\partial_\qusb^\ell \qubb|_{\quSigz},\,\partial_\qusb^\ell\qucb|_{\quSigz},\,\ell \in \{ 0, 1 \},\,\qugamb|_{\quSigz\cup\quUz}
\end{gather}
be given such that each $\quomegb \in \{ \qudlb, \qubb, \qucb, \partial_\quxb \qudlb, \partial_\quvb \qudlb, \partial_\quxb \qubb \}$ satisfies
\begin{gather}
\sum_{\ell = 0}^1 \sum_{\vert I\vert  \leq \qusbn - 1} \left\| \quuprtl^I \partial_\qusb^\ell \quomegb\right\| _{L^2(\quSigs)}^2 \leq \qunu^4 \qusigu^{-1/2} \queTp^{3/2},\label{omegbbaseinitbound}\\
\sum_{\ell = 0}^1 \sum_{|I| \leq \qusbn - 1} \| \quuprtl^I \partial_\qusb^\ell \quomegb\| _{\qusbH^1(\qupbar\quAsigmat)}^2 \leq \qunu^4 \qusigu^{-1},\label{omegbbIbootinf}
\end{gather}
while $\qugamb\vert_{\quSigz\cup\quUz}$ satisfies, for all $\qusigmat \in [0, \qupark\queT]$ and for $\quley, \qulee \leq \qusbn + 1$, $\qules \leq \qusbn$,
\begin{gather}
\partial_\qusb^\quley \partial_\quxb^\qulee \partial_\quvb^\qules \qugamb(0, \quxb, \quvb) = 0\hbox{ for }\quvb \geq 1,\label{gambzIcondendeq}\\
\|(1 + \vert\quxb\vert)^\qusbn \partial_\qusb^\quley \partial_\quxb^\qulee \partial_\quvb^\qules \qugamb\|_{L^2(\quSigz \cup \quUz)}^2 \leq \qunu^2 \queTp^{3/2},\label{gambbaseinitbound}\\
\sum_{\ell, \ell' = 0}^1 \sum_{|I| \leq \qusbn - 1} \sum_{i = 0}^2 \| \quuprtl^I \partial_\qusb^\ell \partial_\quxb^{\ell'} \partial_i \qugamb\| _{\qusbH^1(\qupbar\quAsigmat)}^2 \leq \qunu^2 \qusigu^{-1/2},\label{gambbIbootinf}\\
\|(1 + \vert\quxb\vert)^\qusbn \partial_\qusb^\quley \partial_\quxb^\qulee \partial_\quvb^\qules \qugamb\|_{L^\infty(\quSigz \cap \quUz)}^2 \leq \qunu^2 \queTp^{1/2}.\label{gamblineinitbound}
\end{gather}
(The required transverse derivatives are understood to be a priori values determined from \eqref{nsricone} -- \eqref{swave} as in Section \ref{initdat}.) Then the system \eqref{nsricone} -- \eqref{swave} has a solution, agreeing with the given initial data on $\quSigz\cup\quUz$, which exists throughout the region $\qubulk$ given in \eqref{sbulkdef}, is supported on the Minkowski causal future of $\qusupp\qugamb\vert_{\quSigz \cup \quUz}$, and satisfies the energy bounds, for some universal constant $\enerC_\qusbn > 0$ depending only on $\qusbn$,
\begin{gather}\label{cEboot}
\sum_{\ell = 0}^1 \sum_{|I| \leq \qusbn - 2} \left\|\quuprtl^I \partial_\qusb^\ell \quomegb\right\|_{L^2(\quAsigmat)}^2 \leq \enerC_\qusbn \qunu^4 \qusigu^{-1/2} \queTp^{3/2}\hbox{ for all }\quomegb \in \{ \qudlb, \qubb, \qucb, \partial_\quxb \qudlb, \partial_\quvb \qudlb, \partial_\quxb \qubb \},\\
\quEb_{\qusbn - 1, \quhbar}[\qugamb](\qusigmat) \leq \enerC_\qusbn \qunu^2 \queTp^{3/2}\notag
\end{gather}
for $\qusigmat \in [0, \qupark\queT]$. Further, for $\qunu$ sufficiently small (independent of $\queT$, $\queTp$, $\qupark$), $\qudlb$, $\qubb$, $\qucb$, and $\qugamb$ are unique among the class of solutions to \eqref{nsricone} -- \eqref{swave} satisfying bounds of the form \eqref{cEboot}.
\end{theorem}
\begin{remark} Proposition \ref{initdatbounds} shows that initial data can be constructed which satisfy \eqref{omegbbaseinitbound} -- \eqref{gamblineinitbound} as well as the constraint equations \eqref{scalconsts} -- \eqref{scalconsty}. Further, $\quiota \geq 1/4$ implies that the conditions in \eqref{mainprkestc} can be simultaneously satisfied with values for $\qunu$, $\queT$, $\queTp$ which are independent of $\qupark$.\end{remark}
\begin{remark} Inequalities \eqref{omegbbIbootinf} and \eqref{gambbIbootinf} are used to estimate the boundary term occurring in the Klainerman-Sobolev inequality in Proposition \ref{klainsobolv}.\end{remark}
\begin{remark} Note that the bounds in \eqref{gambbaseinitbound} imply in particular that
\begin{equation}\label{ogambbaseinitbound}
\quib_{\quSigz, \queta, \qusbn}[\qugamb] + \quib_{\quUz, \queta, \qusbn}[\qugamb] \leq \qunu^2 \queTp^{3/2}. 
\end{equation}
The bounds in \eqref{ogambbaseinitbound}, together with \eqref{gambbIbootinf}, are sufficient for the energy estimates we shall use below. The stronger (and in some ways less natural) bounds in \eqref{gambbaseinitbound} and \eqref{gamblineinitbound} are needed only to construct the initial element $\qugamb_0$ in the iteration argument. We believe that it may be possible to replace the bounds in \eqref{gambbaseinitbound} -- \eqref{gamblineinitbound} with more natural bounds, similar to those in \eqref{ogambbaseinitbound}. We do not treat this point both for simplicity and because the initial data given by Proposition \ref{initdatbounds} naturally satisfy bounds of the form in \eqref{gambbaseinitbound} -- \eqref{gamblineinitbound}.\end{remark}
\begin{remark} Performing an isotropic coordinate scaling allows us to relax the fixed upper bound on $\queTp$ in Theorem \ref{seqbound}. The formulation and proof are straightforward and we shall not give them in this paper.\end{remark}
\par
\par
The main idea of the proof is as follows. Suppose given initial data satisfying \eqref{omegbbaseinitbound} -- \eqref{gamblineinitbound}. We begin by obtaining a function $\qugamb_0 : \qubulk \rightarrow \quR^1$ which satisfies, for $\quley \leq \qusbn + 1$, $\qules \leq \qusbn$,
\begin{equation}\label{qugambzcond}
\partial_\qusb^\quley \partial_\quvb^\qules \qugamb_0\vert_{\quSigz \cup \quUz} = \partial_\qusb^\quley \partial_\quvb^\qules \qugamb\vert_{\quSigz \cup \quUz},
\end{equation}
is supported on the {\it Minkowski\/} causal future of $\qusupp \qugamb\vert_{\quSigz \cup \quUz}$, and also satisfies ($\ell \in \{ 0, 1 \}$, $i \in \{ 0, 1, 2 \}$)
\begin{equation}\label{gambzcEboot}
\|\partial_\qusb^\ell\partial_i \qugamb_0\|_{\quDHb^{\qusbn - 1}(\quAsigmat)} \leq \gambC \qunu,\,\|\partial_\quxb^2 \qugamb_0\|_{\quDHb^{\qusbn - 1}(\quAsigmat)} \leq \gambCdx \qunu\hbox{ for $\qusigmat \in [0, \qupark\queT]$}
\end{equation}
for some constants $\gambC, \gambCdx > 0$, where the $\quDHb^{\qusbn - 1}(\quAsigmat)$ norm is introduced in Definition \ref{DHbenerdef} below. (We track the extra bound on $\partial_\quxb^2 \qugamb$ separately for technical convenience.)
We then proceed inductively. Given $\qugamb_\mu$ for $\mu \geq 0$, we construct $\qudlb_{\mu + 1}$, $\qubb_{\mu + 1}$, $\qucb_{\mu + 1}$ by solving \eqref{nsricone} -- \eqref{nsricthree} with the given initial data and $\qugamb$ set equal to $\qugamb_\mu$. It is not hard to see that, for $\quomegb \in \{ \qudlb,\,\qubb,\,\qucb \}$ and $\quomegb_\mu$ the corresponding element of $\{ \qudlb_\mu,\,\qubb_\mu,\,\qucb_\mu \}$,
\begin{equation}\label{quomegbcond}
\partial_\qusb^\quley \partial_\quvb^\qules \quomegb_\mu\vert_{\quSigz \cup \quUz} = \partial_\qusb^\quley \partial_\quvb^\qules \quomegb\vert_{\quSigz \cup \quUz},
\end{equation}
and that the $\quomegb_\mu$ are supported on the Minkowski causal future of $\qusupp\qugamb\vert_{\quSigz\cup\quUz}$ whenever $\qugamb_\mu$ is. We then construct $\qugamb_{\mu + 1}$ by solving \eqref{swave} with $\qudlb$, $\qubb$, $\qucb$ set equal to $\qudlb_{\mu + 1}$, $\qubb_{\mu + 1}$, $\qucb_{\mu + 1}$; again, it is clear that $\qugamb_{\mu + 1}$ is supported on the Minkowski causal future of $\qusupp\qugamb\vert_{\quSigz\cup\quUz}$ whenever $\qudlb_{\mu + 1}$, $\qubb_{\mu + 1}$, $\qucb_{\mu + 1}$ are. ODE Sobolev estimates for \eqref{nsricone} -- \eqref{nsricthree}, together with energy estimates for \eqref{swave}, allow us to obtain bounds on the $\quomegb_\mu$ and $\qugamb_\mu$ as well as the differences $\quomegb_{\mu + 1} - \quomegb_\mu$ and $\qugamb_{\mu + 1} - \qugamb_\mu$ which are sufficient to prove convergence of the $\quomegb_\mu$ and $\qugamb_\mu$ to a solution to \eqref{nsricone} -- \eqref{swave} satisfying the bounds in \eqref{cEboot}.
\par
In Subsection \ref{gambzsval}, we construct the function $\qugamb_0$ and prove bounds on the $\quomegb_\mu$ as well as the differences $\quomegb_{\mu + 1} - \quomegb_\mu$, assuming (inductively) similar bounds on $\qugamb_\mu$ and $\qugamb_{\mu + 1} - \qugamb_\mu$. In Subsection \ref{enerperboot}, we complete the induction by proving bounds on $\qugamb_\mu$ and $\qugamb_{\mu + 1} - \qugamb_\mu$ assuming similar bounds on the $\quomegb_\mu$ and $\quomegb_{\mu + 1} - \quomegb_\mu$. Finally, in Subsections \ref{finishthproof} and \ref{enerboundfin} we complete the proof of Theorem \ref{seqbound}.
\par\noindent
\textbf{Convention.} For convenience, if $\quomegb$ is any element of $\{ \qudlb, \qubb, \qucb, \partial_\quxb \qudlb, \partial_\quvb \qudlb, \partial_\quxb \qubb \}$ and $\mu \in \quN$, we shall (as above) write $\quomegb_\mu$ for the corresponding element of $\{ \qudlb_\mu, \qubb_\mu, \qucb_\mu, \partial_\quxb \qudlb_\mu, \partial_\quvb \qudlb_\mu, \partial_\quxb \qubb_\mu \}$, and vice versa.
\subsection{Boundedness of the metric}\label{gambzsval} We shall use the following weighted Sobolev norms.
\begin{definition}\label{DHbenerdef} Define ($\qusigmat \geq 0$, $m \geq 0$)
\begin{align}\label{DHbnormdef}
\begin{split}
\| f\| _{\quDH^m(\quAsigmat)} &= \queTp^{-3/4} \left[\sum_{\vert I\vert  \leq m} \left\| \quuprtl^I f\right\| _{L^2(\quAsigmat)}^2\right]^{1/2},\\
\| f\| _{\qupbH^{m, 1, 1/4}(\qupbar\quAsigmat)} &= \qusigu^{1/4} \left[\sum_{\vert I\vert  \leq m} \left\| \quuprtl^I f\right\| _{\qusbH^1(\qupbar\quAsigmat)}^2\right]^{1/2},\\
\| f\| _{\quDHb^m(\quAsigmat)} &= \| f\| _{\quDH^m(\quAsigmat)} + \| f\| _{\qupbH^{m, 1, 1/4}(\qupbar\quAsigmat)}
\end{split}
\end{align}
for all functions $f$ on $\quAsigmat$ for which the right-hand sides exist and are finite, and let $\qupbH^{m, 1, 1/4}(\qupbar\quAsigmat)$ and $\quDHb^m(\quAsigmat)$ denote the corresponding Sobolev spaces. If $X = \quAsigmat \cap \{ \quvb \leq \quvb_0 \}$ for some $\quvb_0 > 0$, then we define $\quDHb^m(X)$ by replacing $\quAsigmat$ by $X$ in the foregoing.
\end{definition}
\par
\begin{remark} The $\qupbH^{m, 1, 1/4}(\qupbar\quAsigmat)$ norm is included in $\quDHb^m(\quAsigmat)$ to account for the boundary term in the Klainerman-Sobolev inequality in Proposition \ref{klainsobolv}, which itself arises because of the `thinness' of our spatial sections (cf.\ the discussion at the start of Subsection \ref{PSti}).
\end{remark}
\par
Let initial data and sufficiently many a priori transverse derivatives
\begin{gather}
\partial_\qusb^\quley\partial_\quxb^\qulee\partial_\quvb^\qules \qugamb\vert_{\quSigz\cup\quUz},\label{gambzinitdatlist}\\
\partial_\qusb^\quley\partial_\quxb^\qulee\partial_\quvb^\qules \qudlb\vert_{\quSigz\cup\quUz},\,\partial_\qusb^\quley\partial_\quxb^\qulee\partial_\quvb^\qules \qubb\vert_{\quSigz\cup\quUz},\,\partial_\qusb^\quley\partial_\quxb^\qulee\partial_\quvb^\qules \qucb\vert_{\quSigz\cup\quUz}\label{omegbzinitdatlist}
\end{gather}
be given -- as in Section \ref{initdat}, for example -- such that \eqref{omegbbaseinitbound} -- \eqref{gamblineinitbound} are satisfied. We begin the iteration by constructing $\qugamb_0$.
\par
\begin{lemma}\label{qugambzconstrlem} Given \eqref{gambzinitdatlist}, there is a $\qugamb_0 : \qubulk \rightarrow \quR^1$ satisfying \eqref{qugambzcond} and \eqref{gambzcEboot}, with $\gambC$, $\gambCdx$ constants depending only on $\qusbn$, and supported on the Minkowski causal future of $\qusupp\qugamb\vert_{\quSigz\cup\quUz}$.
\end{lemma}
\par
\begin{proof} Let $\chi \in C^\infty(\quR^1)$ satisfy
\begin{equation}\label{chiCinfprop}
\chi\vert_{[-1, 1]} = 1,\qquad \qusupp\chi \subset (-2, 2).
\end{equation}
For $(\qusb, \quxb, \quvb) \in \qubulk$, set
\begin{equation}\label{fzdef}
f_0(\qusb, \quxb, \quvb) = \Bigl[ \sum_{\quley = 0}^{\qusbn + 1} \sum_{\qules = 0}^\qusbn \frac{\qusb^\quley}{\quley!} \frac{\quvb^\qules}{\qules!} \partial_\qusb^\quley \partial_\quvb^\qules \qugamb(0, \quxb, 0)\Bigr] \cdot \chi(\quvb),
\end{equation}
so that for $\quley \leq \qusbn + 1$, $\qules \leq \qusbn$,
\begin{gather}\label{fzgambzdiffprop}
\partial_\qusb^{\quley} \partial_\quvb^\qules f_0(0, \quxb, 0) = \partial_\qusb^{\quley} \partial_\quvb^\qules \qugamb(0, \quxb, 0).
\end{gather}
Define further ($(\qusb, \quxb, \quvb) \in \qubulk$)
\begin{align}
f_1(\qusb, \quxb, \quvb) &= \sum_{\quley = 0}^{\qusbn + 1} \frac{\qusb^\quley}{\quley!} \partial_\qusb^\quley [\qugamb - f_0] (0, \quxb, \quvb),\label{fydef}\\
f_2(\qusb, \quxb, \quvb) &= \sum_{\qules = 0}^\qusbn \chi(\quvb) \frac{\quvb^\qules}{\qules!} \partial_\quvb^\qules [\qugamb - f_0] (\qusb, \quxb, 0).\label{fedef}
\end{align}
By \eqref{fzgambzdiffprop}, for $\quley \leq \qusbn + 1$, $\qules \leq \qusbn$,
\begin{gather}
\partial_\qusb^\quley \partial_\quvb^\qules f_1(0, \quxb, \quvb) = \partial_\qusb^\quley \partial_\quvb^\qules [\qugamb - f_0](0, \quxb, \quvb),\qquad \partial_\qusb^\quley \partial_\quvb^\qules f_1(\qusb, \quxb, 0) = 0,\\
\partial_\qusb^\quley \partial_\quvb^\qules f_2(0, \quxb, \quvb) = 0,\qquad \partial_\qusb^\quley \partial_\quvb^\qules f_2(\qusb, \quxb, 0) = \partial_\qusb^\quley \partial_\quvb^\qules [\qugamb - f_0](\qusb, \quxb, 0).
\end{gather}
It is then clear that we may satisfy \eqref{qugambzcond} by taking $\qugamb_0 = f_1 + f_2 + f_0$. \eqref{gambzcEboot} follows from \eqref{gambzIcondendeq} -- \eqref{gamblineinitbound} by elementary calculations. (We note that the factors of $(1 + \vert\quxb\vert)^{\qusbn}$ which appear in \eqref{gambbaseinitbound} and \eqref{gamblineinitbound}, together with the support condition in \eqref{gambzIcondendeq}, are necessary to deal with the $\quxb$ and $\quvb$ coefficients, respectively, arising in the differential operators $\quuprtl^I$.) The condition on the support of $\qugamb_0$ holds by construction.\end{proof}
\begin{remark} See \cite{rendall} for a treatment of a similar problem using the Whitney extension theorem. The treatment here -- inspired by Borel's theorem, see \cite{hormander}, Theorem 1.2.6, and also bearing some similarities to the construction in Whitney's original proof, see \cite{whitney} -- allows us to bound $\qugamb_0$, a question which \cite{rendall} does not address.\end{remark}
\par
\begin{definition}\label{indseqdefin} Let $\qugamb_0 : \qubulk \rightarrow \quR^1$ be any smooth function satisfying the statement of Lemma \ref{qugambzconstrlem}. 
For $\mu \in \quZ$, $\mu \geq 1$, define inductively $\qudlb_\mu$, $\qubb_\mu$, $\qucb_\mu$ to be the solutions to \eqref{nsricone} -- \eqref{nsricthree}, with $\qugamb$ replaced by $\qugamb_{\mu - 1}$, for initial data \eqref{omegbzinitdatlist}, and define $\qugamb_\mu$ to be the solution to \eqref{swave} with $\qudlb$, $\qubb$, $\qucb$ replaced by $\qudlb_\mu$, $\qubb_\mu$, and $\qucb_\mu$, for initial data \eqref{gambzinitdatlist}.\end{definition}
\par
\begin{remark} Note that $\qudlb_\mu$, $\qubb_\mu$, $\qucb_\mu$ depend on $\qugamb_{\mu - 1}$ while $\qugamb_\mu$ depends on $\qudlb_\mu$, $\qubb_\mu$, $\qucb_\mu$.\end{remark}
\par
\begin{lemma}\label{qudlbbbcbinitdatag} For $\mu \geq 1$, the transverse derivatives
\begin{gather}
\partial_\qusb^\quley\partial_\quxb^\qulee\partial_\quvb^\qules \qugamb_\mu\vert_{\quSigz\cup\quUz},\\
\partial_\qusb^\quley\partial_\quxb^\qulee\partial_\quvb^\qules \qudlb_\mu\vert_{\quSigz\cup\quUz},\,\partial_\qusb^\quley\partial_\quxb^\qulee\partial_\quvb^\qules \qubb_\mu\vert_{\quSigz\cup\quUz},\,\partial_\qusb^\quley\partial_\quxb^\qulee\partial_\quvb^\qules \qucb_\mu\vert_{\quSigz\cup\quUz}
\end{gather}
agree with the values in \eqref{gambzinitdatlist} -- \eqref{omegbzinitdatlist}. Thus in particular the bounds \eqref{omegbbaseinitbound} -- \eqref{omegbbIbootinf} and \eqref{gambbaseinitbound} -- \eqref{gamblineinitbound} hold with $\qudlb$, $\qubb$, $\qucb$, and $\qugamb$ replaced by $\qudlb_\mu$, $\qubb_\mu$, $\qucb_\mu$, and $\qugamb_\mu$. Moreover, the functions $\qugamb_\mu$, $\qudlb_\mu$, $\qubb_\mu$, and $\qucb_\mu$ are all supported on the Minkowski causal future of $\qusupp\qugamb\vert_{\quSigz\cup\quUz}$.
\end{lemma}
\begin{proof} The first statement follows by applying Lemma \ref{qugambzconstrlem}, together with an induction on the order of derivatives as in the proofs of Propositions \ref{sclinitboundsallgamb} and \ref{gambvbound}. The second statement is then immediate. The condition on the supports follows by induction.\end{proof}
\par
We shall show the following families of estimates ($\ell = 0, 1$, $i = 0, 1, 2$), for constants $\gambC$, $\gambCdx$, $\gambCp$, $\gambCdxp$, $\OmegbC$, and $\OmegbCp$ depending only on $\qusbn$:
\begin{gather}
\|\partial_\qusb^\ell\partial_i \qugamb_\mu\|_{\quDHb^{\qusbn - 1}(\quAsigmat)} \leq \gambC \qunu,\,\|\partial_\quxb^2 \qugamb_\mu\|_{\quDHb^{\qusbn - 1}(\quAsigmat)} \leq \gambCdx \qunu\hbox{ for $\qusigmat \in [0, \qupark\queT]$}\tag{G$_{\mu}$}\label{gambinduct}\\
\|\partial_\qusb^\ell\partial_i (\qugamb_{\mu + 1} - \qugamb_\mu)\|_{\quDHb^{\qusbn - 2}(\quAsigmat)} \leq \frac{1}{2^\mu} \gambCp \qunu,\,\|\partial_\quxb^2 (\qugamb_{\mu + 1} - \qugamb_\mu)\|_{\quDHb^{\qusbn - 2}(\quAsigmat)} \leq \frac{1}{2^\mu} \gambCdxp \qunu\tag{G$_{\mu}'$}\label{gambdiffinduct}\\
\hbox{ for $\qusigmat \in [0, \qupark\queT]$},\notag
\end{gather}
\par\noindent
for $\mu \geq 0$;
\begin{gather}
\|\partial_\qusb^\ell \quomegb_\mu\|_{\quDHb^{\qusbn - 1}(\quAsigmat)} \leq \OmegbC \qunu^2 \qusigu^{-1/4}\tag{O$_{\mu}$}\label{omeginduct}\\
\hbox{ for $\qusigmat \in [0, \qupark\queT]$, $\quomegb_\mu \in \{ \qudlb_\mu, \qubb_\mu, \qucb_\mu, \partial_\quxb \qudlb_\mu, \partial_\quvb \qudlb_\mu, \partial_\quxb \qubb_\mu \}$}\\
\|\partial_\qusb^\ell (\quomegb_{\mu + 1} - \quomegb_\mu)\|_{\quDHb^{\qusbn - 2}(\quAsigmat)} \leq \frac{1}{2^\mu} \OmegbCp \qunu \qusigu^{-1/4}\tag{O$_{\mu}'$}\label{omegdiffinduct}\\
\hbox{ for $\qusigmat \in [0, \qupark\queT]$, $\quomegb_\mu \in \{ \qudlb_\mu, \qubb_\mu, \qucb_\mu, \partial_\quxb \qudlb_\mu, \partial_\quvb \qudlb_\mu, \partial_\quxb \qubb_\mu \}$}.\notag
\end{gather}
\par\noindent
for $\mu \geq 1$.
\par
We proceed by induction. G$_0$ follows from Definition \ref{indseqdefin}; G$_0'$ follows from G$_0$ for an appropriate choice of $\gambCp$. It suffices to show that
\begin{gather}
\hbox{G}_\mu \hbox{ implies } \hbox{O}_{\mu + 1}, \mu \geq 0,\label{GOindimp}\\
\hbox{G}_\mu, \hbox{G}_{\mu + 1} \hbox{ and } \hbox{G}'_\mu \hbox{ imply } \hbox{O}'_{\mu + 1}, \mu \geq 0,\label{GOdiffindimp}\\
\hbox{O}_\mu \hbox{ implies } \hbox{G}_\mu, \mu \geq 1,\label{OGindimp}\\
\hbox{G}_{\mu - 1}, \hbox{O}_\mu \hbox{ and } \hbox{O}_\mu' \hbox{ implies } \hbox{G}_\mu', \mu \geq 1.\label{OGdiffindimp}
\end{gather}
(Note that there is no separate base case for O$_\mu$.) \eqref{GOindimp} -- \eqref{GOdiffindimp} are the simpler of the two and will be shown in this section. \eqref{OGindimp} -- \eqref{OGdiffindimp} will be taken up in the next section.
\par
We first marshall a few preliminary results. The following reordering result follows from Lemma \ref{ishuffleops}:
\par
\begin{corollary}\label{shuffleops} Let $f : \qubulk \rightarrow \quR^1$ be $C^\infty$ on a neighborhood of $\quAsigmat$ ($\qusigmat \geq 0$). Let $(\qucvf_1, \cdots, \qucvf_r)$ denote the ordered set $\quudifset$ (in the order given in Definition \ref{difrerset}, cf.\ \eqref{uprtlIdef}), and let $m \geq 1$, $p \geq 0$. There is a constant $\quCnmi > 0$, depending only on $r$, $m$, and $p$, such that the following holds. For $\qucvf_{r + 1} = \partial_\qusb$, $\qucvf_{r + 2} = \partial_\quvb$ and any $\pi : \{ 1, \cdots, m \} \rightarrow \{ 1, \cdots, r + 2 \}$, letting $\ell_i = \vert \pi^{-1} \{ r + i \}\vert $ ($i = 1, 2$), $m' = m - \ell_1 - \ell_2$,
\begin{equation}\label{shopresester}
\left\| \left[\prod_{i = 1}^m \quops_{\pi(i)}\right] f\right\| _{\quDHb^p(\quAsigmat)} \leq \quCnmi \Biggl[\sum_{\ell'_i \leq \ell_i} \left\| \partial_\qusb^{\ell'_1} \partial_\quvb^{\ell'_2} f\right\| _{\quDHb^{m' + p}(\quAsigmat)}^2\Biggr]^{1/2};
\end{equation}
while for $\qucvf_{r + 1} = \partial_\qutau$ and any $\pi : \{ 1, \cdots, m \} \rightarrow \{ 1, \cdots, r + 1 \}$ satisfying $\ell_1 = \vert \pi^{-1} \{ r + 1 \} \vert \geq 1$, setting $m' = m - \ell_1$,
\begin{equation}\label{shopresestertau}
\left\| \left[\prod_{i = 1}^m \quops_{\pi(i)}\right] f\right\| _{\quDH^p(\quAsigmat)} \leq \quCnmi \Biggl[\sum_{\ell'_1 \leq \ell_1 - 1} \quEb_{m' + p} \left[\partial_\qutau^{\ell'_1} f\right]\Biggr]^{1/2}.
\end{equation}
\end{corollary}
\par
\begin{proof} Taking
\begin{equation}\label{shuffopseqyi}
\quCnmi = (m + p + 1)^{r/2 + 1} (m + p)! (r + 2)^{2(m + p)},
\end{equation}
this follows from Lemma \ref{ishuffleops}\ applied with $\quopsset = \quudifset$ with $q = r + 2$, $q' = r$, $n = m$, noting that we may take $C_\quopsset = 1$, that the exponent $m' + p$ follows from \eqref{lishopIconder}, that $\partial_\quxi$ is the first element of $\quudifset$ and hence of $\quopsset$ (so that we do not need to reposition the extra $\partial_\quxi$ derivative appearing in the $\qupbH^{m, 1, 1/4}$ term), and that the factor $(m + p + 1)^{r/2 + 1}$ in \eqref{shuffopseqyi} comes from replacing the sum in Lemma \ref{ishuffleops}\ by the square root of sums of squares here.\end{proof}
\par
From our work in Section \ref{interlude}\ we obtain the following Klainerman-Sobolev estimate.
\par
\begin{lemma}\label{Asobolevemb} Let $\quCKS$ be the constant defined in \eqref{CKSdefeq}. For all $f \in \quDHb^2(\quAsigmat)$,
\begin{equation}\label{Asembeq}
\| f\| _{L^\infty(\quAsigmat)} \leq \quCKS \qusigu^{-1/4} \| f\| _{\quDHb^2(\quAsigmat)}.
\end{equation}
If $X = \quAsigmat \cap \{ \qusb \leq \qusb_0 \}$ for some $\qusb_0 > 0$, then
\begin{equation}\label{Asembeqthin}
\| f\| _{L^\infty(X)} \leq \quCKS \| f\| _{\quDHb^2(X)}.
\end{equation}
\end{lemma}
\par
\begin{proof} \eqref{Asembeqthin}, and \eqref{Asembeq} for $\qusigmat \leq 4$, follow from Proposition \ref{sobolevemb}. \eqref{Asembeq} for $\qusigmat > 4$ follows from Proposition \ref{klainsobolv}, noting that $\queTp \leq \queTpupp$ and $\qusigmat \geq 3/2$ implies that $\quzeta \leq \queTp\sqrt{2} - \qusigmat \leq -1$, and that $\partial_\quxi, \partial_\quzeta, \quzeta\partial_\quxi - \quxi\partial_\quzeta \in \quudifset$ by Definition \ref{difrerset}, so that Proposition \ref{klainsobolv}\ applies to $\quAsigmat$ for $\qusigmat \geq 3/2$ with $k + \delta = \qusigmat$ and $\delta = \queTp\sqrt{2}$. The factor of $2^{5/8}$ in the definition \eqref{CKSdefeq} of $\quCKS$ comes from the $\sqrt{2}$ in $\delta$, replacing the sum of $L^\infty$ norms in \eqref{kseqero} by the $\qupbH^{2, 1, 1/4}$ norm, and using Proposition \ref{sobolevemb}\ up to $\qusigmat = 4$ (the latter two costing a $\sqrt{2}$ each).\end{proof}
\par
\begin{remark} The right-hand sides of \eqref{Asembeq} and \eqref{Asembeqthin} include one more derivative of $f$ in $\qusbH^1(\qupbar\quAsigmat)$ than is really necessary, but the result as stated is sufficiently sharp for our purposes.\end{remark}
\par
\begin{remark}\label{bigparkrem} In particular, for $\qunu \qupark^{-2\iota} \leq 1/(2\OmegbC \quCKS)$, \eqref{omeginduct} with $\quomegb_\mu = \qudlb_\mu$ implies that $|\quellb_\mu^{-1}| < 2$.\end{remark}
\par
From the Klainerman-Sobolev inequality in Lemma \ref{Asobolevemb}\ we obtain the following multiplicative inequalities. 
\par
\begin{proposition}\label{mfgagniren} Let $m \geq 4$, $N \geq 1$, $\{ I_i \}_{i = 1}^N$ a collection of multiindices satisfying $\sum \vert I_i\vert  \leq m$, and $\phi_i \in \quDHb^m(\quAsigmat)$, $i = 1, \cdots, N$. Then for some constant $\quCsb_N > 0$ depending only on $N$
\begin{align}
\qusigu^{1/4}\left\| \prod_{i = 1}^N \left[\quuprtl^{I_i} \phi_i\right]\right\| _{\qusbH^1(\qupbar\quAsigmat)} &\leq \quCsb_N \qusigu^{-(N - 1)/4} \prod_i \| \phi_i\| _{\quDHb^m(\quAsigmat)},\label{sbHprodpbA}\\
\queTp^{-3/4}\left\| \prod_{i = 1}^N \left[\quuprtl^{I_i} \phi_i\right]\right\| _{L^2(\quAsigmat)} &\leq \quCsb_N \qusigu^{-(N - 1)/4} \prod_i \| \phi_i\| _{\quDHb^m(\quAsigmat)}.\label{DHbprodA}
\end{align}
Further, $\quCsb_N$ can be chosen so that also
\begin{equation}\label{realDHbprodA}
\left\|\prod_{i = 1}^N \phi_i\right\|_{\quDHb^m(\quAsigmat)} \leq \quCsb_N \qusigu^{-(N - 1)/4} \prod_i \|\phi_i\|_{\quDHb^m(\quAsigmat)}
\end{equation}
and so that for $X = \quAsigmat \cap \{ \qusb \leq \qusb_0 \}$ for some $\qusb_0 > 0$
\begin{equation}\label{realDHbprodAthin}
\left\|\prod_{i = 1}^N \phi_i\right\|_{\quDHb^m(X)} \leq \quCsb_N \prod_i \|\phi_i\|_{\quDHb^m(X)}.
\end{equation}
\end{proposition}
\par
\begin{proof} \eqref{sbHprodpbA} holds since $\qusbH^1(\qupbar\quAsigmat)$ is a normed algebra. Now consider \eqref{DHbprodA}. If $N = 1$ then \eqref{DHbprodA} follows from the definition of $\quDHb^m$. Suppose $N > 1$. Reordering if necessary, assume that $\vert I_1\vert  = \max_{1 \leq i \leq N} \vert I_i\vert $. Since $\sum \vert I_i\vert  \leq m$, we must have $\vert I_i\vert  \leq m/2$ for $i > 1$. Since $m \geq 4$, $m/2 \leq m - 2$, and thus $\vert I_i\vert  \leq m - 2$ for all $i > 1$. We then have by \eqref{Asembeq} in Lemma \ref{Asobolevemb}
\begin{multline*}
\queTp^{-3/4} \left\| \prod_{i = 1}^N \left[\quuprtl^{I_i} \phi_i\right]\right\| _{L^2(\quAsigmat)} \leq \queTp^{-3/4} \left\| \quuprtl^{I_1}\phi_1\right\| _{L^2(\quAsigmat)} \prod_{i = 2}^N \left\| \quuprtl^{I_i} \phi_i\right\| _{L^\infty(\quAsigmat)}\\
\leq \quCKS^{N - 1} \qusigu^{-(N - 1)/4} \queTp^{-3/4} \left\| \quuprtl^{I_1} \phi_1\right\| _{L^2(\quAsigmat)} \prod_{i = 2}^N \| \phi_i\| _{\quDHb^m(\quAsigmat)},
\end{multline*}
from which \eqref{DHbprodA} follows from the case $N = 1$ (equivalently, from the definition of $\quDHb^m$). \eqref{realDHbprodA} is clearly a direct consequence of \eqref{sbHprodpbA} and \eqref{DHbprodA}. \eqref{realDHbprodAthin} is shown by using \eqref{Asembeqthin} instead of \eqref{Asembeq} to obtain a modified version of \eqref{DHbprodA}.\end{proof}
\par
The next lemma greatly facilitates our estimates of ODE and wave equation coefficients below.
\par
\begin{lemma}\label{Flipsch} Let $\lipschjc \subseteq \{ 0, 1, 2 \}$, and suppose that $\{ \phi_i \}_{i = 1}^\Flnumphi : \qubulk \rightarrow \quR^1$ are smooth and such that $\partial_j^\ell \phi_i \in \quDHb^{\qusbn - 1}(\quAsigmat)$ for all $\qusigmat \in [0, \qupark\queT]$, $i = 1, \ldots, \Flnumphi$, and $j \in \lipschjc$.
Let $\FlFdom \subset \quR^\Flnumphi$ be a compact convex set containing $\prod_{i = 1}^\Flnumphi \overline{\phi_i(\qubulk)}$. Let $F : \FlFdom \rightarrow \quR^1$ be smooth, and suppose that for some $\FlFexp \in \quN$, $\FlFexp \geq 1$, and all $\Flnumphi$-multiindices $I$ with $\vert I\vert \leq \FlFexp - 1$, 
\begin{equation}\label{FlFbound}
\partial^I F(0) = 0.
\end{equation}
Then there is a constant $\FlCb > 0$, depending only on $\qusbn$, $\Flnumphi$, $\FlFexp$, $\FlFdom$, $F$, and an upper bound on the quantities $\bigl\|\partial_j^\ell \phi_i\bigr\|_{\quDHb^{\qusbn - 1}(\quAsigmat)}$, $j \in \lipschjc$, $\ell \in \{ 0, 1 \}$, $i = 1, \cdots, \Flnumphi$, such that for all $j \in \lipschjc$, $\ell \in \{ 0, 1 \}$,
\begin{equation}\label{FlFres}
\left\|\partial_j^\ell [F(\vecphi)]\right\|_{\quDHb^{\qusbn - 1}(\quAsigmat)} \leq \FlCb \biggr[\sum_{{\ell' \leq \ell}\atop{1 \leq i \leq \Flnumphi}} \bigl\|\partial_j^{\ell'} \phi_i\bigr\|_{\quDHb^{\qusbn - 1}(\quAsigmat)}\biggr]^\FlFexp \qusigu^{-\frac{1}{4} (\FlFexp - 1)}.
\end{equation}
Suppose now that $\{ \tphi_i \}_{i = 1}^\Flnumphi : \qubulk \rightarrow \quR^1$ is another sequence of smooth functions such that also $\partial_j^\ell \tphi_i \in \quDHb^{\qusbn - 1}(\quAsigmat)$ for all $\qusigmat \in [0, \qupark\queT]$, $i = 1, \ldots, \Flnumphi$, and $j \in \lipschjc$, and such that for all $i = 1, \cdots, \Flnumphi$, $\qusigmat \in [0, \qupark\queT]$, $j \in \lipschjc$, $\ell \in \{ 0, 1 \}$,\footnote{This requirement is for convenience and does not seem to be essential.}
\begin{equation}\label{Flphidiffbound}
\left\|\partial_j^\ell (\phi_i - \tphi_i)\right\|_{\qupbH^{m, 1, 1/4}(\qupbar\quAsigmat)} = 0.
\end{equation}
Then there is another constant $\FlCbb > 0$, depending only on $\qusbn$, $\Flnumphi$, $\FlFexp$, $\FlFdom$, $F$, and an upper bound on the quantities $\bigl\|\partial_j^\ell \phi_i\bigr\|_{\quDHb^{\qusbn - 1}(\quAsigmat)}$, $\bigl\|\partial_j^\ell \tphi_i\bigr\|_{\quDHb^{\qusbn - 1}(\quAsigmat)}$, $j \in \lipschjc$, $\ell \in \{ 0, 1 \}$, $i = 1, \cdots, \Flnumphi$, such that for all $j \in \lipschjc$, $\ell \in \{ 0, 1 \}$,
\begin{gather}\label{FlFdiffres}
\left\|\partial_j^\ell [F(\vecphi) - F(\vectphi)]\right\|_{\quDHb^{\qusbn - 1}(\quAsigmat)}\\
\leq \FlCbb \biggl[\sum_{{\ell' \leq \ell}\atop{1 \leq i \leq \Flnumphi}} \bigl\|\partial_j^{\ell'} \phi_i\bigr\|_{\quDHb^{\qusbn - 1}(\quAsigmat)} + \sum_{{\ell' \leq \ell}\atop{1 \leq i \leq \Flnumphi}} \bigl\|\partial_j^{\ell'} \tphi_i\bigr\|_{\quDHb^{\qusbn - 1}(\quAsigmat)}\biggr]^{\FlFexp - 1}\notag\\
\shoveleft{\cdot \biggl[\sum_{{\ell' \leq \ell}\atop{1\leq i\leq\Flnumphi}} \bigl\|\partial_j^{\ell'} (\phi_i - \tphi_i)\bigr\|_{\quDHb^{\qusbn - 1}(\quAsigmat)}\biggr] \qusigu^{-\frac{1}{4} (\FlFexp - 1)}.}\notag
\end{gather}
\end{lemma}
\begin{proof} We use the convention that $|\vecx|_2^0 = 1$ for all $\vecx \in \quR^\Flnumphi$, including $\vecx = 0$. Now by \eqref{FlFbound}, $F$ has a Taylor expansion around $0$ of the form
\begin{equation}\label{Ftaylor}
F(\vecx) = \sum_{|J| = \FlFexp} \partial^J F(0) \vecx^J + O(|\vecx|_2^{\FlFexp + 1});
\end{equation}
from this it is not hard to see that, for any $\Flnumphi$-multiindex $K$ with $|K| \leq \FlFexp$, we have for some constant $C_1$ depending only on $F$, $|K|$, $\FlFexp$, and $\FlFdom$
\begin{equation}\label{FDFbounds}
\left\vert \partial^K F(\vecx)\right\vert \leq C_1 |\vecx|_2^{\FlFexp - |K|},
\end{equation}
while for $|K| > \FlFexp$ we have simply
\begin{equation}\label{FDFboundsl}
|\partial^K F(\vecx)| \leq C_2
\end{equation}
for some constant $C_2$ depending only on $F$, $|K|$, and $\FlFdom$. Now let $I$ be a $\quudifset$-multiindex, $|I| \leq \qusbn - 1$, and let $j \in \lipschjc$, $\ell \in \{ 0, 1 \}$. Then $\quuprtl^I \partial_j^\ell [F(\vecphi)]$ is a sum of terms of the form
\begin{equation}\label{upFsumexp}
(\partial^J F)(\vecphi) \prod_{i = 1}^{|J|} \quuprtl^{K_i} \partial_j^{\ell_i} \phi_{k_i},\quad \sum \ell_i = \ell,\,\sum K_i = I,\,k_i \in \{ 1, \cdots, \Flnumphi \},\,|J| \leq |I| + 1.
\end{equation}
For $|J| \leq \FlFexp$, we obtain by Proposition \ref{mfgagniren} that, for various constants $C_3$ depending only on $F$, $\qusbn$, $\Flnumphi$, $\FlFexp$, and $\FlFdom$,
\begin{align}
\left\|(\partial^J F)(\vecphi) \prod_{i = 1}^{|J|} \quuprtl^{K_i} \partial_j^{\ell_i} \phi_{k_i}\right\|_{L^2(\quAsigmat)} &\leq C_3 \left\| \left(\sum_{i = 1}^\Flnumphi |\phi_i|^{\FlFexp - |J|}\right) \prod_{i = 1}^{|J|} \quuprtl^{K_i} \partial_j^{\ell_i} \phi_{k_i}\right\|_{L^2(\quAsigmat)}\\
&\leq C_3 \sum_{i = 1}^{\Flnumphi} \left\| \phi_i^{\FlFexp - |J|} \prod_{i = 1}^{|J|} \quuprtl^{K_i} \partial_j^{\ell_i} \phi_{k_i}\right\|_{L^2(\quAsigmat)}\\
&\leq C_3 \biggr[\sum_{{\ell' \leq \ell}\atop{1 \leq i \leq \Flnumphi}} \bigl\|\partial_j^\ell \phi_i\bigr\|_{\quDHb^{\qusbn - 1}(\quAsigmat)}\biggr]^\FlFexp \qusigu^{-\frac{1}{4} (\FlFexp - 1)}.\label{FlFboundone}
\end{align}
For $|J| > \FlFexp$ (recall that $|J| \leq |I| + 1 \leq \qusbn$), we have, applying \eqref{FDFboundsl}, and for various constants $C_3$ depending only on $F$, $\qusbn$, $\Flnumphi$, $\FlFexp$, $\FlFdom$, and (now) an upper bound on $\bigl\|\partial_j^\ell \phi_i\bigr\|_{\quDHb^{\qusbn - 1}(\quAsigmat)}$, $j \in \lipschjc$, $\ell \in \{ 0, 1 \}$, $i = 1, \cdots, \Flnumphi$, 
\begin{align}
\left\|(\partial^J F)(\vecphi) \prod_{i = 1}^{|J|} \quuprtl^{K_i} \partial_j^{\ell_i} \phi_{k_i}\right\|_{L^2(\quAsigmat)} &\leq C_3 \left\|\prod_{i = 1}^{|J|} \quuprtl^{K_i} \partial_j^{\ell_i} \phi_{k_i}\right\|_{L^2(\quAsigmat)}\notag\\
&\leq C_3 \quCsb_{|J|} \prod_{i = 1}^{|J|} \left\|\partial_j^{\ell_i} \phi_{k_i}\right\|_{\quDHb^{\qusbn - 1}(\quAsigmat)} \qusigu^{-\frac{1}{4} (|J| - 1)}\notag\\
&\leq C_3 \biggr[\sum_{{\ell' \leq \ell}\atop{1 \leq i \leq \Flnumphi}} \bigl\|\partial_j^\ell \phi_i\bigr\|_{\quDHb^{\qusbn - 1}(\quAsigmat)}\biggr]^\FlFexp \qusigu^{-\frac{1}{4} (\FlFexp - 1)}.\label{FlFboundonep}
\end{align}
These estimates together establish \eqref{FlFres}.
\par
We now consider \eqref{FlFdiffres}. We observe first that, for any $\vecx, \vecxp \in \hbox{dom}\,F$ and any $\Flnumphi$-multiindex $J$,
\begin{equation}\label{Flipschest}
\partial^J F(\vecx) - \partial^J F(\vecxp) = (\vecx - \vecxp) \cdot \int_0^1 (\nabla \partial^J F)((1 - t)\vecxp + t\vecx)\,dt.
\end{equation}
From this and \eqref{FDFbounds} it is not hard to see that, for constants $C'_1$, $C'_2$ depending only on $F$, $\qusbn$, $\Flnumphi$, $\FlFexp$, and $\FlFdom$, we have, for $|J| \leq \FlFexp - 1$,
\begin{equation}\label{Flipschbounds}
\left\vert \partial^J F(\vecx) - \partial^J F(\vecxp)\right\vert \leq C'_1 |\vecx - \vecxp|_2 \left(|\vecx|_2^{\FlFexp - 1 - |J|} + |\vecxp|_2^{\FlFexp - 1 - |J|}\right),
\end{equation}
and for $|J| > \FlFexp - 1$,
\begin{equation}\label{Flipschboundd}
\left\vert \partial^J F(\vecx) - \partial^J F(\vecxp)\right\vert \leq C'_2 |\vecx - \vecxp|_2.
\end{equation}
Our task is to bound the terms analogous to those in \eqref{upFsumexp} which constitute $\quuprtl^I \partial_j^\ell \left[F(\vecphi) - F(\vectphi)\right]$. Noting the elementary algebraic identity
\begin{equation}
\prod_{j = 1}^M x_j - \prod_{j = 1}^M \tilde{x}_j = \sum_{i = 1}^M \prod_{j < i} \tilde{x}_j (x_i - \tilde{x}_i) \prod_{j > i} x_j,
\end{equation}
we see that
\begin{gather}
\left\|(\partial^J F)(\vecphi) \prod_{i = 1}^{|J|} \quuprtl^{K_i} \partial_j^{\ell_i} \phi_{k_i} - (\partial^J F)(\vectphi) \prod_{i = 1}^{|J|} \quuprtl^{K_i} \partial_j^{\ell_i} \tphi_{k_i}\right\|_{L^2(\quAsigmat)}\notag\\
\leq \left\| \left(\partial^J F(\vecphi) - \partial^J F(\vectphi)\right) \prod_{i = 1}^{|J|} \quuprtl^{K_i} \partial_j^{\ell_i} \phi_{k_i}\right\|_{L^2(\quAsigmat)}\notag\\
+ \sum_{i = 1}^{|J|} \left\| \partial^J F(\vectphi) \left(\prod_{m < i} \quuprtl^{K_m} \partial_j^{\ell_m} \tphi_{k_m}\right) \quuprtl^{K_i} \partial_j^{\ell_i} \left(\phi_{k_i} - \tphi_{k_i}\right) \left(\prod_{m > i} \quuprtl^{K_m} \partial_j^{\ell_m} \phi_{k_m}\right)\right\|_{L^2}.\label{FlFboundtwop}
\end{gather}
Let $C_4$ denote various constants depending only on $F$, $\qusbn$, $\Flnumphi$, $\FlFexp$, $\FlFdom$, and (potentially) the quantities $\bigl\|\partial_j^\ell \phi_i\bigr\|_{\quDHb^{\qusbn - 1}(\quAsigmat)}$, $\bigl\|\partial_j^\ell \tphi_i\bigr\|_{\quDHb^{\qusbn - 1}(\quAsigmat)}$, $j \in \lipschjc$, $\ell \in \{ 0, 1 \}$, $i = 1, \cdots, \Flnumphi$. Consider the case $|J| \leq \FlFexp - 1$. Applying \eqref{Flipschbounds} and Proposition \ref{mfgagniren}, we find
\begin{gather*}
\left\| \left(\partial^J F(\vecphi) - \partial^J F(\vectphi)\right) \prod_{i = 1}^{|J|} \quuprtl^{K_i} \partial_j^{\ell_i} \phi_{k_i}\right\|_{L^2(\quAsigmat)}\\
\leq C_4 \sum_{i, i' = 1}^{\Flnumphi} \left\| \left(\phi_{i'} - \tphi_{i'}\right) \left(\phi_i^{\FlFexp - 1 - |J|} + \tphi_i^{\FlFexp - 1 - |J|}\right) \prod_{i'' = 1}^{|J|} \quuprtl^{K_{i''}} \partial_j^{\ell_{i''}} \phi_{k_{i''}}\right\|_{L^2(\quAsigmat)}\\
\leq C_4 \biggl[\sum_{{\ell' \leq \ell}\atop{1 \leq i \leq \Flnumphi}} \bigl\|\partial_j^\ell \phi_i\bigr\|_{\quDHb^{\qusbn - 1}(\quAsigmat)} + \sum_{{\ell' \leq \ell}\atop{1 \leq i \leq \Flnumphi}} \bigl\|\partial_j^\ell \tphi_i\bigr\|_{\quDHb^{\qusbn - 1}(\quAsigmat)}\biggr]^{\FlFexp - 1}\\
\cdot\biggl[\sum_{1\leq i\leq\Flnumphi} \bigl\|\phi_i - \tphi_i\bigr\|_{\quDHb^{\qusbn - 1}(\quAsigmat)}\biggr] \qusigu^{-\frac{1}{4} (\FlFexp - 1)}.
\end{gather*}
By a process analogous to that resulting in \eqref{FlFboundone}, we obtain also the bound
\begin{gather}\label{FlFboundtwopp}
\sum_{i = 1}^{|J|} \left\| \partial^J F(\vectphi) \left(\prod_{m < i} \quuprtl^{K_m} \partial_j^{\ell_m} \tphi_{k_m}\right) \quuprtl^{K_i} \partial_j^{\ell_i} \left(\phi_{k_i} - \tphi_{k_i}\right) \left(\prod_{m > i} \quuprtl^{K_m} \partial_j^{\ell_m} \phi_{k_m}\right)\right\|_{L^2}\\
\shoveleft{\leq C_4 \biggl[\sum_{{\ell' \leq \ell}\atop{1 \leq i \leq \Flnumphi}} \bigl\|\partial_j^\ell \phi_i\bigr\|_{\quDHb^{\qusbn - 1}(\quAsigmat)} + \sum_{{\ell' \leq \ell}\atop{1 \leq i \leq \Flnumphi}} \bigl\|\partial_j^\ell \tphi_i\bigr\|_{\quDHb^{\qusbn - 1}(\quAsigmat)}\biggr]^{\FlFexp - 1}}\\
\shoveleft{\cdot \biggl[\sum_{{\ell' \leq \ell}\atop{1\leq i\leq\Flnumphi}} \bigl\|\partial_j^\ell (\phi_i - \tphi_i)\bigr\|_{\quDHb^{\qusbn - 1}(\quAsigmat)}\biggr] \qusigu^{-\frac{1}{4} (\FlFexp - 1)}.}
\end{gather}
In the case $|J| > \FlFexp - 1$, we write instead
\begin{gather*}
\left\| \left(\partial^J F(\vecphi) - \partial^J F(\vectphi)\right) \prod_{i = 1}^{|J|} \quuprtl^{K_i} \partial_j^{\ell_i} \phi_{k_i}\right\|_{L^2(\quAsigmat)}\\
\leq C_4 \sum_{i = 1}^{\Flnumphi} \left\| \left(\phi_i - \tphi_i\right) \prod_{i'' = 1}^{|J|} \quuprtl^{K_{i''}} \partial_j^{\ell_{i''}} \phi_{k_{i''}}\right\|_{L^2(\quAsigmat)}\\
\leq C_4 \biggl[\sum_{{\ell' \leq \ell}\atop{1 \leq i \leq \Flnumphi}} \bigl\|\partial_j^\ell \phi_i\bigr\|_{\quDHb^{\qusbn - 1}(\quAsigmat)} + \sum_{{\ell' \leq \ell}\atop{1 \leq i \leq \Flnumphi}} \bigl\|\partial_j^\ell \tphi_i\bigr\|_{\quDHb^{\qusbn - 1}(\quAsigmat)}\biggr]^{\FlFexp - 1}\\
\cdot\biggl[\sum_{1\leq i\leq\Flnumphi} \bigl\|\phi_i - \tphi_i\bigr\|_{\quDHb^{\qusbn - 1}(\quAsigmat)}\biggr] \qusigu^{-\frac{1}{4} (\FlFexp - 1)},
\end{gather*}
while \eqref{FlFboundtwopp} still holds by an argument analogous to that giving \eqref{FlFboundonep}. Combining these four estimates gives \eqref{FlFdiffres}.
\end{proof}
\par
Next, we obtain $L^2$ Sobolev bounds on solutions to ordinary differential equations. We introduce the notation (cf.\ \eqref{quusigdef})
\begin{equation}
\usigf(\qusigmat) = \quusig = \qusigmat - 2^{-1/2} \queTp + \frac{1}{2\sqrt{2}} \alpha(\alpha - 2) \left[ 1 - \left(1 - 4\frac{\queTp - \qusigmat\sqrt{2}}{(\alpha - 2)^2}\right)^{1/2}\right];
\end{equation}
recall that the region over which we integrate the Riccati equations \eqref{nsricone} -- \eqref{nsricthree} is precisely (see \eqref{bulksSBdef} and Figure \ref{scalbulk})
\begin{equation}
\qubulks = \{ (\qusb, \quxb, \quvb) \in \qubulk\,\vert \, \quusig\sqrt{2} \leq \quvb \leq \qusigmat\sqrt{2},\,\qutau \leq \qusigmat \},
\end{equation}
and that initial data for \eqref{nsricone} -- \eqref{nsricthree} are bounded on the sets (see \eqref{bulksSBdef}, \eqref{pbarAsigmdef})
\begin{equation}
\qupbar \quAsigmat = \quSigz \cap \{ \qutau = \qusigmat \} = \quSigz \cap \quAsigmat,\quad \quSigs = \qubulks \cap \quSigz.
\end{equation}
\par
\begin{proposition}\label{refgronwallsblv} Let $m \geq 4$, let $\partial_i$ be one of $\partial_\quxb$, $\partial_\quvb$, let $\ell \in \{ 0, 1 \}$, let $\qusigmat \geq 0$, and let $u : \qubulk \rightarrow \quR^1$ satisfy
\begin{equation}\label{udefdiffeqorig}
\partial_\qusb^2 u = \alpha \partial_\qusb u + \beta u + \delta,
\end{equation}
where $\alpha,\,\beta,\,\delta : \qubulk \rightarrow \quR^1$ are in $\quDHb^{m + 1}(\quAsigmatp)$ for $\qusigmat' \in [\usigf(\qusigmat), \qusigmat]$. Then there is a constant $C > 0$, depending only on $m$ and an upper bound on the quantity
\begin{equation}\label{gwalphbetbound}
\sum_{\ell' \leq \ell} \sup_{\qusigmat' \in [\usigf(\qusigmat), \qusigmat]} \left\|\partial_i^{\ell'} \alpha\right\|_{\quDHb^m(\quAsigmatp)} + \sup_{\qusigmat' \in [\usigf(\qusigmat), \qusigmat]} \left\|\partial_i^{\ell'} \beta\right\|_{\quDHb^m(\quAsigmatp)},
\end{equation}
such that for $\ell_1 \in \{ 0, 1 \}$,
\begin{gather}\label{gronwallres}
\|\partial_i^\ell \partial_\qusb^{\ell_1} u\|_{\quDHb^m(\quAsigmat)} \leq C \sum_{\ell' \leq \ell} \Biggl[ \queTp \sup_{\qusigmat' \in [\usigf(\qusigmat), \qusigmat]} \left\|\partial_i^{\ell'} \delta\right\|_{\quDHb^m(\quAsigmatp)}\\
+ \sum_{\ell_1' = 0}^1 \Biggl( \queTp^{-3/4} \left[\sum_{\vert I\vert \leq m} \left\|\quuprtl^I \partial_i^{\ell'} \partial_\qusb^{\ell_1'} u\right\|_{L^2(\quSigs)}^2\right]^{1/2} + \|\partial_i^{\ell'} \partial_\qusb^{\ell_1'} u\|_{\qupbH^{m, 1, 1/4}(\qupbar\quAsigmat)} \Biggr) \Biggr].
\end{gather}
\end{proposition}
\begin{proof} For $\quupsilon \in [\quusig, \qusigmat]$, let $B_\quupsilon = \quAsigmu \cap \qubulks$ (so that, in particular, $B_\quupsilon = \emptyset$ if $\quupsilon < 0$). If $f$ is any $C^\infty$ function on $\qubulk \cap \{ \qutau \in [\quusig, \qusigmat] \}$, then
\begin{align*}
\int_{B_{\quupsilon}} \vert f\vert ^2\,d\quxi\,d\quzeta &= \int_{B_{\quupsilon}} \left\vert \int_0^{\frac{1}{\sqrt{2}} (\quupsilon + \quzeta)} \partial_\qusb f\,d\qusb + f\vert _{\qusb = 0}\right\vert ^2\,d\quxi\,d\quzeta\\
&\leq \int_{B_{\quupsilon}} \sqrt{2} (\quupsilon + \quzeta) \int_0^{\frac{1}{\sqrt{2}} (\quupsilon + \quzeta)} \vert \partial_\qusb f\vert ^2\,d\qusb\,d\quxi\,d\quzeta + 2\sqrt{2} \int_{\quSigs} \left\vert f\right\vert ^2\,d\quxb\,d\quvb\\
&\leq 2\queTp \int_{\quusig}^{\quupsilon} \int_{B_{\quupsilon'}} \vert \partial_\qusb f\vert ^2\,d\quxi\,d\quzeta \,d\quupsilon' + 2\sqrt{2} \int_{\quSigs} \left\vert f\right\vert ^2\,d\quxb\,d\quvb.
\end{align*}
Applying this inequality together with Corollary \ref{shuffleops} and \eqref{realDHbprodAthin} from Proposition \ref{mfgagniren}, we have 
\begin{gather*}
\queTp{}^{-3/2} \sum_{|I| \leq m} \int_{B_{\quupsilon}} \left\vert\quuprtl^I \partial_i^\ell u\right\vert^2\,d\quxb\,d\quzeta \leq \queTp{}^{-3/2} \sum_{|I| \leq m} \Biggl\{ 2\queTp \int_{\quusig}^{\quupsilon} \int_{B_\quupsilon'} \left\vert\partial_\qusb \quuprtl^I \partial_i^\ell u\right\vert^2\,d\quxb\,d\quzeta\,d\quupsilon'\\
+ 2\sqrt{2} \int_{\quSigs} \left\vert\quuprtl^I \partial_i^\ell u\right\vert^2\,d\quxb\,d\quvb\Biggr\}\\
\leq 2\queTp C \int_{\quusig}^{\quupsilon} \left\vert\partial_\qusb \partial_i^\ell u \right\vert_{\quDHb^{m}(B_{\quupsilon'})}^2 + \left\vert\partial_i^\ell u\right\vert_{\quDHb^{m}(B_{\quupsilon'})}^2\,d\quupsilon'\\
+ 2\sqrt{2} \queTp{}^{-3/2} \int_{\quSigs} \left\vert\quuprtl^I \partial_i^\ell u\right\vert^2\,d\quxb\,d\quvb,\\
\queTp{}^{-3/2} \sum_{|I| \leq m} \int_{B_{\quupsilon}} \left\vert\quuprtl^I \partial_\qusb \partial_i^\ell u\right\vert^2\,d\quxb\,d\quzeta \leq 2\queTp C \int_{\quusig}^{\quupsilon} \left\vert\partial_\qusb^2 \partial_i^\ell u \right\vert_{\quDHb^{m}(B_{\quupsilon'})}^2 + \left\vert\partial_\qusb \partial_i^\ell u\right\vert_{\quDHb^{m}(B_{\quupsilon'})}^2\,d\quupsilon'\\
+ 2\sqrt{2} \queTp{}^{-3/2} \int_{\quSigs} \left\vert\quuprtl^I \partial_\qusb \partial_i^\ell u\right\vert^2\,d\quxb\,d\quvb\\
\leq 4\queTp C' \int_{\quusig}^{\quupsilon} \sum_{\ell' \leq \ell} \left\{ \left\vert\partial_i^{\ell'}\alpha\right\vert_{\quDHb^{m}(B_{\quupsilon'})}^2 \left\vert \partial_\qusb\partial_i^{\ell'} u\right\vert_{\quDHb^{m}(B_{\quupsilon'})}^2 + \left\vert\partial_i^{\ell'}\beta\right\vert_{\quDHb^{m}(B_{\quupsilon'})}^2 \left\vert\partial_i^{\ell'} u\right\vert_{\quDHb^{m}(B_{\quupsilon'})}^2\right\}\\
+ \left\vert\partial_\qusb \partial_i^\ell u\right\vert_{\quDHb^{m}(B_{\quupsilon'})}^2 + \left\vert\partial_i^\ell\delta\right\vert_{\quDHb^{m}(B_{\quupsilon'})}^2\,d\quupsilon'\\
+ 2\sqrt{2} \queTp{}^{-3/2} \int_{\quSigs} \left\vert\quuprtl^I \partial_\qusb \partial_i^\ell u\right\vert^2\,d\quxb\,d\quvb,\\
\queTp{}^{-3/2} \sum_{\ell_1' = 0}^1 \sum_{|I| \leq m} \int_{B_\quupsilon} \left\vert\quuprtl^I \partial_\qusb^{\ell_1'} \partial_i^\ell u\right\vert^2\,d\quxb\,d\quzeta\\
\leq 4\queTp C' \Biggl\{ \int_{\quusig}^{\quupsilon} \left(C_1^2 + 2\right) \sum_{{\ell_1' = 0, 1}\atop{\ell' \leq \ell}} \left\vert\partial_\qusb^{\ell_1'}\partial_i^{\ell'} u\right\vert_{\quDHb^{m}(B_{\quupsilon'})}^2\,d\quupsilon' + \int_{\quusig}^{\quupsilon} \left\vert\partial_i^{\ell}\delta\right\vert_{\quDHb^{m}(B_{\quupsilon'})}^2\,d\quupsilon'\Biggr\}\\
+ 2\sqrt{2} \queTp{}^{-3/2} \sum_{\ell_1' = 0, 1} \int_{\quSigs} \left\vert\quuprtl^I \partial_\qusb^{\ell_1'} \partial_i^\ell u\right\vert^2\,d\quxb\,d\quvb,\\
\end{gather*}
where $C_1$ is a constant depending only on an upper bound on the quantity in \eqref{gwalphbetbound}. Thus
\begin{gather*}
\sum_{\ell_1' = 0}^1 \left\vert\partial_\qusb^{\ell_1'} \partial_i^\ell u\right\vert_{\quDHb^{m}(B_{\quupsilon})}^2\\
\leq 8\queTp C' \Biggl\{ \int_{\quusig}^{\quupsilon} \left(C_1^2 + 2\right) \sum_{{\ell_1' = 0, 1}\atop{\ell' \leq \ell}} \left\vert\partial_\qusb^{\ell_1'}\partial_i^{\ell'} u\right\vert_{\quDHb^{m}(B_{\quupsilon'})}^2\,d\quupsilon' + \int_{\quusig}^{\quupsilon} \left\vert\partial_i^\ell\delta\right\vert_{\quDHb^{m}(B_{\quupsilon'})}^2\,d\quupsilon'\Biggr\}\\
+ 4\sqrt{2} \queTp{}^{-3/2} \sum_{\ell_1' = 0}^1 \int_{\quSigs} \left\vert\quuprtl^I \partial_\qusb^{\ell_1'} \partial_i^\ell u\right\vert^2\,d\quxb\,d\quvb + 2\sum_{\ell_1' = 0}^1 \left\vert\partial_\qusb^{\ell_1'} \partial_i^\ell u\right\vert_{\qupbH^{m, 1, 1/4}(\qupbar\quAsigmat)}^2;\\
\end{gather*}
and applying a Gr\"onwall inequality
gives the desired result.\end{proof}
\par
The estimates necessary to obtain \eqref{GOindimp} -- \eqref{GOdiffindimp} can now be obtained as follows. We begin with \eqref{GOindimp}. As in Section \ref{initdat}, we must bound $\qudlb$, $\qubb$, and $\qucb$ (and their derivatives) {\it in that order}, due to the nonlinear dependence of $\qubb$ on $\qudlb$, and $\qucb$ on both $\qudlb$ and $\qubb$, arising from \eqref{nsricone} -- \eqref{nsricthree}. For functions $f$, $g$, $\ldots$, we use the notation $C(f, g, \ldots)$ to indicate a constant which depends only on an upper bound on the quantities in parentheses. We also define the (semi)norms ($\usigf^\Ldeg$, etc., indicate $\Ldeg$ applications of $\usigf$)
\begin{align}\label{gwspecnorm}
\|f\|_{L^\infty_\Ldeg \quDHb^m(\quAsigmat)} &= \sup_{\qusigmat' \in [\usigf^\Ldeg(\qusigmat), \qusigmat]} \|f\|_{\quDHb^m(\quAsigmatp)},\\
\|f\|_{L^\infty_\Ldeg \qupbH^{m, 1, 1/4}(\qupbar\quAsigmat)} &= \sup_{\qusigmat' \in [\usigf^\Ldeg(\qusigmat), \qusigmat]} \|f\|_{\qupbH^{m, 1, 1/4}(\qupbar\quAsigmatp)},\\
\|f\|_{L^\infty_\Ldeg \quDHb^m(\quSigsp)} &= \queTp^{-3/4} \sup_{\qusigmat' \in [\usigf^\Ldeg(\qusigmat), \qusigmat]} \left[\sum_{\vert I\vert \leq m} \left\|f\right\|_{L^2(\quSigsp)}^2\right]^{1/2}
\end{align}
for all functions $f$ for which the right-hand sides are defined. Proposition \ref{refgronwallsblv} then gives the bounds ($\Ldeg \geq 0$)
\begin{gather}\label{gronwallressimp}
\|\partial_i^\ell \partial_\qusb^{\ell_1} u\|_{L^\infty_\Ldeg \quDHb^m(\quAsigmat)} \leq C(\sum_{\ell' \leq \ell} \|\partial_i^{\ell'} \alpha\|_{L^\infty_{\Ldeg + 1} \quDHb^m(\quAsigmat)} + \|\partial_i^{\ell'} \beta\|_{L^\infty_{\Ldeg + 1} \quDHb^m(\quAsigmat)}) \\
\cdot\sum_{\ell' \leq \ell} \Biggl[ \queTp \left\|\partial_i^{\ell'} \delta\right\|_{L^\infty_{\Ldeg + 1} \quDHb^m(\quAsigmat)}\notag\\
+ \sum_{\ell_1' = 0}^1 \Biggl( \left\|\quuprtl^I \partial_i^{\ell'} \partial_\qusb^{\ell_1'} u\right\|_{L^\infty_\Ldeg \quDHb^m(\quSigsp)} + \|\partial_i^{\ell'} \partial_\qusb^{\ell_1'} u\|_{L^\infty_{\Ldeg + 1} \qupbH^{m, 1, 1/4}(\qupbar\quAsigmat)} \Biggr) \Biggr].\notag
\end{gather}
\par
Now $\qudlb_{\mu + 1}$ satisfies (see \eqref{nsricone})
\begin{equation}\label{gwricone}
\partial_\qusb^2 \qudlb_{\mu + 1} = -2 (1 + \qupark^{-2\quiota} \qudlb_{\mu + 1}) (\partial_\qusb\qugamb_\mu)^2,
\end{equation}
so that \eqref{gronwallressimp}, together with the bounds on the initial data in \eqref{omegbbaseinitbound}--\eqref{omegbbIbootinf}, give the estimate ($\partial_i$ one of $\partial_\quxb$, $\partial_\quvb$, $\ell, \ell_1 \in \{ 0, 1 \}$, $\Ldeg \geq 0$)
\begin{gather}\label{qudlbgwbound}
\|\partial_i^\ell \partial_\qusb^{\ell_1} \qudlb_{\mu + 1}\|_{L^\infty_\Ldeg \quDHb^m(\quAsigmat)} \leq C(\sum_{\ell' \leq \ell} \|\partial_i^{\ell'} \partial_\qusb \qugamb_\mu\|_{L^\infty_{\Ldeg + 1} \quDHb^m(\quAsigmat)})\\
\cdot\Biggl\{ \queTp \sum_{\ell' \leq \ell} \Biggl[ \left\|\partial_i^{\ell'} \partial_\qusb \qugamb_\mu\right\|_{L^\infty_{\Ldeg + 1} \quDHb^m(\quAsigmat)}^2 \widehat{\usigf^{\Ldeg + 1}(\qusigmat)}^{-1/4} \Biggr] + \qunu^2 \widehat{\usigf^\Ldeg(\qusigmat)}^{-1/4} (1 + [\Ldeg + 1]^{1/2})\Biggr\}.\notag
\end{gather}
In particular, when $\qunu \qupark^{-2\iota} \leq 1/(2C(\gambC) \quCKS)$, we have $|\quellb_\mu| \geq 1/2$ (see Remark \ref{bigparkrem}). We assume this lower bound on $|\quellb_\mu|$ throughout the rest of this section. Next, $\qubb_{\mu + 1}$ satisfies (see \eqref{nsrictwo})
\begin{equation}\label{gwrictwo}
\partial_\qusb^2 \qubb_{\mu + 1} = \frac{1}{\quellb_{\mu + 1}} \qupark^{-2\quiota} (\partial_\qusb \qudlb_{\mu + 1}) (\partial_\qusb\qubb_{\mu + 1}) - 4 \partial_\qusb\qugamb_\mu (\partial_\quxb \qugamb_\mu + \qupark^{-2\quiota} \qubb_{\mu + 1} \partial_\qusb\qugamb_\mu),
\end{equation}
so that we obtain from Lemma \ref{Flipsch}, \eqref{gronwallressimp}, \eqref{omegbbaseinitbound}--\eqref{omegbbIbootinf}, and \eqref{qudlbgwbound} that ($\partial_i = \partial_\quxb$, $\ell, \ell_1 \in \{ 0, 1 \}$, $\Ldeg \geq 0$)
\begin{gather}\label{qubbgwbound}
\|\partial_i^\ell \partial_\qusb^{\ell_1} \qubb_{\mu + 1}\|_{L^\infty_\Ldeg \quDHb^m(\quAsigmat)}\\
\leq C\biggl(\sum_{\ell' \leq \ell} \biggl[ \|\partial_i^{\ell'} \partial_\qusb \qugamb_\mu\|_{L^\infty_{\Ldeg + 1} \quDHb^m(\quAsigmat)} + \sum_{\ell_1' = 0}^1 \|\partial_i^{\ell'} \partial_\qusb^{\ell_1'} \qudlb\|_{L^\infty_{\Ldeg + 1} \quDHb^m(\quAsigmat)}\biggr]\biggr)\\
\cdot \Biggl\{ \queTp \left[ \|\partial_\qusb \partial_i^{\ell'} \qugamb_\mu\|_{L^\infty_{\Ldeg + 1} \quDHb^m(\quAsigmat)} + \|\partial_\quxb \partial_i^{\ell'} \qugamb_\mu\|_{L^\infty_{\Ldeg + 1} \quDHb^m(\quAsigmat)}\right]^2 \widehat{\usigf^{\Ldeg + 1}(\qusigmat)}^{-1/4}\notag\\
+ \qunu^2\widehat{\usigf^\Ldeg(\qusigmat)}^{-1/4} (1 + [\Ldeg + 1]^{1/2})\Biggr\}\notag\\
\leq C(\sum_{\ell' \leq \ell} \|\partial_i^{\ell'} \partial_\qusb \qugamb_\mu\|_{L^\infty_{\Ldeg + 2} \quDHb^m(\quAsigmat)}, r)\notag\\
\cdot \Biggl\{ \queTp \left[ \|\partial_\qusb \partial_i^{\ell'} \qugamb_\mu\|_{L^\infty_{\Ldeg + 1} \quDHb^m(\quAsigmat)}^2 + \|\partial_\quxb \partial_i^{\ell'} \qugamb_\mu\|_{L^\infty_{\Ldeg + 1} \quDHb^m(\quAsigmat)}^2\right] \widehat{\usigf^{\Ldeg + 1}(\qusigmat)}^{-1/4}\notag\\
+ \qunu^2\widehat{\usigf^\Ldeg(\qusigmat)}^{-1/4} (1 + [\Ldeg + 1]^{1/2})\Biggr\}.\notag
\end{gather}
Finally, $\qucb_{\mu + 1}$ satisfies (see \eqref{nsricthree})
\begin{gather}\label{gwricthree}
\partial_\qusb^2 \qucb_{{\mu} + 1} = -4 \partial_\qusb \qugamb_{\mu} \partial_\quvb \qugamb_{\mu} - \frac{2}{\quab_{{\mu} + 1}} (\partial_\quxb\qugamb_{\mu})^2\\
\kern -0.125in\notag + \qupark^{-2\quiota} \left( \frac{1}{2\quab_{{\mu} + 1}} (\partial_\qusb \qubb_{{\mu} + 1})^2 - 2\qucb_{{\mu} + 1}(\partial_\qusb\qugamb_{\mu})^2 - \frac{4}{\quab_{{\mu} + 1}} \qubb_{{\mu} + 1} \partial_\quxb \qugamb_{\mu} \partial_\qusb \qugamb_{\mu} \right)\\
\kern-0.125in\notag- \qupark^{-4\quiota} \frac{2}{\quab_{{\mu} + 1}} \qubb_{{\mu} + 1}^2 (\partial_\qusb\qugamb_{\mu})^2,
\end{gather}
so that $\qucb_{\mu + 1}$ satisfies the bounds ($\ell_1 \in \{ 0, 1 \}$, $\Ldeg \geq 0$)
\begin{gather}\label{qucbgwbound}
\|\partial_\qusb^{\ell_1} \qucb_{\mu + 1}\|_{L^\infty_\Ldeg \quDHb^m(\quAsigmat)} \leq C(\|\partial_\qusb \qugamb_\mu\|_{L^\infty_{\Ldeg + 1} \quDHb^m(\quAsigmat)}) \Biggl\{ \queTp\\
\cdot\Biggl[ \|\partial_\qusb \qugamb_\mu\|_{L^\infty_{\Ldeg + 1} \quDHb^m(\quAsigmat)} + \|\partial_\quxb\qugamb_\mu\|_{L^\infty_{\Ldeg + 1}\quDHb^m(\quAsigmat)} + \|\partial_\quvb\qugamb_\mu\|_{L^\infty_{\Ldeg + 1}\quDHb^m(\quAsigmat)}\notag\\
+ \|\qubb_{\mu + 1}\|_{L^\infty_{\Ldeg + 1}\quDHb^m(\quAsigmat)} + \|\partial_\qusb\qubb_{\mu + 1}\|_{L^\infty_{\Ldeg + 1}\quDHb^m(\quAsigmat)} + \|\qudlb_{\mu + 1}\|_{L^\infty_{\Ldeg + 1}\quDHb^m(\quAsigmat)}\Biggr]^2 \widehat{\usigf^{\Ldeg + 1}(\qusigmat)}^{-1/4}\notag\\
+ \qunu^2\widehat{\usigf^\Ldeg(\qusigmat)}^{-1/4} (1 + [\Ldeg + 1]^{1/2})\Biggr\}\notag\\
\kern-12pt \leq C(\|\partial_\qusb \qugamb_\mu\|_{L^\infty_{\Ldeg + 3} \quDHb^m(\quAsigmat)}, \|\partial_\quxb \qugamb_\mu\|_{L^\infty_{\Ldeg + 2} \quDHb^m(\quAsigmat)}, r)\notag\\
\cdot \Biggl\{ \queTp \Biggl[ \|\partial_\qusb \qugamb_\mu\|_{L^\infty_{\Ldeg + 2} \quDHb^m(\quAsigmat)}^2 + \|\partial_\quxb\qugamb_\mu\|_{L^\infty_{\Ldeg + 2}\quDHb^m(\quAsigmat)}^2 + \|\partial_\quvb\qugamb_\mu\|_{L^\infty_{\Ldeg + 1}\quDHb^m(\quAsigmat)}^2\Biggr]\widehat{\usigf^{\Ldeg + 1}(\qusigmat)}^{-1/4}\notag\\
+ \qunu^2\widehat{\usigf^\Ldeg(\qusigmat)}^{-1/4} (1 + [\Ldeg + 1]^{1/2})\Biggr\}.\notag
\end{gather}
Now it is clear that there is a constant $C(\Ldeg)$ such that for all $\qusigmat \geq 0$
\begin{equation}\label{Lsigest}
\widehat{\usigf^\Ldeg(\qusigmat)}^{-1/4} \leq C(\Ldeg) \qusigu^{-1/4}.
\end{equation}
It is thus clear that the induction hypothesis \eqref{gambinduct},
\begin{equation}
\|\partial_\qusb^\ell\partial_i \qugamb_\mu\|_{\quDHb^{\qusbn - 1}(\quAsigmat)} \leq \gambC \qunu,\,\|\partial_\quxb^2 \qugamb_\mu\|_{\quDHb^{\qusbn - 1}(\quAsigmat)} \leq \gambCdx \qunu\hbox{ for $\qusigmat \in [0, \qupark\queT]$},\label{concrgambinduct}\\
\end{equation}
gives the estimates ($\ell \in \{ 0, 1 \}$, $\qusigmat \in [0, \qupark\queT]$)
\begin{equation}\label{quomegbgwbound}
\|\partial_\qusb^\ell \quomegb\|_{\quDHb^{\qusbn - 1}(\quAsigmat)} \leq \qOmegbC(\gambC, \gambCdx) \qunu^2 \qusigu^{-1/4}
\end{equation}
for all $\quomegb \in \{ \qudlb, \partial_\quxb \qudlb, \partial_\quvb \qudlb, \qubb, \partial_\quxb\qubb, \qucb \}$. Note that the bound on $\partial_\quxb^2 \qugamb_\mu$ in \eqref{concrgambinduct} is needed only in bounding $\partial_\quxb \qubb$.
\par
It now remains to consider the bounds \eqref{omegdiffinduct}, i.e., we are to prove
\begin{gather}
\|\partial_\qusb^\ell (\quomegb_{\mu + 2} - \quomegb_{\mu + 1})\|_{\quDHb^{\qusbn - 2}(\quAsigmat)} \leq \frac{1}{2^{\mu + 1}} \OmegbCp \qunu \qusigu^{-1/4}\tag{O$_{\mu}'$}\\
\hbox{ for $\qusigmat \in [0, \qupark\queT]$, $\quomegb_{\mu'} \in \{ \qudlb_{\mu'}, \qubb_{\mu'}, \qucb_{\mu'}, \partial_\quxb \qudlb_{\mu'}, \partial_\quvb \qudlb_{\mu'}, \partial_\quxb \qubb_{\mu'} \}$}.\notag
\end{gather}
We begin with the following general considerations: suppose we have functions $u$, $u'$ which satisfy
\begin{align}
\partial_\qusb^2 u &= \alpha \partial_\qusb u + \beta u + \delta,\\
\partial_\qusb^2 u' &= \alpha' \partial_\qusb u' + \beta' u' + \delta',
\end{align}
and also have identical initial data; then the difference satisfies
\begin{equation}
\partial_\qusb^2 (u - u') = \alpha \partial_\qusb (u - u') + \beta (u - u') + (\alpha - \alpha') \partial_\qusb u' + (\beta - \beta') u' + \delta - \delta',
\end{equation}
so that by \eqref{gronwallressimp}, schematically, $u - u'$ satisfies bounds of the form
\begin{gather*}
\|\partial_i^\ell \partial_\qusb^{\ell_1} (u - u')\|_{L^\infty_\Ldeg \quDHb^m(\quAsigmat)} \leq C(\sum_{\ell' \leq \ell} \|\partial_i^{\ell'} \alpha\|_{L^\infty_{\Ldeg + 1} \quDHb^m(\quAsigmat)} + \|\partial_i^{\ell'} \beta\|_{L^\infty_{\Ldeg + 1} \quDHb^m(\quAsigmat)})\\
\cdot\sum_{\ell' \leq \ell} \queTp \Biggl[ \left\|\partial_i^{\ell'} (\delta - \delta')\right\|_{L^\infty_{\Ldeg + 1} \quDHb^m(\quAsigmat)} + \left\|\partial_i^{\ell'} [(\alpha - \alpha')\partial_\qusb u']\right\|_{L^\infty_{\Ldeg + 1} \quDHb^m(\quAsigmat)}\\
+ \left\|\partial_i^{\ell'} [(\beta - \beta') u']\right\|_{L^\infty_{\Ldeg + 1} \quDHb^m(\quAsigmat)}\Biggr]\\
\leq C(\sum_{\ell' \leq \ell} \|\partial_i^{\ell'} \alpha\|_{L^\infty_{\Ldeg + 1} \quDHb^m(\quAsigmat)} + \|\partial_i^{\ell'} \beta\|_{L^\infty_{\Ldeg + 1} \quDHb^m(\quAsigmat)})\\
\cdot \Biggl\{ \queTp \sum_{\ell' \leq \ell} \left\|\partial_i^{\ell'} (\delta - \delta')\right\|_{L^\infty_{\Ldeg + 1} \quDHb^m(\quAsigmat)}\\
+ \biggl[ \sum_{\ell' \leq \ell} \bigl( \|\partial_i^{\ell'} (\alpha - \alpha')\|_{L^\infty_{\Ldeg + 1} \quDHb^m(\quAsigmat)} + \left\|\partial_i^{\ell'} (\beta - \beta')\right\|_{L^\infty_{\Ldeg + 1} \quDHb^m(\quAsigmat)}\bigr) \biggr]\\
\sum_{{\ell' \leq \ell}\atop{\ell_1' = 0, 1}} \|\partial_i^{\ell'} \partial_\qusb^{\ell_1} u'\|_{L^\infty_{\Ldeg + 1} \quDHb^m(\quAsigmat)} \Biggr\}.
\end{gather*}
Note that the only substantial difference with \eqref{gronwallressimp} is the additional forcing terms involving $\alpha - \alpha'$ and $\beta - \beta'$. These terms -- as well as those involving $\delta - \delta'$ -- can be bounded by the second half of Lemma \ref{Flipsch}. From \eqref{gwricone}, then, we obtain the following bounds: for $\qudlb$ (as before, $\partial_i$ is one of $\partial_\quxb$, $\partial_\quvb$, $\ell, \ell_1 \in \{ 0, 1 \}$, and $\Ldeg \geq 0$),
\begin{gather}\label{qudlbdiffgwbound}
\|\partial_i^\ell \partial_\qusb^{\ell_1} (\qudlb_{\mu + 2} - \qudlb_{\mu + 1})\|_{L^\infty_\Ldeg \quDHb^m(\quAsigmat)} \leq C(\sum_{\ell' \leq \ell} \|\partial_i^{\ell'} \partial_\qusb \qugamb_{\mu + 1}\|_{L^\infty_{\Ldeg + 1} \quDHb^m(\quAsigmat)})\\
\cdot \Biggl\{ \sum_{\ell' \leq \ell} \Biggl[ \bigl( \left\|\partial_i^{\ell'} \partial_\qusb \qugamb_{\mu + 1}\right\|_{L^\infty_{\Ldeg + 1}\quDHb^m(\quAsigmat)} + \left\|\partial_i^{\ell'} \partial_\qusb \qugamb_\mu\right\|_{L^\infty_{\Ldeg + 1}\quDHb^m(\quAsigmat)}\bigr)\notag\\
\cdot\left\|\partial_i^{\ell'} \partial_\qusb (\qugamb_{\mu + 1} - \qugamb_\mu)\right\|_{L^\infty_{\Ldeg + 1} \quDHb^m(\quAsigmat)} \widehat{\usigf^{\Ldeg + 1}(\qusigmat)}^{-1/4} \Biggr]\notag\\
\cdot \Biggl[ \queTp + \sum_{{\ell' \leq \ell}\atop{\ell_1' \leq \ell_1}} \|\partial_i^{\ell'} \partial_\qusb^{\ell_1} \qudlb_{\mu + 1}\|_{L^\infty_{\Ldeg + 1} \quDHb^m(\quAsigmat)} \Biggr] \Biggr\}\notag\\
\leq C(\sum_{\ell' \leq \ell} \|\partial_i^{\ell'} \partial_\qusb \qugamb_{\mu + 1}\|_{L^\infty_{\Ldeg + 1} \quDHb^m(\quAsigmat)}, \sum_{\ell' \leq \ell} \|\partial_i^{\ell'} \partial_\qusb \qugamb_\mu\|_{L^\infty_{\Ldeg + 2} \quDHb^m(\quAsigmat)})\notag\\
\cdot\left\|\partial_i^{\ell'} \partial_\qusb (\qugamb_{\mu + 1} - \qugamb_\mu)\right\|_{L^\infty_{\Ldeg + 1} \quDHb^m(\quAsigmat)} \widehat{\usigf^{\Ldeg + 1}(\qusigmat)}^{-1/4};\notag
\end{gather}
for $\qubb$ ($\partial_i = \partial_\quxb$, $\ell, \ell_1 \in \{ 0, 1 \}$, $\Ldeg \geq 0$),
\begin{gather*}
\|\partial_i^\ell \partial_\qusb^{\ell_1} (\qubb_{\mu + 2} - \qubb_{\mu + 1})\|_{L^\infty_\Ldeg \quDHb^m(\quAsigmat)}\\
\leq C(\sum_{{\ell' \leq \ell, j = 0, 1}\atop{\mu' = \mu, \mu + 1}} \|\partial_i^{\ell'} \partial_j \qugamb_{\mu'}\|_{L^\infty_{\Ldeg + 1} \quDHb^m(\quAsigmat)}, \sum_{{\ell' \leq \ell, \ell_1' = 0, 1}\atop{\mu' = \mu, \mu + 1}} \|\partial_i^{\ell'} \partial_\qusb^{\ell'_1} \qudlb_{\mu'}\|_{L^\infty_{\Ldeg + 1} \quDHb^m(\quAsigmat)}, r)\\
\cdot \Biggl\{ \Biggl[ \queTp \sum_{{\ell' \leq \ell, j = 0, 1}\atop{\mu' = \mu, \mu + 1}} \|\partial_i^{\ell'} \partial_j \qugamb_{\mu'}\|_{L^\infty_{\Ldeg + 1}\quDHb^m(\quAsigmat)}\\
\sum_{\ell' \leq \ell, j = 0, 1} \bigl( \|\partial_i^{\ell'} (\partial_j \qugamb_{\mu + 1} - \partial_j \qugamb_\mu)\|_{L^\infty_{\Ldeg + 1}\quDHb^m(\quAsigmat)}\bigr) \widehat{\usigf^{\Ldeg + 1}(\qusigmat)}^{-1/4}\Biggr]\notag\\
+ \Biggl[ \sum_{{\ell' \leq \ell, \ell'_1 = 0, 1}\atop{\mu' = \mu, \mu + 1}} \|\partial_i^{\ell'} \partial_\qusb^{\ell'_1} \qudlb_{\mu'}\|_{L^\infty_{\Ldeg + 1} \quDHb^m(\quAsigmat)} \sum_{{\ell' \leq \ell}\atop{\ell'_1 = 0, 1}} \|\partial_i^{\ell'} \partial_\qusb^{\ell'_1} (\qudlb_{\mu + 1} - \qudlb_\mu)\|_{L^\infty_{\Ldeg + 1} \quDHb^m(\quAsigmat)}\notag\\
+ \sum_{{\ell' \leq \ell}\atop{\mu' = \mu, \mu + 1}} \|\partial_i^{\ell'} \partial_\qusb \qugamb_{\mu'}\|_{L^\infty_{\Ldeg + 1} \quDHb^m(\quAsigmat)} \sum_{\ell' \leq \ell} \|\partial_i^{\ell'} \partial_\qusb (\qugamb_{\mu + 1} - \qugamb_\mu)\|_{L^\infty_{\Ldeg + 1} \quDHb^m(\quAsigmat)}\Biggr]\notag\\
\cdot \left[ \queTp \sum_{\ell' \leq \ell, j = 0, 1} \|\partial_i^{\ell'} \partial_j \qugamb_\mu\|_{L^\infty_{\Ldeg + 2}\quDHb^m(\quAsigmat)}^2 + 1\right] \widehat{\usigf^{\Ldeg + 2}(\qusigmat)}^{-1/4}\Biggr\}\notag\\
\leq C(\sum_{{\ell' \leq \ell, j = 0, 1}\atop{\mu' = \mu, \mu + 1}} \|\partial_i^{\ell'} \partial_j \qugamb_{\mu'}\|_{L^\infty_{\Ldeg + 2} \quDHb^m(\quAsigmat)}, \sum_{\ell' \leq \ell} \|\partial_i^{\ell'} \partial_\qusb \qugamb_\mu\|_{L^\infty_{\Ldeg + 3} \quDHb^m(\quAsigmat)}, r)\notag\\
\cdot \biggl[ \sum_{\ell' \leq \ell} \|\partial_i^{\ell'} \partial_\quxb (\qugamb_{\mu + 1} - \qugamb_\mu)\|_{L^\infty_{\Ldeg + 1} \quDHb^m(\quAsigmat)} \widehat{\usigf^{\Ldeg + 1}(\qusigmat)}^{-1/4}\notag\\
+ \sum_{\ell' \leq \ell} \|\partial_i^{\ell'} \partial_\qusb (\qugamb_{\mu + 1} - \qugamb_\mu)\|_{L^\infty_{\Ldeg + 2} \quDHb^m(\quAsigmat)} \widehat{\usigf^{\Ldeg + 2}(\qusigmat)}^{-1/4}\biggr];\notag
\end{gather*}
and for $\qucb$, leaving the details to the reader ($\ell \in \{ 0, 1 \}$, $\Ldeg \geq 0$),
\begin{gather*}
\|\partial_\qusb^\ell (\qucb_{\mu + 2} - \qucb_{\mu + 1})\|_{L^\infty_\Ldeg \quDHb^m(\quAsigmat)}\\
\leq C(\sum_{\mu' = \mu, \mu + 1} \|\partial_\qusb \qugamb_{\mu'}\|_{L^\infty_{\Ldeg + 4} \quDHb^m(\quAsigmat)}, \sum_{\mu' = \mu, \mu + 1} \|\partial_\quvb \qugamb_{\mu'}\|_{L^\infty_{\Ldeg + 3} \quDHb^m(\quAsigmat)},\\
\sum_{\mu' = \mu, \mu + 1} \|\partial_\quvb \qugamb_{\mu'}\|_{L^\infty_{\Ldeg + 2} \quDHb^m(\quAsigmat)}, r)\notag\\
\cdot \Biggl[ \|\partial_\qusb (\qugamb_{\mu + 1} - \qugamb_\mu)\|_{L^\infty_{\Ldeg + 3}\quDHb^m(\quAsigmat)} \widehat{\usigf^{\Ldeg + 3}(\qusigmat)}^{-1/4} + \|\partial_\quxb (\qugamb_{\mu + 1} - \qugamb_\mu)\|_{L^\infty_{\Ldeg + 2}\quDHb^m(\quAsigmat)} \widehat{\usigf^{\Ldeg + 2}(\qusigmat)}^{-1/4}\notag\\
+ \|\partial_\quvb (\qugamb_{\mu + 1} - \qugamb_\mu)\|_{L^\infty_{\Ldeg + 1}\quDHb^m(\quAsigmat)} \widehat{\usigf^{\Ldeg + 1}(\qusigmat)}^{-1/4} \Biggr].\notag
\end{gather*}
It is clear that the foregoing estimates are sufficient to establish \eqref{omegdiffinduct} when G$_{\mu}$, G$_{\mu + 1}$, and G$'_\mu$ hold, thus completing the proof of \eqref{GOdiffindimp}.
\subsection{Boundedness of $\qugamb_\mu$}\label{enerperboot} We now complete the induction by proving \eqref{OGindimp} -- \eqref{OGdiffindimp}, that is, that for any $\mu \geq 1$ (see \eqref{gambinduct}, \eqref{omeginduct})
\begin{gather}
\|\partial_\qusb^\ell \quomegb_\mu\|_{\quDHb^{\qusbn - 1}(\quAsigmat)} \leq \OmegbC \qunu^2 \qusigu^{-1/4}\label{ciomeginduct}\\
\hbox{ for $\qusigmat \in [0, \qupark\queT]$, $\quomegb_\mu \in \{ \qudlb_\mu, \qubb_\mu, \qucb_\mu, \partial_\quxb \qudlb_\mu, \partial_\quvb \qudlb_\mu, \partial_\quxb \qubb_\mu \}$}\notag
\end{gather}
for $\ell = 0, 1$ implies
\begin{equation}\label{cigambinduct}
\|\partial_\qusb^\ell\partial_i \qugamb_\mu\|_{\quDHb^{\qusbn - 1}(\quAsigmat)} \leq \gambC \qunu,\,\|\partial_\quxb^2 \qugamb_\mu\|_{\quDHb^{\qusbn - 1}(\quAsigmat)} \leq \gambCdx \qunu\hbox{ for $\qusigmat \in [0, \qupark\queT]$}
\end{equation}
for $\ell = 0, 1$, $i = 0, 1, 2$, and (see \eqref{gambdiffinduct}, \eqref{omegdiffinduct})
\begin{gather}
\|\partial_\qusb^\ell (\quomegb_{\mu + 1} - \quomegb_\mu)\|_{\quDHb^{\qusbn - 2}(\quAsigmat)} \leq \frac{1}{2^\mu} \OmegbCp \qunu \qusigu^{-1/4}\label{ciomegdiffinduct}\\
\hbox{ for $\qusigmat \in [0, \qupark\queT]$, $\quomegb_\mu \in \{ \qudlb_\mu, \qubb_\mu, \qucb_\mu, \partial_\quxb \qudlb_\mu, \partial_\quvb \qudlb_\mu, \partial_\quxb \qubb_\mu \}$}\notag
\end{gather}
for $\ell = 0, 1$ implies
\begin{gather}
\|\partial_\qusb^\ell\partial_i (\qugamb_{\mu + 1} - \qugamb_\mu)\|_{\quDHb^{\qusbn - 2}(\quAsigmat)} \leq \frac{1}{2^\mu} \gambCp \qunu,\,\|\partial_\quxb^2 (\qugamb_{\mu + 1} - \qugamb_\mu)\|_{\quDHb^{\qusbn - 2}(\quAsigmat)} \leq \frac{1}{2^\mu} \gambCdxp \qunu\label{cigambdiffinduct}\\
\hbox{ for $\qusigmat \in [0, \qupark\queT]$}\notag
\end{gather}
for $\ell = 0, 1$, $i = 0, 1, 2$.
\par
We begin as in the previous section by marshalling some preliminary results. The following estimates are elementary.
\par
\begin{proposition}\label{siguprop} (i) If $\quvb \geq 0$ and $\qusb \in [0, 1/(2\sqrt{2})]$, then
\begin{equation}\label{vbjysigu}
\frac{1}{3} (1 + \quvb) \leq \qutauu \leq \frac{5}{4} (1 + \quvb)
\end{equation}
(where $\qutauu = \max \{ 1, 2^{-1/2} (\qusb + \quvb) \}$).
\par
(ii) For $p \in (0, 1)$,
\begin{equation*}
\int_0^\qusigmat \quupsu^{-p}\,d\quupsilon \leq \frac{1}{1 - p} \qusigu^{1 - p}.
\end{equation*}
\end{proposition}
\par
We next have a divergence theorem which is basic to the energy inequalities we wish to prove. Note that we write the boundary terms as contractions with one-forms rather than inner products with normal vectors.
\par
\begin{proposition}\label{divth} Let $\qusigmat \in [0, \qupark\queT]$. If $\quaaph$ is any Lorentzian metric on $\qubulk\cap\{\qutau\leq\qusigmat\}$ and $X^i$ is any $C^1$ vector field on $\qubulk\cap\{\qutau\leq\qusigmat\}$ such that $d\quxb_i X^i \vert \qudet\quaaph\vert ^{1/2} \in L^1(\qubulk\cap\{\qutau\leq\qusigmat\})$ and all of the integrals below are absolutely convergent, then 
\begin{multline}\label{findivth}
\int_{\qubulk \cap \{ \qutau \leq \qusigmat \} } X^i_{;i} \vert \qudet\quaaph\vert ^{1/2}\,d\qutau\,d\quxi\,d\quzeta = \int_{\quAsigmat} d\qutau_i X^i \vert \qudet\quaaph\vert ^{1/2} \,d\quxi\,d\quzeta\\
- \int_{\quSigz \cap \{ \qutau \leq \qusigmat \} } d\qusb_i X^i \vert \qudet\quaaph\vert ^{1/2}\,d\quxb\,d\quvb - \int_{\quUz \cap \{ \qutau \leq \qusigmat \} } d\quvb_i X^i \vert \qudet\quaaph\vert ^{1/2}\,d\qusb\,d\quxb\\
+ \int_{\quBsigfb \cap \{ \qutau \leq \qusigmat \} } \left(d\qusb_i - \quBsigffb'(\quvb) d\quvb_i\right) X^i \vert \qudet\quaaph\vert ^{1/2}\,d\quxb\,d\quvb.
\end{multline}
\end{proposition}
\par
\begin{proof} By multiplying $X^i$ by a suitable $C^\infty$ cutoff in the $\quxb$ direction and using $d\quxb_i X^i \vert \qudet\quaaph\vert ^{1/2} \in L^1(\qubulk\cap\{\qutau\leq\qusigmat\})$ and a density argument, it suffices to consider $X^i$ which (when restricted to $\qubulk \cap \{ \qutau \leq \qusigmat\}$) have support bounded in the $\quxi$ direction. Let $\quvf_{ij\ell} = \vert \qudet\quaaph\vert ^{1/2} \epsilon_{ij\ell}$ denote the volume element on $\qubulk$ corresponding to $\quaaph$, and define the two-form
\begin{equation*}
\qufX_{ij} = \quvf_{\ell ij} X^\ell.
\end{equation*}
Then (see \cite{wald}, Appendix B.2) we have
\begin{equation*}
d\qufX_{ij\ell} = X^m_{;m} \quvf_{ij\ell},
\end{equation*}
so by Stokes's Theorem (see \cite{schouten}, II, (8.13); there are no boundary terms in the $\quxi$ direction because of the assumption on the support of $X^i$) 
\begin{multline}\label{primdiv}
\int_{\qubulk \cap \{ \qutau\leq\qusigmat \} } X^i_{;i} \vert \qudet\quaaph\vert ^{1/2}\,d\qutau\,d\quxi\,d\quzeta = \int_{\qubulk \cap \{ \qutau\leq\qusigmat \}} d\qufX_{ij\ell}\\
= \int_{\quSigz\cap\{\qutau\leq\qusigmat\}} \qufX_{ij} + \int_{\quUz\cap\{\qutau\leq\qusigmat\}} \qufX_{ij} + \int_{\quBsigfb\cap\{\qutau\leq\qusigmat\}} \qufX_{ij} + \int_{\quAsigmat} \qufX_{ij}.
\end{multline}
To evaluate the first two integrals we work in the $\qusb,\,\quxb,\,\quvb$ coordinate system. It is not hard to see that on $\quSigz$ and $\quUz$, respectively, choices of ordered bases consistent with the orientation given by $\{\partial_\qusb, \partial_\quxb, \partial_\quvb \}$ on $\qubulk$ are $\{ \partial_\quvb, \partial_\quxb \}$ and $\{ \partial_\quxb, \partial_\qusb \}$. Further, on $\quSigz\cap\{\qutau\leq\qusigmat\}$ and $\quUz\cap\{\qutau\leq\qusigmat\}$, respectively,
\begin{equation*}
\quvf_{\ell ij} X^\ell = \quvf_{0ij} X^0,\qquad \quvf_{\ell ij} X^\ell = \quvf_{2ij} X^2,
\end{equation*}
so
\begin{align}\label{nullsurf}
\int_{\quSigz\cap\{\qutau\leq\qusigmat\}} \qufX_{ij} &= \int_{\quSigz\cap\{\qutau\leq\qusigmat\}} X^0\quvf_{021}\,d\quxb\,d\quvb\\
&= -\int_{\quSigz\cap\{\qutau\leq\qusigmat\}} d\qusb_i X^i \vert \qudet\quaaph\vert ^{1/2}\,d\quxb\,d\quvb,\notag\\
\int_{\quUz\cap\{\qutau\leq\qusigmat\}} \qufX_{ij} &= \int_{\quUz\cap\{\qutau\leq\qusigmat\}} X^2 \quvf_{210}\,d\qusb\,d\quxb\notag\\
&= -\int_{\quUz\cap\{\qutau\leq\qusigmat\}} d\quvb_i X^i \vert \qudet\quaaph\vert ^{1/2}\,d\qusb\,d\quxb.\notag
\end{align}
For the integral over $\quAsigmat$ we work with respect to coordinates $\qutau,\,\quzeta,\,\quxi$ (which have the same orientation as $\qusb,\,\quxb,\,\quvb$). A choice of basis on $\quAsigmat$ consistent with the orientation on $\qubulk$ is given by $\{ \partial_\quzeta, \partial_\quxi \}$, so that (indices $0, 1, 2$ now refer to $\qutau,\,\quzeta,\,\quxi$!)
\begin{equation}\label{spacsurf}
\int_{\quAsigmat} \qufX_{ij} = \int_{\quAsigmat} X^0\quvf_{012}\,d\quxi\,d\quzeta = \int_{\quAsigmat} d\qutau_i X^i \vert \qudet\quaaph\vert ^{1/2}\,d\quxi\,d\quzeta.
\end{equation}
For the final integral, we note that a bit of algebra shows that if $\Sigma$ is a surface given by $\qusb = S(\quxb, \quvb)$ and $\widetilde{X}$, $\widetilde{V}$ are the projections on $T\Sigma$ of $\partial_\quxb$ and $\partial_\quvb$, respectively, then the basis $\{ \widetilde{X}, \widetilde{V} \}$ is consistent with the orientation on $\qubulk$ and
\begin{equation}\label{Sigsurf}
\int_\Sigma \qufX_{ij} = \int_\Sigma \left[d\qusb_i X^i - \frac{\partial S}{\partial\quxb} d\quxb_i X^i - \frac{\partial S}{\partial\quvb} d\quvb_i X^i\right]\vert \qudet\quaaph\vert ^{1/2}\,d\quxb\,d\quvb.
\end{equation}
Applying this with $S(\quxb, \quvb) = \quBsigffb(\quvb)$ and combining with \eqref{primdiv}, \eqref{nullsurf} and \eqref{spacsurf} gives the result.\end{proof}
\par
Note that we do not need any bounds on $\| d\quxb_i X^i \vert \qudet\quaaph\vert ^{1/2}\| _{L^1(\qubulk\cap\{\qutau\leq\qusigmat\})}$ -- only that it be finite -- so that it is permissible for it to depend on parameters such as $\qupark$. 
\par
Now fix some $\mu \geq 1$, and assume that \eqref{ciomeginduct} and \eqref{ciomegdiffinduct} hold for this value of $\mu$. Note that this implies that, for $\ell = 0, 1$,
\begin{gather}
\|\partial_\qusb^\ell \quomegb_{\mu + 1}\|_{\quDHb^{\qusbn - 2}(\quAsigmat)} \leq (\OmegbC + \OmegbCp) \qunu \qusigu^{-1/4}\label{quomegbsuppbound}\\
\hbox{ for $\qusigmat \in [0, \qupark\queT]$, $\quomegb_\mu \in \{ \qudlb_\mu, \qubb_\mu, \qucb_\mu, \partial_\quxb \qudlb_\mu, \partial_\quvb \qudlb_\mu, \partial_\quxb \qubb_\mu \}$}.\notag
\end{gather}
We obtain bounds on various quantities derived from the metric $\quhbar_\mu$. We define metric perturbations $\qutdh_\mu$ and $\qutdhi_\mu$ by (see \eqref{hbardef})
\begin{align}\label{hbarmudef}
\begin{split}
(\quhbar_\mu)_{ij} &= \queta_{ij} + \qupark^{-2\quiota} \qutdh_\mu\\
&= \begin{pmatrix}
0&0&-1\\0&1&0\\-1&0&0
\end{pmatrix} + \qupark^{-2\quiota} \begin{pmatrix}
0&0&0\\0&\qudab_\mu &\qubb_\mu\\0&\qubb_\mu&\qucb_\mu
\end{pmatrix},\\
(\quhbar_\mu^{-1})^{ij} &= \eta^{ij} + \qupark^{-2\quiota} \qutdhi_\mu\\
&= \begin{pmatrix}
0&0&-1\\0&1&0\\-1&0&0
\end{pmatrix} + \frac{\qupark^{-2\quiota}}{\quab_\mu} \begin{pmatrix}
\qupark^{-2\quiota} \qubb_\mu^2 - \quab_\mu\qucb_\mu & \qubb_\mu & 0\\ \qubb_\mu & -\qudab_\mu & 0 \\ 0 & 0 & 0 
\end{pmatrix}.\\
\end{split}
\end{align}
\par
\begin{lemma}\label{bootstrapimp}
There is a constant $\quuCnmi > 0$ such that the following holds. For $\qusigmat \in [0, \qupark\queT]$, $\ell = 0, 1$, and $\quomegb_\mu \in \{ \qudlb_\mu, \qubb_\mu, \qucb_\mu, \partial_\quxb \qudlb_\mu, \partial_\quvb \qudlb_\mu, \partial_\quxb \qubb_\mu \}$,
\begin{equation}\label{bigambb}
\| \partial_\qutau^\ell \quomegb_\mu\| _{L^\infty(\quAsigmat)} < [(1 + \sqrt{2})\quuCnmi]^\ell\OmegbC\quCKS\qunu^2\qusigu^{-1/2}.
\end{equation}
\end{lemma}
\par
\begin{proof} Since $\qusbn \geq \qusbnmin$ and $\partial_\qutau = \sqrt{2} \partial_\qusb - \partial_\quzeta$, this is an immediate consequence of \eqref{ciomeginduct}, Corollary \ref{shuffleops}, and Lemma \ref{Asobolevemb}.\end{proof}
\par
\begin{corollary}\label{bootstrapimpcor} Suppose that
\begin{equation}\label{parkCKSnu}
\qunu \qupark^{-2\quiota} \leq 1/(12\OmegbC\quCKS).
\end{equation}
Then for $\qusigmat \in [0, \qupark\queT]$ 
\begin{gather}\label{ablowboundeq}
\left\| \qupark^{-2\quiota} \qudlb_\mu\right\| _{L^\infty(\quAsigmat)}, \left\| \qupark^{-2\quiota} \qubb_\mu\right\| _{L^\infty(\quAsigmat)}, \left\| \qupark^{-2\quiota}\qucb_\mu\right\| _{L^\infty(\quAsigmat)} \leq \frac{1}{12}\qunu\qusigu^{-1/2},\\
\ququad \left\| \quab_\mu^{-1}\right\| _{L^\infty(\quAsigmat)} < \frac{5}{4},\notag\\
\left\| \qupark^{-2\quiota} \qudab_\mu\right\| _{L^\infty(\quAsigmat)} < \frac{1}{5}\qunu\qusigu^{-1/2},\ququad \left\| \qupark^{-2\quiota} \qudaib_\mu\right\| _{L^\infty(\quAsigmat)} < \frac{1}{4}\qunu\qusigu^{-1/2}.\notag
\end{gather}
Thus in particular, on $\quAsigmat$,
\begin{gather}
\qupark^{-2\iota} \left\vert \qutdhi_\mu\right\vert_{HS} \leq \qunu\qusigu^{-1/2},\quad \left\vert \quhbar_\mu\right\vert_{HS},\, \left\vert\quhbar^{-1}_\mu\right\vert_{HS} \leq 4.
\end{gather}
\end{corollary}
\par
\begin{proof} This follows from \eqref{bigambb} with $\ell = 0$, recalling (see \eqref{fscalcdefii}) that $\qudaib = -\qudab/\quab$.
\end{proof}
\par
This allows us to compare volume elements and various geometric seminorms for the solution metric $\quhbar_\mu$ and the Minkowski metric $\eta$, as follows. 
\par
\begin{lemma}\label{amsmalldet} 
For $\qusigmat \in [0, \qupark\queT]$,
\begin{equation}\label{asdparkeqii}
\left\| \vert \qudet\quth_\mu\vert ^{1/2} - \vert \qudet\queta\vert ^{1/2}\right\| _{L^\infty(\quAsigmat)} \leq \qupark^{-2\quiota}\OmegbC\quCKS\qunu^2\qusigu^{-1/2},
\end{equation}
and, in particular, if \eqref{parkCKSnu} holds, then $\vert \qudet\quth_\mu\vert ^{1/2} \in (11/12, 13/12)$ on $\qubulk \cap \{ \qutau \leq \quepsb \}$. 
\end{lemma}
\par
\begin{proof} Since $\qudet\quth_\mu = \quab_\mu = (1 + \qupark^{-2\quiota}\qudlb_\mu)^2$, this follows from Lemma \ref{bootstrapimp}\ and Corollary \ref{bootstrapimpcor}.\end{proof}
\par
\begin{proposition}\label{hbarmubd}
There is a constant $\hbmdC(\OmegbC, \qusbn)$, depending only on $\OmegbC$ and $\qusbn$, such that for all $\qusigmat \in [0, \qupark\queT]$,
\begin{gather}
\left\|\quhbar_\mu^{-1} - \eta^{-1} \right\|_{\quDHb^{\qusbn - 1}(\quAsigmat)} \leq \hbmdC(\OmegbC, \qusbn)\qupark^{-2\iota}\qunu^2\qusigu^{-1/4},\\
\left\|\partial_\qutau \quhbar_\mu^{-1}\right\|_{\quDHb^{\qusbn - 1}(\quAsigmat)} \leq \hbmdC(\OmegbC, \qusbn) \qupark^{-2\iota} \qunu^2 \qusigu^{-1/4},\notag
\end{gather}
and hence
\begin{gather}
\left\|\quhbar_\mu^{-1} - \eta^{-1}\right\|_{L^\infty(\quAsigmat)} \leq \quCKS \hbmdC(\OmegbC, \qusbn) \qupark^{-2\iota}\qunu^2\qusigu^{-1/2},\\
\left\|\partial_\qutau \quhbar_\mu^{-1}\right\|_{L^\infty(\quAsigmat)} \leq \quCKS \hbmdC(\OmegbC, \qusbn) \qupark^{-2\iota} \qunu^2 \qusigu^{-1/2}.\notag
\end{gather}
\end{proposition}
\par
\begin{proof} By \eqref{hbarmudef},
\begin{equation}
(\quhbar_\mu^{-1})^{ij} = \begin{pmatrix}
0&0&-1\\0&1&0\\-1&0&0
\end{pmatrix} + \frac{\qupark^{-2\quiota}}{\quab} \begin{pmatrix}
\qupark^{-2\quiota} \qubb^2 - \quab\qucb & \qubb & 0\\ \qubb & -\qudab & 0 \\ 0 & 0 & 0 
\end{pmatrix};
\end{equation}
thus the result follows from Lemma \ref{Asobolevemb}, Lemma \ref{Flipsch}, and Lemma \ref{bootstrapimp}.\end{proof}
\par
\begin{proposition}\label{Bsigfbexist} Suppose that \eqref{parkCKSnu} holds and that $\qupark$, $\queT$, and $\queTp$ satisfy
\begin{equation}\label{parkeTeTp}
\qunu \qupark^{1/2 - 2\iota} \leq 2^{13/4} \OmegbC\quCKS \queTp\queT^{-1/2}.
\end{equation}
Then the surfaces $\quAsigmat$ ($\qusigmat \in [0, \qupark\queT]$) and
\begin{equation*}
\quBsigfb = \{ (\qusb, \quxb, \quvb) \in \qubulk \,\vert \,\qusb = \quBsigffb(\quvb) \}
\end{equation*}
are spacelike for $\quhbar_\mu$.
\end{proposition}
\par
\begin{proof} That $\quAsigmat$ is spacelike follows from Corollary \ref{bootstrapimpcor}\ and \eqref{hbarmudef}:
\begin{equation*}
\quhbar_\mu^{-1}(d\qutau, d\qutau) = -1 + \frac{1}{2} \qupark^{-2\quiota} \left( \qupark^{-2\quiota} \frac{\qubb^2}{\quab} - \qucb\right) \leq -1 + \frac{1}{2} \left[ 2 \left(\frac{1}{4}\right)^2 + \frac{1}{4}\right] < -\frac{1}{2}.
\end{equation*}
Next, note that \eqref{parkeTeTp} implies that $\quBsigffb$ maps $[0, \qupark\queT\sqrt{2}]$ into $[\queTp/2, \queTp]$.
It suffices to show that $\quhbar_\mu^{-1} (d[\qusb - \quBsigffb(\quvb)], d[\qusb - \quBsigffb(\quvb)]) < 0$ for $\quvb \in [0, \qupark\queT\sqrt{2}]$. Recalling (see \eqref{Bsigffbdefeq}) that
\begin{equation*}
\quBsigffb(\quvb) = \queTp - 4\OmegbC\quCKS\qunu^2\qupark^{-2\quiota} \left[(1 + \quvb)^{1/2} - 1\right],
\end{equation*}
we have
\begin{equation*}
d[\qusb - \quBsigffb(\quvb)] = d\qusb + 2\OmegbC\quCKS\qunu^2\qupark^{-2\quiota} (1 + \quvb)^{-1/2} d\quvb,
\end{equation*}
so that, using \eqref{hbardef}, Lemma \ref{bootstrapimp}, Proposition \ref{siguprop}, and \eqref{vbjysigu},
\begin{align*}
\quhbar_\mu^{-1} (d[\qusb - \quBsigffb(\quvb)], d[\qusb - \quBsigffb(\quvb)])\kern -1in&\\
&= \qupark^{-2\quiota} \left[\qupark^{-2\quiota} \frac{\qubb^2}{\quab} - \qucb\right] - 4\OmegbC\quCKS\qunu^2\qupark^{-2\quiota} (1 + \quvb)^{-1/2}\\
&\leq \frac{53}{48} \OmegbC\quCKS\qunu^2\qupark^{-2\quiota}\qutauu^{-1/2} - 4\sqrt{3}\OmegbC\quCKS\qunu^2\qupark^{-2\quiota} \qutauu^{-1/2} < 0.
\end{align*}
This completes the proof.\end{proof}
\par
We next obtain bounds on the coefficients in the wave operator $\quwbox_{\quhbar_\mu}$. Here and below, we define coefficients $\qudAc_\mu[\partial]$ by (see \eqref{iswave}) 
\begin{align}\label{dAcdef}
\quwbox_{\quhbar_\mu} f &= \left(-2\partial_\qusb\partial_\quvb + \partial_\quxb^2\right) f + \qupark^{-2\quiota} \sum_{\partial\in\qudifopset} \qudAc_\mu[\partial] \partial f\\
&= \Biggl\{\left(-2\partial_\qusb\partial_\quvb + \partial_\quxb^2\right) + \qupark^{-2\quiota} \Biggl[\qudaib_\mu \partial_\quxb^2 - \qucb_\mu \partial_\qusb^2 + 2\frac{\qubb_\mu}{\quab_\mu} \partial_\qusb \partial_\quxb\notag\\
&\kern 2pt- \left(\partial_\qusb \qucb_\mu - \frac{1}{\quab_\mu} \partial_\quxb \qubb_\mu + \frac{\partial_\quvb\qudlb_\mu}{\qulb_\mu}\right)\partial_\qusb + \left(\frac{1}{\quab_\mu} \partial_\qusb \qubb_\mu - \frac{\qulb_\mu\partial_\quxb\qudlb_\mu}{\quab_\mu^2}\right)\partial_\quxb - \frac{\partial_\qusb \qudlb_\mu}{\qulb_\mu}\partial_\quvb\Biggr]\notag\\
&\kern 2pt\qupark^{-4\quiota} \left(\frac{\qubb_\mu^2}{\quab_\mu} \partial_\qusb^2 - \left(\qucb_\mu \frac{\partial_\qusb \qudlb_\mu}{\qulb_\mu} - 2\frac{\qubb_\mu\partial_\qusb \qubb_\mu}{\quab_\mu} + \frac{\qubb_\mu\,\quellb_\mu \partial_\quxb \qudlb_\mu}{\quab_\mu^2}\right) \partial_\qusb - \frac{\qubb_\mu\,\qulb_\mu\partial_\qusb\qudlb_\mu}{\quab_\mu^2}\partial_\quxb\right)\notag\\
&\kern 2pt - \qupark^{-6\quiota} \frac{\qubb_\mu^2\qulb_\mu\partial_\qusb\qudlb_\mu}{\quab_\mu^2}\partial_\qusb\Biggr\} f,\notag
\end{align}
where
\begin{equation}
\qudifopset = \{ \partial_\qusb^2, \partial_\qusb\partial_\quxb, \partial_\quxb^2, \partial_\qusb, \partial_\quxb, \partial_\quvb \}.
\end{equation}
\par
\begin{proposition}\label{dAcbounds} Suppose that \eqref{parkCKSnu} holds. Then there are constants $C^{\qudAc}(\OmegbC, \qusbn) > 0$, $\overline{C^{\qudAc}}(\OmegbC, \OmegbCp, \qusbn) > 0$, depending only on $\OmegbC$, $\qusbn$ and $\OmegbC$, $\OmegbCp$, $\qusbn$, respectively, such that ($\ell = 0, 1$)
\begin{gather}
\|\partial_\qusb^\ell \qudAc_\mu[\partial]\|_{\quDHb^{\qusbn - 1}(\quAsigmat)} \leq C^{\qudAc}(\OmegbC, \qusbn) \qunu^2 \qusigu^{-1/4}\,\,\,\hbox{for all $\partial\in\qudifopset$,}\label{dAcboundseq}\\
\|\partial_\qusb^\ell (\qudAc_{\mu + 1} - \qudAc_\mu)[\partial]\|_{\quDHb^{\qusbn - 2}(\quAsigmat)} \leq \frac{1}{2^\mu} \overline{C^{\qudAc}}(\OmegbC, \OmegbCp, \qusbn) \qunu \qusigu^{-1/4}\,\,\,\hbox{for all $\partial\in\qudifopset$.}\label{dAcdiffboundseq}
\end{gather}
\end{proposition}
\par
\begin{proof} Since $\partial_\qusb$, $\partial_\quxb$, and $\partial_\quvb$ are all (linear combinations of) elements of $\quudifset$, these estimates follow immediately from Lemma \ref{Flipsch}, \eqref{ciomeginduct}, \eqref{ciomegdiffinduct}, \eqref{quomegbsuppbound}, Corollary \ref{bootstrapimpcor}, and \eqref{dAcdef}.\end{proof}
\par
\begin{corollary}\label{gambdxest} Suppose that \eqref{parkCKSnu} holds. Then there are constants $C(\gambC, \qusbn) > 0$, $\overline{C}(\gambC, \OmegbCp, \qusbn) > 0$, depending only on $\gambC$, $\qusbn$ and $\gambC$, $\OmegbCp$, $\qusbn$, respectively, such that for $\ell = 0, 1$
\begin{gather}
\|\partial_\qusb^\ell \partial_\quxb^2 \qugamb_\mu\|_{\quDHb^{\qusbn - 1 - \ell}(\quAsigmu)} \leq C(\gambC, \qusbn) \left( \qunu\qusigu^{-1/4} + \sum_{{\ell' = 0, 1, \ell'' \leq \ell}\atop{i = 0, 1, 2}} \|\partial_\qusb^{\ell' + \ell''} \partial_i \qugamb_\mu\|_{\quDHb^{\qusbn - 1 - \ell}(\quAsigmu)}\right),\label{cquxbtwogambb}\\
\|\partial_\qusb^\ell (\partial_\quxb^2 \qugamb_{\mu + 1} - \partial_\quxb^2 \qugamb_\mu)\|_{\quDHb^{\qusbn - 2 - \ell}(\quAsigmat)} \leq \overline{C}(\gambC, \OmegbCp, \qusbn)\notag\\
\cdot \left( \nu \qusigu^{-1/4} 2^{-\mu} + \sum_{{\ell' = 0, 1, \ell'' \leq \ell}\atop{i = 0, 1, 2}} \|\partial_\qusb^{\ell' + \ell''} \partial_i (\qugamb_{\mu + 1} - \qugamb_\mu)\|_{\quDHb^{\qusbn - 2 - \ell}(\quAsigmu)}\right).\label{cquxbtwogambdiffb}
\end{gather}
\end{corollary}
\par
\begin{proof}
By $\quwbox_{\quhbar_\mu} \qugamb_\mu = 0$, we have
\begin{equation}
\partial_\quxb^2 \qugamb_\mu = \frac{1}{\quab_\mu} \left[ 2\partial_\qusb\partial_\quvb\qugamb_\mu - \qupark^{-2\iota} \sum_{\partial \in \qudifopset\backslash \{ \partial_\quxb^2 \}} \qudAc_\mu[\partial] \partial\qugamb_\mu\right],
\end{equation}
so by Lemma \ref{Flipsch} and Proposition \ref{dAcbounds}, we have, for various constants $C(\gambC, \qusbn)$ depending only on $\gambC$ and $\qusbn$,
\begin{gather}
\|\partial_\qusb^\ell \partial_\quxb^2 \qugamb_\mu\|_{\quDHb^{\qusbn - 1 - \ell}(\quAsigmu)} \leq C(\gambC, \qusbn) \Biggl[ \OmegbC \qunu\qusigu^{-1/4} + \|\partial_\qusb^{1 + \ell}\partial_\quvb\qugamb_\mu\|_{\quDHb^{\qusbn - 1 - \ell}(\quAsigmu)}\\
+ \qupark^{-2\iota} \left(\qunu\qusigu^{-1/4} + \sum_{\partial \in \qudifopset\backslash\{\partial_\quxb^2\}} \|\partial_\qusb^\ell \partial\qugamb_\mu\|_{\quDHb^{\qusbn - 1 - \ell}(\quAsigmu)}\right)\Biggr]\\
\leq C(\gambC, \qusbn) \left( \qunu\qusigu^{-1/4} + \sum_{{\ell' = 0, 1}\atop{i = 0, 1, 2}} \|\partial_\qusb^{\ell' + \ell} \partial_i \qugamb_\mu\|_{\quDHb^{\qusbn - 1 - \ell}(\quAsigmu)}\right),
\end{gather}
showing \eqref{cquxbtwogambb}. \eqref{cquxbtwogambdiffb} follows analogously, recalling \eqref{quomegbsuppbound}.
\end{proof}
\par
By Corollary \ref{gambdxest}, it is sufficient to show only the first inequalities in \eqref{cigambinduct} and \eqref{cigambdiffinduct}.
\par
\par
Recall that we have defined, for $\quaph$ any smooth Lorentzian metric on $\qubulk$ and $f$ any smooth function on $\qubulk$ (see \eqref{QTaphdef}, \eqref{mffdef}, \eqref{ibounddef}, and \eqref{epsdef}),
\begin{gather}
\quQT_\quaph[f]_{ij} = \partial_i f\,\partial_j f - \frac{1}{2} \quaph_{ij} \quaph^{\ell m} \partial_\ell f\,\partial_m f,\notag\\
\qumff_\quaph[f]^i = -\quaph^{i\ell} \quQT_\quaph[f]_{j\ell} \partial_\qutau^j = -\quaph^{i\ell} \partial_\ell f\,\partial_\qutau f + \frac{1}{2} \quaph^{j\ell} \partial_j f\,\partial_\ell f\,\partial_\qutau^i,\notag\\
\quIb_{\quSigz, \quaph}[f] = \int_{\quSigz} d\qusb_i \qumff_\quaph[f]^i \vert \qudet\quaph\vert ^{1/2}\,d\quvb\,d\quxb,\ququad \quIb_{\quUz, \quaph}[f] = \int_{\quUz} d\quvb_i \qumff_\quaph[f]^i \vert \qudet\quaph\vert ^{1/2}\,d\qusb\,d\quxb,\label{ribounddef}\\
\quepsn_\quaph[f](\qusigmat) = \int_{\quAsigmat} d\qutau_i \qumff_\quaph[f]^i \vert \qudet\quaph\vert ^{1/2} \,d\quxi\,d\quzeta.\label{repsdef}
\end{gather}
\par
Given the bounds on $\quth_\mu$ derived above, we may bound the seminorms in \eqref{ribounddef} and \eqref{repsdef} for $\quaph = \quth_\mu$ in terms of their values for $\quaph = \eta$, as follows.
\begin{proposition}\label{initenernormeq} Suppose that \eqref{parkCKSnu} holds. Then
\begin{align}\label{Ibsim}
\begin{split}
\frac{1}{2} \quIb_{\quSigz, \queta}[f]^{1/2} \leq &\quIb_{\quSigz, \quth_\mu}[f]^{1/2} \leq  2\quIb_{\quSigz, \queta}[f]^{1/2},\\
\frac{1}{2} \quIb_{\quUz, \queta}[f]^{1/2} \leq &\quIb_{\quUz, \quth_\mu}[f]^{1/2} \leq 2\quIb_{\quUz, \queta}[f]^{1/2},
\end{split}
\end{align}
and for all $\qusigmat \in [0, \qupark\queT]$, 
\begin{equation}\label{epssim}
\frac{1}{2} \quepsn_\queta[f](\sigma)^{1/2} \leq \quepsn_{\quth_\mu}[f](\sigma)^{1/2} \leq 2\quepsn_\queta[f](\sigma)^{1/2} \leq 2\left[\queTp^{3/2} \| f\| _{\quDH^1(\quAsigmat)}^2 + \| \partial_\qutau f\| _{L^2(\quAsigmat)}^2\right]^{1/2}.
\end{equation}
In particular,
\begin{equation}\label{Ebmsim}
\frac{1}{2} \quEb_{n, \queta}[f] \leq \quEb_{n, \quth_\mu}[f] \leq 2\quEb_{n, \queta}[f].
\end{equation}
\end{proposition}
\par
\begin{proof} Working in $\qusb,\,\quxb,\,\quvb$ coordinates, note that (for $\ell = 0, 1, 2$), on $\quSigz$, $\qutdhi_\mu^{0\ell} = 0$, while on $\quUz$, $\qutdhi_\mu^{2\ell} = 0$. Thus we have, by \eqref{mffdef}, in $\qusb,\,\quxb,\,\quvb$ coordinates,
\begin{equation*}
\qumff_{\quth_\mu}[f]^0 = \qumff_\queta[f]^0 + \frac{1}{2\sqrt{2}}\qupark^{-2\quiota}\sum_{j, \ell = 1}^2 \qutdhi_\mu^{j\ell} \partial_j f \partial_\ell f,
\end{equation*}
and so by
Lemma \ref{HSinequal},
\begin{equation}\label{mffthetazdiff}
\vert \qumff_{\quth_\mu}[f]^0 - \qumff_\queta[f]^0\vert  \leq \qumff_\queta[f]^0 \qupark^{-2\quiota} \qubHS{\qutdhi_\mu}.
\end{equation}
The first line in \eqref{Ibsim} follows from Corollary \ref{bootstrapimpcor}. The proof of the second line is analogous.
To show \eqref{epssim} we work in $\qutau,\,\quxi,\,\quzeta$ coordinates, obtaining (since in this coordinate system $\qumff_\queta[f]^0$ bounds all three derivatives of $f$)
\begin{align*}
\qumff_{\quth_\mu}[f]^0 &= \qumff_\queta[f]^0 + \qupark^{-2\quiota}\left[-\qutdhi_\mu^{0\ell} \partial_\ell f \partial_\qutau f + \frac{1}{2} \qutdhi_\mu^{j\ell} \partial_j f \partial_\ell f\right]\\
&= \qumff_\queta[f]^0 + \qupark^{-2\quiota}\left[ -\frac{1}{2} \qutdhi_\mu^{00} (\partial_\qutau f)^2 + \frac{1}{2} \sum_{j, \ell = 1}^2 \qutdhi_\mu^{j\ell} \partial_j f\partial_\ell f\right]
\end{align*}
and so
\begin{equation*}
\vert \qumff_{\quth_\mu}[f]^0 - \qumff_\queta[f]^0\vert  \leq \qumff_\queta[f]^0 \qupark^{-2\quiota} \qubHS{\qutdhi_\mu}.
\end{equation*}
As before, this establishes the equivalence in \eqref{epssim}. The second part of \eqref{epssim} follows from
Definition \ref{DHbenerdef}. Finally, \eqref{Ebmsim} follows directly from the equivalence in \eqref{epssim} and the definition of $\quEb$ (see \eqref{eenerdefi}).\end{proof}
\par
We may now obtain the following first-order energy inequality for $\quepsn_{\quhbar_\mu}[f]$. As is usual in the study of wave equations on curved spacetimes, there will be an error term in the bulk integral arising from the deformation tensor of the metric ${\quhbar_\mu}$ along the unit normal $\partial_\qutau$ to $\quAsigmat$.
\par
\begin{proposition}\label{epsenerbound} Let $\qusigmat \in [0, \qupark\queT]$, let $f$ be a smooth function on $\qubulk \cap \{ \qutau \leq \qusigmat \}$ satisfying $\quepsn_{\quth_\mu}[f](\qusigmat') < \infty$ for $\qusigmat' \in [0, \qusigmat]$, and suppose that
\eqref{parkCKSnu} and \eqref{parkeTeTp} hold. Then we have
\begin{multline}\label{epsenerboundresineqyi}
\quepsn_{\quth_\mu}[f](\qusigmat) \leq 2\int_0^\qusigmat \biggl[\left(\int_{\quAsigmu} \left\vert \partial_\qutau f\quwbox_{\quth_\mu} f\right\vert \,d\quxi\,d\quzeta\right)\\
+ 18 \quCKS \hbmdC(\OmegbC, \qusbn) \qupark^{-2\quiota} \qunu^2 \quupsu^{-1/2} \quepsn_{\quth_\mu}[f](\quupsilon)\biggr]\,d\quupsilon\\
+ 2\quIb_{\quSigz\cap\{\qutau\leq\qusigmat\}, \queta}[f] + 2\quIb_{\quUz\cap\{\qutau\leq\qusigmat\}, \queta}[f].
\end{multline}
where $\hbmdC(\OmegbC, \qusbn)$ is the constant in Proposition \ref{hbarmubd}.
\end{proposition}
\par
\begin{proof} We note the elementary calculation
\begin{multline}\label{enerineqerrtermexp}
\qumff_{\quth_\mu}[f]^i_{;i} = -\left(\quQT_{\quth_\mu}[f]^i_j \partial_\qutau^j\right)_{;i}\\
= -\quwbox_{\quth_\mu} f \partial_\qutau f - \quQT_{\quth_\mu}[f]^{ij}\quL_{\partial_\qutau} (\quth_\mu)_{ij} = -\quwbox_{\quth_\mu} f \partial_\qutau f - \qupark^{-2\quiota} \quQT_{\quth_\mu}[f]_{ij}\partial_\qutau \qutdhi_\mu^{ij}.
\end{multline}
Now on $\quAsigmat$ we have, by Lemma \ref{HSinequal}, Lemma \ref{Asobolevemb}, and Proposition \ref{hbarmubd},
\begin{align}\label{quQTbound}
\left\vert \quQT_{\quth_\mu}[f]_{ij} \partial_\qutau \qutdhi_\mu^{ij}\right\vert  &= \left\vert  \partial_i f \partial_j f \partial_\qutau \qutdhi_\mu^{ij} - \frac{1}{2} (\quth_\mu)_{ij} \partial_\qutau \qutdhi_\mu^{ij} \quth_\mu^{\ell m} \partial_\ell f \partial_m f\right\vert \\
&\kern -0.5in\leq 2d\qutau_i \qumff_\queta[f]^i \quHS{\partial_\qutau \qutdhi_\mu} + d\qutau_i \qumff_\queta[f]^i \quHS{\quth_\mu} \quHS{\partial_\qutau \qutdhi_\mu} \quHS{\quth_\mu^{-1}}\\
&\kern-0.5in\leq 18 \quCKS \hbmdC(\OmegbC, \qusbn) d\qutau_i \qumff_\queta[f]^i \qunu^2 \qusigu^{-1/2}.
\end{align}
Next, as in the proof of Proposition \ref{initenernormeq}, we have for $\qutau \leq \qusigmat$
\begin{multline}\label{mffthmffetasmallepsn}
\left\vert \qumff_{\quth_\mu}[f]^1\right\vert  \leq \left\vert \qumff_\queta[f]^1\right\vert  + \qupark^{-2\quiota} \left\vert \qutdhi_\mu^{1\ell} \partial_\ell f \partial_\qutau f\right\vert  \leq 2\qumff_\queta[f]^0 \left[ 1 + \vert\qutdhi_\mu\vert_{HS}\right]\\
\leq 4\qumff_{\quth_\mu}[f]^0,
\end{multline}
from which $\qumff_{\quth_\mu}[f]^1\vert \qudet{\quth_\mu}\vert ^{1/2} \in L^1(\qubulk\cap\{\qutau\leq\qusigmat\})$ follows from $\quepsn_{\quth_\mu}[f](\qusigmat') < \infty$ for $\qusigmat' \in [0, \qusigmat]$. We may thus apply Proposition \ref{divth}\ with $X^i = \qumff_{\quth_\mu}[f]^i$, \eqref{enerineqerrtermexp}, Corollary \ref{bootstrapimpcor}, and Lemma \ref{amsmalldet} to obtain
\begin{gather}
\quepsn_{\quth_\mu}[f](\qusigmat) = \int_{\quAsigmat} d\qutau_i \qumff_{\quth_\mu}[f]^i \vert \qudet{\quth_\mu}\vert ^{1/2} \,d\quxi\,d\quzeta\notag\\
= \int_{\qubulk \cap \{ \qutau \leq \qusigmat\} } \qumff_{\quth_\mu}[f]^i_{;i} \vert \qudet{\quth_\mu}\vert ^{1/2}\,d\qutau\,d\quxi\,d\quzeta\notag\\
+ \int_{\quSigz \cap \{ \qutau\leq\qusigmat \}} d\qusb_i \qumff_{\quth_\mu}[f]^i \vert \qudet{\quth_\mu}\vert ^{1/2}\,d\quxb\,d\quvb\notag + \int_{\quUz\cap\{\qutau\leq\qusigmat\}} d\quvb_i \qumff_{\quth_\mu}[f]^i \vert \qudet{\quth_\mu}\vert ^{1/2}\,d\qusb\,d\quxb\notag\\
- \int_{\quBsigfb\cap\{\qutau\leq\qusigmat\}} \left(d\qusb_i - \quBsigffb'(\quvb) d\quvb_i\right) \qumff_{\quth_\mu}[f]^i \vert \qudet{\quth_\mu}\vert ^{1/2}\,d\quxb\,d\quvb \label{epsenerboundeqi}\\
\leq 2\int_0^\qusigmat \int_{\quAsigmu} \left[\left\vert \quwbox_{\quth_\mu} f\partial_\qutau f\right\vert + 18 \quCKS \hbmdC(\OmegbC, \qusbn) d\qutau_i \qumff_\queta[f]^i \qunu^2\quupsu^{-1/2} \qupark^{-2\iota} \right]\,d\quxi\,d\quzeta\,d\quupsilon\notag\\
+ \quIb_{\quSigz\cap\{\qutau\leq\qusigmat\}, {\quth_\mu}}[f] + \quIb_{\quUz\cap\{\qutau\leq\qusigmat\}, {\quth_\mu}}[f]\notag\\
\leq 2\int_0^\qusigmat \left[\left(\int_{\quAsigmu} \left\vert \partial_\qutau f\quwbox_{\quth_\mu} f\right\vert \,d\quxi\,d\quzeta\right) + 18 \quCKS \hbmdC(\OmegbC, \qusbn) \qupark^{-2\quiota} \qunu^2 \quupsu^{-1/2} \quepsn_{\quth_\mu}[f](\quupsilon)\right]\,d\quupsilon\notag\\
+ 2\quIb_{\quSigz\cap\{\qutau\leq\qusigmat\}, \queta}[f] + 2\quIb_{\quUz\cap\{\qutau\leq\qusigmat\}, \queta}[f],\notag
\end{gather}
as claimed.\end{proof}
\par
We note the following bounds.
\par
\begin{proposition}\label{sthreegambderiv} Suppose that \eqref{parkCKSnu} holds. Then there are constants $C(\gambC, \qusbn)$, $\overline{C}(\gambC, \OmegbCp, \qusbn)$, depending only on $\gambC$, $\qusbn$ and $\gambC$, $\OmegbCp$, $\qusbn$, respectively, such that, for all $\partial\in\qudifopset\cup [\quudifset, \qudifopset]$, $\ell \in \{ 0, 1 \}$,
\begin{gather}\label{partgambbdeq}
\|\partial \partial_\qusb^\ell \qugamb_\mu\|_{\quDHb^{\qusbn - 1 - \ell}(\quAsigmat)} \leq C(\gambC, \qusbn) \biggl[ \nu\qusigu^{-1/4} + \sum_{{\ell' = 0, 1, \ell' \leq \ell}\atop{i = 0, 1, 2}} \|\partial_\qusb^{\ell + \ell'} \partial_i \qugamb_\mu\|_{\quDHb^{\qusbn - 1 - \ell}(\quAsigmat)}\biggr],\\
\|\partial \partial_\qusb^\ell (\qugamb_{\mu + 1} - \qugamb_\mu)\|_{\quDHb^{\qusbn - 2 - \ell}(\quAsigmat)} \leq \overline{C}(\gambC, \OmegbCp, \qusbn)\notag\\
\cdot\left( \nu \qusigu^{-1/4} 2^{-\mu} + \sum_{{\ell' = 0, 1, \ell' \leq \ell}\atop{i = 0, 1, 2}} \|\partial_\qusb^{\ell + \ell'} \partial_i (\qugamb_{\mu + 1} - \qugamb_\mu)\|_{\quDHb^{\qusbn - 2 - \ell}(\quAsigmu)}\right).\notag
\end{gather}
\end{proposition}
\par
\begin{proof} We note that for all $i$, $j$, $\ell$, $m$,
\begin{equation}
[ \partial_i \partial_j, x_\ell \partial_m ] = \delta_{j\ell} \partial_i \partial_m + \delta_{i\ell} \partial_j \partial_m,
\end{equation}
from which it is straightforward to see that
\begin{equation}
[\quudifset, \qudifopset] \subset \hbox{span }\{ \partial_\qusb^2, \partial_\qusb\partial_\quxb, \partial_\qusb\partial_\quvb, \partial_\quxb^2, \partial_\quxb\partial_\quvb, \partial_\qusb, \partial_\quxb, \partial_\quvb \},
\end{equation}
and the result then follows from Corollary \ref{gambdxest}.\end{proof}
\par
Finally, we bound commutators of $\quwbox_{\quhbar_\mu}$ with differential operators $\quuprtl^I \partial_\qusb^\ell$ to obtain the following bounds.
\par
\begin{proposition}\label{Dfboxcomm} There are constants $C(\qusbn)$ and $C'(\gambC, \OmegbCp, \qusbn)$, depending only on $\qusbn$ and $\gambC$, $\OmegbCp$, $\qusbn$, respectively, such that for $\qusigmat \in [0, \qupark\queT]$, if $I$ is a multiindex with $|I| \leq \qusbn - 1$,
\begin{gather}
\left\| \quwbox_{\quhbar_\mu} \quuprtl^I \qugamb_\mu\right\|_{L^2(\quAsigmat)} \leq C(\qusbn) \left(\sum_{\partial\in\qudifopset} \|\qudAc_\mu[\partial]\|_{\quDHb^{\qusbn - 1}(\quAsigmat)}\right)\notag\\
\cdot \left( \sum_{\partial \in \qudifopset \cup \qudifopsett} \|\partial\qugamb_\mu\|_{\quDHb^{\qusbn - 2}(\quAsigmat)} \right) \queTp^{3/4} \qusigu^{-1/4} \qupark^{-2\iota},\label{Dfboxcommresyii}\\
\left\| \quwbox_{\quhbar_\mu} \quuprtl^I \partial_\qusb \qugamb_\mu\right\|_{L^2(\quAsigmat)} \leq C(\qusbn) \left(\sum_{{\partial\in\qudifopset}\atop{\ell' = 0, 1}} \|\partial_\qusb^{\ell'} \qudAc_\mu[\partial]\|_{\quDHb^{\qusbn - 1}(\quAsigmat)}\right)\label{Dfboxcommreseri}\\
\left( \sum_{\partial \in \qudifopset \cup \qudifopsett} \|\partial\partial_\qusb \qugamb_\mu\|_{\quDHb^{\qusbn - 2}(\quAsigmat)} + \sum_{\partial \in \qudifopset} \|\partial\qugamb_\mu\|_{\quDHb^{\qusbn - 1}(\quAsigmat)} \right) \queTp^{3/4} \qusigu^{-1/4} \qupark^{-2\iota},\notag
\end{gather}
and if $I$ is a multiindex with $|I| \leq \qusbn - 2$,
\begin{gather}
\left\| \quwbox_{\quhbar_{\mu + 1}} \quuprtl^I \partial_\qusb^\ell (\qugamb_{\mu + 1} - \qugamb_\mu)\right\|_{L^2(\quAsigmat)} \leq C'(\gambC, \OmegbCp, \qusbn) \queTp^{3/4} \qusigu^{-1/4} \qupark^{-2\iota}\label{Dfboxcommdiffres}\\
\cdot\Biggl[ \biggl(\sum_{{\partial\in\qudifopset}\atop{\ell' = 0, 1}} \|\partial_\qusb^{\ell'} \qudAc_{\mu + 1}[\partial]\|_{\quDHb^{\qusbn - 2}(\quAsigmat)}\biggr)\notag\\
\biggl(\sum_{{\partial' \in \qudifopset \cup \qudifopsett}\atop{\ell' = 0, 1}} \|\partial_\qusb^{\ell'} \partial' (\qugamb_{\mu + 1} - \qugamb_\mu)\|_{\quDHb^{\qusbn - 3}(\quAsigmat)} + \sum_{\partial\in\qudifopset} \|\partial(\qugamb_{\mu + 1} - \qugamb_\mu)\|_{\quDHb^{\qusbn - 2}(\quAsigmat)}\biggr)\notag\\
+ \biggl(\sum_{{\partial\in\qudifopset}\atop{\ell' = 0, 1}} \|\partial_\qusb^{\ell'} (\qudAc_{\mu + 1}[\partial] - \qudAc_\mu[\partial])\|_{\quDHb^{\qusbn - 2}(\quAsigmat)}\biggr) \biggl(\sum_{{\partial\in\qudifopset}\atop{\ell' = 0, 1}} \|\partial_\qusb^{\ell'} \partial\qugamb_\mu\|_{\quDHb^{\qusbn - 2}(\quAsigmat)}\biggr)\Biggr].\notag
\end{gather}
\end{proposition}
\par
\begin{proof} We first note that, for $\ell \in \{ 0, 1 \}$,
\begin{gather}
\quwbox_{\quhbar_\mu} \quuprtl^I \partial_\qusb^\ell \qugamb_\mu = [\quwbox_{\quhbar_\mu}, \quuprtl^I \partial_\qusb^\ell] \qugamb_\mu.
\end{gather}
Let $I$ be a multiindex with $|I| \leq \qusbn - 1$, $\ell \in \{ 0, 1 \}$. Then
\begin{gather}\label{splitbmucomm}
[\qubmu, \quuprtl^I \partial_\qusb^\ell] \qugamb_\mu = [\qubeta, \quuprtl^I \partial_\qusb^\ell] \qugamb_\mu + [\qubmu - \qubeta, \quuprtl^I\partial_\qusb^\ell] \qugamb_\mu.
\end{gather}
Now
\begin{gather}
\|[\qubeta, \quuprtl^I\partial_\qusb^\ell] \qugamb_\mu\|_{L^2(\quAsigmat)} \leq C(\qusbn) \qupark^{-2\iota} \queTp^{3/4} \sum_{\partial\in\qudifopset} \|\partial_\qusb^\ell [ \qudAc[\partial] \partial\qugamb_\mu]\|_{\quDHb^{\qusbn - 2}(\quAsigmat)}\label{betacommestyi}\\
\leq C(\qusbn) \queTp^{3/4} \qupark^{-2\iota} \qusigu^{-1/4} \left(\sum_{{\partial\in\qudifopset}\atop{\ell' = 0, 1}} \|\partial_\qusb^{\ell'} \qudAc_\mu[\partial]\|_{\quDHb^{\qusbn - 2}(\quAsigmat)}\right) \left(\sum_{{\partial\in\qudifopset}\atop{\ell' = 0, 1}} \|\partial_\qusb^{\ell'} \partial\qugamb_\mu\|_{\quDHb^{\qusbn - 2}(\quAsigmat)}\right),\notag
\end{gather}
where $C(\qusbn)$ is some constant depending only on $\qusbn$. Further,
\begin{equation}
[\qubmu - \qubeta, \quuprtl^I] \qugamb_\mu = \qupark^{-2\iota} \sum_{\partial\in\qudifopset} [\qudAc_\mu[\partial]\partial, \quuprtl^I]\qugamb_\mu,
\end{equation}
and, for any $\partial \in \qudifopset$,
\begin{gather}
\|[\qudAc_\mu[\partial]\partial, \quuprtl^I]\qugamb_\mu\|_{L^2(\quAsigmat)} \leq \|\qudAc_\mu[\partial] [\partial, \quuprtl^I] \qugamb_\mu\|_{L^2(\quAsigmat)} + \|[\qudAc[\partial], \quuprtl^I] \partial\qugamb_\mu\|_{L^2(\quAsigmat)},\\
\|\qudAc_\mu[\partial] [\partial, \quuprtl^I] \qugamb_\mu\|_{L^2(\quAsigmat)}  \leq C(\qusbn) \qusigu^{-1/4} \|\qudAc_\mu[\partial]\|_{\quDHb^{\qusbn - 1}(\quAsigmat)} \sum_{\partial' \in \qudifopsett} \|\partial' \qugamb_\mu\|_{\quDHb^{\qusbn - 2}(\quAsigmat)},\\
\|[\qudAc_\mu[\partial], \quuprtl^I] \partial\qugamb_\mu\|_{L^2(\quAsigmat)} \leq C(\qusbn) \sum_{{J + K = I}\atop{|J| < |I|}} \|\quuprtl^K\{ \qudAc_\mu[\partial]\} \quuprtl^J \partial\qugamb_\mu\|_{L^2(\quAsigmat)}\\
\leq C(\qusbn) \qusigu^{-1/4} \sum_{\partial' \in I} \|\partial' \qudAc_\mu[\partial]\|_{\quDHb^{\qusbn - 2}(\quAsigmat)} \|\partial\qugamb_\mu\|_{\quDHb^{\qusbn - 2}(\quAsigmat)}\\
\leq C(\qusbn) \qusigu^{-1/4} \|\qudAc_\mu[\partial]\|_{\quDHb^{\qusbn - 1}(\quAsigmat)} \|\partial\qugamb_\mu\|_{\quDHb^{\qusbn - 2}(\quAsigmat)}.
\end{gather}
Similarly,
\begin{gather}
[\qudAc_\mu[\partial]\partial, \quuprtl^I\partial_\qusb]\qugamb_\mu = \qudAc_\mu[\partial] [\partial, \quuprtl^I] \partial_\qusb\qugamb_\mu + [\qudAc_\mu[\partial], \quuprtl^I \partial_\qusb] \partial\qugamb_\mu,\\
[\qudAc_\mu[\partial], \quuprtl^I \partial_\qusb] \partial\qugamb_\mu = [\qudAc_\mu[\partial], \quuprtl^I]\partial_\qusb \partial\qugamb_\mu - \quuprtl^I \{ \partial_\qusb \qudAc_\mu[\partial] \partial\qugamb_\mu \},
\end{gather}
from which we obtain
\begin{gather}\label{dAcusbound}
\|[\qudAc_\mu[\partial]\partial, \quuprtl^I\partial_\qusb]\qugamb_\mu\|_{L^2(\quAsigmat)} \leq C(\qusbn) \qusigu^{-1/4} \Biggl( \|\qudAc_\mu[\partial]\|_{\quDHb^{\qusbn - 1}(\quAsigmat)}\\
\biggl[ \|\partial_\qusb \partial \qugamb_\mu\|_{\quDHb^{\qusbn - 2}(\quAsigmat)} + \sum_{\partial' \in \qudifopsett} \|\partial' \partial_\qusb \qugamb_\mu\|_{\quDHb^{\qusbn - 2}(\quAsigmat)} \biggr]\notag\\
+ \|\partial_\qusb \qudAc_\mu[\partial]\|_{\quDHb^{\qusbn - 1}(\quAsigmat)} \|\partial\qugamb_\mu\|_{\quDHb^{\qusbn - 1}(\quAsigmat)}\Biggr).\notag
\end{gather}
\eqref{Dfboxcommresyii} and \eqref{Dfboxcommreseri} follow readily. Note that \eqref{dAcusbound} holds with $\qugamb_\mu$ replaced with any function for which the right-hand side exists and is finite.
\par
\eqref{Dfboxcommdiffres} can be shown as follows. Let $I$ be a multiindex with $|I| \leq \qusbn - 2$. First,
\begin{equation}
\qubmup (\qugamb_{\mu + 1} - \qugamb_\mu) = -(\qubmup - \qubmu)\qugamb_\mu = \sum_{\partial\in\qudifopset} (\qudAc_\mu[\partial] - \qudAc_{\mu + 1}[\partial]) \partial\qugamb_\mu,
\end{equation}
so that
\begin{equation}
\qubmup \quuprtl^I \partial_\qusb^\ell (\qugamb_{\mu + 1} - \qugamb_\mu) = [\qubmup, \quuprtl^I \partial_\qusb^\ell] (\qugamb_{\mu + 1} - \qugamb_\mu) + \quuprtl^I \partial_\qusb^\ell \{ \qubmup (\qugamb_{\mu + 1} - \qugamb_\mu)\}.
\end{equation}
The first term on the right-hand side can be bounded as above, using the analogue of \eqref{splitbmucomm}, though some extra work is needed for the $[\qubeta, \quuprtl^I \partial_\qusb^\ell] (\qugamb_{\mu + 1} - \qugamb_\mu)$ term:
\begin{gather*}
\|[\qubeta, \quuprtl^I \partial_\qusb^\ell] (\qugamb_{\mu + 1} - \qugamb_\mu)\|_{L^2(\quAsigmat)} \leq C(\qusbn) \queTp^{3/4} \qupark^{-2\iota} \|\qubeta\partial_\qusb^\ell (\qugamb_{\mu + 1} - \qugamb_\mu)\|_{\quDHb^{\qusbn - 3}(\quAsigmat)}\\
\leq C(\qusbn) \queTp^{3/4} \qupark^{-2\iota} \sum_{\partial\in\qudifopset} \|\partial_\qusb^\ell (\qudAc_{\mu + 1}[\partial] \partial\qugamb_{\mu + 1} - \qudAc_\mu[\partial]\partial\qugamb_\mu)\|_{\quDHb^{\qusbn - 3}(\quAsigmat)}\\
\leq C(\qusbn) \queTp^{3/4} \qupark^{-2\iota} \qusigu^{-1/4} \Biggl[ \biggl( \sum_{{\partial\in\qudifopset}\atop{\ell' = 0, 1}} \|\partial_\qusb^{\ell'} \qudAc_{\mu + 1}[\partial]\|_{\quDHb^{\qusbn - 3}(\quAsigmat)}\biggr)\\
\cdot\biggl( \sum_{{\partial\in\qudifopset}\atop{\ell' = 0, 1}} \|\partial_\qusb^{\ell'} \partial(\qugamb_{\mu + 1} - \qugamb_\mu)\|_{\quDHb^{\qusbn - 3}(\quAsigmat)}\biggr)\\
+ \biggl(\sum_{{\partial\in\qudifopset}\atop{\ell' = 0, 1}} \|\partial_\qusb^{\ell'} (\qudAc_{\mu + 1}[\partial] - \qudAc_\mu[\partial]) \|_{\quDHb^{\qusbn - 3}(\quAsigmat)} \biggr) \biggl( \sum_{{\partial\in\qudifopset}\atop{\ell' = 0, 1}} \|\partial_\qusb^{\ell'} \partial\qugamb_\mu\|_{\quDHb^{\qusbn - 3}(\quAsigmat)}\biggr) \Biggr].
\end{gather*}
Finally,
\begin{gather*}
\|\quuprtl^I \partial_\qusb^\ell \qubmup (\qugamb_{\mu + 1} - \qugamb_\mu)\|_{L^2(\quAsigmat)} \leq \queTp^{3/4} \qupark^{-2\iota}\\
\cdot\sum_{\partial\in\qudifopset} \bigl\|\partial_\qusb^\ell \bigl\{ (\qudAc_{\mu + 1}[\partial] - \qudAc_\mu[\partial]) \partial\qugamb_\mu\bigr\}\bigr\|_{\quDHb^{\qusbn - 2}(\quAsigmat)}\\
\leq C(\qusbn) \queTp^{3/4} \qupark^{-2\iota} \qusigu^{-1/4} \biggl(\sum_{{\partial\in\qudifopset}\atop{\ell' = 0, 1}} \|\partial_\qusb^{\ell'} (\qudAc_{\mu + 1}[\partial] - \qudAc_\mu[\partial])\|_{\quDHb^{\qusbn - 2}(\quAsigmat)}\biggr)\\
\cdot\biggl(\sum_{{\partial\in\qudifopset}\atop{\ell' = 0, 1}} \|\partial_\qusb^{\ell'} \partial\qugamb_\mu\|_{\quDHb^{\qusbn - 2}(\quAsigmat)}\biggr).
\end{gather*}
\eqref{Dfboxcommresyii} now follows. \end{proof}
\par
\eqref{cigambinduct} and \eqref{cigambdiffinduct} can now be shown as follows. From Proposition \ref{epsenerbound} and the definition of $\quEb_{\qusbn, \quhbar}[\qugamb]$ (see \eqref{eenerdefi}), we have for $\qusigmat \in [0, \qupark\queT]$
\begin{gather}\label{quEbmainbound}
\quEb_{\qusbn, \quhbar_\mu}[\qugamb_\mu](\qusigmat) \leq C(\qusbn) \int_0^\qusigmat \Biggl[\sum_{{|I| \leq \qusbn - 1}\atop{\ell = 0, 1}} \left(\int_{\quAsigmu} \left\vert \quuprtl^I \partial_\qusb^\ell \partial_\qutau \qugamb_\mu \quwbox_{\quth_\mu} \quuprtl^I \partial_\qusb^\ell \qugamb_\mu \right\vert \,d\quxi\,d\quzeta\right)\\
+ 18 \quCKS \hbmdC(\OmegbC, \qusbn) \qupark^{-2\quiota} \qunu^2 \quupsu^{-1/2} \quEb_{\qusbn, \quhbar_\mu}[\qugamb_\mu](\quupsilon)\Biggr]\,d\quupsilon\notag\\
+ 2\quib_{\quSigz\cap\{\qutau\leq\qusigmat\}, \queta, \qusbn}[\qugamb_\mu] + 2\quib_{\quUz\cap\{\qutau\leq\qusigmat\}, \queta, \qusbn}[\qugamb_\mu].\notag
\end{gather}
By \eqref{ogambbaseinitbound}, the initial data terms are bounded by $2\qunu^2\queTp^{3/2}$. Further, from \eqref{eenerdefi}, \eqref{Ebmsim}, and Definition \ref{DHbenerdef}, together with \eqref{gambbIbootinf}, it is clear that
\begin{gather}
\sum_{{\ell = 0, 1}\atop{i = 0, 1, 2}} \|\partial_\qusb^\ell \partial_i \qugamb_\mu\|_{\quDHb^{\qusbn - 1}(\quAsigmat)} \leq 2 \queTp^{-3/4} \left[\quEb_{\qusbn, \quhbar_\mu}[\qugamb_\mu]\right]^{1/2} + \qunu,\label{DHbquEbbound}\\
\queTp^{-3/4} \left[\quEb_{\qusbn, \quhbar_\mu}[\qugamb_\mu]\right]^{1/2} \leq \sum_{{\ell = 0, 1}\atop{i = 0, 1, 2}} \|\partial_\qusb^\ell \partial_i \qugamb_\mu\|_{\quDHb^{\qusbn - 1}(\quAsigmat)}.\label{quEbDHbbound}
\end{gather}
Now
\begin{gather}\label{locmultineq}
\sum_{{|I| \leq \qusbn - 1}\atop{\ell = 0, 1}} \left(\int_{\quAsigmu} \left\vert \quuprtl^I \partial_\qusb^\ell \partial_\qutau \qugamb_\mu \quwbox_{\quth_\mu} \quuprtl^I \partial_\qusb^\ell \qugamb_\mu \right\vert \,d\quxi\,d\quzeta\right) \leq \left(\sum_{\ell = 0, 1} \|\partial_\qusb^\ell \partial_\qutau \qugamb_\mu\|_{\quDHb^{\qusbn - 1}(\quAsigmu)}\right)\\
\cdot \left(\sum_{{|I| \leq \qusbn - 1}\atop{\ell = 0, 1}} \|\quwbox_{\quth_\mu} \quuprtl^I \partial_\qusb^\ell \qugamb_\mu\|_{L^2(\quAsigmu)}\right)\queTp^{3/4}\notag\\
\leq C(\qusbn) \left(\sum_{\ell = 0, 1} \|\partial_\qusb^\ell \partial_\qutau \qugamb_\mu\|_{\quDHb^{\qusbn - 1}(\quAsigmu)}\right) \left(\sum_{{\partial\in\qudifopset}\atop{\ell = 0, 1}} \|\partial_\qusb^\ell \qudAc_\mu[\partial]\|_{\quDHb^{\qusbn - 1}(\quAsigmu)}\right)\notag\\
\cdot \left( \sum_{{\partial \in \qudifopset \cup \qudifopsett}\atop{\ell = 0, 1}} \|\partial\partial_\qusb^\ell\qugamb_\mu\|_{\quDHb^{\qusbn - 2}(\quAsigmu)} + \sum_{\partial\in\qudifopset} \|\partial\qugamb_\mu\|_{\quDHb^{\qusbn - 1}(\quAsigmu)} \right) \queTp^{3/2} \quupsu^{-1/4} \qupark^{-2\iota}.\notag
\end{gather}
For $\qusigmat \in [0, \qupark\queT]$, define
\begin{equation}\label{gambMdefeq}
\gambM(\qusigmat) = \sup \{ \sum_{{\ell, \ell' = 0, 1}\atop{i = 0, 1, 2}} \|\partial_\qusb^{\ell + \ell'} \partial_i \qugamb_\mu\|_{\quDHb^{\qusbn - 1 - \ell}(\quAsigmu)}\,\vert\,\quupsilon \in [0, \qusigmat] \}.
\end{equation}
Then by \eqref{locmultineq}, Proposition \ref{dAcbounds}, and Corollary \ref{gambdxest}, we obtain
\begin{multline}\label{multgambMbound}
\sum_{{|I| \leq \qusbn - 1}\atop{\ell = 0, 1}} \left(\int_{\quAsigmu} \left\vert \quuprtl^I \partial_\qusb^\ell \partial_\qutau \qugamb_\mu \quwbox_{\quth_\mu} \quuprtl^I \partial_\qusb^\ell \qugamb_\mu \right\vert \,d\quxi\,d\quzeta\right)\\
\leq C(\gambC, \qusbn) \queTp^{3/2} \quupsu^{-1/2} \qupark^{-2\iota} \qunu^2 \gambM(\quupsilon) \bigl[\qunu + \gambM(\quupsilon)\bigr].
\end{multline}
Inserting \eqref{DHbquEbbound} -- \eqref{quEbDHbbound} and \eqref{multgambMbound} into \eqref{quEbmainbound}, we thus obtain the inequality
\begin{multline}
\gambM^2(\qusigmat) \leq 10\qunu + C(\gambC, \qusbn) \int_0^\qusigmat \quupsu^{-1/2} \qupark^{-2\iota} \qunu^2 \gambM(\quupsilon) (\qunu + \gambM(\quupsilon))\\
\shoveright{+ \quupsu^{-1/2} \qupark^{-2\iota} \qunu^2 \gambM^2(\quupsilon)\,d\quupsilon}\\
\leq 10\qunu + C(\gambC, \qusbn) (\qusigu^{1/2} \qupark^{-2\iota} \qunu^2) [\gambM^2(\qusigmat) + \qunu\gambM(\qusigmat)],\label{gambMestCdef}
\end{multline}
so that when $\qusigmat \leq \qupark\queT$ and $\queT^{1/2} \qunu \leq 1/(2C(\gambC, \qusbn))$, we have
\begin{equation}
\gambM(\qusigmat) \leq 42\qunu.
\end{equation}
Requiring $\gambC \geq 42$ and $\gambCdx \geq C(\gambC, \qusbn)(1 + \gambC)$ (which are permissible since we have not put any upper bound on $\gambC$ or $\gambCdx$), we obtain \eqref{cigambinduct}. Since the induction showing \eqref{cigambinduct} does not rely in any way on \eqref{cigambdiffinduct}, we have now shown that the sequences $\{ \quomegb_\mu \}_\mu$ and $\{ \qugamb_\mu \}_\mu$ are bounded.
\par
The proof of \eqref{cigambdiffinduct} is analogous, and we only point out the differences. All initial data terms vanish; the analogues of \eqref{DHbquEbbound} -- \eqref{quEbDHbbound} are
\begin{gather}
\sum_{{\ell = 0, 1}\atop{i = 0, 1, 2}} \|\partial_\qusb^\ell \partial_i (\qugamb_{\mu + 1} - \qugamb_\mu)\|_{\quDHb^{\qusbn - 2}(\quAsigmat)} \leq 2 \queTp^{-3/2} \left[\quEb_{\qusbn - 1, \quhbar_\mu}[\qugamb_{\mu + 1} - \qugamb_\mu]\right]^{1/2},\label{diffDHbquEbbound}\\
\queTp^{-3/4} \left[\quEb_{\qusbn - 1, \quhbar_\mu}[\qugamb_{\mu + 1} - \qugamb_\mu]\right]^{1/2} \leq \sum_{{\ell = 0, 1}\atop{i = 0, 1, 2}} \|\partial_\qusb^\ell \partial_i (\qugamb_{\mu + 1} - \qugamb_\mu)\|_{\quDHb^{\qusbn - 2}(\quAsigmat)}.\label{diffquEbDHbbound}
\end{gather}
For $\qusigmat \in [0, \qupark\queT]$, define
\begin{equation}
\gambMdiff(\qusigmat) = \sup \{ \sum_{{\ell, \ell' = 0, 1}\atop{i = 0, 1, 2}} \|\partial_\qusb^{\ell + \ell'} \partial_i (\qugamb_{\mu + 1} - \qugamb_\mu)\|_{\quDHb^{\qusbn - 2 - \ell}(\quAsigmu)}\,\vert\,\quupsilon \in [0, \qusigmat] \}.
\end{equation}
The analogue of \eqref{multgambMbound} is (recall that we have already shown \eqref{ciomeginduct} for {\it all\/} $\mu \geq 0$)
\begin{gather}\label{diffmultgambMbound}
\sum_{{|I| \leq \qusbn - 2}\atop{\ell = 0, 1}} \left(\int_{\quAsigmu} \left\vert \quuprtl^I \partial_\qusb^\ell \partial_\qutau (\qugamb_{\mu + 1} - \qugamb_\mu) \quwbox_{\quth_\mu} \quuprtl^I \partial_\qusb^\ell (\qugamb_{\mu + 1} - \qugamb_\mu) \right\vert \,d\quxi\,d\quzeta\right)\\
\leq C(\gambC, \OmegbCp, \qusbn) \queTp^{3/4} \quupsu^{-1/2} \qupark^{-2\iota} \qunu^2 \gambMdiff(\quupsilon) \bigl[2^{-\mu} + \gambMdiff(\quupsilon)\bigr].\notag
\end{gather}
We thus obtain the inequality
\begin{multline*}
\gambMdiff{}^2 (\qusigmat) \leq C(\qusbn, \gambC, \OmegbCp, \gambCp) \int_0^\qusigmat \quupsu^{-1/2} \qupark^{-2\iota} \qunu^2 [ \gambMdiff(\quupsilon) + 2^{-\mu}]\gambMdiff(\quupsilon)\\
+ \quupsu^{-1/2} \qupark^{-2\iota} \qunu^2 \gambMdiff{}^2 (\quupsilon)\,d\quupsilon,
\end{multline*}
from which \eqref{cigambdiffinduct} follows as above. We omit the details.
\par
\subsection{Proof of Theorem \ref{seqbound}}\label{finishthproof} We may now complete the proof of Theorem \ref{seqbound}, as follows.
\par
\begin{proof}[Proof of Theorem \ref{seqbound}.] We apply Lemma \ref{qugambzconstrlem} to obtain a smooth function $\qugamb_0 : \qubulk \rightarrow \quR^1$ and constants $\gambC$, $\gambCdx$ such that \eqref{cigambinduct}, \eqref{cigambdiffinduct} hold with $\mu = 0$. Let $\OmegbC = \OmegbC(\gambC, \qusbn)$ be as constructed in Subsection \ref{gambzsval}, let $C(\gambC, \qusbn)$ be the constant in \eqref{gambMestCdef}, and assume that
\begin{equation}
\gambC \geq 42,\qquad \gambCdx \geq C(\gambC, \qusbn) (1 + \gambC).
\end{equation}
Suppose that \eqref{mainprkestc} -- \eqref{qunuqueTcondc} hold with $C_1 = \OmegbC$, $C_2 = 2C(\gambC, \qusbn)$. Then Subsections \ref{gambzsval} and \ref{enerperboot} apply to show that the sequences $\{ \quomegb_\mu \}$ ($\quomegb \in \{ \qudlb, \qubb, \qucb, \partial_\quxb \qudlb, \partial_\quvb \qudlb, \partial_\quxb \qubb \}$), $\{ \qugamb_\mu \}$ are bounded and moreover satisfy the bounds
\begin{gather*}
\|\partial_\qusb^\ell (\quomegb_{\mu + 1} - \quomegb_\mu)\|_{\quDHb^{\qusbn - 2}(\quAsigmat)} \leq \frac{1}{2^\mu} \OmegbCp \qunu \qusigu^{-1/4}\\
\hbox{ for $\qusigmat \in [0, \qupark\queT]$, $\quomegb_\mu \in \{ \qudlb_\mu, \qubb_\mu, \qucb_\mu, \partial_\quxb \qudlb_\mu, \partial_\quvb \qudlb_\mu, \partial_\quxb \qubb_\mu \}$},\\
\|\partial_\qusb^\ell\partial_i (\qugamb_{\mu + 1} - \qugamb_\mu)\|_{\quDHb^{\qusbn - 2}(\quAsigmat)} \leq \frac{1}{2^\mu} \gambCp \qunu,\,\|\partial_\quxb^2 (\qugamb_{\mu + 1} - \qugamb_\mu)\|_{\quDHb^{\qusbn - 2}(\quAsigmat)} \leq \frac{1}{2^\mu} \gambCdxp \qunu\\
\hbox{ for $\qusigmat \in [0, \qupark\queT]$}.
\end{gather*}
Thus the sequences $\{ \quomegb_\mu \}$ and $\{ \qugamb_\mu \}$ are Cauchy in $\quDHb^{\qusbn - 2}(\quAsigmat)$, and hence converge there. Since $\{ \partial_\qusb, \partial_\quxb, \partial_\quvb \} \subset \quudifset$, it follows that for all $\qusigmat \in (0, \qupark\queT)$, $\{ \quomegb_\mu \}$ and $\{ \qugamb_\mu \}$ converge in the ordinary Sobolev space $H^{\qusbn - 2}(\quAsigmat \times (-\epsilon, \epsilon))$, necessarily to a solution of \eqref{nsricone} -- \eqref{swave} satisfying the bounds \eqref{cEboot}. The condition on the support of $\qugamb$ and the $\quomegb$ follows from Lemma \ref{qudlbbbcbinitdatag}.
\par
Finally, we sketch a proof of uniqueness. Let $\qudlb$, $\qubb$, $\qucb$, $\qugamb$ be the solution to \eqref{nsricone} -- \eqref{swave} just constructed, and suppose that $\qudlbp$, $\qubbp$, $\qucbp$, $\qugambp$ form another solution to \eqref{nsricone} -- \eqref{swave} with the same initial data which satisfy the bounds (cf.\ \ref{cEboot})
\begin{gather}\label{cEbootp}
\sum_{\ell = 0}^1 \sum_{|I| \leq \qusbn - 2} \left\|\quuprtl^I \partial_\qusb^\ell \quomegbp\right\|_{L^2(\quAsigmat)}^2 \leq \enerCp\hbox{ for all }\quomegbp \in \{ \qudlbp, \qubbp, \qucbp, \partial_\quxb \qudlbp, \partial_\quvb \qudlbp, \partial_\quxb \qubbp \},\\
\quEb_{\qusbn - 1, \quhbar}[\qugamb](\qusigmat) \leq \enerCp\notag
\end{gather}
for some constant $\enerCp > 0$. We allow the constants below to depend on $\enerCp$. Define
\begin{gather*}
\alpha(\qusigmat) = \sup \{ \|\partial_\qusb^\ell \partial_i (\qugamb - \qugambp)\|_{\quDHb^{\qusbn - 3}(\quAsigmatp)}\,|\,\ell \in \{ 0, 1 \}, i \in \{ 0, 1, 2 \},\qusigmat' \in [0, \qusigmat] \},\\
\beta(\qusigmat) = \sup \biggl\{ \|\partial_\qusb^\ell (\quomegb - \quomegbp)\|_{\quDHb^{\qusbn - 3}(\quAsigmatp)}\,|\,\ell \in \{ 0, 1 \},\qusigmat' \in [0, \qusigmat],\\
\quomegb \in \{ \qudlb, \qubb, \qucb, \partial_\quxb \qudlb, \partial_\quvb \qudlb, \partial_\quxb \qubb \},\,\quomegbp \in \{ \qudlbp, \qubbp, \qucbp, \partial_\quxb \qudlbp, \partial_\quvb \qudlbp, \partial_\quxb \qubbp \} \biggr\}.
\end{gather*}
$\alpha$ and $\beta$ are bounded and nondecreasing on $[0, \qupark\queT]$. Applying an analogue of \eqref{cquxbtwogambdiffb}, we find for some constant $C > 0$
\begin{equation}\label{basequxbsqugamb}
\|\partial_\quxb^2 (\qugamb - \qugambp)\|_{\quDHb^{\qusbn - 3}(\quAsigmat)} \leq C(\alpha(\qusigmat) + \beta(\qusigmat)\qunu).
\end{equation}
Differencing \eqref{nsricone} -- \eqref{nsricthree} for $\quomegb$ and $\quomegbp$, applying a formula like \eqref{Flipschest}, and using Proposition \ref{refgronwallsblv} and \eqref{basequxbsqugamb}, we find that for some constant $C > 0$ and all $\qusigmat \in [0, \qupark\queT]$,
\begin{equation}
\beta(\qusigmat) \leq C (\alpha(\qusigmat) + \beta(\qusigmat)\qunu);
\end{equation}
thus for $\qunu \leq 1/(2C)$, we have
\begin{equation}\label{basebetabound}
\beta(\qusigmat) \leq C \alpha(\qusigmat).
\end{equation}
Since the initial data for $\qudlb$, $\qubb$, $\qucb$, $\qugamb$ and $\qudlbp$, $\qubbp$, $\qucbp$, $\qugambp$ coincide, the differences $\qudlb - \qudlbp$, $\qubb - \qubbp$, $\qucb - \qucbp$, $\qugamb - \qugambp$ must vanish in $\quDHb^{\qusbn - 3}(A_0)$, where $A_0 = \quSigz \cap \quUz$. Suppose that these differences vanish in $\quDHb^{\qusbn - 3}(A_\qusigmatp)$ for $\qusigmatp \in [0, \qusigmatz]$ for some $\qusigmatz \in [0, \qupark\queT]$. By an argument similar to that in the proof of Proposition \ref{epsenerbound} but with $\quAsigmat \cup (\quUz \cap \{ \tau \geq \qusigmat \})$ instead of $\quUz$, together with an analogue of \eqref{Dfboxcommdiffres} with $\qugamb_{\mu + 1}$, etc., replaced by $\qugamb$, etc., and $\qugamb_\mu$, etc., replaced by $\qugambp$, etc., we obtain for some constant $C > 0$ and all $\qusigmat \in [\qusigmatz, \qupark\queT]$ the inequality
\begin{equation}\label{basealphabound}
\alpha^2(\qusigmat) \leq C (\qusigmat - \qusigmatz) \alpha(\qusigmat) (\alpha(\qusigmat) + \beta(\qusigmat)).
\end{equation}
Applying \eqref{basebetabound}, we find that for $\qusigmat - \qusigmatz$ sufficiently small
\begin{equation}
\alpha^2(\qusigmat) \leq \frac{1}{2} \alpha^2(\qusigmat),
\end{equation}
so that $\alpha^2(\qusigmat) = 0$. We thus obtain that $\alpha(\qusigmat) = 0$ for all $\qusigmat \in [0, \qupark\queT]$, and elementary ODE theory then implies that $\beta(\qusigmat) = 0$ for all $\qusigmat \in [0, \qupark\queT]$ as well.
\end{proof}
\par
\begin{remark} By Sobolev embedding, we have $\quomegb, \partial_i \qugamb \in C^{\qusbn - 4}(\qubulk)$ for $\quomegb \in \{ \qudlb, \qubb, \qucb, \partial_\quxb \qudlb, \partial_\quvb \qudlb, \partial_\quxb \qubb \}$, $i \in \{ 0, 1, 2 \}$.\end{remark}
\par
\subsection{Flux bounds in Theorem \ref{iseqbound}}\label{enerboundfin} Finally, we prove the flux bounds in \eqref{fluxboundsone} -- \eqref{fluxboundstwo}. Existence and smoothness of the null foliation $\Sigma_{\qutcds}$ depends on Propositions \ref{foccoordsmth} and \ref{divthSigy} from the next section; we note that nothing in the proofs of those propositions depends on the flux bounds here.
\par
\begin{proposition}\label{seqboundfluxbounds} Let $\qudlb$, $\qubb$, $\qucb$, $\qugamb$ be any solution to \eqref{nsricone} -- \eqref{swave} given by Theorem \ref{seqbound} with initial data as constructed in Section \ref{initdat} and satisfying \eqref{omegbbaseinitbound} -- \eqref{gamblineinitbound} with the set of commutation vector fields $\quudifset$ set equal to $\quudifseta$, and let $\quhbar$ be the metric on $\qubulk$ determined by $\qudlb$, $\qubb$, $\qucb$. Then $U_{\qutvb} = \{ (\qusb, \quxb, \qutvb)\,|\, \qusb \in [0, \quBsigffb(\qutvb)], \quxb\in\quR^1 \}$, $\qutvb \in [0, \qupark\queT\sqrt{2}]$, forms an $\quhbar$-null foliation of $\qubulk$ passing through $\{ \qusb = 0, \quvb = \qutvb \}$. Assume also that $\queT$, $\queTp$, $\qupark$, and $\qunu$ satisfy
\begin{equation}\label{sbfbparmcond}
\qunu^2 \qupark^{-2\iota} \leq 1/\quCGeod,\quad \qupark\queT \geq 2,\quad \qunu^2\queT^{1/2} \queTp^{-1} \leq 1/(8\quCGeod)
\end{equation}
where $\quCGeod > 0$ is the constant in Proposition \ref{foccoordsmth}. Then there is a family of smooth functions $\quSigsbf_\qutsb : \quR^1 \times [0, \qupark\queT\sqrt{2} - \qutsb] \rightarrow [0, \queTp/4]$, $\qutsb \in [0, \queTp/8]$, whose graphs $\Sigma_{\qutsb} = \{ (\quSigsbf_\qutsb(\quxb, \quvb), \quxb, \quvb)\,|\,\quxb \in \quR^1, \quvb \in [0, \qupark\queT\sqrt{2} - \qutsb] \}$ form an $\quhbar$-null foliation of a neighborhood of $\quSigz$ in $\qubulk$ passing through $\{ \qusb = \qutsb, \quvb = \qupark\queT\sqrt{2} - \qutsb \}$ and transverse to $U_{\qutvb}$. Moreover, the solution $\qugamb$ satisfies the following flux bounds, for any $\qutsb \in [0, \queTp/8]$, $\qutvb \in [0, \qupark\queT\sqrt{2}]$, all multiindices $I$ with $|I| \leq \qusbn - 5$, and some constant $C > 0$:
\begin{gather}
\int_0^{\qupark\queT\sqrt{2} - \qutsb} \int_{-\infty}^\infty \biggl[ d\qusb_i \qumff_\quhbar[\qugamb]^i (\quSigsbf_\qutsb(\quxb, \quvb), \quxb, \quvb) - \frac{\partial\quSigsbf_\qutsb}{\partial\quxb}(\quxb, \quvb) d\quxb_i \qumff_\quhbar[\qugamb]^i (\quSigsbf_\qutsb(\quxb, \quvb), \quxb, \quvb)\label{sbfbi}\\
- \frac{\partial\quSigsbf_\qutsb}{\partial\quvb}(\quxb, \quvb) d\quvb_i \qumff_\quhbar[\qugamb]^i(\quSigsbf_\qutsb(\quxb, \quvb), \quxb, \quvb)\biggr]|\det\quhbar|^{1/2}(\quSigsbf_\qutsb(\quxb, \quvb), \quxb, \quvb)\,d\quxb\,d\quvb \leq C\qunu^2,\notag\\
\int_{U_{\qutvb}} d\quvb_i \qumff_\quhbar[\qugamb]^i|\det\quhbar|^{1/2}\,d\qusb\,d\quxb \leq C\qunu^2.\label{sbfbii}
\end{gather}
\end{proposition}
\begin{proof} That $U_{\qutvb}$ is an $\quhbar$-null foliation follows from the form of $\quhbar$ in \eqref{gaugemetform} (see Proposition \ref{gaugemetii}); the stated properties of $\Sigma_{\qutsb}$ follow from Proposition \ref{foccoordsmth}. The proofs of \eqref{sbfbi} and \eqref{sbfbii} are similar and we only prove \eqref{sbfbi}. Fix $\qutsb \in [0, \queTp/8]$. Define (see Proposition \ref{divthSigy})
\begin{multline*}
\qubulkF' = \{ (\qusb, \quxb, \quvb) \in \qubulk\,\vert \,0 \leq \qusb \leq \quSigsbf_\qutsb(\quxb, \quvb) \}\\
\cup \{ (\qusb, \quxb, \quvb) \in \qubulk\,\vert \, \quvb \geq \qupark\queT\sqrt{2} - \qutsb,\,0\leq \qusb \leq \qupark\queT\sqrt{2} - \quvb\} \subset \qubulkF.
\end{multline*}
Let $I$ be a multiindex, $|I| \leq \qusbn - 5$. Then by Proposition \ref{divthSigy} with $X^i = \qumff_\quhbar[\quuprtl^I \qugamb]^i$ we have ($\qumff_\quhbar[\quuprtl^I \qugamb]^i$ is $C^1$ since $|I| \leq \qusbn - 5$, and all integrals are absolutely convergent since the functions $\qudlb$, $\qubb$, $\qucb$, $\qugamb$ are compactly supported)
\begin{gather}
\int_0^{\qupark\queT\sqrt{2} - \qutsb} \int_{-\infty}^\infty \biggl[ d\qusb_i \qumff_\quhbar[\quuprtl^I \qugamb]^i (\quSigsbf_\qutsb(\quxb, \quvb), \quxb, \quvb) - \frac{\partial\quSigsbf_\qutsb}{\partial\quxb}(\quxb, \quvb) d\quxb_i \qumff_\quhbar[\quuprtl^I \qugamb]^i (\quSigsbf_\qutsb(\quxb, \quvb), \quxb, \quvb)\\
- \frac{\partial\quSigsbf_\qutsb}{\partial\quvb}(\quxb, \quvb) d\quvb_i \qumff_\quhbar[\quuprtl^I \qugamb]^i (\quSigsbf_\qutsb(\quxb, \quvb), \quxb, \quvb)\biggr] \vert \qudet\quhbar\vert ^{1/2} (\quSigsbf_\qutsb(\quxb, \quvb), \quxb, \quvb)\,d\quxb\,d\quvb\notag\\
= \int_{\qubulkF'} \qumff_\quhbar[\quuprtl^I \qugamb]^i_{;i} \vert \qudet\quhbar\vert ^{1/2}\,d\qutau\,d\quxi\,d\quzeta\notag\\
+ \int_{\quSigz} d\qusb_i \qumff_\quhbar[\quuprtl^I \qugamb]^i \vert \qudet\quhbar\vert ^{1/2}\,d\quxb\,d\quvb + \int_{\quUz \cap \qubulkF'} d\quvb_i \qumff_\quhbar[\quuprtl^I \qugamb]^i \vert \qudet\quhbar\vert ^{1/2}\,d\qusb\,d\quxb\notag\\
- \int_{A_{\qupark\queT} \cap \qubulkF'} d\qutau_i \qumff_\quhbar[\quuprtl^I \qugamb]^i \vert \qudet\quhbar\vert ^{1/2} \,d\quxi\,d\quzeta.\notag
\end{gather}
\eqref{sbfbi} follows by applying \eqref{gambzIcondendeq} -- \eqref{gamblineinitbound} to the initial data terms and bounding the bulk integral using \eqref{quQTbound} and \eqref{enerineqerrtermexp}, as in \eqref{gambMestCdef}.\end{proof}
\par
\section{CONCENTRATED SOLUTIONS}\label{focsol}
\subsection{Introduction}\label{fsintro} As an application of the results of Sections \ref{initdat} and \ref{enerineq}, we construct the concentrated solutions to the Einstein vacuum equations \eqref{intRicwav} described in Theorem \ref{irbthm}. In detail, we shall prove the following theorem.
\par
\begin{theorem}\label{rbthm} Let $\qusbn \geq 7$. Let $C_1$, $C_2$ be the constants from Theorem \ref{seqbound} corresponding to $\qusbn$, and let $\quCGeod$ be the constant in Proposition \ref{foccoordsmth}. Assume that $\queT > 0$, $\queTp \in (0, 1/(2\sqrt{2})]$, $\qupark \geq 1$, and $\qunu \in (0, 1)$ satisfy \eqref{mainprkestc} -- \eqref{qunuqueTcondc} and \eqref{sbfbparmcond}; it is sufficient to require
\begin{gather}
\qunu \qupark^{-2\quiota} \leq \min\{ 1/(12C_1\quCKS), 1/\quCGeod \},\qquad \qunu \qupark^{1/2 - 2\iota} \leq 2^{13/4} C_1 \quCKS\queTp\queT^{-1/2},\\
\qunu \queT^{1/2} \queTp^{-1} \leq \min\{\frac{1}{C_2}, \frac{1}{8\quCGeod} \},\qquad \qupark\queT \geq 2.\label{queTnuboundrb}
\end{gather}
Let $\quupM$, $\quupN \in \quN$, $\quupM$, $\quupN \geq 1$. Then there are constants $C$, $C'$, $C'_1 > 0$ such that the following holds. Suppose that $\quparr$ satisfies
\begin{equation}
\quparr \geq C'_1,
\end{equation}
and that $\qupark$ satisfies also
\begin{equation}\label{rbthmsqpb}
\qupark^{-2\iota} \leq \quparr^{\quupM - \quupN}.
\end{equation}
Then there is a solution $\quing$ to the Einstein vacuum equations $\quRic(\quing) = 0$ of the form \eqref{ingexp}, with $\qumeth$ and $\qugamma$ corresponding, via \eqref{fscalcdefii} -- \eqref{fscalgamdef} and \eqref{gaugemetform}, to a solution $\qudlb_\quparr$, $\qubb_\quparr$, $\qucb_\quparr$, $\qugamb_\quparr$ to \eqref{nsricone} -- \eqref{swave} given by Theorem \ref{seqbound} which satisfies the following properties:
\begin{description}
\item{(i)} the bounds in \eqref{omegbbaseinitbound} -- \eqref{cEboot} hold with $\qunu = \quparr^{-\quupM}$;
\item{(ii)} the metric $\quhbar_\rho$ given by \eqref{hbardef} admits a null foliation $\quSigsb$, $\qutsb \in [0, \queTp/8]$ of a neighborhood of $\quSigz$ in $\qubulk$ of the form described in Proposition \ref{foccoordsmth};
\item{(iii)} there is a function $\quappfri$ on $\qubulk$, supported on $\quQS = \{ (\qusb, \quxb, \quvb) \in \qubulk\,\vert\,\quxb,\,\quvb \in (0, 1)\}$, such that
\begin{align}
C' \quparr^{-\quupM - \qusbn - 4} \geq \left\| \quappfri\right\| _{\quDH^{\qusbn - 2}(\quHyp{\qutsb})} &\geq \frac{1}{C'} \quparr^{-\quupM - \qusbn - 4},\label{firmbdleq}\\
\left\| \qugamb_\quparr - \quappfri\right\| _{\quDH^{\qusbn - 2}(\quHyp{\qutsb})} &\leq C \quparr^{-\min\{\frac{3}{2} \quupM, \quupM + \frac{1}{2} (\quupN + \qusbn + 3)\}},\label{firmbdeq}
\end{align}
where the norm $\|\cdot\|_{\quDH^{\qusbn - 2}(\quHyp{\qutsb})}$ is defined in \quEEEnullHbspace.
\end{description}
\end{theorem}
\par
Note that Theorem \ref{irbthm} follows from Theorem \ref{rbthm}.
\par
\begin{remark} Note that \eqref{firmbdleq} -- \eqref{firmbdeq} imply, for $\quupN + \qusbn + 3 \geq \quupM$ and $\quparr$ sufficiently large, a bound on the relative error of the form
\begin{equation*}
\| \qugamb_\quparr - \quappfri\| _{\quDH^\qusbn(\quHyp{\qutsb})}/\| \qugamb_\quparr\| _{\quDH^\qusbn(\quHyp{\qutsb})} \leq C''\quparr^{-\quupM/2 - \qusbn - 4}
\end{equation*}
for some constant $C'' > 0$. Since the functions $\quappfr$ are all supported on the parallelepiped $\quQS = \{ (\qusb, \quxb, \quvb) \in \qubulk\,\vert \,\quxb, \quvb \in (0, 1) \}$, this implies in particular that
\begin{equation*}
\| \qugamb_\quparr\| _{\quDH^\qusbn(\quHyp{\qutsb}\backslash\quQS)}/\| \qugamb_\quparr\| _{\quDH^\qusbn(\quHyp{\qutsb})} \leq C'' \quparr^{-\quupM/2 - \qusbn - 4},
\end{equation*}
which implies that the percentage of $\qugamb_\quparr$ supported off of $\quQS$ (measured in $\quDH^\qusbn$) can be made arbitrarily small by taking $\quparr$ sufficiently large. The solutions $\qugamb_\quparr$ are thus {\it \quconadj\/} on $\quQS$. We emphasize, though, that the existence time $\queT$ depends on $\quparr$ through (i) and \eqref{queTnuboundrb}.
\end{remark}
\par
The construction of the solutions $\qugamb$ in Theorem \ref{rbthm} involves two main ideas, one standard and the other (to our knowledge) new. The standard procedure for obtaining {\it exact\/} localized solutions to a linear wave equation $\quwbox_\quaaph f = 0$ for a given metric $\quaaph$ consists of the following three steps, which we describe on $\qubulk$ for concreteness (see, e.g., \cite{sbierski}):
\begin{description}
\item{(i)} We construct a suitably localized function $f_0$ which, for some large parameter $\quparr$, some exponent $\quupN$, and a suitable Sobolev norm $\|\cdot\|$, satisfies
\begin{equation}\label{genfzwsmall}
\|\quwbox_\quaaph f_0\| \sim \quparr^{-\quupN}.
\end{equation}
\item{(ii)} Next, we use the standard existence theorem for wave equations to obtain an $f$ satisfying
\begin{equation}
f|_{\quSigz\cup\quUz} = f_0|_{\quSigz\cup\quUz},\qquad \quwbox_\quaaph f = 0.
\end{equation}
\item{(iii)} Finally, we use energy estimates based on $\quaaph$ and the bound
\begin{equation}
\|\quwbox_\quaaph (f - f_0)\| \sim \quparr^{-\quupN}
\end{equation}
to conclude that $f$ must be close (in some suitable norm) to $f_0$, and hence (approximately) localized.
\end{description}
This procedure as it stands does not make sense for us since our wave equation is nonlinear: we must solve $\quwbox_\quhbar \qugamb = 0$, and the metric $\quhbar$ itself depends on the solution $\qugamb$. On the other hand,
\begin{equation}\label{quhbaretaeq}
\quwbox_\quhbar \qugamb = \quwbox_\eta \qugamb + \qupark^{-2\iota} \sum_{\partial \in \qudifopset} \qudAc[\partial] \partial\qugamb.
\end{equation}
Now the coefficients $\qudAc[\partial]$ corresponding to any solution $\qudlb$, $\qubb$, $\qucb$, $\qugamb$ to \eqref{nsricone} -- \eqref{swave} obtained from Theorem \ref{seqbound} must satisfy bounds such as those in Proposition \ref{dAcbounds}. Thus, a function $f_0$ which satisfies \eqref{genfzwsmall} with $\quaaph = \eta$ will, by \eqref{quhbaretaeq}, satisfy (schematically)
\begin{equation}
\|\quwbox_\quhbar f_0\| \sim \quparr^{-\quupN} + \qupark^{-2\iota},
\end{equation}
which gives \eqref{genfzwsmall} for $\quaaph = \quhbar$ when $\qupark$ is sufficiently large. Steps (ii) and (iii) may then be performed as in the standard procedure. The key point is that the lower bound on $\qupark$ is, by Proposition \ref{dAcbounds}, {\it independent\/} of the choice of solution $\qudlb$, $\qubb$, $\qucb$, $\qugamb$, as long as the initial data conditions \eqref{omegbbaseinitbound} -- \eqref{gamblineinitbound} hold.
To put it another way, Theorem \ref{seqbound} ensures that {\it any\/} solution to \eqref{nsricone} -- \eqref{swave} with initial data satisfying \eqref{omegbbaseinitbound} -- \eqref{gamblineinitbound} will also satisfy \eqref{cEboot}, giving us, in some sense, `a priori' (relative to the choice of $f_0$) bounds on the coefficients $\qudAc[\partial]$. 
\par
This procedure is not strongly tied to a particular choice of approximate solution, and we use a geometric optics ansatz.\footnote{The preliminary version of our results in \cite{thesis}, Chapter 7, used a (more complicated) Gaussian beam ansatz; we are indebted to Jan Sbierski for suggesting that we use a geometric optics ansatz instead.} We shall give only that part of the theory which is necessary for our immediate purposes and refer the reader to, e.g., \cite{couranthilbert} or \cite{whitham} for a more general discussion.
\par
We proceed as follows. In \quEEEsmoothnull\ we describe the foliation in (ii).
In \quEEEapproxsol\ we construct geometric optics approximate solutions to the Minkowski wave equation and
apply our work in Section \ref{initdat}\ to obtain and bound initial data, and in \quEEEremainbound\ we apply our work in Section \ref{enerineq}\ to obtain and bound the solutions $\qudlb_\quparr, \qubb_\quparr, \qucb_\quparr, \qugamb_\quparr$. Finally, in \quEEEfocapp\ we sketch the proof of existence and $C^1$ smoothness of the foliation in (ii).
\par
As a guide to the arguments below, we note that we continue to use $\qusbn$ to denote the Sobolev exponent fixed in Section \ref{enerineq}; recall that $\qusbn \geq \qusbnmin$. We use $\quupM$ to control the overall size of the solution (cf.\ \eqref{firmbdeq} -- \eqref{firmbdleq}), and $\quupN$ to denote the order of the geometric optics approximation. See \eqref{gambNdefeq}.
\par
\subsection{Auxiliary null foliation}\label{smoothnull}
The null foliation of $\qubulk$ transverse to $\{ \quvb = \hbox{const} \}$ which appears in Proposition \ref{seqboundfluxbounds} and Theorem \ref{rbthm} can be characterized in greater detail as follows. See \quEEEnullfolfig\ for a rough sketch of the situation.
\begin{figure}[h]\centering
\includegraphics[keepaspectratio]{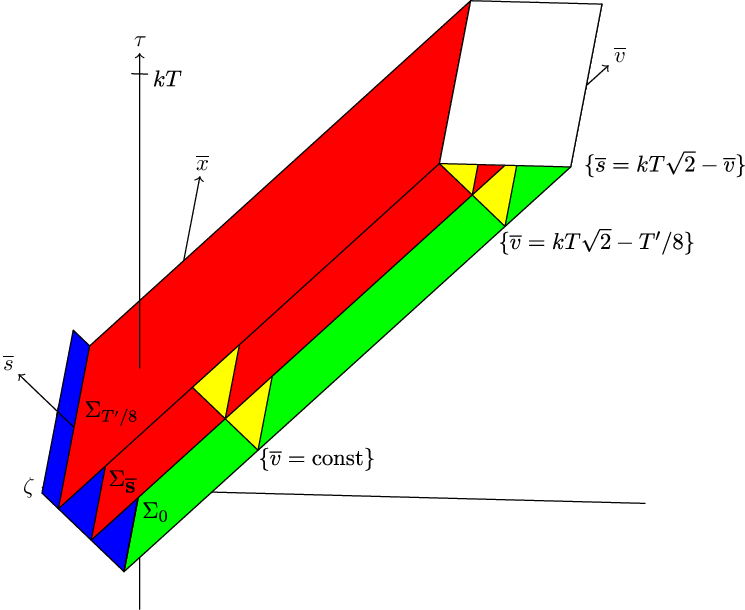}
\caption{Schematic of the auxiliary null foliation}\label{nullfolfig}
\end{figure}
\par
\begin{proposition}\label{foccoordsmth} There is a constant $\quCGeod > 0$ such that the following holds. Let $\queT > 0$, $\queTp \in (0, \queTpupp]$, $\qupark \geq 1$, and $\qunu \in (0, 1)$ be such that \eqref{mainprkestc} -- \eqref{qunuqueTcondc} hold, and that also 
\begin{equation}\label{foccoordparkcond}
\qunu^2 \qupark^{-2\iota} \leq 1/\quCGeod,\quad \qupark\queT \geq 2,\quad \qunu^2\queT^{1/2} \queTp^{-1} \leq 1/(8\quCGeod).
\end{equation}
Let $\qudlb$, $\qubb$, $\qucb$, $\qugamb$ be any solution to \eqref{nsricone} -- \eqref{swave} given by Theorem \ref{seqbound} with initial data as constructed in Section \ref{initdat} and satisfying \eqref{omegbbaseinitbound} -- \eqref{gamblineinitbound} with $\quudifset = \quudifseta$, and let $\quhbar$ be the metric on $\qubulk$ determined by $\qudlb$, $\qubb$, $\qucb$. 
Then we can foliate a portion of the bulk region $\qubulk$ by $C^1$ $\quhbar$-null hypersurfaces, obtained as graphs of $C^1$ functions.
In detail, there is a family of $C^1$ functions of $(\quxb, \quvb)$, $\quSigsbf_\qutsb : \quR^1 \times [0, \qupark\queT\sqrt{2} - \qutsb] \rightarrow [0, \queTp/4]$, $\qutsb \in [0, \queTp/8]$ with the following properties:
\begin{description}
\item{(i)} $\quSigsbf_0(\quxb, \quvb) = 0$ for all $(\quxb, \quvb) \in \quR^1 \times [0, \qupark\queT\sqrt{2}]$, i.e., the graph of $\quSigsbf_0$ coincides with $\quSigz$;
\item{(ii)} $\quSigsbf_\qutsb(\quxb, \qupark\queT\sqrt{2} - \qutsb) = \qutsb$, i.e., the graph of $\quSigsbf_\qutsb$ contains the line $A_{\qupark\queT} \cap \{ \qusb = \qutsb \}$;
\item{(iii)} the graphs of the $\quSigsbf_\qutsb$, i.e., the family of $C^1$ hypersurfaces
\begin{equation}\label{Hyptsbdef}
\quHyp{\qutsb} = \{ (\qusb, \quxb, \quvb) \in \qubulk\,\vert \,\quvb \in [0, \qupark\queT\sqrt{2} - \qutsb],\,\qusb = \quSigsbf_\qutsb(\quxb, \quvb) \},\,\, \qutsb \in [0, \queTp/8],
\end{equation}
gives a $C^1$ $\quhbar$-null foliation of (see Figure \ref{nullfolfig})
\begin{multline}\label{bulkFfcdefi}
\qubulkF = \{ (\qusb, \quxb, \quvb) \in \qubulk\,\vert \,\quvb \in [0, \qupark\queT\sqrt{2} - \queTp/8],\,0 \leq \qusb \leq \quSigsbf_{\queTp/8}(\quxb, \quvb) \}\\
\cup \{ (\qusb, \quxb, \quvb) \in \qubulk\,\vert \,\quvb \geq \qupark\queT\sqrt{2} - \queTp/8,\,0\leq \qusb \leq \qupark\queT\sqrt{2} - \quvb\};
\end{multline}
\item{(iv)} the functions $\quSigsbf_\qutsb$ satisfy the following bounds: for all $\qutsb \in [0, \queTp/8]$, $(\quxb, \quvb) \in \quR^1 \times [0, \qupark\queT\sqrt{2} - \qutsb]$, $i \in \{ 0, 1 \}$,
\begin{equation*}
\left\vert \quSigsbf_\qutsb(\quxb, \quvb) - \qutsb\right\vert  \leq \quCGeod\qunu\queT^{1/2},\qquad \left\vert \partial_i \quSigsbf_\qutsb(\quxb, \quvb)\right\vert  \leq \quCGeod\qunu\queT^{1/2}.
\end{equation*}
\end{description}
\end{proposition}
\par
\begin{proof} See \quEEEfocapp.\end{proof}
\par \noindent
\par
\begin{remark} The null foliation in Proposition \ref{foccoordsmth}, together with the null foliation $U_{\qutvb}$, would allow us to introduce a `double-null' coordinate system similar in some respects to that used elsewhere (see, e.g., \cite{christodoulou}, \cite{klainlukrod}, \cite{klainrod}, \cite{klainrodii}, \cite{lukrod}, \cite{lukrodii}, \cite{lukrodiii}). It differs in that the leaves $\quHyp{\qutsb}$ are ruled by geodesics developed from a {\it terminal}, rather than initial, hypersurface. (This is reflected in property (ii) above.) This is forced on us for the following reason. Generically, null vectors orthogonal to $\partial_\quxb$ and transverse to $U_{\qutvb}$ differ from (the Minkowskian version) $\partial_\quvb$ by terms of size $\sim \nu\qupark^{-2\iota} \widehat{\qutvb}^{-1/2}$, which suggests that, generically, null geodesics transverse to $U_{\qutvb}$ remain within $\qubulk$ only up to (at most) null parameter $\lambda \sim \qunu^{-1} \qupark^{2\iota} \widehat{\qutvb}^{1/2}$. For a foliation ruled by such geodesics to satisfy Proposition \ref{foccoordsmth}, we require $\lambda \sim \qupark$, so that when $\iota = 1/4$, we are forced to take $\qutvb \sim \qupark$.\end{remark}
\begin{remark}The condition that \eqref{omegbbaseinitbound} -- \eqref{gamblineinitbound} hold with $\quudifset = \quudifseta$, rather than $\quudifset = \quudifsetr$, is necessary to obtain sufficient decay of the metric $\quhbar$. See \quEEEfocapp.\end{remark}
\par
We define energy fluxes through, and Sobolev (semi)norms on, the leaves of the foliation given by Proposition \ref{foccoordsmth}\ as follows, cf.\ \eqref{ibounddef} and \eqref{DHbnormdef}. 
\par
\begin{definition}\label{nullHbspace} Let $\quSigsb$, $\qutsb \in [0, \queTp/8]$ be the foliation given by Proposition \ref{foccoordsmth}, and let $\quaph$ be one of $\quhbar$ and $\queta$. For $f$ any function for which the right-hand sides below exist and are finite and $\qumff_\quaph[f]^i$ as defined in \eqref{mffdef}, define the flux
\begin{multline}\label{IbSigsbdef}
\quIb_{\quSigsb, \quaph}[f] = \int_0^{\qupark\queT\sqrt{2} - \qutsb} \int_{-\infty}^{\infty} \Biggl[\left(d\qusb_i \qumff_\quaph[f]^i\right)(\quSigsbf_\qutsb(\quxb, \quvb), \quxb, \quvb)\\
- \partial_\quxb \quSigsbf_\qutsb(\quxb, \quvb) \left(d\quxb_i \qumff_\quaph[f]^i\right)(\quSigsbf_\qutsb(\quxb, \quvb), \quxb, \quvb)\\
- \partial_\quvb \quSigsbf_\qutsb(\quxb, \quvb) \left(d\quvb_i \qumff_\quaph[f]^i\right)(\quSigsbf_\qutsb(\quxb, \quvb), \quxb, \quvb)\Biggr] \vert \qudet\quaph\vert ^{1/2}(\quSigsbf_\qutsb(\quxb, \quvb), \quxb, \quvb)\,d\quxb\,d\quvb
\end{multline}
and for $m \geq 0$ the seminorm
\begin{equation}\label{nDHbnormdef}
\| f\| _{\quDH^m(\quSigsb)} = \left[\sum_{\vert I\vert  \leq m - 1} \quIb_{\quSigsb,\quhbar}[\quuprtl^I f]\right]^{1/2}.
\end{equation}
\end{definition}
\par
We have finally the following version of the divergence theorem, cf.\ Proposition \ref{divth}\ above.
\par
\begin{proposition}\label{divthSigy}
Let $\qutsb \in (0, \queTp/8]$, set (cf.\ \eqref{bulkFfcdefi}) 
\begin{multline*}
\qubulkF' = \{ (\qusb, \quxb, \quvb) \in \qubulk\,\vert \,\quvb \in [0, \qupark\queT\sqrt{2} - \qutsb],\,0 \leq \qusb \leq \quSigsbf_\qutsb(\quxb, \quvb) \}\\
\cup \{ (\qusb, \quxb, \quvb) \in \qubulk\,\vert \, \quvb \geq \qupark\queT\sqrt{2} - \qutsb,\,0\leq \qusb \leq \qupark\queT\sqrt{2} - \quvb\} \subset \qubulkF,
\end{multline*}
and let $\quaaph$ be any Lorentzian metric on $\qubulkF'$. Let $X^i$ be any $C^1$ vector field on $\qubulkF' \cap \{ \qutau \leq \qusigmat \}$ such that $d\quxb_i X^i \vert \qudet\quaaph\vert ^{1/2} \in L^1(\qubulkF')$ and all of the integrals below are absolutely convergent,
\begin{multline}\label{findiv}
\int_{\qubulkF' } X^i_{;i} \vert \qudet\quaaph\vert ^{1/2}\,d\qutau\,d\quxi\,d\quzeta\\
= \int_0^{\qupark\queT\sqrt{2} - \qutsb} \int_{-\infty}^\infty \biggl[ d\qusb_i X^i(\quSigsbf_\qutsb(\quxb, \quvb), \quxb, \quvb) - \frac{\partial\quSigsbf_\qutsb}{\partial\quxb}(\quxb, \quvb) d\quxb_i X^i(\quSigsbf_\qutsb(\quxb, \quvb), \quxb, \quvb)\\
- \frac{\partial\quSigsbf_\qutsb}{\partial\quvb}(\quxb, \quvb) d\quvb_i X^i(\quSigsbf_\qutsb(\quxb, \quvb), \quxb, \quvb)\biggr] \vert \qudet\quaaph\vert ^{1/2}(\quSigsbf_\qutsb(\quxb, \quvb), \quxb, \quvb)\,d\quxb\,d\quvb\\
-\int_{\quSigz} d\qusb_i X^i \vert \qudet\quaaph\vert ^{1/2}\,d\quxb\,d\quvb - \int_{\quUz \cap \qubulkF'} d\quvb_i X^i \vert \qudet\quaaph\vert ^{1/2}\,d\qusb\,d\quxb\\
+ \int_{A_{\qupark\queT} \cap \qubulkF'} d\qutau_i X^i \vert \qudet\quaaph\vert ^{1/2} \,d\quxi\,d\quzeta.
\end{multline}
\end{proposition}
\par
\begin{proof} The proof is analogous to that of Proposition \ref{divth}; the first term on the right-hand side follows from \eqref{Sigsurf}.\end{proof}
\par
\subsection{Geometric optics approximation; initial data bounds}\label{approxsol} We next bound suitable approximate solutions to the Minkowski wave equation. Set (see the statement of Theorem \ref{rbthm})
\begin{align}
\quQS &= \{ (\qusb, \quxb, \quvb) \in \qubulk\,\vert \,\qusb \in [0, \queTp],\,\quxb, \quvb \in (0, 1)\},\label{QSdefeq}\\
\qupbar\quQS &= (0, 1) \times (0, 1).
\end{align}
We now fix an arbitrary function $A_0 \in C^\infty_0(\qupbar\quQS)$; by abuse of notation we also write $A_0$ for the function on $\quQS$ taking $(\qusb, \quxb, \quvb)$ to $A_0(\quxb, \quvb)$. Further, fix some $\quparr \geq 1$, $\quupN \in \quN$. The approximate solutions we use can be written as follows.
\begin{definition}\label{Ioperator} Define $\quI : C^\infty_0(\quQS) \rightarrow C^\infty_0(\quQS)$ by
\begin{equation}\label{Idefeq}
[\quI f](\qusb, \quxb, \quvb) = \partial_\quvb f (\qusb, \quxb, \quvb) - \partial_\quvb f (0, \quxb, \quvb) - \frac{1}{2} \int_0^{\qusb} \partial_\quxb^2 f(u, \quxb, \quvb)\,du,
\end{equation}
and define
\begin{equation}\label{gambNdefeq}
\qugambGN = \quRe \left[e^{-\qui\quparr\quvb} \sum_{j = 0}^\quupN (\qui\quparr)^{-j} (\quI^j A_0)(\qusb, \quxb, \quvb)\right],
\end{equation}
where $\qui = \sqrt{-1}$ and $\quRe z$ is the real part of $z \in \quC$.
\end{definition}
The functions $\qugambGN$ have the following properties.
\begin{lemma}\label{gambGappr}
Let $\qules \in \quN$. There is a constant $C = C(\quupN, \qules)$, depending only on $\quupN$ and $\qules$, such that for all $\quley$, $\qulee \in \quN$,
\begin{gather}
\|\partial_\qusb^\quley \partial_\quxb^\qulee \partial_\quvb^\qules \qugambGN\|_{L^\infty(\quQS)} \leq C(\quupN, \qules)\|A_0\|_{W^{2\quupN - \quley + \qulee + \qules}(\quQS)}\quparr^{\qules - \quley},\label{frJKLeq}\\
\|\partial_\qusb^\quley \partial_\quxb^\qulee \partial_\quvb^\qules \quwbox_\eta \qugambGN\|_{L^\infty(\quQS)} \leq C(\quupN, \qules) \|A_0\|_{W^{2\quupN + \qulee + \qules + 2}(\quQS)} \quparr^{-\quupN + \qules}.\label{frwboxJKLeq}
\end{gather}
\end{lemma}
\par
\begin{proof} We note that $\quI A_0 = -\frac{1}{2} \qusb \partial_\quxb^2 A_0(\quxb, \quvb)$. It is then clear that, for all $j \geq 0$, $\quI^j A_0$ is a polynomial in $\qusb$ (with coefficients depending on $\quxb$, $\quvb$), and that moreover, when $j \geq 1$, $\quI^j A_0|_{\qusb = 0} = 0$. Let us define $\quIi$ to be the operator taking $s^\ell f(\quxb, \quvb)$ to $s^{\ell + 1}/(\ell + 1) f(\quxb, \quvb)$ for any smooth $f$; then for all $j \geq 1$
\begin{multline*}
\quI [\quI^j A_0] = \partial_\quvb [\quI^j A_0] (\qusb, \quxb, \quvb) - \frac{1}{2} \partial_\quxb^2 [\quIi \quI^j A_0] (\qusb, \quxb, \quvb)\\
= (\partial_\quvb - \partial_\quvb|_{\qusb = 0} - \frac{1}{2} \partial_\quxb^2 \quIi) [\quI^j A_0] (\qusb, \quxb, \quvb).
\end{multline*}
Now the operators $\partial_\quvb$ and $-\frac{1}{2} \partial_\quxb^2 \quIi$ commute, so that we may apply the binomial expansion theorem to obtain, for any $j \geq 2$,
\begin{equation}\label{Ijeq}
[\quI^j A_0](\qusb, \quxb, \quvb) = [\quI^{j - 1} \quI A_0](\qusb, \quxb, \quvb) = \sum_{\ell = 1}^j \left({j - 1}\atop {\ell - 1}\right) \frac{(-1)^\ell}{2^\ell \ell!} \qusb^\ell \partial_\quvb^{j - \ell} \partial_\quxb^{2\ell} A_0(\quxb, \quvb);
\end{equation}
a posteriori, \eqref{Ijeq} holds for $j = 1$ as well.
From \eqref{Ijeq} we obtain that, for $j \geq 1$ and any $\ell \geq 1$, $\ell \leq j$,
\begin{equation}\label{qusbderivIjA}
\partial_\qusb^\ell \quI^j A_0 = \sum_{\ell' = \ell}^j \left({j - 1}\atop{\ell' - 1}\right) \frac{(-1)^{\ell'}}{2^{\ell'} (\ell' - \ell)!} \qusb^{\ell' - \ell} \partial_\quvb^{j - \ell'} \partial_\quxb^{2\ell'} A_0,
\end{equation}
while $\partial_\qusb^\ell \quI^j A_0$ vanishes identically for $\ell > j$. Also, clearly $\partial_\qusb^\ell \quI^0 A_0 = \partial_\qusb^\ell A_0 = 0$ for all $\ell > 0$. Note that \eqref{qusbderivIjA} implies that $\partial_\qusb^\ell \quI^j A_0$ involves at most $2j$ $\quxb$ derivatives and at most $j - \ell$ $\quvb$ derivatives of $A_0$. Next, we have
\begin{gather*}
\quwbox_\eta \qugambGN = \quRe \biggl[ 2\qui\quparr e^{-\qui\quparr\quvb} \sum_{j = 0}^\quupN (\qui\quparr)^{-j} \partial_\qusb \quI^j A_0 + e^{-\qui\quparr\quvb} \sum_{j = 0}^\quupN (\qui\quparr)^{-j} \quwbox_\eta \quI^j A_0 \biggr]\\
= \quRe \biggl[ e^{-\qui\quparr\quvb} (\qui\quparr)^{-\quupN} \quwbox_\eta \qui^\quupN A_0\\
+ e^{-\qui\quparr\quvb} \biggl\{ 2\qui\quparr\sum_{j = 1}^\quupN (\qui\quparr)^{-j} \bigl(-\frac{1}{2} \quwbox_\eta \quI^{j - 1} A_0\bigr) + \sum_{j = 0}^{\quupN - 1} (\qui\quparr)^{-j} \quwbox_\eta \quI^j A_0\biggr\}\biggr]\\
= \quRe \biggl[ e^{-\qui\quparr\quvb} (\qui\quparr)^{-\quupN} \quwbox_\eta \quI^{\quupN} A_0\biggr].
\end{gather*}
\eqref{frJKLeq} and \eqref{frwboxJKLeq} follow by counting derivatives:
\begin{gather*}
\|\partial_\qusb^\quley \partial_\quxb^\qulee \partial_\quvb^\qules \qugambGN\|_{L^\infty(\quQS)}\\
= \Biggl\|\quRe \biggl[ e^{-\qui\quparr\quvb} \sum_{j = 0}^\quupN \sum_{\qules' + \qules'' = \qules} \left(\qules\atop{\qules'}\right) (-\qui\quparr)^{\qules'} (\qui\quparr)^{-j} \partial_\qusb^\quley\partial_\quxb^\qulee\partial_\quvb^{\qules''} \quI^j A_0\biggr]\Biggr\|_{L^\infty(\quQS)}\notag\\
\leq C(\quupN, \qules)\|A_0\|_{W^{2\quupN - \quley + \qulee + \qules}(\quQS)}\quparr^{\qules - \quley},\\
\|\partial_\qusb^\quley\partial_\quxb^\qulee\partial_\quvb^\qules\qugambGN\|_{L^\infty(\quQS)}\\
= \Biggl\|\quRe \biggl[ (\qui\quparr)^{-\quupN} e^{-\qui\quparr\quvb} \sum_{\qules' + \qules'' = \qules} \left(\qules\atop{\qules'}\right) (-\qui\quparr)^{\qules'} \quwbox_\eta \partial_\qusb^\quley\partial_\quxb^\qulee\partial_\quvb^{\qules''} \quI^\quupN A_0\biggr]\Biggr\|_{L^\infty(\quQS)}\notag\\
\leq C(\quupN, \qules) \|A_0\|_{W^{2\quupN + \qulee + \qules + 2}(\quQS)}\quparr^{-\quupN + \qules}.
\end{gather*}
\end{proof}
\par
Initial data obtained from $\qugambGN$ satisfy the following bounds.
\par
\begin{lemma}\label{gambkinitbound} There is a constant $C_0(\quupN, \qusbn, A_0) > 0$ depending only on $\quupN$, $\qusbn$, and $A_0$ such that the following holds. Define, for $(\qusb, \quxb, \quvb) \in \quQS$,
\begin{equation}\label{qugambGNfdef}
\qugambGNf(\qusb, \quxb, \quvb) = \frac{1}{2C_0(\quupN, \qusbn, A_0)^2} \quparr^{-\quupM - 2\qusbn - 2} \queTp^{3/4} \qugambGN(\qusb, \quxb, \quvb);
\end{equation}
extend $\qugambGNf$ to $\qubulk$ by $0$. Define further, for $(\quxb, \quvb) \in \quSigz$,
\begin{equation}
\quginit(\quxb, \quvb) = \qugambGNf(0, \quxb, \quvb) = \frac{1}{2C_0(\quupN, \qusbn, A_0)^2} \quparr^{-\quupM - 2\qusbn - 2} \queTp^{3/4} \cos (\quparr\quvb) A_0(\quxb, \quvb).
\end{equation}
Then the values for $\qudlb$, $\qubb$, $\qucb$, $\qugamb$ and their transverse derivatives on $\quSigz \cup \quUz$ constructed from $\quginit$ as in Proposition \ref{initdatbounds} satisfy the bounds in \eqref{omegbbaseinitbound} -- \eqref{gamblineinitbound} with $\qunu = \quparr^{-\quupM}$, and furthermore
\begin{equation}\label{gambIdiff}
\quib_{\quSigz, \queta, \qusbn - 2}[\qugamb - \qugambGNf] + \quib_{\quUz, \queta, \qusbn - 2}[\qugamb - \qugambGNf] < C(\quupN, \qusbn) \cdot (\qupark^{-4\quiota}\quparr^{-4\quupM} + \quparr^{-2(\quupM + \quupN + \qusbn + 2)})
\end{equation}
for some constant $C(\quupN, \qusbn) > 0$ depending only on $\quupN$ and $\qusbn$.
\end{lemma}
\par
\begin{proof} We first note that, for $C_0(\quupN, \qusbn, A_0)$ sufficiently large (we need at least $C_0(\quupN, \qusbn, A_0) \geq C(\quupN, 2\qusbn + 2) \|A_0\|_{\qusbW^{2\quupN + 9\qusbn + 10}(\quQS)}$, where $C$ is the constant in the statement of Lemma \ref{gambGappr}), Lemma \ref{gambGappr} gives
\begin{equation}\label{fqugbound}
\|\quginit\|_{\qusbW_\quxb^{7\qusbn + 8} \qusbW_\quvb^{2\qusbn + 2}((0, 1)\times(0, 1))} \leq \frac{1}{2C_0(\quupN, \qusbn, A_0)} \quparr^{-\quupM} \queTp^{3/4}.
\end{equation}
Next, let $C$ be the constant in Proposition \ref{initdatbounds} corresponding to $\quupey = \qusbn + 2$, $\quupee = 3\qusbn + 3$, $\quupes = \qusbn + 1$. We may assume that $C_0(\quupN, \qusbn, A_0) \geq C$. \eqref{omegbbaseinitbound} -- \eqref{gamblineinitbound} with $\qunu = \quparr^{-\quupM}$ then follow from \eqref{fqugbound} and \eqref{stuffdelbound} and item (iii) in Proposition \ref{initdatbounds} for $C_0(\quupN, \qusbn, A_0)$ sufficiently large, since $\qusupp\quginit \subset (0, 1) \times (0, 1)$.
\par
We now note that, for any $\quley$, $\qulee$, $\qules \in \quN$, $1 \leq \quley \leq \qusbn - 1$, $\qulee \leq \qusbn - 2$, $\qules \leq \qusbn - 2$, we have on $\quSigz$ (here and below, all transverse derivatives of $\qugamb$ are a priori values)
\begin{gather}
\partial_\qusb^{\quley - 1} \partial_\quxb^\qulee \partial_\quvb^\qules \quwbox_\eta (\qugamb - \qugambGNf) = (-2\partial_\qusb^\quley \partial_\quxb^\qulee \partial_\quvb^{\qules + 1} + \partial_\qusb^{\quley - 1} \partial_\quxb^{\qulee + 2} \partial_\quvb^\qules) (\qugamb - \qugambGNf)\\
= - \partial_\qusb^{\quley - 1} \partial_\quxb^\qulee \partial_\quvb^\qules \quwbox_\eta \qugambGNf - \partial_\qusb^{\quley - 1} \partial_\quxb^\qulee \partial_\quvb^\qules \qupark^{-2\iota} \sum_{\partial \in \qudifopset} \qudAc[\partial] \partial\qugamb,
\end{gather}
so integrating $2\partial_\qusb^\quley \partial_\quxb^\qulee \partial_\quvb^{\qules + 1} (\qugamb - \qugambGNf)$ from $1$ to any $\qusb \in [0, 1]$, and bounding $L^1$ norms by $L^\infty$ norms since $\qusb \in [0, 1]$, gives
\begin{gather}\label{quleyindbase}
2\|\partial_\qusb^\quley \partial_\quxb^\qulee \partial_\quvb^\qules (\qugamb - \qugambGNf)\|_{L^\infty(\quSigz)} \leq \|\partial_\qusb^{\quley - 1} \partial_\quxb^{\qulee + 2} \partial_\quvb^{\qules} (\qugamb - \qugambGNf)\|_{L^\infty(\quSigz)}\\
+ \|\partial_\qusb^{\quley - 1} \partial_\quxb^\qulee \partial_\quvb^\qules \quwbox_\eta \qugambGNf\|_{L^\infty(\qupbar\quQS)} + \qupark^{-2\iota} \sum_{\partial\in\qudifopset} \|\partial_\qusb^{\quley - 1} \partial_\quxb^\qulee \partial_\quvb^\qules (\qudAc[\partial]\partial\qugamb)\|_{L^\infty(\quSigz)}.\notag
\end{gather}
Recalling that $\qudifopset = \{ \partial_\qusb^2, \partial_\qusb\partial_\quxb, \partial_\quxb^2, \partial_\qusb, \partial_\quxb, \partial_\quvb \}$, that $\qudAc[\partial]$ is a rational function in $\{ \qudlb, \qubb, \qucb, \partial_\qusb\qudlb, \partial_\qusb\qubb, \partial_\qusb\qucb, \partial_\quxb\qudlb, \partial_\quvb\qudlb, \partial_\quxb\qubb \}$, and that $\quley \leq \qusbn - 1$, $\qulee \leq \qusbn - 2$, $\qules \leq \qusbn - 2$, we see that \eqref{stuffdelbound} with $\quupey = \qusbn + 2$, $\quupee = 3\qusbn + 3$, $\quupes = \qusbn + 1$ implies that for some constant $C'(\quupN, \qusbn, A_0) > 0$
\begin{equation}
\qupark^{-2\iota} \sum_{\partial\in\qudifopset} \|\partial_\qusb^{\quley - 1} \partial_\quxb^\qulee \partial_\quvb^\qules (\qudAc[\partial]\partial\qugamb)\|_{L^\infty(\quSigz)} \leq C'(\quupN, \qusbn, A_0) \qupark^{-2\iota} \quparr^{-2\quupM};
\end{equation}
further, by Lemma \ref{gambGappr},
\begin{equation}
\|\partial_\qusb^{\quley - 1} \partial_\quxb^\qulee \partial_\quvb^\qules \quwbox_\eta \qugambGNf\|_{L^\infty(\qupbar\quQS)} \leq C'(\quupN, \qusbn, A_0) \quparr^{-\quupN - \quupM - 2\qusbn - 2 + \qules}
\end{equation}
for some (new) constant $C'(\quupN, \qusbn, A_0) > 0$. Performing an induction on $\quley$ in \eqref{quleyindbase} and recalling that, by definition,
\begin{equation}
\|\partial_\quxb^{\qulee'} \partial_\quvb^{\qules'} (\qugamb - \qugambGNf)\|_{L^\infty(\quSigz)} = 0
\end{equation}
for all $\qulee', \qules' \in \quN$, we obtain
\begin{equation}\label{quibSigzbound}
\quib_{\quSigz, \queta, \qusbn - 2}[\qugamb - \qugambGNf] < C'(\quupN, \qusbn, A_0) \cdot (\qupark^{-4\iota}\quparr^{-4\quupM} + \quparr^{-2(\quupM + \quupN + \qusbn + 2)})
\end{equation}
for some constant $C'(\quupN, \qusbn, A_0) > 0$. This establishes half of \eqref{gambIdiff}. The bound
\begin{equation}
\quib_{\quUz, \queta, \qusbn - 2}[\qugamb - \qugambGNf] < C'(\quupN, \qusbn, A_0) \cdot (\qupark^{-4\quiota}\quparr^{-4\quupM} + \quparr^{-2(\quupM + \quupN + \qusbn + 2)})
\end{equation}
can be shown in the same fashion, by interchanging the roles of $\qusb$ and $\quvb$, integrating from $\qusb = 0$, and using \eqref{quibSigzbound} on $\quSigz\cap\quUz$. We omit the details.
\end{proof}
\par
\subsection{Bounds on the remainder}\label{remainbound} The desired concentrated solutions can now be obtained from Theorem \ref{seqbound} applied to the initial data constructed in Subsection \ref{approxsol}.
\par
\begin{proposition}\label{rbprop} Let $\queT > 0$, $\queTp \in (0, 1/(2\sqrt{2})]$, $\qupark \geq 1$, $\quparr \geq 1$, and $\qunu \in (0, 1)$ be such that \eqref{mainprkestc} -- \eqref{qunuqueTcondc} and \eqref{foccoordparkcond} hold. Let $\qudlb$, $\qubb$, $\qucb$, $\qugamb$ be the solution to \eqref{nsricone} -- \eqref{swave} given by Theorem \ref{seqbound}, with initial data that given by Lemma \ref{gambkinitbound}. Then there are constants $C_1, C_2, C_3 > 0$, depending only on $\quupN$, $\quupM$, $\qusbn$, and $A_0$, such that, if
\begin{equation}\label{parkparrbd}
\quparr \geq C_1,\qquad \qupark^{-2\iota} \leq \quparr^{\quupM - \quupN},\qquad \qutsb \in [0, \queTp/8],
\end{equation}
then for $\quupP = \min \{ 3\quupM/2, \quupM + (\quupN + \qusbn + 3)/2 \}$,
\begin{equation}\label{rmbdeq}
\left\| \qugamb - \qugambGNf\right\| _{\quDH^{\qusbn - 2}(\quHyp{\qutsb})} \leq C_2 \quparr^{-\quupP},
\end{equation}
and moreover
\begin{equation}\label{rmbdleq}
C_3 \quparr^{-\quupM - \qusbn - 4} \geq \| \qugambGNf\| _{\quDH^{\qusbn - 2}(\quHyp{\qutsb})} \geq \frac{1}{C_3} \quparr^{-\quupM - \qusbn - 4}.
\end{equation}
\end{proposition}
\begin{remark} Theorem \ref{rbthm} follows.\end{remark}
\par
\begin{proof} By an argument analogous to that in the proof of Proposition \ref{epsenerbound}, we find the estimate, for any $\qutsb \in [0, \queTp/8]$ and any multiindex $I$ for $\quudifset$ satisfying $|I| \leq \qusbn - 2$ and some constant $C > 0$,
\begin{gather}
\quib_{\Sigma_\qutsb, \quhbar}[\quuprtl^I (\qugamb - \qugambGNf)] \leq C\Biggl[\int_\qubulk \biggl|\partial_\qutau \quuprtl^I (\qugamb - \qugambGNf) \quwbox_\quhbar \quuprtl^I (\qugamb - \qugambGNf)\biggr|\,d\qusb\,d\quxb\,d\quvb\\
+ \int_0^{\qupark\queT} \qupark^{-2\iota} \quupsu^{-1/2} \qunu (\quepsn_\quhbar[\quuprtl^I \qugamb] + \quepsn_\quhbar[\quuprtl^I \qugambGNf])\,d\quupsilon\Biggr]\notag\\
+ 2\quIb_{\quSigz, \queta, \qusbn - 2}[\qugamb - \qugambGNf] + 2\quIb_{\quUz, \queta, \qusbn - 2}[\qugamb - \qugambGNf].\notag
\end{gather}
Now by Proposition \ref{Dfboxcomm}, Proposition \ref{sthreegambderiv}, Proposition \ref{dAcbounds}, Lemma \ref{gambGappr}, and the fact that $\quuprtl^I$ is composed of conformal Killing fields for the Minkowski metric $\eta$, we have the bounds, for some constant $C > 0$ (we can add factors of $\qusigu^{-2}$ to the bounds on $\qugambGNf$ since it is supported on $\quvb \leq 1$)
\begin{gather*}
\|\partial_\qutau \quuprtl^I \qugamb\|_{L^2(\quAsigmat)} \leq C\qunu,\qquad \|\quwbox_\quhbar \quuprtl^I \qugamb\|_{L^2(\quAsigmat)} \leq C\qunu^2 \qupark^{-2\iota} \qusigu^{-1/2},\\
\|\partial_\qutau \quuprtl^I \qugambGNf\|_{L^2(\quAsigmat)} \leq C\quparr^{-\quupM - \qusbn - 3},\\
\|\quwbox_\quhbar \quuprtl^I \qugambGNf\|_{L^2(\quAsigmat)} \leq C (\quparr^{-\quupM - \quupN - \qusbn - 3} + \qupark^{-2\iota} \quparr^{-2\quupM - \qusbn - 3}) \qusigu^{-2},\\
\quepsn_\quhbar[\quuprtl^I \qugamb] \leq C\qunu^2,\\
\quepsn_\quhbar[\quuprtl^I \qugambGNf] \leq C\quparr^{-\quupM - \qusbn - 3}.
\end{gather*}
Applying Lemma \ref{gambkinitbound}, and recalling that $\qunu = \quparr^{-\quupM}$, we thus find that
\begin{equation}
\quib_{\Sigma_\qutsb, \quhbar}[\quuprtl^I (\qugamb - \qugambGNf)] \leq C \quparr^{-\min\{2\quupM + \quupN + \qusbn + 3, 3\quupM\}}.
\end{equation}
\eqref{rmbdeq} follows. \eqref{rmbdleq} follows from Lemma \ref{gambGappr} by noting that $\|\qugambGNf\|_{\quDH^{\qusbn - 2}(\quHyp{\qutsb})}$ has only one top-order (in $\quparr$) term (with a constant coefficient), of order $\quparr^{-\quupM - \qusbn - 4}$, and that for any continuous function $f$ on $[0, 1] \times [0, 1]$ independent of $\quparr$, $\|\cos(\quparr\quvb) f\|_{L^2(\qupbar\quQS)}$ converges to a constant multiple of $\|f\|_{L^2(\qupbar\quQS)}$ as $\quparr \rightarrow\infty$. We omit the details.
\end{proof}
\par
\subsection{Appendix: Proof of Proposition \ref{foccoordsmth}}\label{focapp} In this appendix we prove Proposition \ref{foccoordsmth}. We begin with some notation and preliminary results.
\par
\par
\begin{definition}\label{natchardef} For $i \in \quZ$, we define
\begin{equation*}
\qunatchar_i = \begin{cases}
1,&\ququad i \geq 0\\ 0,&\ququad i < 0.
\end{cases}
\end{equation*}
\end{definition}
\par
Recall (see \eqref{hatfuncdef}) that for $u \in \quR^1$ we define
\begin{equation*}
\qubsighat{u} = \max \{ 1, u \}.
\end{equation*}
\par
\begin{lemma}\label{siguintlem} (i) Let $u_1$, $u_2 \in \quR^1$. Then
\begin{equation*}
\qubsighat{u_1 + u_2} \leq \qubsighat{u_1} + \qubsighat{u_2},
\end{equation*}
while if $u_1 \in [0, 1]$
\begin{equation}\label{sigubaseineq}
\frac{1}{\sqrt{2}}\qubsighat{u_2} \leq \qubsighat{\frac{1}{\sqrt{2}} (u_1 + u_2)} \leq \sqrt{2} \qubsighat{u_2},\qquad \qubsighat{u_2 - u_1} \geq \frac{1}{1 + u_1} \qubsighat{u_2}.
\end{equation}
(ii) Let $L \in \quR^1$, $L > 1$. For $x, p \in \quR^1$, $-1 \leq x \leq L$,
\begin{equation}\label{siguintineq}
\int_x^L \qubsighat{u}^p\,du \leq \begin{cases}
 \frac{2\vert p\vert }{\left\vert 1 + p\right\vert } \qubsighat{x}^{p + 1},&\ququad p < -1\\
2 + \qulog L,&\ququad p = -1\\
2 + \frac{1}{1 + p} L^{p + 1},&\ququad p > -1.
\end{cases}
\end{equation}
\end{lemma}
\par
\begin{proof} These follow by straightforward calculations. For (ii), note that for $x \leq 1$
\begin{equation*}
\int_x^L \qubsighat{u}^p\,du = \int_1^L u^p\,du + 1 - x.
\end{equation*}
We omit the details.\end{proof}
\par
The following global injectivity result was inspired by an unpublished note of Taylor (see \cite{taylor}, Proposition 2).
\par
\begin{lemma}\label{injfunc} Let $n \geq 1$ and $U \subset \quR^n$ be a convex set, and suppose that $F : U \rightarrow \quR^n$ satisfies
\begin{equation}\label{injfuncHScond}
\qubHS{DF - \quido} < 1
\end{equation}
on $U$, where $DF^i_j = \partial_j F^i$ and $\quido$ is the $n \times n$ identity matrix. Then $F$ is injective.
\end{lemma}
\par
\begin{proof} Let $\qux_1$, $\qux_2 \in U$ be such that $F(\qux_1) = F(\qux_2)$. Since \eqref{injfuncHScond} is invariant under all translations, and also under simultaneous, equal rotation of the domain and range of $F$, we may assume that $\qux_1 = 0$, $\qux_2^i - \qux_1^i = \qux_2^i = \vert \qux_2\vert  \delta^i_1$. Note next that \eqref{injfuncHScond} implies in particular that on $U$
\begin{equation}\label{Fyineqz}
\left\vert \partial_1 F^1 - 1\right\vert  < 1,\qquad\hbox{so}\qquad \partial_1 F^1 \neq 0\hbox{ on $U$.}
\end{equation}
Define $f : [0, 1] \rightarrow \quR^1$ by
\begin{equation*}
f(t) = F^1(\qux_1 + t(\qux_2 - \qux_1)) = F^1(t\vert \qux_2\vert , 0, \cdots, 0).
\end{equation*}
$F^1(\qux_2) = F^1(\qux_1)$ implies $f(1) = f(0)$, so that by Rolle's Theorem there must be a $t^* \in (0, 1)$ such that $f'(t^*) = 0$. But then
\begin{equation*}
\vert \qux_2\vert  \partial_1 F^1(t^*\vert \qux_2\vert , 0, \cdots, 0) = f'(t^*) = 0,
\end{equation*}
so by \eqref{Fyineqz} $\vert \qux_2\vert  = 0$ and $\qux_2 = 0 = \qux_1$.\end{proof}
\par
\begin{proof}[Proof of Proposition \ref{foccoordsmth}] We first obtain sufficiently fast decay of $\qubb$ and $\qucb$. Define
\begin{gather}
\quBxi = \quxi\partial_\qutau + \qutau\partial_\quxi = 2^{-1/2}\left[ \quxb \left(\partial_\qusb + \partial_\quvb\right) + (\qusb + \quvb) \partial_\quxb\right],\notag\\
\quBzet = \quzeta\partial_\qutau + \qutau\partial_\quzeta = \qusb\partial_\qusb - \quvb\partial_\quvb,\notag\\
\quXvec = \quBxi - \quxb \partial_\qutau,\hbox{ so } \partial_\quxb = \qutau^{-1} \quXvec,\notag\\
\quVvec = \qusb \partial_\qusb - \quBzet,\hbox { so } \partial_\quvb = \quvb^{-1} \quVvec.\notag
\end{gather}
Noting the commutators
\begin{gather}
[\quXvec, \frac{1}{\qutau}] = [\quXvec, \frac{1}{\quvb} ] = 0,\quad [\quVvec, \frac{1}{\quvb^\ell}] = -\frac{\ell}{\quvb^\ell},
\end{gather}
we obtain by induction the formulas, for some collection of sets of multiindices $\{ \{ I \}_{\qulee_1 \qulee_2} \}$ and various constants $c(\qules')$, $c(I, \qulee_1, \qulee_2)$ depending only on the indicated parameters,
\begin{gather}
\biggl(\frac{1}{\quvb} \quVvec\biggr)^\qules = \frac{1}{\quvb^\qules} \sum_{\qules' = 1}^\qules c(\qules') \quVvec^{\qules'},\notag\\
\quXvec^\qulee = \sum_{\qulee_1 + 2\qulee_2 \leq \qulee} \quxb^{\qulee_1} \qutau^{\qulee_2} \sum_I c(I, \qulee_1, \qulee_2) (\quBxi, \partial_\quxb)^I.\notag
\end{gather}
From this it is not hard to see that there is a constant $C(\qulee, \qules) > 0$, depending only on $\qulee$, $\qules$, such that for any function $f \in \quDHb^{\qulee + \qules + 2}(\quAsigmat)$
\begin{equation}\label{decaybound}
\|\partial_\quxb^\qulee \partial_\quvb^\qules f\|_{L^\infty(\quAsigmat)} \leq C(\qulee, \qules) \qusigu^{-1/4 - \qulee/2 - \qules} \|f\|_{\quDHb^{\qulee + \qules + 2}(\quAsigmat)}.
\end{equation}
Further, suppose that on $\qugamb$ we have functions $u \in \quDHb^{\qusbn - 2}(\quAsigmat)$, $M$, and $b$ such that
\begin{equation}\label{focappbode}
\partial_\qusb u = Mu + b,
\end{equation}
and that there are quantities $\mxp{M}, \mxp{b}, \dxp{M}, \dxp{b} > 0$ independent of $\qusigu$ such that, for $\qulee + \qules \leq \qusbn - 4$,
\begin{equation}
\|\partial_\quxb^\qulee \partial_\quvb^\qules M\|_{L^\infty(\quAsigmat)} \leq \mxp{M} \qusigu^{-\dxp{M} - \qulee/2 - \qules},\quad \|\partial_\quxb^\qulee \partial_\quvb^\qules b\|_{L^\infty(\quAsigmat)} \leq \mxp{b} \qusigu^{-\dxp{b} - \qulee/2 - \qules}.
\end{equation}
Differentiating \eqref{focappbode} $\qulee$ times with respect to $\quxb$ and $\qules$ times with respect to $\quvb$ and applying \eqref{decaybound} to $u$, it is not hard to show that there is a constant $C(\mxp{M}, \mxp{b}, \qulee, \qules)$ such that, for all $\qusigmat \in [0, \qupark\queT]$,
\begin{equation}
\|\partial_\quxb^\qulee \partial_\quvb^\qules u\|_{L^\infty(\quAsigmat)} \leq C(\mxp{M}, \mxp{b}, \qulee, \qules) \qusigu^{-\min\{\dxp{M} + 1/4, \dxp{b}\} - \qulee/2 - \qules},
\end{equation}
from which we find that for some other constant $C(\mxp{M}, \mxp{b}, \qulee, \qules)$,
\begin{equation}\label{focappbodebound}
\|\partial_\quxb^\qulee \partial_\quvb^\qules u\|_{L^\infty(\quAsigmat)} \leq C(\mxp{M}, \mxp{b}, \qulee, \qules) \qusigu^{-\dxp{b} - \qulee/2 - \qules}.
\end{equation}
Lemma \ref{Asobolevemb} and \eqref{decaybound} applied to \eqref{cEboot} give immediately, for $\qulee + \qules \leq \qusbn - 4$ and some constant $C(\qusbn, \gambC, \OmegbC)$ depending only on the indicated parameters,
\begin{equation}
\|\partial_\quxb^\qulee \partial_\quvb^\qules \qudlb\|_{L^\infty(\quAsigmat)} \leq C(\qusbn, \gambC, \OmegbC) \qunu^2 \qusigu^{-1/2 - \qulee/2 - \qules},
\end{equation}
while applying \eqref{focappbodebound} to \eqref{nsrictwo} -- \eqref{nsricthree}, using Lemma \ref{Flipsch} to bound terms of the form $\quellb^{-1} f$, a straighforward calculation gives, for $\qulee + \qules \leq \qusbn - 4$ and a similar constant $C(\qusbn, \gambC, \OmegbC)$,
\begin{gather}
\|\partial_\quxb^\qulee \partial_\quvb^\qules \qubb\|_{L^\infty(\quAsigmat)} \leq C(\qusbn, \gambC, \OmegbC) \qunu^2 \qusigu^{-1 - \qulee/2 - \qules},\\
\|\partial_\quxb^\qulee \partial_\quvb^\qules \qucb\|_{L^\infty(\quAsigmat)} \leq C(\qusbn, \gambC, \OmegbC) \qunu^2 \qusigu^{-3/2 - \qulee/2 - \qules}.
\end{gather}
From \eqref{hbardef} we obtain next, again on $\quAsigmat$, and for $\quley + \qulee + \qules \leq \qusbn - 4$ and a similar constant $C(\qusbn, \gambC, \OmegbC)$,
\begin{gather}\label{quhbardiffbd}
|\partial_\qusb^\quley \partial_\quxb^\qulee \partial_\quvb^\qules \bigl(\quhbar_{ij} - \eta_{ij}\bigr)|_{HS} \leq C(\qusbn, \gambC, \OmegbC) \qunu^2 \qupark^{-2\iota} \qusigu^{-(i + j - 1)/2 - \qulee/2 - \qules} \chi_{i + j - 2},\\
|\partial_\qusb^\quley \partial_\quxb^\qulee \partial_\quvb^\qules \bigl((\quhbar^{-1}){}^{ij} - \eta^{ij}\bigr)|_{HS} \leq C(\qusbn, \gambC, \OmegbC) \qunu^2 \qupark^{-2\iota} \qusigu^{-(3 - i - j)/2 - \qulee/2 - \qules} \chi_{2 - i - j},\\
\end{gather}
from which we obtain, finally, that the Christoffel symbols $\Gamma^i_{jk}$ satisfy (for various constants $C = C(\qusbn, \gambC, \OmegbC)$)
\begin{gather}
\|\partial_\ell \Gamma^i_{jk}\|_{L^\infty(\quAsigmat)} = \frac{1}{2} \|\partial_\ell \bigl[(\quhbar^{-1}){}^{im} (\partial_j \quhbar_{mk} + \partial_k \quhbar_{mj} + \partial_m \quhbar_{jk})\bigr]\|_{L^\infty(\quAsigmat)}\notag\\
\leq C \qusigu^{-\ell/2} \|[|\eta^{i m}| + \qunu^2\qupark^{-2\iota} \qusigu^{-(3 - i - m)/2} \chi_{2 - i - m}] \nu^2 \qupark^{-2\iota} \qusigu^{-(j + k + m - 1)/2}\|_{L^\infty(\quAsigmat)}\notag\\
\leq C(\qusbn, \gambC, \OmegbC) \qunu^2 \qupark^{-2\iota} \qusigu^{-\frac{1}{2} (j + k - i + \ell - 1)}.\label{chrsymmainbd}
\end{gather}
For convenience, we let $\quChhbarC = \quChhbarC(\qusbn, \gambC, \OmegbC)$ denote the maximum of the constants in \eqref{quhbardiffbd} and \eqref{chrsymmainbd}.
\par
Define auxiliary subsets of $\quR^3$
\begin{align}
\qubulkGt &= \{ (\qutsb, \qutxb, \qutvb) \in \quR^3\,\vert \,\qutsb \in [0, \queTp/8],\,\qutxb \in \quR^1,\,\qutvb \in [-1, \qupark\queT\sqrt{2} - \qutsb] \},\label{bulkGtdef}\\
\qubulkex &= \{ (\qusb, \quxb, \quvb)\in\quR^3\,\vert \,\qusb \in [0, \queTp/2],\,\quxb \in \quR^1,\,\quvb \in [-2, \qupark\queT\sqrt{2} - \qusb] \}.\label{bulkexdefeq}
\end{align}
Points $(\qutsb, \qutxb, \qutvb)$ will always be assumed to lie in $\qubulkGt$. Extend the Christoffel symbols $\Gamma^i_{jk}$ to $\qubulkex$ in such a way that \eqref{chrsymmainbd} holds on $\qubulkex$, possibly with a new constant $C(\qusbn, \gambC, \OmegbC)$. Let $\qunL$ be the unique $\quhbar$-null vector field on $A_{\qupark\queT} \cap \qubulkGt$ satisfying
\begin{equation}\label{nLcompdefeq}
\qunL^2 = 1,\ququad\quhbar_{ij} \qunL^i \partial_\quxb^j = 0,\quad\hbox{so}\quad \qunL^0 = \frac{1}{2} \qupark^{-2\iota} \qucb - \qupark^{-4\iota} \frac{\qubb^2}{2\quab}, \qunL^1 = -\qupark^{-2\iota} \frac{\qubb}{\quab}, \qunL^2 = 1;
\end{equation}
using \eqref{quhbardiffbd}, it is not hard to show that for some constant $\nLC(\qusbn, \gambC, \OmegbC)$
\begin{equation}\label{Geodbyi}
|\qunL^i - \delta^i_2| \leq \nLC(\qusbn, \gambC, \OmegbC) \qunu^2 \qupark^{-2\iota} (\qupark\queT)^{-\frac{1}{2}(3 - i)}.
\end{equation}
Let $\quido$ denote the identity operator. Define $\quGeod(\qutsb, \qutxb, \qutvb)$ to be the solution to the final-value problem
\begin{gather}
\quGeod\vert _{\qutA_{\qupark\queT}} = \quido,\qquad \partial_\qutvb\quGeod\vert _{\qutA_{\qupark\queT}} = \qunL,\label{Geoddefcond}\\
\partial_\qutvb^2 \quGeod^i (\qutsb, \qutxb, \qutvb) = \Gamma^i_{jk} (\quGeod(\qutsb, \qutxb, \qutvb)) \partial_\qutvb \quGeod^j(\qutsb, \qutxb, \qutvb) \partial_\qutvb \quGeod^k(\qutsb, \qutxb, \qutvb)\label{Geoddefeq}
\end{gather}
for all $(\qutsb, \qutxb, \qutvb) \in \qubulkGt$ for which the solution exists. 
It can be shown that \eqref{nLcompdefeq} and \eqref{Geoddefcond} -- \eqref{Geoddefeq} imply that there is a constant $C(\qusbn, \gambC, \OmegbC)$ such that on $A_{\qupark\queT} \cap \qubulkGt$ 
\begin{align}\label{dGeodinitbound}
\begin{split}
\left\vert \partial_i \quGeod^j - \delta^j_i\right\vert  &\leq C(\qusbn, \gambC, \OmegbC)\qunu\qupark^{-2\quiota} (\qupark\queT)^{-\frac{1}{2} (1 - j)},\\
\left\vert \partial_\qutvb \partial_i \quGeod^j\right\vert  &\leq C(\qusbn, \gambC, \OmegbC)\qunu\qupark^{-2\quiota} (\qupark\queT)^{-\frac{1}{2}(3 - j)}.
\end{split}
\end{align}
Now fix some $\qutsb \in [0, \queTp/8]$, $\qutxb \in \quR^1$, and set $\qutvbz = \qupark\queT\sqrt{2} - \qutsb$. Let $\qutvbs$ be the least element of $[-1, \qutvbz]$ such that on $\{ \qutsb \} \times \{\qutxb\} \times [\qutvbs, \qutvbz]$, $\quGeod$ exists and satisfies
\begin{gather}
\left\vert\quGeod^i - \quido\right\vert \leq C_1 \qunu^2 \queT^{1/2},\quad \left\vert\partial_\qutvb \quGeod^i - \delta^i_2\right\vert \leq C'_1 \qunu^2 \qupark^{-2\iota} \qutvbu^{-\frac{1}{2} (3 - i)},\label{Geodbootb}\\
\left\vert \partial_i \quGeod^j - \delta^j_i\right\vert  \leq C_2 \qunu^2\qupark^{-2\quiota} (\qupark\queT)^{-\frac{1}{2} (1 - j)},\quad\left\vert \partial_\qutvb \partial_i \quGeod^j\right\vert  \leq C'_2\qunu^2\qupark^{-2\quiota} \qutvbu^{-\frac{1}{2}(3 - j)}\label{dGeodbootb}
\end{gather}
for constants $C_1$, $C'_1$, $C_2$, $C'_2$. Here $\partial_i$ is differentiation with respect to $\qutsb$, $\qutxb$, $\qutvb$. By \eqref{Geodbyi}, \eqref{Geoddefcond}, and \eqref{dGeodinitbound}, $\qutvbs < \qutvbz$. We claim that $\qutvbs = -1$. We may assume that $\quCGeod \geq C_1$, so that \eqref{Geodbootb} at $\qutvb = \qutvbs$ together with \eqref{foccoordparkcond} imply that if $\qutvbs > -1$, $\quGeod$ can be extended to some interval $[-\qutvbs - \delta, \qutvbz]$. Thus either $\qutvbs = -1$ or one of the estimates in \eqref{Geodbootb} -- \eqref{dGeodbootb} must saturate at $\qutvb = \qutvbs$. Noting that for $p$ one of $-5/2$, $-2$, $-3/2$, $-1$, and $-1/2$, and any $\qutvb \in [-1, \qutvbz]$, Lemma \ref{siguintlem}(ii) implies that
\begin{equation}\label{sghupint}
\int_\qutvb^\qutvbz \qusighat{u}^p\,du \leq 2 + 6\qutvbz^{1/2} \leq 8(\qupark\queT)^{1/2},
\end{equation}
it is clear that, for appropriate choices of $C_1$ and $C_2$ depending on $C'_1$ and $C'_2$ respectively, the first inequalities in each of \eqref{Geodbootb} -- \eqref{dGeodbootb} follow from the respective second inequalities, so that either $\qutvbs = -1$ or one of
\begin{equation}\label{satineq}
\left\vert\partial_\qutvb \quGeod^i - \delta^i_2\right\vert \leq C'_1 \qunu^2 \qupark^{-2\iota} \qutvbu^{-\frac{1}{2} (3 - i)},\,\left\vert \partial_\qutvb \partial_i \quGeod^j\right\vert  \leq C'_2\qunu\qupark^{-2\quiota} (\qupark\queT)^{-\frac{1}{2}(3 - j)}
\end{equation}
saturates at $\qutvb = \qutvbs$. We shall show that, for an appropriate choice of $C'_1$, $C'_2$, neither of the inequalities in \eqref{satineq} can saturate at $\qutvb = \qutvbs$, showing that $\qutvbs = -1$. First, note that, at any point $\quGeod(\qutsb, \qutxb, \qutvb)$,
\begin{equation}\label{siguGeodrel}
\qusigu = \qubsighat{\frac{1}{\sqrt{2}} \left(\quGeod^0(\qutsb, \qutxb, \qutvb) + \quGeod^2(\qutsb, \qutxb, \qutvb)\right)};
\end{equation}
Lemma \ref{siguintlem}(i), \eqref{Geodbootb}, and \eqref{foccoordparkcond} then imply that for some numerical constant $C > 0$
\begin{equation}
\frac{1}{C} \qutvbu \leq \qusigu \leq C \qutvbu.
\end{equation}
From \eqref{chrsymmainbd}, we thus obtain, for $\qusigmat$ as in \eqref{siguGeodrel} and some constant $C_3(\qusbn, \gambC, \OmegbC)$,
\begin{equation}\label{chrsymtvbb}
\|\partial_\ell \Gamma^i_{jk}\|_{L^\infty(\quAsigmat)} \leq C_3(\qusbn, \gambC, \OmegbC) \qunu^2 \qupark^{-2\iota} \qutvbu^{-\frac{1}{2} (j + k - i + \ell - 1)}.
\end{equation}
\eqref{satineq}, \eqref{chrsymmainbd}, and \eqref{Geoddefeq} imply that for any $\qutvb \in [\qutvbs, \qutvbz]$ we have (by abuse of notation, we use $k$ both to denote the scaling parameter and an index), requiring $\quCGeod \geq C'_2$ so that \eqref{foccoordparkcond} gives $C'_2 \qunu^2 \qupark^{-2\iota} \leq 1$,
\begin{multline*}
\left| \partial_\quvb \quGeod^i(\qutsb, \qutxb, \qutvb) - \qunL^i(\qutsb, \qutxb, \qutvbz)\right| \leq \int_{\qutvb}^{\qutvbz} \left|\Gamma^i_{jk}(\qutsb, \qutxb, \qutvb')\right| \left|\quGeod^j(\qutsb, \qutxb, \qutvb')\right| \left|\quGeod^k(\qutsb, \qutxb, \qutvb')\right|\\
\leq C_3 \qunu^2 \qupark^{-2\iota} \sum_{j, k = 0}^2 \int_{\qutvb}^{\qutvbz} \qutvbpu^{-\frac{1}{2} (j + k - i + 1)} \bigl[ \delta^j_2 + C'_2 \qunu^2\qupark^{-2\iota} \qutvbpu^{-\frac{1}{2} (3 - j)}\bigr]\\
\shoveright{\bigl[ \delta^k_2 + C'_2 \qunu^2 \qupark^{-2\iota} \qutvbpu^{-\frac{1}{2}(3 - k)}\bigr]\,d\qutvb'}\\
\leq 36 C'_3 \qunu^2 \qupark^{-2\iota} \int_{\qutvb}^{\qutvbz} \qutvbpu^{-\frac{1}{2}(5 - i)}\,d\qutvb' \leq 360 C_3 \qunu^2 \qupark^{-2\iota} \qutvbu^{-\frac{1}{2}(3 - i)},
\end{multline*}
so that, by \eqref{Geodbyi}, the first inequality in \eqref{satineq} will not saturate if we choose $C'_2 \geq C_3(\qusbn, \gambC, \OmegbC) + \nLC(\qusbn, \gambC, \OmegbC)$. Differentiating \eqref{Geoddefeq} and proceeding in a similar fashion, we obtain that for some numerical constant $c$
\begin{equation}
\left|\partial_\qutvb^2 \partial_i \quGeod^j\right| \leq c C_3 \qunu^2 \qupark^{-2\iota} \qutvbu^{-\frac{1}{2}(5 - i)},
\end{equation}
and find that the second inequality in \eqref{satineq} also cannot saturate if we choose $C'_2 \geq cC_3(\qusbn, \gambC, \OmegbC) + C(\qusbn, \gambC, \OmegbC)$, where $C(\qusbn, \gambC, \OmegbC)$ is the constant in \eqref{dGeodinitbound}. Thus $\quGeod$ is defined and satisfies \eqref{Geodbootb} -- \eqref{dGeodbootb} everywhere on $\qubulkGt$.
\par
Assuming now that $\quCGeod \geq C_2$, Lemma \ref{injfunc} and \eqref{dGeodbootb} show that $\quGeod$ is injective on $\qubulkGt$. By the inverse function theorem it is therefore a $C^1$ diffeomorphism onto its image. By \eqref{Geodbootb} and \eqref{foccoordparkcond}, and standard ODE theory, $\quGeod(\qubulkGt)$ must contain a $C^1$-smooth hypersurface lying entirely in $\qubulkex \backslash \qubulk$. Now fix $\qutsbz \in [0, \queT/8]$, and define
\begin{align}\label{Geodsdefeq}
\begin{split}
\quGeods : \quR^1 \times [-1, \qupark\queT\sqrt{2} - \qutsbz] &\rightarrow \quR^1 \times [-2, \qupark\queT\sqrt{2} - \qutsbz]\\
(\qutxb, \qutvb) &\mapsto (\quGeod^1(\qutsbz, \qutxb, \qutvb), \quGeod^2(\qutsbz, \qutxb, \qutvb)).
\end{split}
\end{align}
The same logic applied to $\quGeod$ shows that $\quGeods$ is injective, with $C^1$ inverse. We see further that its image contains $\quR^1 \times [0, \qupark\queT\sqrt{2} - \qutsbz]$. Thus the function
\begin{align}\label{Sigsbfmotdefeq}
\begin{split}
\quSigsbf_\qutsbz : \quR^1 \times [0, \qupark\queT\sqrt{2} - \qutsbz] &\rightarrow [0, \queTp]\\
(\quxb, \quvb) &\mapsto \quGeod^0(\qutsbz, \quGeods^{-1}(\quxb, \quvb))
\end{split}
\end{align}
is $C^1$. It is then clear from basic homotopy considerations that $\quGeod(\qubulkGt)$ must contain
\begin{equation}
\{ (\qusb, \quxb, \quvb) \in \qubulk\,|\,0 \leq \qusb \leq \quSigsbf_{\queTp/8}(\quxb, \quvb) \}.
\end{equation}
The surfaces $\quSigsbf_\qutsb$, $\qutsb \in [0, \queTp/8]$, give the desired foliation. (Note that the graph of $\quSigsbf_0$ is the surface $\quSigz$ by our gauge choice.) The bounds in (iv) in the statement of Proposition \ref{foccoordsmth} follow from \eqref{Geodbootb} -- \eqref{dGeodbootb} and the chain rule.
\end{proof}
\bibliographystyle{amsplain}
\bibliography{paper_lt_working_noc}
\end{document}